\documentclass{aa}  
\def\mbh{$M_{\rm BH}$\/}
\def\nh{$n_{\mathrm{H}}$\/}
\def\lledd{$L_\mathrm{bol}/L_\mathrm{Edd}$}

\def\rfe{$R_{\rm FeII}$}

\def\siiv{{Si\sc{iv+Oiv]}}\/}
\def\ciii{{\sc{Ciii]}}\/}
\def\civ{{\sc{Civ}}\/}
\def\aliii{{Al\sc{iii}}\/}
\def\siiii{{Si\sc{iii]}}\/}

\def\feiiq{\rm Fe{\sc ii}$\lambda$4570\/}

\def\ltsima{$\; \buildrel < \over \sim \;$}
\def\ltsim{\lower.5ex\hbox{\ltsima}}  
\def\gtsima{$\; \buildrel > \over \sim \;$}

\def\gtsim{\lower.5ex\hbox{\gtsima}}

\def\oiii{[\ion{O}{iii}]$\lambda\lambda$4959,5007\/}

\def\hb{{\sc{H}}$\beta$\/}

\def\feii{{Fe\sc{ii}}\/}
\def\fe{{\sc{Fe}}\/}
\def\feiii{{Fe\sc{iii}}\/}

\def\fe76087{{\sc [Fe vii]}$\lambda$6087\/}
\def\kms{km~s$^{-1}$}

\def\heii{{{\sc H}e{\sc ii}}$\lambda$4686\/}

\def\o4959{{\sc{[Oiii]}}$\lambda$4959\/}
\defcitealias{Sniegowska2021}{S21}
\defcitealias{Garnica2022}{G22}
\defcitealias{Marziani2023}{M23}

\usepackage{xcolor} 
\usepackage{graphicx}
\usepackage{txfonts}
\begin{document}

   \title{Chemical abundances along the quasar main sequence}
   \author{A. Floris 
   \inst{1,2}
   \and
   P. Marziani 
   \inst{1} 
   \and
   S. Panda \inst{3,}\thanks{CNPq Fellow}
   \and
   M. Sniegowska \inst{4}
          \and \\
          M. D'Onofrio\inst{2}
          \and A. Deconto-Machado\inst{5}
          \and A. del Olmo\inst{5} \and B. Czerny\inst{6}
          }
           
   \institute{National Institute for Astrophysics (INAF), Astronomical Observatory of Padova, IT-35122 Padova, Italy \\
    \email{alberto.floris@inaf.it, paola.marziani@inaf.it}
    \and
     Dipartimento di Fisica e Astronomia, Universit\`a di Padova, Vicolo dell' Osservatorio 3, IT-35122 Padova, Italy\\       
    \email{mauro.donofrio@unipd.it}
    \and
     Laborat\'{o}rio Nacional de Astrofísica, MCTI, R. dos Estados Unidos 154, Na\c{c}\~{o}es, CEP 37504-364, Itajub\'a, Brazil\\
    \email{spanda@lna.br}
    \and
     School of Physics and Astronomy, Tel Aviv University, Tel Aviv, 69978, Israel\\
    \email{msniegowska@tauex.tau.ac.il}
    \and
    Instituto de Astrof\'\i sica de Andaluc\'\i a\ (IAA-CSIC), Glorieta de Astronom\'\i a s/n, 38038 Granada, Spain\\
    \email{adeconto@iaa.es, chony@iaa.es}
    \and 
    Center for Theoretical Physics, Polish Academy of Sciences, Al. Lotnik\'ow 32/46, 02-668 Warsaw, Poland\\
    \email{bcz@cft.edu.pl}
    }
     
   \date{Received ; accepted }


  \abstract
   {The 4D eigenvector 1 (E1) sequence has emerged as a powerful tool for organizing the observational and physical characteristics of type-1 active galactic nuclei (AGNs). }
   {In this study, we present a comprehensive analysis of the metallicity of the broad line region gas, incorporating both new data and previously published findings, to assess the presence of any trend along the sequence. }
   {We perform a multi-component analysis on the strongest UV and optical emission lines,  compute $\sim 10$\ diagnostic ratios, and compare them with the prediction of CLOUDY photoionization simulations, identifying a photoionization solution closest to the data.}
   {Our investigation reveals a consistent pattern along the optical plane of the E1. We observe a systematic progression in metallicity, ranging from sub-solar values to metallicity levels several times higher than solar values}
   {These findings underscore the role of metallicity as a fundamental correlate of the 4DE1/main sequence. Extreme values of metallicity, at least several tens solar, are confirmed in low-$z$ AGNs radiating at a high Eddington ratio, although the origin of the extreme enrichment remains open to debate.}


\authorrunning{A. Floris et al.}

   \maketitle
   
%
\section{Introduction}

In recent years the 4DE1 (4 Dimensional Eigenvector 1) correlation space, initially defined in \cite{Boroson1992,Sulentic2000}, has proved to be a crucial tool for unraveling the spectral properties of type 1 Active Galactic Nuclei (AGN, e.g., \citealt{Panda2019,pandasniegowska22,fuetal22,wangetal22,nagoshifumihilde22}, among the most recent works), even at the highest redshifts \citep[e.g.,][]{yiangetal23,loiaconoetal24}. The optical plane, referred to as the Main Sequence (MS) of quasars \citep{Marziani2001, Marziani2010,Shen2014} is the parameter space defined by \rfe\ (the ratio between the intensity of the \feii\  emission at $\lambda 4570$\ to H$\beta$) and the FWHM(H$\beta$) \citep{gaskell85,Boroson1992}. The MS allows for the definition of two main populations of quasars occupying different regions of this parameter space \citep{sulenticetal00a,sulenticetal11}: 
\begin{itemize}
    \item {\bf Population A} found below 4000 km s$^{-1}$ of FWHM(H$\beta$),
    \item {\bf Population B} found above 4000 km s$^{-1}$ of FWHM(H$\beta$).
\end{itemize}

A subpopulation of Population A AGNs, characterized by a \rfe \gtsim 1, is the Extreme Population A (xA), associated with extreme-accreting AGNs \citep{marzianisulentic14,duetal16a,Panda2023}. 
These populations exhibit different spectral features associated with an array of physical conditions in the Broad Line Region (BLR) of these objects \citep{Fraix2017,Marziani2018,Panda2019}. 

Several parameters vary along the MS \citep[see e.g.,][for summaries]{sulenticetal11,Fraix2017,Panda2021PhD}.   Trends involve parameters associated with line-formation, such as the ionization parameter $U$, which has been shown to gradually increase moving from xA to Population B objects \citep{Sulentic2000,Marziani2010}. The gas density $n_{\rm H}$ seems to follow an inverse trend to $U$, with xA objects associated with much higher densities than Population B objects \citep{Marziani2001,matsuokaetal08,Negrete2012,martinez-aldamaetal15,Panda2018}. Notably, \rfe\ is believed to be primarily driven by the Eddington ratio $L/L_\mathrm{Edd}$ \ \citep{Boroson1992,Marziani2001,sunshen15, Donofrio2021}. The position of objects along the MS is thus connected to the radiation emitted by the AGN continuum source, itself connected to the rate of accretion of matter onto the SMBH. 

The FWHM(H$\beta$) is strongly affected by the viewing angle even if the objects considered in the Eigenvector sequence are all type 1  AGNs. There is increasing evidence that the low-ionization lines (LILs) are emitted in a flattened configuration in correspondence of the accretion disk \citep[e.g.,][]{decarlietal11,mejia-restrepoetal18a,negreteetal18,Panda2021}, as suggested since long \citep{collinsouffrinetal88}. The effect of the viewing angle on the line width (roughly proportional to $\sin \theta$, where $\theta$ is the angle between the line of sight and the accretion disk axis) is compounded with the effect of the supermassive black hole (SMBH) mass that is weak but becomes significant if AGNs in luminosity ranges of several orders of magnitudes are considered \citep{marzianietal18a}. 
If no correction for viewing angle is applied, the 
Population B objects,  by definition with larger FWHM, are characterized by systematically more massive SMBHs than their Population A counterparts \citep{marzianietal03b}. 

The role of the chemical content of the BLR is, however, not yet completely understood in the process of line formation. Previous studies like \cite{Sniegowska2021} and \cite{Garnica2022}, focused on the metal content of xA quasars and determined high metallicity $Z$ in the BLR of these objects, reaching extreme values. High $Z$ is a potential explanation for the observed trends of density and ionization parameter, as resonant scattering becomes more efficient when metals are more abundant \citep[e.g.,][]{Murrayetal1995}.
Additionally, various studies have indicated the systematic enrichment of the BLR of AGNs compared to their host galaxies, nearly always reaching super-solar values \citep[e.g., ][and references therein]{Hamann1999, Nagao2006, Xu2018, Maiolino2019}. High $Z$ values have been confirmed through independent techniques, such as iron emission and absorption features detected in the nuclear region \citep{Jiangetal2018, Maiolino2019}. 
Some proposed explanations include selective enrichment of the circumnuclear region of the AGN \citep{Sniegowska2021}. Metallicities around solar are possible for xA sources only under the most extreme conditions \citep{templetal21}. In addition,  there is still the possibility that the adopted empirical and photoionization models are inadequate. 

Recent works suggest that enrichment may not occur for the BLR gas of all AGNs: toward the opposite extreme of the sequence, estimation of metallicity indicate solar (for radio-quiet Population B) or sub-solar (for radio-loud Population  B; \citealt{Punsly2018,Marziani2022}). Previous studies involved a few objects and samples in different redshift and luminosity domains \citepalias{Sniegowska2021,Garnica2022, Marziani2023}, focusing on specific spectral bins or populations of the MS. It is necessary to understand what happens between the extremes of the MS, attempting to uncover any trend in metallicity along the MS that could help unveil possible mechanisms behind the evolution of quasars and their host galaxies: evidence of blueshifts \citep{Zamanov2002,komossaetal08,boroson05,aokietal05,zhangetal11,negreteetal18} in the lines of the spectra of most AGNs hints at the possibility that outflows could chemically enrich the gas found in the Interstellar Medium (ISM) or even the Intergalactic Medium (IGM) \citep[][]{choietal24,sabhloketal24}. Indeed,  \citet{Harrisonetal14} pointed out the far-reaching impact of such outflows beyond the BLR of these objects, suggesting an interplay between the enriched material produced in the regions close to the active nucleus and expelled over time in the direction of the host galaxy.

In this work, we present the systematic study of the metallicity of a sample of objects covering most spectral types along the MS at low redshift (Section \ref{sample}). Section \ref{methods} describes the methodology and techniques used. Results for the objects of the sample are provided in Section \ref{results}. In Section \ref{validat} we address the potential issues of our methodology.  The discussion of the results in Section \ref{disc} focuses on the correlation of the metallicity of objects with accretion parameters and the possible enrichment mechanisms that may explain our results. The conclusions are summarized in Section \ref{conclusions}.

\section{Sample}
\label{sample}

\begin{table*}
\caption{Basic sample properties}
\label{tab:objbasic}
\centering
\begin{tabular}{l c c c  c c c c c}
\hline\hline
Common Name & IAU Code &  $z$ & FWHM(H$\beta$) & R$_\mathrm{FeII}$ & Spectral type  & $R_\mathrm{K}$ &    UV   & Optical\\
(1) & (2) & (3) & (4) & (5) & (6) & (7) & (8) & (9)   \\
\hline
Mrk 335 	 & J00063+2012      & 0.0256 & $2209\pm220$ & $0.21\pm0.04$ & A1     & 0.42   &   COS & M03 \\
TXS 0042+101 & J00449+1026      & 0.587  & $7279\pm728$ & $0.04\pm0.06$ & B1/RL  & 197.37 &   FOS & J91 \\ 
I Zw 1 		 & J00535+1241      & 0.061  & $1506\pm151$ & $1.83\pm0.28$ & A4     & 0.40    &  FOS, COS & M03 \\ 
Fairall 9    & J01237$-$5848    & 0.046  & $5958\pm596$ & $0.43\pm0.08$ & B1     & 0.92  &     FOS, COS & M03 \\ 
PHL 1092     & J01399+0619      & 0.3965 & $2494\pm249$ & $1.76\pm0.39$ & A4     & 0.42   &   STIS, M20 & M20 \\ 
Ark 120      & J05161$-$0008    & 0.033  & $5863\pm586$ & $0.49\pm0.04$ & B1     & 0.30    &  FOS & M03 \\ 
Mrk 110      & J09252+5217      & 0.036  & $2048\pm205$ & $0.19\pm0.04$ & A1     & 2.18   &   COS & M03 \\
Ton 28       & J10040+2855     & 0.329  & $2398\pm240$ & $1.04\pm0.18$ & A3      & $<0.40$ &  FOS & M03 \\ 
NGC 3783     & J11390$-$3744   & 0.009  & $3514\pm351$ & $0.20\pm0.04$ & B1      & 0.30   &   FOS, COS & M03 \\ 
LB 2522      & J13012+5902   &  0.472  & $3999\pm400$ & $1.30\pm0.32$ & A3      & $<0.25$ &  FOS & M03 \\ 
LEDA 51016   & J14170+4456     & 0.114  & $2720\pm272$ & $1.01\pm0.16$ & A3      & 0.17   &   FOS, COS & M03 \\ 
Mrk 478      & J14421+3526     & 0.077  & $1322\pm132$ & $1.04\pm0.10$ & A3      & 0.20  &    FOS & M03 \\
Mrk 509      & J20441$-$1043   & 0.035  & $3309\pm331$ & $0.02\pm0.02$ & A1      & 0.35   &   COS & M03 \\
Ark 564      & J22426+2943     & 0.025  & $1011\pm101$ & $0.67\pm0.22$ & A2      & 0.88   &   FOS, COS & M03 \\ 
\hline
\end{tabular}
\tablefoot{(1) Object common name. (2) IAU coordinate name. (3) Redshift. (4) FWHM of H$\beta$, in units of km s$^{-1}$. (5) \rfe= \feiiq/\hb\ measured in this work. (6) Spectral type.  (7) Kellerman ratio, from NED, or from the following sources: \cite{Ailing2023}  for Mrk 110, \cite{Jarvela2022} for  Ark 564; \cite{Capetti2021}, \cite{Baldi2022} and \cite{Kellerman1994}, for Ton 28, LB 2522 and LEDA 51016 respectively. (8) HST camera used for collecting UV spectra of the object, using the abbreviation FOS for the Faint Object Spectrograph, and COS for the Cosmic Origins Spectrograph. (9) Optical data source, using the abbreviations M03, M20 and J91 for the papers \cite{Marziani2003}, \cite{Marinello2020}, \cite{Jackson1991}, respectively.}
\end{table*}

\subsection{Sample definition}
\label{sampledef}

The sample of objects used in this work was selected based on a few criteria:

\begin{enumerate}
    \item Availability of ground observations in the optical wavelength range of objects below a redshift $z$ of 0.8, where H$\beta$ can be observed.
    \item Availability of archival UV data from HST covering the $\lambda$1300-2100 \AA\ rest-frame range of the object.
    \item Requirement of Radio Quiet (RQ) objects, except for TXS 0042+101 that is Radio Loud (RL), to study the properties of the BLR without the influence of a collimated jet \citep{Baker1994, Punsly2018, Gaur2019}.
\end{enumerate}

Table \ref{tab:objbasic} reports the main observational parameters,  in the following order:  common name, IAU coordinate name, adopted redshift, the \hb\ FWHM,  the FeII-to-H$\beta$  intensity ratio \rfe, with \feii\ emission integrated over the $\lambda4434$-$4684$ \AA\ range, and \hb\  measured on the full broad profile; spectral type (ST) based on \hb\ FWHM and \rfe, and the Kellerman's ratio \citep{Kellerman1989}, necessary to distinguish RQ from RL objects, defined as
\begin{equation}
\label{Rk}
    {\rm R}_K = \frac{f_{\text{Radio}}}{f_B},
\end{equation}
using data from the NED database. Here, $f_{\text{Radio}}$ is the specific flux at a wavelength of $\lambda=6$ cm (5 GHz) and $f_{B}$ is the specific flux at $\lambda=4400$ \AA\ (680 THz) in the $B$ band. When the NED data was not available, the flux density used in the calculation was obtained from other works.  The last two columns summarize  the UV and optical data sources for each object.

Most objects that satisfied these selection criteria were included in the sample of objects in \cite{Marziani2003}.
The spectra of two objects, PHL 1092 and TXS 0042+101, which are found in more extreme regions of the MS, were obtained from \cite{Marinello2020} and \cite{Jackson1991}, respectively, with the latter being digitized from \cite{Jackson1991} due to the unavailability of the original spectrum.

\begin{figure}[h!]
\centering
\includegraphics[width=9.15cm]{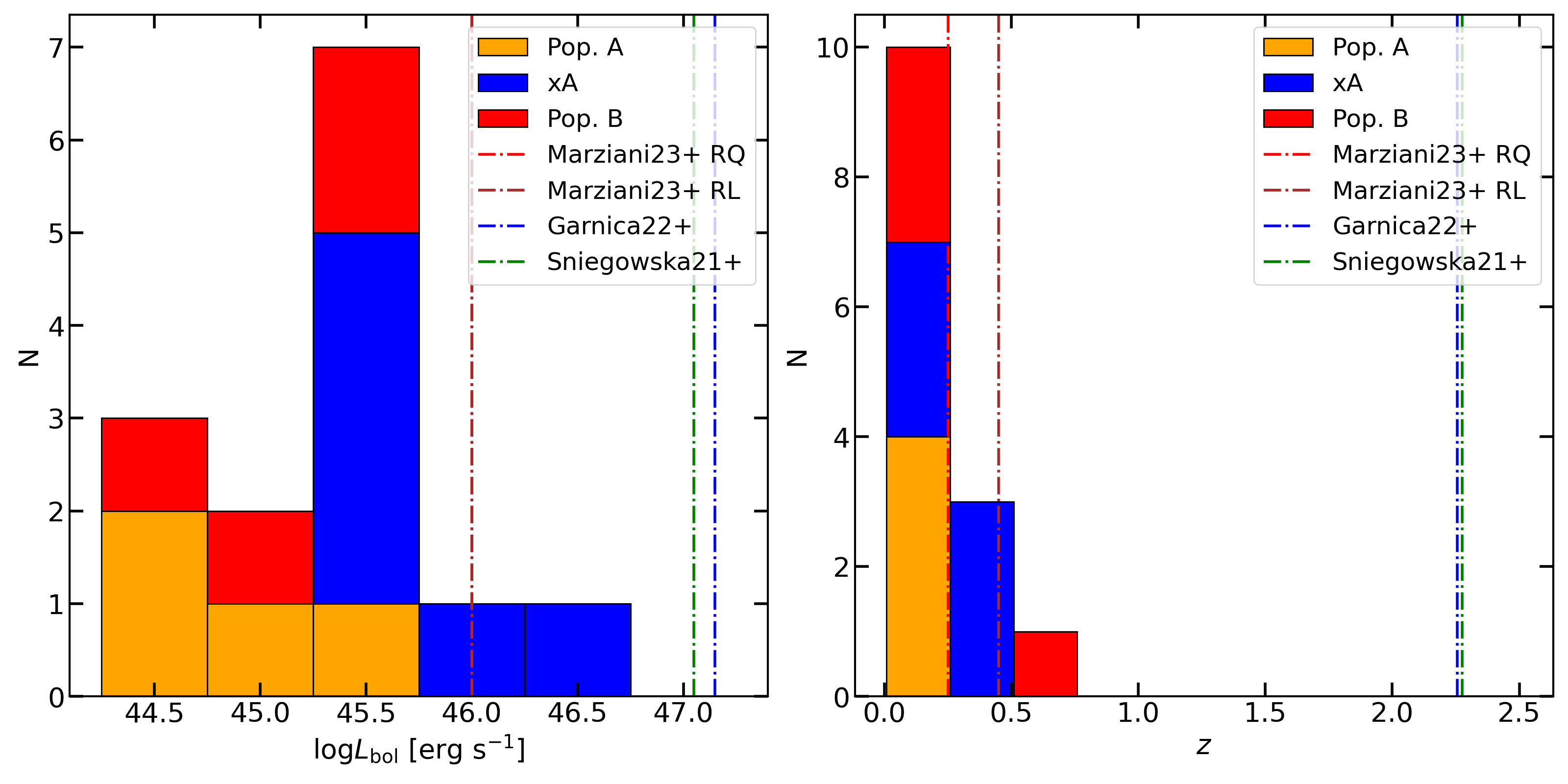}
\caption{Stacked histograms depicting the distribution of sample objects in $\log(L_{\rm bol})$ and $z$, categorized by the respective population. Assigned colors correspond to the legend, while vertical lines denote average values from the objects used in the referenced works \citetalias{Sniegowska2021}, \citetalias{Garnica2022} and \citetalias{Marziani2023}.}
\label{zlumplot}
\end{figure}

Figure \ref{zlumplot} describes the distribution of the objects of the sample in bolometric luminosity (ad computed in Section \ref{accrpar}) and $z$, confronting them with average values from the samples used in the works of \cite{Sniegowska2021}, \cite{Garnica2022} and \cite{Marziani2023} (hereafter referred to as \citetalias{Sniegowska2021}, \citetalias{Garnica2022}, \citetalias{Marziani2023}, respectively).

\section{Methodology}
\label{methods}

\subsection{Subdivision of spectral types}

An advantage of the MS/E1 approach is the possibility of defining spectral types along the optical plane of the sequence with simple selection criteria \citep{Sulentic2002,Shen2014,pandasniegowska22}. The expectation is that the effect of systematic trends along the sequence is minimized in each spectral type, so that an average spectrum of the sources in each bin is representative of all or most of the individual sources, not only phenomenologically, but also in terms of physical properties. This has been shown to be the case \citep{Sulentic2002,bachevetal04,Marziani2010}, even with some limitations  \citep{Marziani2001,Marziani2018,marzianietal18a,Panda2019}, as the 2D optical E1/MS is not a simple 1D trend \citep{wildyetal19,MLMA2021}, but rather the projection of a trend in a 4D space \citep{Sulentic2000}. 

In this paper we follow the naming convention and the subdivision of \citet{Sulentic2002} in regular steps of $\delta$\rfe=0.5 (increasing numerals with increasing \rfe),  and  $\delta$FWHM \hb = 4000 \kms (A, B, and afterward B$^+$, B$^{++}$, etc.).    

\subsection{Emission line interpretation and empirical modelling}
\label{empmodel}

In this work, we adopt an interpretation of the emission line profiles of Type-I AGNs based on three main components:

\begin{itemize}
    \item {\bf Broad Component (BC)}: A virialized component that constitutes the bulk of the line profiles of the main emission lines. It follows a Lorentzian profile in Population A objects, and a Gaussian profile in Population B objects \citep{Marziani2010}. It is a distinctive component of Type-I AGN line profiles, identifying the BLR of these objects.
    
    \item {\bf Blueshifted Component (BLUE)}: A blue-shifted outflow component present in the lines of almost all AGNs, affecting broad and narrow lines \citep{bachevetal04,Zamanov2002}. It is typically modeled by a skewed Gaussian that exhibits an asymmetry towards the blue \citep{zhangetal11}. This component is predominant in xA objects that are characterized by strong winds, being mainly present in High Ionization Lines (HILs) like \civ, that require high column density and density of the gas, as well as low $U$ in the BLR \citep{Negrete2012, Negrete2013}. 
    
    
    \item {\bf Very Broad Component (VBC)}: A virialized component with larger FWHM compared to the BC, as it models regions of the BLR that are closer to the SMBH \citep[a Very Broad Line Region (VBLR), ][]{petersonferland86,sulenticetal00c,sneddengaskell07}. It is a component exclusive of Population B objects that are modeled with a Gaussian that is redshifted by several km s$^{-1}$ to the rest-frame wavelength of the emission line. It is a simplification introduced to account for the gradients more properly modeled in the "Locally optimally emitting clouds" (LOCs)    scheme \citep{Baldwin1995} that accounts for the stratification of properties in the BLR assuming that there is a gradual gradient of variation of properties in the BLR that follows a power-law.
    
\end{itemize}

An additional component that can be encountered is the Narrow Component (NC) which is, however, not associated with the BLR and thus modeled with narrow lines and disregarded for the estimation of the metallicity of the gas in the BLR.


\subsection{Multi-component fitting technique}
\label{multicomp}

Our sample was analyzed using the {\tt specfit} \citep{Kriss1994} task from {\tt IRAF}, inspecting the regions of our spectra centered on the \siiv$\lambda1400$, \civ$\lambda1549$, \ciii$\lambda1909$ (hereafter, \siiv, \civ, \ciii, respectively) and \hb\ lines. The laboratory vacuum wavelengths of these lines, as well as all other lines measured during the analysis are   from   \citet{VandenBerk2001}. The {\tt specfit} task runs on the {\it Space Telescope Science Data Analysis System} ({\tt STSDAS}) {\tt contrib} package of {\tt IRAF}, which was developed by the {\it National Optical Astronomy Observatory} (NOAO) in Tucson, Arizona \citep{tody86,tody93}.

{\tt Specfit} allows users to interactively fit a wide range of emission and absorption lines on the continuum model of the input spectrum. To achieve this, it employs input ASCII files created using the {\tt dbcreate} task, containing the necessary functions to model various components of the spectrum. Gaussian and Lorentzian functions, widely used in multi-component fitting, are each described by values of line intensity, central wavelength, FWHM, and asymmetry parameter.

In addition to the standard functions provided, {\tt specfit} allows the use of user-provided functions. In particular, we utilized the {\tt usercont} and {\tt userline} functions to model the complex \feiii\ (found in the UV range) and \feii\ (found also in optical ranges) emissions respectively. 

The {\tt specfit} multi-component fitting procedure is interactive, based
on the input file described in the previous paragraphs. The initial parameter values in the input file serve as starting guesses for modeling the components. Each parameter can be refined using a minimization fit provided by the routine. When the algorithm converges successfully, it provides the best-fit parameters that minimize the $\chi^2$  between the model and observed data, resulting in an optimal solution. This fitting is carried out with  two different algorithms (out of the five available in {\tt specfit}):

\begin{itemize}
    \item {\bf Marquardt} is based on the Levenberg-Marquardt (LM) algorithm, which iteratively minimizes the $\chi^2$ using the Jacobian matrix. This matrix contains the partial derivatives of the model function fitting the data to each fit parameter. The algorithm combines a combination of gradient descent and Gauss-Newton methods, adjusting the step size at each iteration. It takes larger steps when the fitting improves, and smaller steps when it does not. 

    \item The downhill {\bf Simplex}  algorithm utilizes an iterative method for fitting, and adjusting values for user-provided components starting from initial guesses and constraints. It iteratively improves the solution toward the minimum $\chi^2$\ with operations on an N-dimensional (the number of free parameters) simplex \citep{pressetal92}. Unlike the Levenberg-Marquardt method, the simplex algorithm requires only function evaluations and no derivatives. Even if it is not efficient in terms of iteration number, it is very robust and never causes the {\tt specfit} task to crash. 
\end{itemize}


We start by fitting the continuum emission in the wavelength range with a power-law function, carefully selecting regions of the spectrum unaffected by the presence of emission or absorption lines. The flux values of the continuum at various reference wavelengths are provided in Table \ref{table:rms}.

In the optical wavelength range, we incorporate a \feii\ template to correct for \feii\ contamination, which can act as a pseudo-continuum on the modeled spectrum, particularly problematic in objects with strong \feii\ emission. Our \feii\ template is derived from a European Southern Observatory (ESO) spectrum of I Zw 1 \citep{Marziani2009, Marziani2010,marzianietal21}. To address \feiii\ contamination, prevalent in the $\lambda$1800-2100 \AA\ wavelength range, we utilize a template from \cite{Vestergaard2001}, based on the \feiii\ emission of I Zw 1. In cases where the \feiii\ emission is strong, and there is evidence of a stronger feature at 1914 \AA\ to the template, we model the 1914 \AA\ feature with a Gaussian profile, even if it is severely blended with \ciii. This step is typically necessary for xA objects.

When fitting emission lines, we allow them to vary while imposing the NC of lines to remain fixed at the rest frame of the object. Additionally, we ensure that the FWHM of  \siiv\ is similar to those measured for \civ\ and \hb\ \citep{Marziani2010}. This is because lines associated with the virialized component of H$\beta$, \ciii, \civ, and \siiv\ are produced in approximately the same regions. By the same token,  BLUE and VBC widths and asymmetries are also made consistent in the fit.


All measurements reported in the tables along the paper and those employed for the metallicity computations have been corrected for galactic extinction for each object using the equations of \cite{Cardelli1989}.


\begin{table*}[h!]
\caption{Flux and uncertainty measurements}             
\label{table:rms}
\centering                          
\begin{tabular}{l c c c c c c c r}        
\hline\hline                 
\multicolumn{1}{l}{Object name} & \multicolumn{2}{c}{$\lambda$1300-1450\AA} & \multicolumn{3}{c}{$\lambda$1450-2100\AA}  &  \multicolumn{2}{c}{Optical range} & \multicolumn{1}{c}{$A(V)$}\\
& $f_{\lambda1350}$ & ${\rm RMS}_{\rm SiIV}$ & $f_{\lambda1700}$ & ${\rm RMS}_{\rm CIV}$ & ${\rm RMS}_{\rm CIII]}$ & $f_{\lambda5100}$ & ${\rm RMS}_{\rm opt}$ & \\
(1) & (2) & (3) & (4) & (5) & (6) & (7) & (8) & (9) \\
\hline                        
Mrk 335      & 22.11 & 0.65 & 14.83 & 0.70 &  1.30 & 5.25 & 0.12 & 0.096\\
TXS 0042+101 & 0.17 & 0.05 & 0.15 & 0.04 & 0.03 & 0.08 & 0.01 &    0.186\\ 
I Zw 1       & 10.15 & 0.38 & 16.43 & 0.70 & 0.37 & 6.65 & 0.12 &  0.176\\ 
Fairall 9    & 35.19 & 0.91 & 26.36 & 1.63 & 1.19 & 2.72 & 0.09 &  0.071\\ 
PHL 1092 & 3.40 & 0.08 & 2.77 & 0.06 & 0.07 & 0.28 & 0.01 &        0.110\\ 
Ark 120 & 33.13 & 2.32 & 13.81 & 1.00 & 0.88 & 9.67 & 0.15 &       0.349\\ 
Mrk 110 & 24.01 & 1.57 & 16.02 & 1.74 & ... & 0.76 & 0.03 &        0.034\\
Ton 28 & 10.36 & 3.11 & 7.98 & 1.38 &1.25 & 1.92 & 0.04 &          0.060\\ 
NGC 3783 & 79.16 & 2.05 & 63.26 & 2.84 & 2.09 & 10.36 & 0.20 &     0.332\\ 
LB 2522 & 6.25 & 2.08 & 5.89 & 1.67 & 1.27 & 0.54 & 0.02 &         0.022\\ 
LEDA 51016 & 8.90 & 0.24 & 7.22 & 0.45 & 0.26 & 0.79 & 0.02 &      0.024\\ 
Mrk 478 & 36.23 & 2.06 & 26.93 & 0.93 & 0.52 & 2.00 & 0.02 &       0.039\\
Mrk 509 & 84.18 & 1.71 & 73.99 & 2.10 & 1.91 & 8.87 & 0.31 &       0.157\\
Ark 564 & 6.86 & 0.52 & 6.06 & 0.84 & 0.27 & 0.75 & 0.04 &         0.166\\ 
\hline
\end{tabular}
\tablefoot{(1) Object common name. (2), (4), (7) Specific flux of the continuum emission measured at 1350 \AA, 1700 \AA\ and 5100 \AA, respectively, and measured in units of $10^{-15}$ erg s$^{-1}$ cm$^{-2}$ ${\rm\AA}^{-1}$. (3), (5), (6), (8) RMS of the flux associated with the continuum emission, measured in units of $10^{-15}$ erg s$^{-1}$ cm$^{-2}$ ${\rm\AA}^{-1}$, in the ranges of the various lines inspected in this work. (9) Absorption parameter in the V band obtained from the NED database.}
\end{table*}

\subsection{Estimation of uncertainties}
\label{errors}

The placement of the continuum is one of the main sources of uncertainty in measuring emission-line intensities.  In particular, the continuum placement strongly influences extended features such as the \feiii\ and \heiiuv\ emissions in the UV range, and the \feii\ emission in the H$\beta$ range.


The uncertainty associated with continuum placement was estimated as follows  \citep[cf. ][]{Sniegowska2021}: initially, the intensity of the continuum was measured using {\tt IRAF} in each range where the best fit was conducted, yielding the Root Mean Square (RMS) of the flux (measured in units of $10^{-15}$ erg s$^{-1}$ cm$^{-2}$ ${\rm\AA}^{-1}$), as shown in Table \ref{table:rms}. Subsequently, the intensity of each line was measured by fixing the continuum's intensity at both maximum and minimum values, by adding or subtracting the RMS, to obtain the line's intensity at its minimum and maximum, respectively. When the continuum is higher (lower), the line intensity decreases (increases), compensating for the flux change, resulting in lower (upper) limits that are asymmetric and represent the range of emission line intensities.

To handle these uncertainties, we assume that the error distribution follows a triangular distribution \citep{D'Agostini2003}. This simplifies error computation and is the preferred method for dealing with asymmetric uncertainties. For each line measurement, the variance $\sigma^2(X)$ of any parameter $X$   is calculated using the formula for the triangular distribution: 

\begin{equation}
    \sigma^2(X) = \frac{\Delta^2x_+ + \Delta^2x_- + \Delta x_+ + \Delta x_-}{18}
\end{equation}

where $\Delta x_+$ and $\Delta x_-$ represent the differences between measurements with the maximum and best continuum, and with the best and minimum continuum, respectively. 
Finally, errors in spectral diagnostic ratios were calculated using standard error propagation formulas.

\subsection{CLOUDY photoionization simulations}
\label{cloudy}

To constrain the properties of the gas in the BLR, we conducted photoionization simulations using the {\tt CLOUDY} spectral synthesis code \citep{Ferland2017}.
{\tt CLOUDY} is a {\tt C++} code that models the ionization, chemical, and thermal state of the material that may be exposed to an external radiation field or other sources of heating. It predicts observables, including emission and absorption spectra, by solving the statistical and thermal equilibrium equations for many chemical species, including ions. The code accounts for collisional excitations, and radiative processes, and treats radiation transfer through an escape probability formalism.

Photoionization simulations require input parameters that define ionization and level populations, along with electron temperature and optical thickness as a function of the geometrical depth within the emitting gas cloud. The assumed geometry is open and plane-parallel, representing a cloud of emitting gas as a slab exposed to a radiation field only on one side.
The input parameters for a complete computation of the gas state, used to predict emission line intensities, include:

\begin{enumerate}
\item The ionization parameter $U$, defined by the equation:
\begin{equation}
\label{ionization}
    U=\frac{\int_{\nu_0}^{\infty} \frac{L_\nu}{h\nu} \,d\nu}{4\pi r_{\rm BLR}^2 c n_{\rm H}} = \frac{Q({\rm H})}{4\pi r_{\rm BLR}^2 c n_{\rm H}},
\end{equation}
 where $Q({\rm H})$ is the number of ionizing photons, $r_{\rm BLR}$ is the distance between the continuum source and the line emitting gas. It provides the ratio between the Hydrogen-ionizing photons and the Hydrogen number density.
\item Gas Hydrogen number density $n_{\rm H}$ [cm$^{-3}$].
\item Metallicity $Z$, to solar metallicity $Z_\odot$, defined as $\log Z = \log[\frac{Z}{H}] - \log[\frac{Z}{H}]_\odot$, where $[\frac{Z}{H}]$ and $[\frac{Z}{H}]_\odot$ are the ratio of the measured number density of metals to Hydrogen, and the same quantity for the Sun, respectively.
\item Spectral energy distribution (SED) of the quasar.
\item Column density $N_{\rm c}$.
\item Micro-turbulence parameter. 
\end{enumerate}

This resulting six-dimensional parameter space is not typically fully explored, and in our case, we examined trends using arrays of simulations that were organized as follows: 3 different SEDs, one for each of the following cases: Pop. B RQ, a SED for Population A sources \citep{Mathews1987} and one for xA quasars \citep{Ferland2020}. $Z$ was assumed to scale as solar ($Z_\odot$), with 12 values ranging between $0.01$\ and 1000 $Z_\odot$ for Pop. A and xA sources, and 14 values between $0.001$\ and 20 $Z_\odot$ for Pop. B. The micro-turbulence parameter was set to 0 km s$^{-1}$. This is negligible for resonance UV lines as outlined in \cite{Sniegowska2021} but is expected to lead to an under-prediction of \feii\ emission \citep{Sigut2003, bruhweilerverner08, Panda2018, Panda_etal_2019_WC, Sniegowska2021, Panda2021}. For each metallicity value, we considered an array of simulations covering the $n_H$ and $U$ parameter plane in the range $7 \le \log n_{\rm H} \le 14$ cm$^{-3}$, $-4.5 \le \log U \le 1$ for Pop. A (667 simulations), and $7 \le \log  n_{\rm H} \le 13$ cm$^{-3}$, $-3 \le \log U \le 1$\ for Pop. B (425 simulations). 

The separation of the three components (BC, VBC, BLUE) can be heuristically associated with three different regions expected to be in different physical conditions, BLR, VBLR, BLUE (see also the scheme in Fig. 2  \citealt{deconto-machadoetal23}). A source of uncertainty comes from the fact that photoionization computations are carried out under the assumption of single-zone emission, albeit separately for each of the three regions. While this assumption seems a good one for the BLR of extreme Population A, where density and ionization tend toward limiting values \citep{Marziani2001,Panda2019,Panda2021}, this might not be the case for Pop. B sources \citep{Korista_Goad_2000}, or less-extreme Population A. 

\subsection{Diagnostic ratios}
\label{diagratios}

Diagnostic ratios of lines provide constraints on the ionization, metallicity, and density of the emitting gas in the BLR. The following diagnostic ratios were used for computing the metallicity of our low $z$ quasar sample. 

\begin{figure}[h!]
    \centering
    \includegraphics[width=6.5cm]{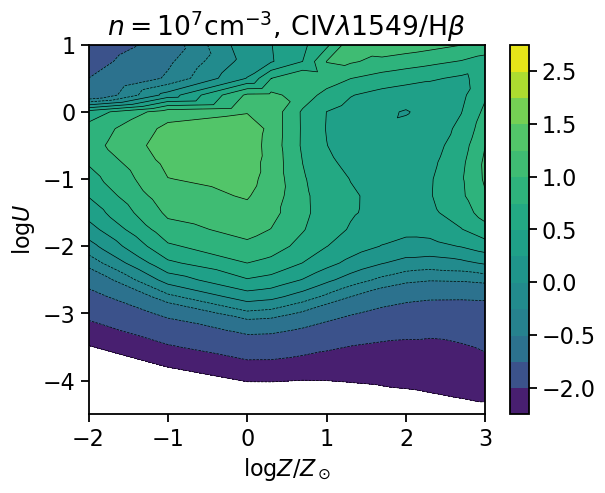} 
    \includegraphics[width=6.5cm]{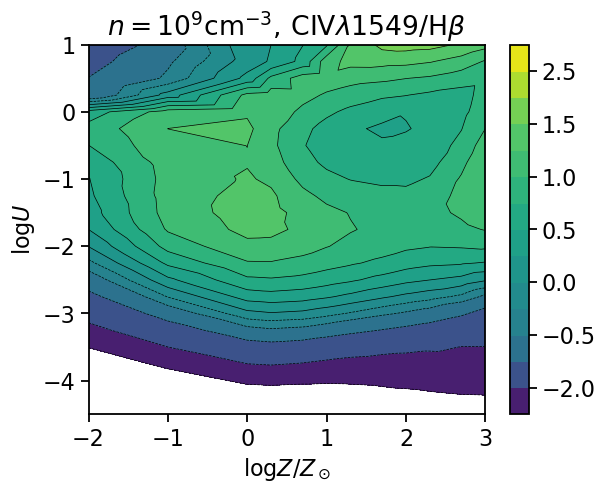}   
    \includegraphics[width=6.5cm]{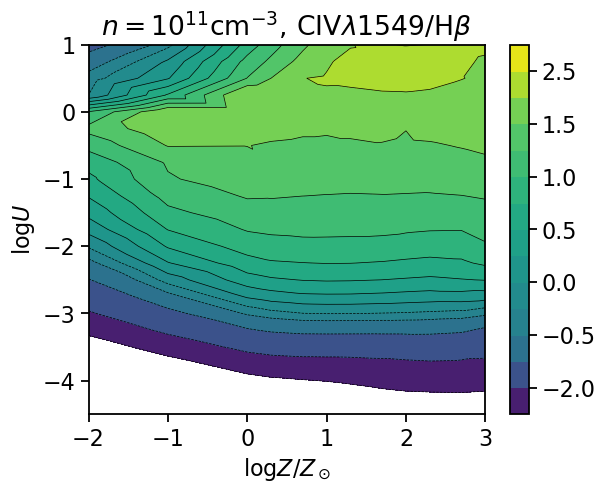} 
    \includegraphics[width=6.5cm]{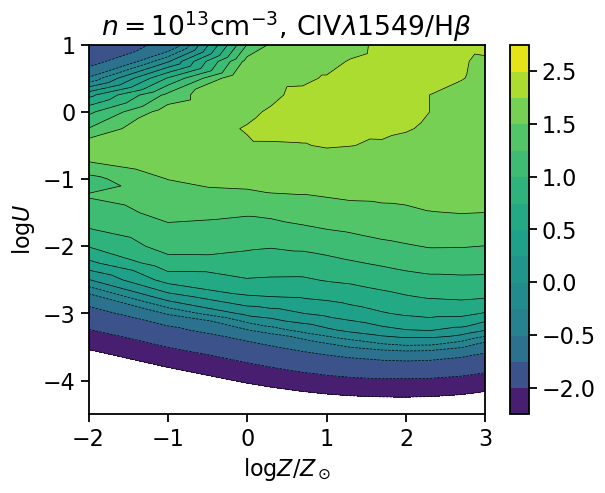}
    \caption{Behavior of the diagnostic ratio \civ/\hb\ (shown in logarithmic scale) as a function of ionization parameter $U$ and metallicity $Z$, for four values of Hydrogen density, $\log$ \nh = 7,9,11,13 [cm$^{-3}$], using the \cite{Mathews1987} Population A SED.}
    \label{fig:c4hbzu}
\end{figure}

\begin{enumerate}
    \item {\bf (SiIV+OIV])/CIV}: widely used as a metallicity indicator  \citep{Nagao2006, Shin2013}. An excellent explanation of why this ratio is a reliable metallicity indicator is provided by \cite{Huang2023}. 
   
    
    \item {\bf (SiIV+OIV])/HeII}: sensitive to metallicity, assuming that He abundance relative to Hydrogen is constant. The OIV] line has a critical density $n_{\rm c}$$\sim$ 10$^{11}$ cm$^{-3}$ \citep{Marziani2020}, making this diagnostic ratio affected by density.   
    
    \item {\bf CIV/HeII}: diagnostic ratio sensitive to metallicity, again assuming constant He abundance relative to Hydrogen. The reasons why the \siiv/\heiiuv\ and \civ/\heiiuv\ are very reliable metallicity indicators are similar. First, there is the linear increase of the abundance of Si and C to He with increasing $Z$. In the \civ/\heiiuv\ case,  the ionization potentials of C$^{2+}$ and He$^+$ are almost coincident; however, the main difference is that \heiiuv\ is a recombination line and the \civ\ is due to collisional excitation \citep{Marziani2020, Sniegowska2021}. As the metallicity $Z$\ increases, the electron temperature $T_\mathrm{e}$\ decreases  \citep[e.g.,][]{Pagel1979} and hence   metal lines excitation is disfavored. Conversely, the lower $T_\mathrm{e}$\ increases the rate of He$^{++}$ recombination, reinforcing the increase of the ratios with $Z$. The ratio \civ/\heiiuv\ has been used   \citep{Shin2013,sulenticetal14} despite of the \heiiuv\ weakness. However, with high S/N and resolution measurements from FOS and especially COS, we can accurately estimate the line intensity of \heiiuv, previously very difficult to assess, and self-consistently deblend it from the much stronger \civ\ emission. 
    
    \item {\bf CIV/H$\beta$}. Unlike all other ratios employed in this work, the \civ/\hb\ ratio is subject to major observational uncertainty in the photometric precision of the optical data, and lack of concurrent optical and UV observations. The depth of absorption features close to the rest frame in the \civ\ profile is difficult to assess and may significantly lower the \civ/\hb\ (one of the most serious cases could be PHL 1092, Table \ref{table:civhbratio}). The diagnostic value of the \civ/\hb\ ratio is primarily due to its strong dependence on the ionization parameter. Fig. \ref{fig:c4hbzu} shows the behavior of the \civ/\hb\ intensity ratio in the plane $\log Z$, $\log U$: the isophotal contours are almost parallel to the $Z$ axis.  
    
    In the presence of a very low \civ/\hb\ ($\sim 1)$ ratio, the $U$\ must be very low, with $\log U \lesssim -2.5$. At higher ionization parameters $-1 \lesssim \log U \lesssim 0$, and moderate density, the \civ\ emission to the Balmer line is optimized at relatively low values of $Z$. At higher $Z$ ($\gtrsim 10 Z_\odot$), for the same $U$, the electron temperature is significantly lowered, decreasing the collisional excitation rate, and therefore lowering the line emission. There are two implications relevant to this work: a very weak \civ\ requires very low $U$,\ with the lowest \civ\ prominence obtained for low $Z$ (Fig. \ref{fig:c4hbzu}). Conversely, for $\log U \sim -1 \div 0$, where \civ\ is efficiently produced, \civ/\hb\ $\gtrsim 10$\ can be achieved for $\log Z \sim -2$, for density $\log$\nh $\sim 11 \div 13$.    
    
    \item {\bf AlIII/CIV}: a diagnostic ratio sensitive to the ionization parameter. This is because it relates lines of elements with different ionization potentials. It is potentially useful to constrain the effect of feedback from supernov\ae\ since \aliii\ is overproduced in the explosion of type Ibc supernovae\ to the \civ\ solar abundances \citep{chieffilimongi13}. This diagnostic ratio may tend to bias the estimated $Z$ toward higher values.
    
    \item {\bf AlIII/SiIII]}: a diagnostic ratio sensitive to density, since it involves an intercombination line with a well-defined critical density: $n_{\rm c}$$\sim$ 10$^{11}$ cm$^{-3}$ for \siiii\ \citep{Negrete2013}. If the ratio \aliii/\siiii\ is $\gg 1$, the resulting density will be much greater than the critical density. 
    Historically, diagnostic ratios employing \aliii\ have rarely been used, since the measurement of the intensity of \aliii\ was quite uncertain, but with high-quality COS and FOS spectra, we have been able to measure it accurately.
    
    \item {\bf SiIII]/CIII]}: a diagnostic ratio sensitive to density, similar to \aliii/\siiii\ but around lower values, $n_{\rm c}$$\sim$ 10$^{10}$ cm$^{-3}$ for \ciii\ \citep{Hamann2002}. As the ratio described above, if \siiii/\ciii\ is $\gg 1$, the resulting density will be much greater than the critical density of \ciii\ as well, and is a very powerful density estimator,  constraining density very well. The inclusion of strong \ciii\ may however bias the ($U$, \nh, $Z$) solution toward lower densities. The issue is analyzed in Section \ref{validat}.  
    
    \item {\bf CIII]/CIV}: it is a diagnostic ratio sensitive to the ionization parameter as it compares the intensity of different ionization degrees of the same atomic species. However, since \ciii\ is an intercombination line, it is also sensitive to density. For xA sources the ratio \ciii/\civ\ is subject to higher uncertainties due to \feiii\ contamination, making it difficult to measure the intensity of \ciii\ accurately. However, since xA \ciii\ is weakest along the MS \citep{bachevetal04,aokiyoshida99}, a fairly high-density solution is anyway suggested despite the uncertainties.   
    
    \item {\bf FeII/H$\beta$}: it is a diagnostic ratio sensitive to metallicity, as it involves a metal and Hydrogen. However, this diagnostic ratio is also sensitive to the density, ionization parameter, and column density of the line-emitting gas \citep{verneretal99,verneretal04,bruhweilerverner08,Panda2018, Panda2019,Panda2021}. High \feii/\hb\ is only possible if conditions of low ionization, high density ($n_{\rm H}$$\gtrsim 10^{11}$ cm$^{-3}$) and large column density ($N_{\rm c}$ $\gtrsim 10^{23}$ cm$^{-2}$) are satisfied \citep{Panda2018, Panda2019, Panda2021}. The ratio \feii/\hb\ is relatively easy to measure with good precision ($\sim 0.1$\ at 1$\sigma$  confidence level). However, the mean escape probability formalism is known to perform poorly in the prediction of the emission line strength for gas irradiated by strong X-ray sources \citep[e.g.,][and references therein] {dumontetal03}. In the case of this particular ratio, there is therefore considerable uncertainty associated with the photoionization computations. 
    
    \item {\bf HeII$_{\rm opt}$/H$\beta$}: it is a diagnostic ratio sensitive to the ionization parameter. Assuming the abundance of He relative to Hydrogen is constant, it is the ratio between two species with very different ionization potential: \heii\ requires $\sim 50$ eV to be ionized, while Hydrogen requires $\sim 10$ eV.  The \heii/\hb\ ratio is relatively insignificant in the $Z$\ estimates. 

\end{enumerate}

These diagnostic ratios are used differently depending on the quasar population. Notably, not all components are present in all objects. Although most objects can be modeled assuming a BLUE component for HILs, it is not always present for LILs. By the same token, the VBC appears only in Population B sources. Population A objects allow reliable estimation only of physical parameters associated with the BC. Population B quasars allow for the estimation of physical parameters of gas located in the VBLR and BLR, while xA quasars allow for the estimation of the properties of the BLUE and BC. In this study, we exclusively examined the BLUE of PHL 1092, since it was the only object with a sufficient number of diagnostic ratios enabling an accurate constraint of its physical conditions.

The solution for the single-zone model, represented by a single point in the 3D space (\nh, $U$, $Z$) is determined by minimizing the $\chi^2$\ computed from the difference  between the observed line ratios and the predicted line ratios across the entire 3D space \citep{Marziani24}, applying the following relation:
\begin{equation}
    \chi^2_{\rm kc}(n_{\rm H}, U, Z) = \sum_{\rm i} \left(\frac{R_{\rm kci}-R_{\rm kci, mod}(n_{\rm H}, U, Z)}{\delta R_{\rm kci}}\right)^2,
    \label{chi2}
\end{equation}

with $R_{\rm kc}$ identifying the diagnostic ratio considered for the object and component, and $\delta R_{\rm kc}$ identifying the error associated with each component. The "mod" subscript identifies the value of the specific diagnostic ratio predicted by the {\tt CLOUDY} simulations for each value in the 3D grid of the (\nh, $U$, $Z$) parameter space. 

\subsection{FeII and AlIII Upper limits}
\label{upperlim}

\begin{figure*}[h!]
\centering
\includegraphics[width=6.05cm]{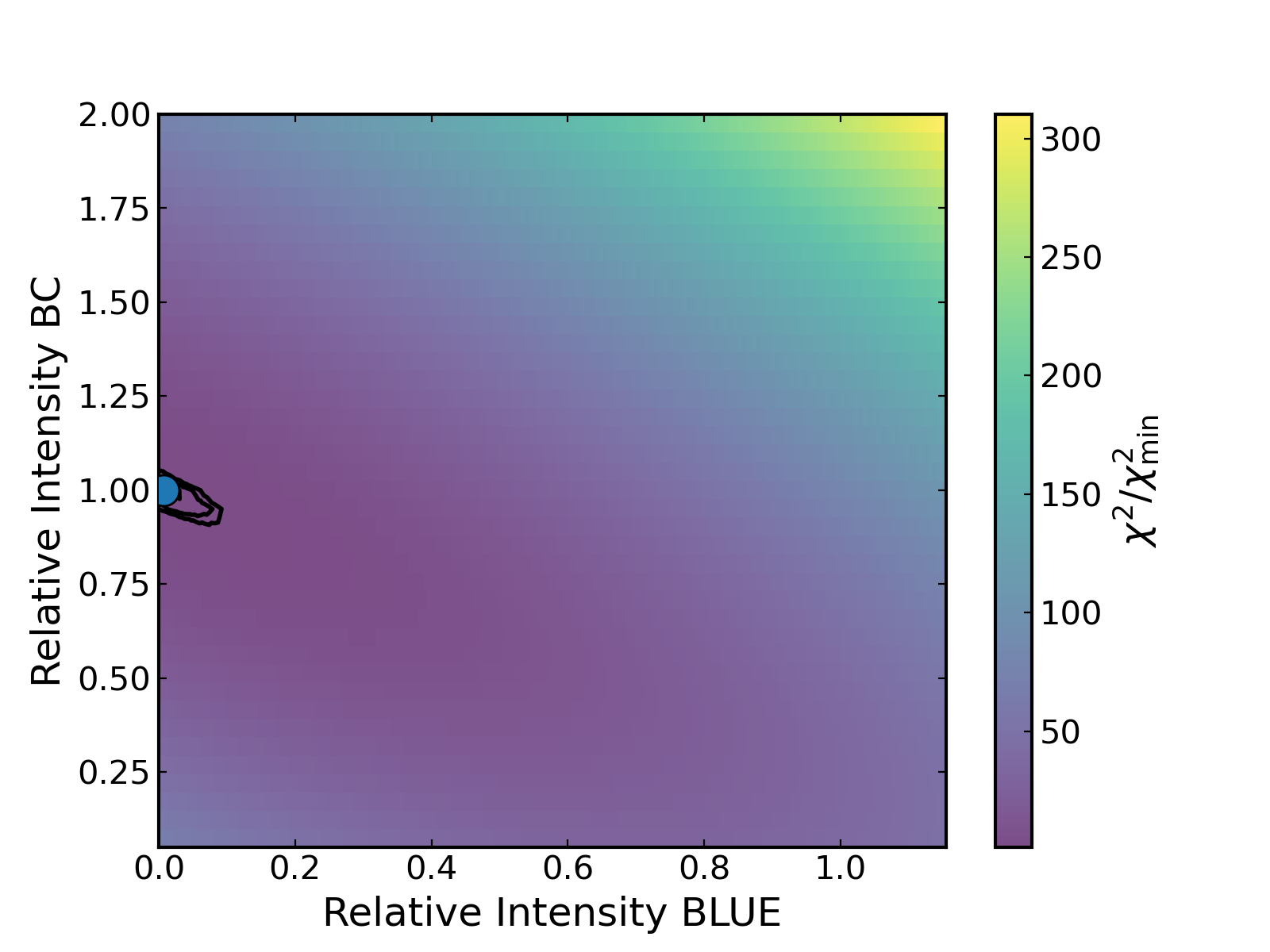}
\includegraphics[width=6.05cm]{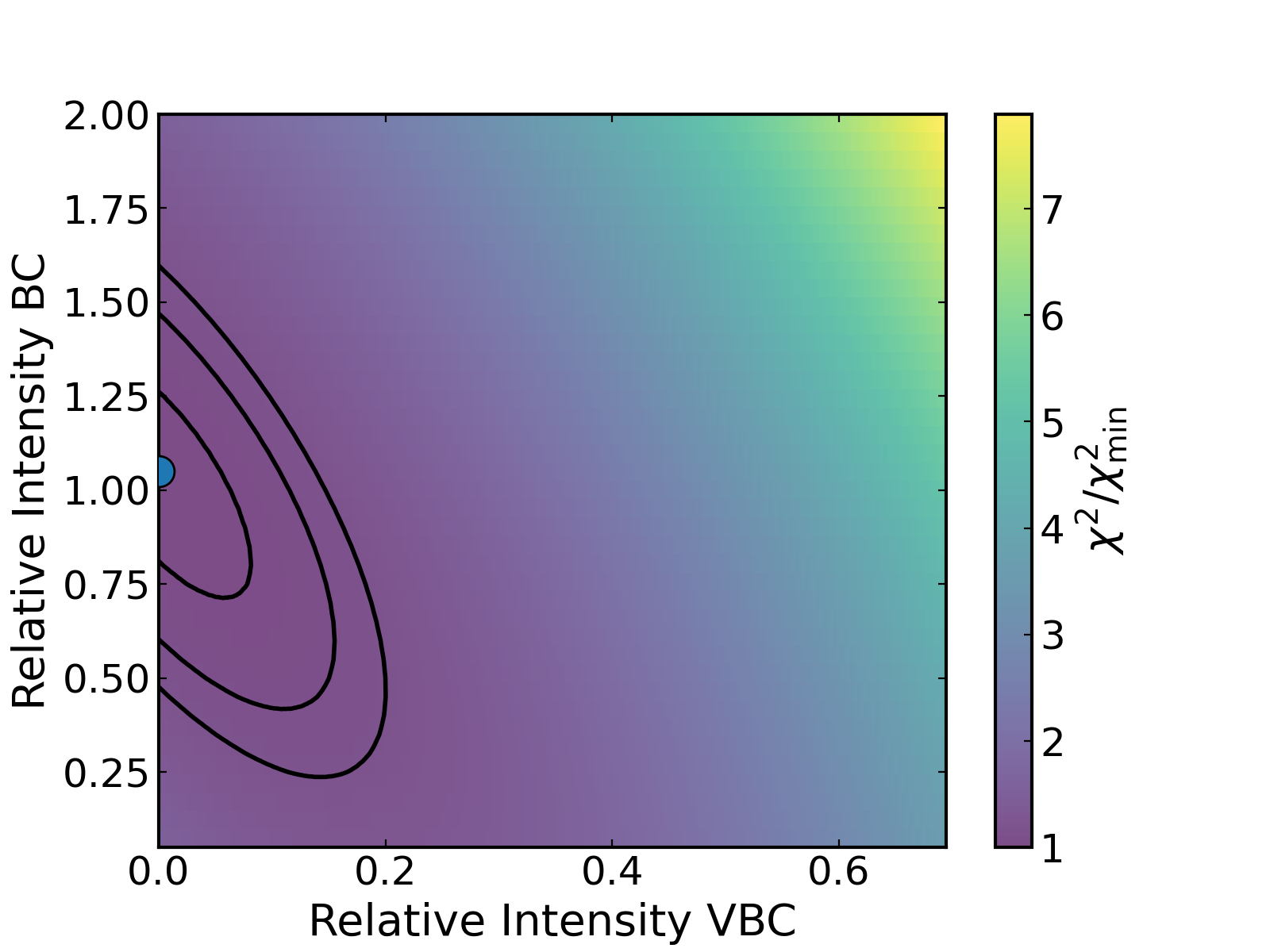}
\includegraphics[width=6.05cm]{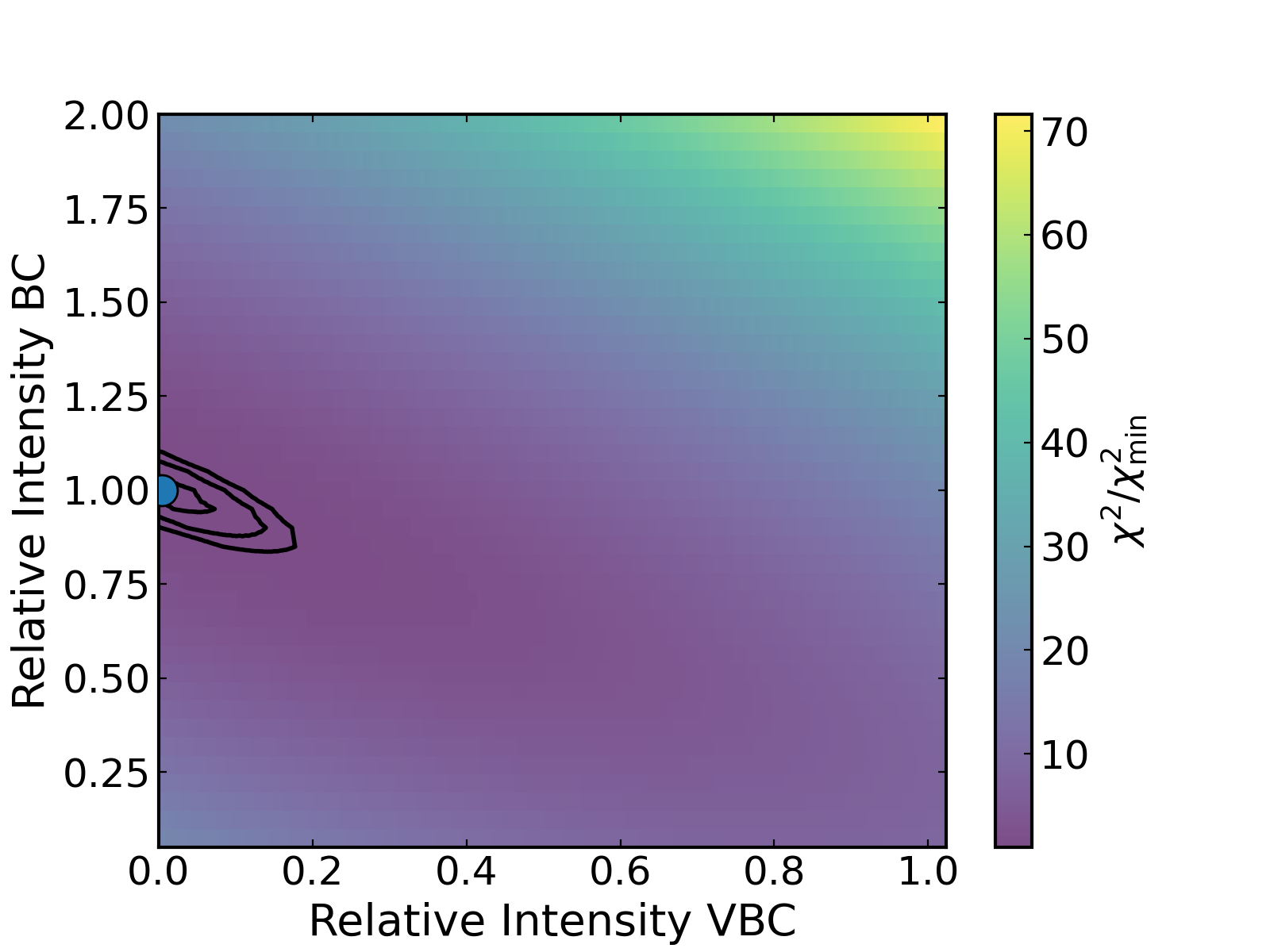}
\caption{{\it Left}: Plot of the $\chi^2/\chi^2_{\rm min}$ distribution as a function of the relative intensity of the FeII$_{\rm BC}$ of PHL 1092 with respect to the {\tt specfit} retrieved intensity, versus the relative intensity of the FeII$_{\rm BLUE}$ with respect to the original intensity of FeII$_{\rm BC}$. {\it Middle}: Same, but for the AlIII$_{\rm VBC}$ emission of Fairall 9. {\it Right}: Same, but for the FeII$_{\rm VBC}$ emission of Fairall 9.}
\label{fig:upperfig}
\end{figure*}

{\tt CLOUDY} photoionization simulations have previously hinted at the possibility of a contribution to the intensity of \aliii\ and \feii\ emission from the VBC for Pop. B Quasars and from the BLUE for xA Quasars \citep{zhouetal23}. This possibility was thus investigated by damping the intensity of \aliii\ emission obtained from the multi-component fitting of our spectra, adding a Gaussian component to the fit with the same blueshift (redshift) for xA Quasars (for Population B quasars) correspondent to the shift between the H$\beta$ BC and the H$\beta$ BLUE (H$\beta$ VBC), and the same FWHM as the BLUE (VBC). After including this additional component the $\chi^2$ associated with the fit was calculated for varying intensity values of the Gaussian line. As a result, regions associated with 1$\sigma$, 2$\sigma$, and 3$\sigma$ compatibility were drawn in the plots shown in Figure \ref{fig:upperfig}. The same procedure was applied to the \feii\ emission in the optical wavelength range, with the difference that instead of adding a Gaussian component, the additional \feii\ emission was modeled from the ESO \feii\ template \citep{Marziani2009, Marziani2010}. The results of this computation were then added to the known diagnostic ratios associated with the VBC or the BLUE of our objects, by using the 1$\sigma$ upper limit as the diagnostic ratio. 

\section{Results}
\label{results}

\subsection{Results on individual sources}
\label{ressources}

\begin{table*}[h!]
\caption{\siiv\ blend line measurements.}
\label{table:siiv}      
\centering                          
\begin{tabular}{l c c c}        
\hline\hline                 
\multicolumn{1}{c}{Object name} & \multicolumn{3}{c}{\siiv}\\
 & BLUE & BC & VBC\\
(1) & (2) & (3) & (4)\\
\hline
Mrk 335 & $85.2\pm25.4$ & $634.9\pm74.6$ & ...\\
Mrk 110 & ... & $706.3\pm104.9$ & ... \\
Mrk 509 & $251.7\pm98.9$ & $1859\pm300$ & ...\\
Ark 564 & $11.5\pm10.5$ & $115.5\pm24.5$ & ...\\
Mrk 478 & $6.9\pm22.8$ & $541\pm122$ & ...\\
Ton 28 & $63.5\pm40.7$ & $200\pm139$ & ...\\
LB 2522 & $146\pm67$ & $76.9\pm113.5$ & ...\\
LEDA 51016 & $34.3\pm12.2$ & $305\pm35$ & ...\\
I Zw 1 & $104.9\pm22.8$ & $481\pm56$ & ...\\
PHL 1092 & $27.1\pm5.6$ & $7.31\pm3.55$ & ...\\
Fairall 9 & $108\pm26$ & $363\pm32$ & $390\pm70$\\
Ark 120 & $220\pm114$ & $881\pm155$ & $501\pm188$ \\
NGC 3783 & $404\pm141$ & $1013\pm159$ & $539\pm145$\\
TXS 0042+101 & $0.6\pm1.1$ & $3.4\pm1.7$ & $3.5\pm3.3$\\

\hline                                   

\end{tabular}
\tablefoot{(1) Object common name. (2) Intensity of the line associated with the BLUE, with associated errors, measured in units of $10^{-15}$ erg s$^{-1}$ cm$^{-2}$. (3) Intensity of the line associated with the BC, with associated errors, measured in units of $10^{-15}$ erg s$^{-1}$ cm$^{-2}$. (4) Intensity of the line associated with the VBC, with associated errors, measured in units of $10^{-15}$ erg s$^{-1}$ cm$^{-2}$.}
\end{table*}

\begin{table*}[h!]
\caption{\civ\ and \heiiuv\ line measurements.}
\label{table:civ}      
\centering                          
\begin{tabular}{l c c c c c c}        
\hline\hline                 
\multicolumn{1}{c}{Object name} & \multicolumn{3}{c}{\civ} & \multicolumn{3}{c}{\heiiuv}\\
 & BLUE & BC & VBC & BLUE & BC & VBC\\
(1) & (2) & (3) & (4) & (5) & (6) & (7)\\
\hline
Mrk 335 & $24.9\pm16.9$ & $3943\pm274$ & ... & $153\pm33$ & $454\pm77$ & ...\\
Mrk 110 & $68.2\pm40.2$ & $5526\pm427$ & ... & ... & $413\pm99$ & ...\\
Mrk 509 & $266\pm130$ & $11100\pm994$ & ... & $394\pm154$ & $1096\pm300$ & ...\\
Ark 564 & $29.5\pm21.5$ & $409\pm54$ & ... & $42.1\pm24.9$ & $140\pm30$ & ...\\
Mrk 478 & $213\pm30$ & $820\pm100$ & ... & $124\pm25$ & $102\pm35$ & ...\\
Ton 28 & $203\pm65$ & $221\pm88$ & ... & $63.9\pm46.5$ & $36.4\pm62.7$ & ...\\
LB 2522 & $201\pm83$ & $87.6\pm99.2$ & ... & $53.5\pm59.8$ & $31.8\pm95.1$ & ...\\
LEDA 51016 & $151\pm25$ & $446\pm67$ & ... & $58.4\pm17.1$ & $42.5\pm25.0$ & ...\\
I Zw 1 & $987\pm 97$ & $678\pm96$ & ... & $197\pm47$ & $143\pm45$ & ...\\
PHL 1092 & $26.2\pm4.4$ & $5.23\pm2.84$ & ... & $3.17\pm2.06$ & $0.56\pm2.23$ & ...\\
Fairall 9 & $448\pm84$ & $2354\pm185$ & $1772\pm295$ & $14.7\pm33.2$ & $172\pm59$ & $127\pm86$\\
Ark 120 & $1500\pm155$ & $2500\pm169$ & $1002\pm154$ & ... & $150\pm48$ & $401\pm107$ \\
NGC 3783 & $2318\pm282$ & $8720\pm534$ & $6119\pm867$ & $85.6\pm86.0$ & $794\pm175$ & $2180\pm570$ \\
TXS 0042+101 & $5.0\pm1.6$ & $18.7\pm2.8$ & $33.0\pm6.9$ & ... & $1.5\pm1.1$ & $7.5\pm4.2$\\

\hline                                   

\end{tabular}
\tablefoot{(1) Object common name. (2),(5) Intensity of the line associated with the BLUE, with associated errors, measured in units of $10^{-15}$ erg s$^{-1}$ cm$^{-2}$. (3),(6) Intensity of the line associated with the BC, with associated errors, measured in units of $10^{-15}$ erg s$^{-1}$ cm$^{-2}$. (4),(7) Intensity of the line associated with the VBC, with associated errors, measured in units of $10^{-15}$ erg s$^{-1}$ cm$^{-2}$.}
\end{table*}

\begin{table*}[t!]
\caption{$\lambda1900$ blend line measurements.}
\setlength{\tabcolsep}{1.5pt}
\label{table:1900}      
\centering                          
\begin{tabular}{l c c c c c c c}        
\hline\hline                 
\multicolumn{1}{c}{Object name} & \multicolumn{2}{c}{AlIII} & \multicolumn{2}{c}{SiIII]} & \multicolumn{3}{c}{CIII]}\\
& BLUE & BC & BLUE & BC & BLUE & BC & VBC\\
(1) & (2) & (3) & (4) & (5) & (6) & (7) & (8)\\
\hline
Mrk 335 & ... & $42.5\pm45.9$ & ... & $186\pm68$ & ... & $705\pm125$ & ...\\
Mrk 110 & ... &... & ... & ... & ... & ... & ...\\
Mrk 509 & ... & $441\pm210$ & ... & $916\pm277$ & ... & $2034\pm401$ & ...\\
Ark 564 & ... & $14.6\pm7.4$ & ... & $36.5\pm10.5$ & ... & $231\pm25$ & ...\\
Mrk 478 & ... & $101\pm26$ & ... & $217\pm38$ & ... & $298\pm46$ & ...\\
Ton 28 & ... & $135\pm81$ & ... & $84.8\pm74.3$ & ... & $51.2\pm51.8$ & ...\\
LB 2522 & ... & $60.2\pm87.4$ & ... & $13.1\pm89.4$ & ... & $85.7\pm92.0$ & ...\\
LEDA 51016 & ... & $30.9\pm14.6$ & ... & $93.9\pm23.0$ & ... & $94.0\pm23.1$ & ...\\
I Zw 1 & ... & $153\pm34$ & ... & $270\pm46$ & ... & $162\pm35$ & ...\\
PHL 1092 & $3.9\pm1.6$ & $9.6\pm4.1$ & $4.1\pm1.7$ & $9.7\pm4.2$ & $3.3\pm1.5$ & $1.2\pm3.9$ & ...\\
Fairall 9 & ... & $137\pm43$ & ... & $440\pm70$ & ... & $294\pm59$ & $189\pm51$\\
Ark 120 & ... & $116\pm51$ & ... & $396\pm82$ & ... & $457\pm77$ & $301\pm93$\\
NGC 3783 & ... & $499\pm151$ & ... & $661\pm170$ & ... & $1160\pm201$ & $1024\pm344$\\
TXS 0042+101 & ... & $2.1\pm1.1$ & ... & $2.3\pm1.2$ & ... & $5.4\pm1.5$ & $6.0\pm3.0$\\

\hline                                   

\end{tabular}
\tablefoot{(1) Object common name. (2),(4),(6) Intensity of the line associated with the BLUE, with associated errors, measured in units of $10^{-15}$ erg s$^{-1}$ cm$^{-2}$. (3),(5),(7) Intensity of the line associated with the BC, with associated errors, measured in units of $10^{-15}$ erg s$^{-1}$ cm$^{-2}$. (8) Intensity of the line associated with the VBC, with associated errors, measured in units of $10^{-15}$ erg s$^{-1}$ cm$^{-2}$.}
\end{table*}

\begin{table*}[t!]
\caption{Optical line measurements.}
\setlength{\tabcolsep}{1.5pt}
\label{table:hbeta}      
\centering                          
\begin{tabular}{l c c c c c c}        
\hline\hline                 
\multicolumn{1}{c}{Object name} & \multicolumn{2}{c}{HeII$\lambda4686$} & \multicolumn{3}{c}{H$\beta$} & \multicolumn{1}{c}{FeII}\\
 & BC & VBC &  BLUE & BC & VBC & BC\\
(1) & (2) & (3) & (4) & (5) & (6) & (7)\\
\hline
Mrk 335 & $199\pm46$ & ... & ... & $669\pm72$ & ... & $139\pm25$\\
Mrk 110 & $24.1\pm4.3$ & &... & $147\pm16$ & ... & $29.2\pm5.6$\\
Mrk 509 & $447\pm92$ & ... &... & $1877\pm230$ & ... & $38.6\pm28.7$\\
Ark 564 & $10.2\pm3.7$ & ... &... & $36.5\pm6.4$ & ... & $24.3\pm6.8$\\
Mrk 478 & $5.35\pm0.99$ & ... & $6.15\pm1.53$ & $157\pm12$ & ... & $169\pm10$\\
Ton 28 & ... & ... &... & $124\pm17$ & ... & $129\pm12$\\
LB 2522 & ... & ... & ... & $35.3\pm7.8$ & ... & $45.7\pm5.3$\\
LEDA 51016 & ... & ... & $8.1\pm1.6$ & $58.7\pm8.3$ & ... & $67.4\pm6.4$\\
I Zw 1 & ... & ... & $29.2\pm6.2$ & $298\pm40$ & ... & $599\pm50$\\
PHL 1092 & ... & ... & $2.3\pm0.7$ & $12.7\pm2.7$ & ... & $26.2\pm3.0$\\
Fairall 9 & ... & ... & ... & $187.0\pm20.7$ & $239\pm42$ & $184\pm24$\\
Ark 120 & ... & $154\pm39$ & $261\pm26$ & $809\pm56$ & $991\pm103$ & $1160\pm86$\\
NGC 3783 & ... & $231\pm55$ & ... & $1142\pm73$ & $521\pm112$ & $332\pm56$\\
TXS 0042+101 & ... & ... & ... & $3.1\pm1.1$ & $8.8\pm4.0$ & $0.5\pm0.7$\\

\hline                                   

\end{tabular}
\tablefoot{(1) Object common name. (2),(5),(7) Intensity of the line associated with the BC, with associated errors, measured in units of $10^{-15}$ erg s$^{-1}$ cm$^{-2}$. (3),(6) Intensity of the line associated with the VBC, with associated errors, measured in units of $10^{-15}$ erg s$^{-1}$ cm$^{-2}$. (4) Intensity of the line associated with the BLUE, with associated errors, measured in units of $10^{-15}$ erg s$^{-1}$ cm$^{-2}$.}
\end{table*}

\begin{sidewaystable*}
\setlength{\tabcolsep}{4pt}
\caption{Diagnostic intensity ratios}
\label{tab:ratios}
\centering
\begin{tabular}{l c c c c c c c c c c c r}
\hline
\hline
Object & Component & $\frac{{\rm CIV}}{{\rm HeII}_{\rm UV}}$ & $\frac{{\rm CIV}}{\rm H\beta}$ & $\frac{\rm SiIV+OIV]}{{\rm CIV}}$ & $\frac{\rm SiIV+OIV]}{{\rm HeII}_{\rm UV}}$ & $\frac{{\rm AlIII}}{{\rm CIV}}$ & $\frac{{\rm AlIII}}{{\rm SiIII]}}$ & $\frac{{\rm SiIII]}}{{\rm CIII]}}$ & $\frac{\rm FeII}{{\rm H}\beta}$ & $\frac{{\rm HeII}_{\rm opt}}{\rm H\beta}$ & $\frac{{\rm CIII]}}{{\rm CIV}}$ \\
(1) & (2) & (3) & (4) & (5) & (6) & (7) & (8) & (9) & (10) & (11) & (12) \\
\hline
Mrk 335 & BC & $8.69\pm1.60$ & --- & $0.16\pm0.02$ & $1.40\pm0.29$ & $0.01\pm0.01$ & $0.23\pm0.26$& $0.26\pm0.11$ & $0.21\pm0.04$ & $0.30\pm0.08$ & $0.18\pm0.03$ \\
Mrk 110 & BC & $13.39\pm3.37$ & --- & $0.13\pm0.02$ & $1.71\pm0.48$ & --- & --- & --- & $0.20\pm0.04$ & $0.16\pm0.03$ & --- \\
Mrk 509 & BC & $10.16\pm2.93$ & --- & $0.17\pm0.03$ & $1.70\pm0.54$ & $0.04\pm0.02$ & $0.48\pm0.27$ & $0.45\pm0.16$ & $0.02\pm0.02$ & $0.24\pm0.06$ & $0.18\pm0.04$ \\
Ark 564 & BC & $2.91\pm0.74$ & $11.19\pm2.45$ & $0.28\pm0.07$ & $0.82\pm0.25$ & $0.04\pm0.02$ & $0.40\pm0.23$ & $0.16\pm0.05$ & $0.67\pm0.22$ & $0.28\pm0.11$ & $0.57\pm0.10$ \\
Mrk 478 & BC & $8.04\pm2.94$ & $5.21\pm0.75$ & $0.66\pm0.17$ & $5.31\pm2.18$ & $0.12\pm0.03$ & $0.46\pm0.14$ & $0.73\pm0.17$ & $1.08\pm0.11$ & $0.03\pm0.01$ & $0.36\pm0.07$ \\
Ton 28 & BC & $6.08\pm10.72$ & --- & $0.90\pm0.72$ & $5.47\pm10.15$ & $0.61\pm0.44$ & $1.59\pm1.69$ & $1.66\pm2.22$ & $1.04\pm0.18$ & --- & $0.23\pm0.25$ \\
LB 2522 & BC & $2.75^{+8.79}_{-2.75}$ & --- & $1.78^{+2.40}_{-1.78}$ & $4.91^{+15.08}_{-4.91}$ & $0.69^{+1.27}_{-0.69}$ & $4.59^{+32.04}_{-4.59}$ & $0.15^{+1.06}_{-0.15}$ & $1.30\pm0.32$ & --- & $0.98^{+1.53}_{-0.98}$ \\
LEDA 51016 & BC & $10.50\pm6.37$ & --- & $0.68\pm0.13$ & $7.17\pm4.30$ & $0.07\pm0.03$ & $0.33\pm0.18$ & $1.00\pm0.35$ & $1.15\pm0.19$ & --- & $0.21\pm0.06$ \\
I Zw 1 & BC & $4.73\pm1.63$ & --- & $0.71\pm0.13$ & $3.36\pm1.12$ & $0.23\pm0.06$ & $0.57\pm0.16$ & $1.66\pm0.46$ & $2.01\pm0.32$ & --- & $0.24\pm0.06$ \\
PHL 1092 & BC & $9.27^{+36.94}_{-9.27}$ & --- & $1.40\pm1.02$ & $12.97^{+51.60}_{-12.97}$ & $1.84\pm1.27$ & $0.99\pm0.60$ & $7.84^{+24.72}_{-7.84}$ & $2.07\pm0.49$ & --- & $0.24^{+0.75}_{-0.24}$ \\
 & BLUE & $8.26\pm5.56$ & --- & $1.04\pm0.27$ & $8.55\pm5.84$ & $0.15\pm0.07$ & $0.96\pm0.56$ & $1.24\pm0.76$ & $\leq1.28$ & --- & $0.13\pm0.06$ \\
Fairall 9 & BC & $13.69\pm4.80$ & $12.60\pm1.71$ & $0.15\pm0.02$ & $2.11\pm0.75$ & $0.06\pm0.02$ & $0.31\pm0.11$ & $1.49\pm0.38$ & $0.98\pm0.17$ & --- & $0.12\pm0.03$ \\
 & VBC & $13.92\pm9.73$ & $7.41\pm1.79$ & $0.17\pm0.03$ & $3.06\pm2.15$ & $\leq0.02$ & --- & --- & $\leq0.15$ & --- & $0.11\pm0.03$ \\
Ark 120 & BC & $16.66\pm5.41$ & --- & $0.35\pm0.07$ & $5.87\pm2.13$ & $0.05\pm0.02$ & $0.29\pm0.14$ & $0.87\pm0.23$ & $0.84\pm0.08$ & --- & $0.18\pm0.03$ \\
& VBC & $2.50\pm0.77$ & --- & $0.50\pm0.20$ & $1.25\pm0.58$ & $\leq0.05$ & --- & --- & $\leq0.17$ & $0.16\pm0.04$ & $0.30\pm0.10$ \\
NGC 3783 & BC & $10.99\pm2.51$ & $7.64\pm0.68$ & $0.12\pm0.02$ & $1.28\pm0.34$ & $0.06\pm0.02$ & $0.75\pm0.30$ & $0.57\pm0.18$ & $0.29\pm0.05$ & --- & $0.13\pm0.02$ \\
& VBC & $2.80\pm0.83$ & $11.74\pm3.02$ & $0.06\pm0.02$ & $0.25\pm0.09$ & $\leq0.03$ & --- & --- & $\leq0.67$ & $0.44\pm0.14$ & $0.12\pm0.04$ \\
TXS 0042+101 & BC & $12.56\pm9.78$ & $6.01\pm2.24$ & $0.18\pm0.09$ & $2.29\pm2.07$ & $0.11\pm0.07$ & $0.90\pm0.66$ & $0.43\pm0.24$ & $0.16\pm0.24$ & --- & $0.29\pm0.09$ \\
& VBC & $4.40\pm2.63$ & $3.75\pm1.89$ & $0.19\pm0.18$ & $0.46\pm0.51$ & $\leq0.05$ & --- & --- & $\leq0.20$ & --- & $0.32\pm0.17$ \\
\hline
\end{tabular}
\tablefoot{(1) Object common name. (2) Component considered for the estimation of metallicity. (3)-(12) Diagnostic ratios used for the estimation of metallicity, with the associated error. ---: not used or not available. Upper limits for some diagnostic ratios as described in Section \ref{upperlim} are shown.} 
\end{sidewaystable*}

Tables \ref{table:siiv}, \ref{table:civ},   \ref{table:1900}, \ref{table:hbeta} report the intensity of each different line component for the objects of the sample measured from the multi-component fitting, with associated errors, for the four spectral ranges that have been fit separately, namely the \siiv, \civ+\heiiuv, \aliii+\siiii+\ciii, \hb+\feii. The fits are shown in the Figures of Appendix \ref{app:fits}, and the projections of the ($U$, \nh, $Z$) parameter space in Appendix \ref{app:isoph}. Table \ref{tab:ratios} summarizes the diagnostic intensity ratios used for the estimation of the physical parameters, while estimates for the metallicity and physical conditions in the BLR for our sample are shown in Table \ref{tab:z}. Here follow considerations on the results obtained for each of the individual sources:

\begin{itemize}
    \item {\bf Mrk 335} is a Population A object. It has been a target of several reverberation mapping campaigns in the optical \citep{petersonetal04}: \citet{grieretal12} obtain 13.9 $\pm$ 0.9 days for \hb. The resulting \mbh\ is $(2.6 \pm 0.8) \times 10^7$ M$_\odot$, for a luminosity $\approx 7.7 \cdot 10^{44}$ erg s$^{-1}$. Mrk 335 is therefore considered a Population A source radiating at a modest Eddington ratio (\lledd$\lesssim 0.2$).    
    \citep[although see][]{Mastroserio2020}. Its BLR was spatially resolved in the infrared, obtaining an upper limit for its size of 0.155 pc \citep{Gravity2020}. The source might presently be close to its lowest state \citep{grupeetal18}, although there has been a recent upturn \citep{komossaetal20}. Both \civ\ and \hb\ do not reveal any significant blueshifted excess to the Lorentzian profile \citep[cf.][]{vangroningen87}, implying a very weak wind component.  
    
    The estimation of the physical conditions in the BLR yielded a result characterized by subsolar metallicity and very low density. The low metallicity reported in the paper is a lower limit since the result is likely due to the lack of {\tt CLOUDY} simulations with 0.2 $Z_\odot$ or 0.5 $Z_\odot$\ (the $Z$ of Mrk 110 and 509, with similar spectra, is around solar).  Additionally,  the gas density value is the lowest in the sample ($\log$\nh\ $\sim 8$ cm$^{-3}$). A low-density value is supported by very strong \ciii\ emission and a very low \aliii/\ciii\ ratio. Interestingly, the \oiii\ doublet can be fit by a narrow component with a semi-broad component with a width similar to the one of \hb$_{\rm BC}$. The estimated BLR density is higher than the critical density of \oiii. However,  some  \oiii\ emission is predicted by CLOUDY simulations to be $\sim 0.05$ the one of \hb\ in the BLR. Considering the low BLR \nh, a radial stratification  of \nh\ 
   \citep{Baldwin1995} may facilitate \oiii\ emission. 
     
    \item {\bf Mrk 110} is a Population A object of spectral type A1 that, not surprisingly,  exhibits spectral features remarkably similar to Mrk 335. A peculiarity is the presence of a weak asymmetry towards the red in the profiles of both \civ\ and \hb. The asymmetry was modeled adding a faint redshifted component affecting \civ\  intensity by about 20\%, with no significant effect on the \hb\ flux. A redshifted asymmetry had been already detected in the Balmer, He{\sc i}, and \heii\ lines and has been interpreted as due to gravitational redshift \citep{kollatschny04,mullerwold06,gavrilovicetal07}. Consistently, the BLR of Mrk 110 appears to be compact: $\tau \approx 20- 30$\ ld\ \citep{petersonetal98,kollatschnyetal01,lietal13}. If the \mbh $\approx 2 \cdot 10^7$ M$_\odot$\ derived by Kollatschny et al. is corrected for a viewing angle $\theta \approx 30^{\circ{}}$  the central mass could reach $\sim 10^8$. A photoionization solution following the locally optimized cloud model indicates emission from the BLR region starting at $\log r \approx 15.3$ \citep{juranova24}, implying an inner radius $\sim 100$\ gravitational radii. Gravitational and transverse shifts could then explain the shift amplitude of the additional components of \civ\ and \hb\ included in the fit ($\approx 4500$ \kms). If this interpretation is correct, the compact BLR and an overmassive black hole compared to other A1 sources could be at the origin of the redshifted excess \citep{gavrilovicetal07,marziani23}.    
    
    The   \civ/\hb\ $\sim 37$ was deemed inappropriate and thus excluded from the measurement of the metallicity. Indeed, similar measurements of the intensity of the lines from \cite{juranova24} returned a value of $\sim 13$, much more appropriate, indicating a light loss in our optical spectrum as a possible cause of the mismatch. The solution obtained is very well-constrained around solar metallicity (Fig. \ref{fig:proj1}, Tab. \ref{tab:z}). 
    
    \item {\bf Mrk 509} is a  Population A1 object, with very weak or absent \feii\ lines in its optical spectrum. It has been monitored in the optical, UV, and X-ray domain   \citep{Kriss19}, and more recently in IR \citep{Mitchell_etal_2024}. Any blueshifted excess is barely visible in \siiv\ and \heiiuv, and very faint in both \civ\ and \hb, although spectropolarimetry reveals the presence of an accretion disk wind \citep{lira21}. \hb\ reverberation mapping yields a time delay of $\approx$ 79.6$^{+6.1}_{-5.4}$ ld \citep{Bentz2013,duetal15}, and a mass $\sim 2 \cdot 10^8$ M$_\odot$\ \citep{yuetal20}.  Additionally, its BLR was recently spatially resolved in the infrared, revealing an upper limit to its size of 0.249 pc \citep{Gravity2020}. 
    
    Measurement of the intensity of its lines, especially \civ\ and \siiv, has been affected by the presence of many absorption lines. However, the measured line ratios constrain very well  $Z$, $U$, and $n_{\rm H}$,  yielding about solar metallicity, a value that could be considered typical of spectral type A1. 
    
    \item {\bf Ark 564} is a Population A object belonging to the A2 spectral bin of the MS. Long-term monitoring of the object has been performed over time in the optical range \citep{Shapovalova12}, yielding a rather short time delay between continuum and \hb\ variation, around $6.7\pm 10$ days. There is evidence of an excess emission on the blue side of the of \civ\ and \heiiuv\ and a high-ionization, the relativistic wind has been detected in the  X-ray domain via the discovery of lines of highly ionized oxygen shifted by $\sim 0.1 c$ \citep{guptaetal13}. These properties are fairly typical of A2 sources with quite strong \feii\ emission and higher $L/L_\mathrm{Edd}$ compared to the objects of spectral type A1.
    
    The ionization parameter is consistent with   $\log U \sim -1$. and the presence of strong \ciii\ emission drives the solution toward low-density values $\log$ \nh $\sim 7$, as in the case of Mrk 335. 
    Repeating the estimation of the physical parameters without the use of the diagnostic ratios including \ciii\, returned a somewhat different scenario, with higher density ($\log n_{\rm H} \sim 11$) and solar metallicity, in better agreement with the reverberation delay (see the discussion in Section \ref{revphot}).
    The rather high $\chi^2$\ value (Tab. \ref{tab:z}) is due to the contributions of the \civ/\heiiuv\ and \ciii/\civ, which are in turn due mainly to the small uncertainties associated with these two ratios, rather than to a disagreement between the observed and the predicted value from the minimum $\chi^2$\ solution (\civ/\heiiuv\ $\approx$ 4.3 vs. 2.9, and \ciii/\civ\ $\approx$ 0.87 vs. 0.57 predicted vs. observed).   
The result obtained for the metallicity,  5 $Z_\odot$ is thus stable, also compared to other Population A objects. A detached 1$\sigma$ region (Fig. \ref{fig:proj1}) would imply a slightly lower $Z$\ and a significantly higher \nh\ ($\log n_\mathrm{H}\sim 10$ [cm$^{-3}$]). 
    
    \item {\bf Mrk 478} is a Population A object, a NLSy1 with prominent but not extreme \feii\ emission. A very recent study indicates an \hb\ lag of $\approx 45 - 55$ days to continuum changes \citep{wooetal24}. Not unlike Ark 564, Mrk 478 could be regarded as a bridge between the A1 and xA spectral types, since it exhibits some spectral characteristics that are typical of both. Its \rfe\ $\sim$ 1 makes Mrk 478 a borderline source  \citep[][see also Table \ref{tab:ratios}]{Marziani2010}.  The high $Z$ value obtained, along with $U$ and $n_{\rm H}$ are plausible given our measurements.  
    A relevant question is what drives the metallicity to a value that is higher by a factor of 10 to an object that is not so dissimilar in the optical spectral range to Ark 564? As a matter of fact, \civ/\heiiuv, \siiv/\civ\ and \siiv/\heiiuv, all $Z$\ indicators, are $\approx 2.8$, 2.4, and 6.5 times higher in Mrk 478. The \siiv/\civ\ close to unity requires, under the physical conditions of the BLR, very high $Z \sim \mathcal{O}(100)$\ \citep{Sniegowska2021,Garnica2022}. The high $\chi^2$\ (actually the highest of the sample) is due to the \feii/\hb\ ratio: the predicted value is too low, by a factor of 3. As in the case of Ark 564, the high intensity of \ciii\ measurement returns a lower density of the gas compared to that obtained for other xA sources. The relatively low density and high ionization are not conducive to strong \feii\ emission, but the observed prominence of \feii\ is certainly consistent with a high metallicity solution.
    
    \item {\bf Ton 28} Even if the \rfe\ is not as high, Ton 28 shares several extreme properties with I Zw 1, such as extreme blueshifts in both \civ\ and \oiii\ \citep{Zamanov2002}.  Estimates of the time-lag $\tau$\ between continuum variations and \hb\ response are dependent on the method applied for the cross-correlation analysis; the Javelin-based \citep{zuetal11} estimate is $c\tau \approx 48.2^{+3.3}_{-3.1}$\ ld, larger than the CCF centroid $c\tau \approx 37.3_{-6.0}^{+6.9}$\ ld \citep{lietal21}, consistent with a previous estimate affected by a large uncertainty \citep[$32.2_{-4.2}^{+43.5}$\ ld,][]{duetal18}.  Velocity-resolved reverberation maps of \hb\ reveal a signature of Keplerian disk-like motion \citep{lietal21}. Spectro-polarimetric observations indicate a large H$\alpha$\  line width in polarized light ($\approx 4220$ \kms) compared to the width in natural light ($\approx 1770$ \kms, \citealt{Capetti2021}), implying an   almost face-on view \citep[$\theta \approx 15^{\circ{}}$, ][]{sniegowskaetal23}. Despite the lower S/N to other sources used in this work, the results obtained during the estimation of physical parameters are well constrained and suggest a solution of $\approx 50 Z_\odot$, with high density.
    
    \item {\bf LB 2522 $\equiv$ PG 1259+593} represents a typical xA object in our sample \citep{negreteetal18}, belonging to the A3 spectral type. LB 2522 has been subject to a reverberation study in the MIR wavelength range \citep{Lyu19} that yields a delay time in the W2 band\footnote{\url{https://www.astro.ucla.edu/~wright/WISE/passbands.html}} of $\tau \approx 794$ days. These values set an upper limit for $r_\mathrm{BLR}$, with the delay time of the H$\beta$ emission being estimated around $\tau = 225\pm75$ days (Du, private communication).  
    
    Although it is one of the most luminous objects of the sample, due to its large distance its FOS spectra are subject to more noise ($S/N \approx 5$) than other sources studied in this work. Nevertheless, the line intensity yielded a fairly precise measurement of the physical parameters of the BLR, with a metallicity of about 20 $Z_\odot$ and an average density of the gas   (Tab. \ref{tab:z}, Fig. \ref{fig:proj3}). 
    
    \item {\bf LEDA 51016 $\equiv$ PG 1415+451} is a typical xA object, exhibiting a very strong \feii\ emission in the optical wavelength range, and a prominent blueshifted component in \civ, appreciable also in \hb\ (Fig. \ref{fig:fits2}). Optical monitoring for reverberation mapping yields a time delay of \hb\ to continuum at 5100 \AA\ of $\tau \approx 30.0_{-4.6}^{+5.4}$ days \citep{huetal21}, to be compared with $\tau \approx 433$\ days in the MIR \citep{Lyu19}. 
    
    The COS and FOS spectra of LEDA 51016 appear with very high S/N, and the results are consistent with other xA sources of the sample, including well-constrained supersolar metallicity ($\log Z \approx 1.3^{+0.4}_{-0.3}$) and relatively high gas density. In this case, as for the prototypical source I Zw 1, the low ionization, high-density solution retrieved by \citet{Negrete2012} is possible within the 1$\sigma$ uncertainty range (Tab. \ref{tab:z}, Fig. \ref{fig:proj2}). 
    

    \item {\bf I Zw 1} is a narrow-line Seyfert 1 galaxy with a strong \feii\ emission \citep{phillips78}. Emission lines are narrow, with \hb\ FWHM estimates in the range between 1100--1500 \kms. The line narrowness has made I Zw 1  an ideal source for the extraction of a template of \feii\ emission in the optical \citep{Boroson1992,laoretal97a,Marziani2009}, IR \citep{garciarissmannetal12}, and UV \citep{Vestergaard2001} domains. The source has been hailed as a prototype for Population A \citep{sulenticetal00a} and, more recently, for extreme Population A \citep{marzianisulentic14}, due to its strong \feii, weak \ciii, high \siiv/\civ\ ratio, and weak \civ\ equivalent width.  I Zw 1 has been monitored for reverberation mapping: the rest-frame delay for \hb\ is $\tau \approx 37.2^{+4.9}_{-4.6}$ \ days \citep{huangetal19}. More recently, the amplitude of flux variation has been so small to make it is impossible to estimate a time lag for \hb\ \citep{uetal22}.
    
    Previous studies focused on the physical conditions in the virialized BLR of I Zw 1 suggested a low ionization solution, with high density \citep{Marziani2010,Panda2021} and a possible metallicity enhancement \citep{Negrete2012}.  The estimation of the physical parameters in this work leads to a solution of $\approx 20 Z_\odot$, with medium-high density.

    \item {\bf PHL 1092} has been hailed as one of the strongest \feii\ emitters ever discovered \citep[][and references therein]{Marinello2020, Panda2020, MLMA2021}. Our measurements confirm the extremely strong \rfe $\approx 1.75$\, placing PHL 1092 in the spectral type A4, where only $\approx$3\%\ of optically selected quasars are found \citep{Marziani2013}. The measurements of line intensities and FWHM for the BC and BLUE in both the optical and UV ranges are consistent with those reported in \cite{Marinello2020} (the slightly higher value of \rfe\ reported by these authors is due to their removal of the BLUE of \hb). The extreme attributes of this object are highlighted by its distinction as the sole source in the sample displaying a BLUE in the $\lambda1900$ blend, a rare occurrence even among xA objects \citep{Marziani2022,buendia-riosetal23}.

    The metallicity estimated for this source ($Z \sim 1000 Z_\odot$) is apparently out of scale, higher than the highest estimates for xA quasars at high luminosity, seldom exceeding $Z \sim 100  Z_\odot$\ \citep{Garnica2022}. 
    However, this result aligns with the low ionization, high-density solution of xA sources \citep{Negrete2012}: low \civ, high \aliii/\siiii, high \siiv/\heii\ prominent \feii, and very low \ciii/\siiii\ imply low $U$, high \nh, and high $Z$. The inordinately high value of the $Z$ is driven by the  \siiv/\civ\ $\gtrsim 1$, which excludes $Z$ values only a few times above solar.   In the case of PHL 1092, the \civ/\heii\ and  \civ/\hb\ ratios are highly uncertain, in part because of absorptions. The ionization parameter $\sim 10^{-3}$\ accounts for the weak \civ: the \civ\ BC equivalent width is just $\lesssim 2 \AA$, with a \civ/\hb\ ratio $\sim 1$. The  $W$ \civ\ BC+BLUE $\approx 8 \AA$, qualifies PHL 1092 among the weak-lined quasars (WLQ; \citealt{diamond-stanicetal09,shemmeretal10,shemmerlieber15}), that are xA sources by the wide majority \citep{marzianietal16a, martinez-aldama18}. 
     
    A major difficulty is introduced by the large errors associated with most measured ratios (Tab. \ref{tab:ratios}). Most ratios should be considered upper limits.  In addition, the out-of-scale value $Z \sim 1000 Z_\odot$, is greatly reduced to $\sim 50$\ $Z_\odot$ if a microturbulence $\approx $ 10 \kms\ is introduced in the computations which act as a metallicity controller \citep[see][for an in-depth analysis]{Panda2021}. 

   Therefore, the metallicity estimated for the BLUE ($Z\sim 50Z_\odot$) could provide a representative lower limit for this object, since it accounts for most of the observed flux in the HILs. BLUE and the BC revision with turbulence will still place the source on par with the highest $Z$ estimates in the sample. 
     

\item {\bf Fairall 9} is a prototypical Population B source, with its spectrum almost coinciding with the composite spectrum of Population B objects \citep[][\citetalias{Marziani2023}]{Jiang2021}. 
    Reverberation mapping of the object has been performed in the past both in the optical, yielding a well-defined delay $\tau \approx 17.4^{+3.2}_{-4.3}$ days \citep{Bentz2013}. Dust-reverberation mapping yielded a delay in the W1 and W2 maps $\approx$ 250 days in the W1 band at $3.4\mu$m, and between $\approx 280$ and $\approx 370$\ days for the W2 band at 4.6 $\mu$m \citep{chenetal23}. The Fairall 9 \hb\   red wing extends   up to $\sim 18,000$ \kms, beyond \oiii\ (Fig. \ref{fig:fits3}). This is not uncommon among Pop. B sources. The only peculiarity is the occurrence of a plateau on the red side of \hb\ NC, leaving a faint semi-broad residual at $v_\mathrm{r}\approx 2500$ \kms. Analogous features are seen in Pop. B objects, albeit rarely. One case in point is Ark 120, discussed below. A spectropolarimetric analysis of Fairall 9 \citep{Jiang2021} suggested the presence of a warped geometry of the accretion disk, but it is not obvious if the residual feature detected in natural light might be connected to the warped disk. 
 
   The estimations of the metallicity of the BC and VBC return about 2 $Z_\odot$ for the BC, and 1 $Z_\odot $ \ for the VBC, consistent within the uncertainties (Fig. \ref{fig:proj3}). The $1\sigma$ confidence range of $Z$ for the BC  is rather broad ranging from -1 to 0.7. The ionization parameter is constrained within an order of magnitude, and the \nh\ is only within 4 orders of magnitude.  The VBC parameters appear better constrained. The wide range of confidence intervals might be the indication of radial stratification in the BLR gas emitting the BC. The VBC, as associated with gas in the innermost region, may have intrinsically better-defined physical parameters. 

        
    \item {\bf Ark 120} stands out as the most peculiar object in the sample. The main peculiarity is its unusual double-peaked H$\beta$ profile, which was modeled in this work using a BC and two satellite components, one blueshifted and the other redshifted, roughly symmetrically displaced by $\pm 2000$ \kms. 
    This profile is not unique, and likely hints toward some reproducible structure or dynamical process within the BLR \citep{Marziani1992,korista92,Korista1997}. Even after extensive monitoring, the nature of its double-peaked profile has not been completely understood, although periodic variations have been suggested    \citep{Li2019}. Fairall 9 might undergo a similar phenomenon. 
    Ark 120  exhibits a rather strong \feii\ emission for a Population B object (\rfe{} = 0.84, see Table \ref{tab:ratios}), likely originating in a region further from the one in which originates H$\beta$, which exhibits a time delay of $\tau = 57.3\pm23.0$ days \citep{Kuehn2008}.
    
    The presence of a mismatch between the UV and optical fluxes after correcting the flux for galactic extinction indicates the presence of a stronger extinction than expected. This leads to an underestimation of the \civ/\hb\ ratio, which reduces the resulting metallicity. We therefore excluded the \civ/\hb\ diagnostic ratio during the computation. 
    Once excluded from the inordinately low \civ/\hb, the $Z$ solution converges to a well-defined metallicity value.  The moderate \siiv/\civ\ ratio indicates abundances at least solar or a few times above solar. The value obtained ($Z \sim 10 Z_\odot$) for the BC is consistent with the one for the VBC ($Z \sim 5 Z_\odot$). This is in the case that the satellite lines are included in the total flux of the BC. Removing them, and using only the central core of the \hb\ profile (Fig. \ref{fig:fits4}) would imply an \rfe $\approx 1.5$ that is out-of-scale to the other line ratios. 
    
    \item {\bf NGC 3783} is the closest object of the sample, located at a redshift, z = 0.009. It is a well-known object, extensively monitored for reverberation mapping, with an H$\beta$-emitting region having a median radius delay $r_{\rm median} = 10.07^{+1.10}_{-1.21}$ lt-day \citep{Bentz21}. \cite{lira21} performed a spectropolarimetric study of the object, revealing evidence of a nuclear wind whose presence is also revealed by the excess of the blue side of \civ\ and \siiv\ (Fig. \ref{fig:fits4}), most likely due to outflowing gas.  The BLR of NGC 3783 was spatially resolved in the infrared, obtaining an upper limit for its size of 0.110 pc \citep{Gravity2020}. 
    The spectrum of this object almost coincides with the composite spectrum obtained for the RQ Population B. 
    
Confronting the diagnostic ratios of the analyzed spectra with the {\tt CLOUDY} simulations, the BC is associated with a slightly sub-solar metallicity, with the VBC having a compatible result of solar metallicity. The ionization parameter gradient obtained for the BC and VBC of this object is consistent with a more ionized VBC, being closer to the ionizing source.

    
    \item {\bf TXS 0042+101} is the farthest object of the sample. We measured $z \approx$ 0.587,  in agreement with \cite{Tang2012}. It is the only RL object of the sample. Like many other RL sources, it is well studied in the radio wavelength range, with its FRII radio morphology described in \cite{Hutchings98}. Optical observations are poor.  The VBC is relatively prominent, allowing for an independent $Z$\ estimate, even if the S/N is among the worst in the sample. 
    
    The $Z$\ results are reliable, with sub-solar metallicity for the BC and VBC, consistent among each other, and with a low minimum $\chi^2$. The spectrum of TXS 0042+101 and its derived properties are consistent with the median composite of RL quasars in the B1 spectral type \citepalias{Marziani2023}. 
    
\end{itemize}
\subsection{The trend along the MS}

\begin{figure*}[ht!]
    \centering
    \includegraphics[width=15cm]{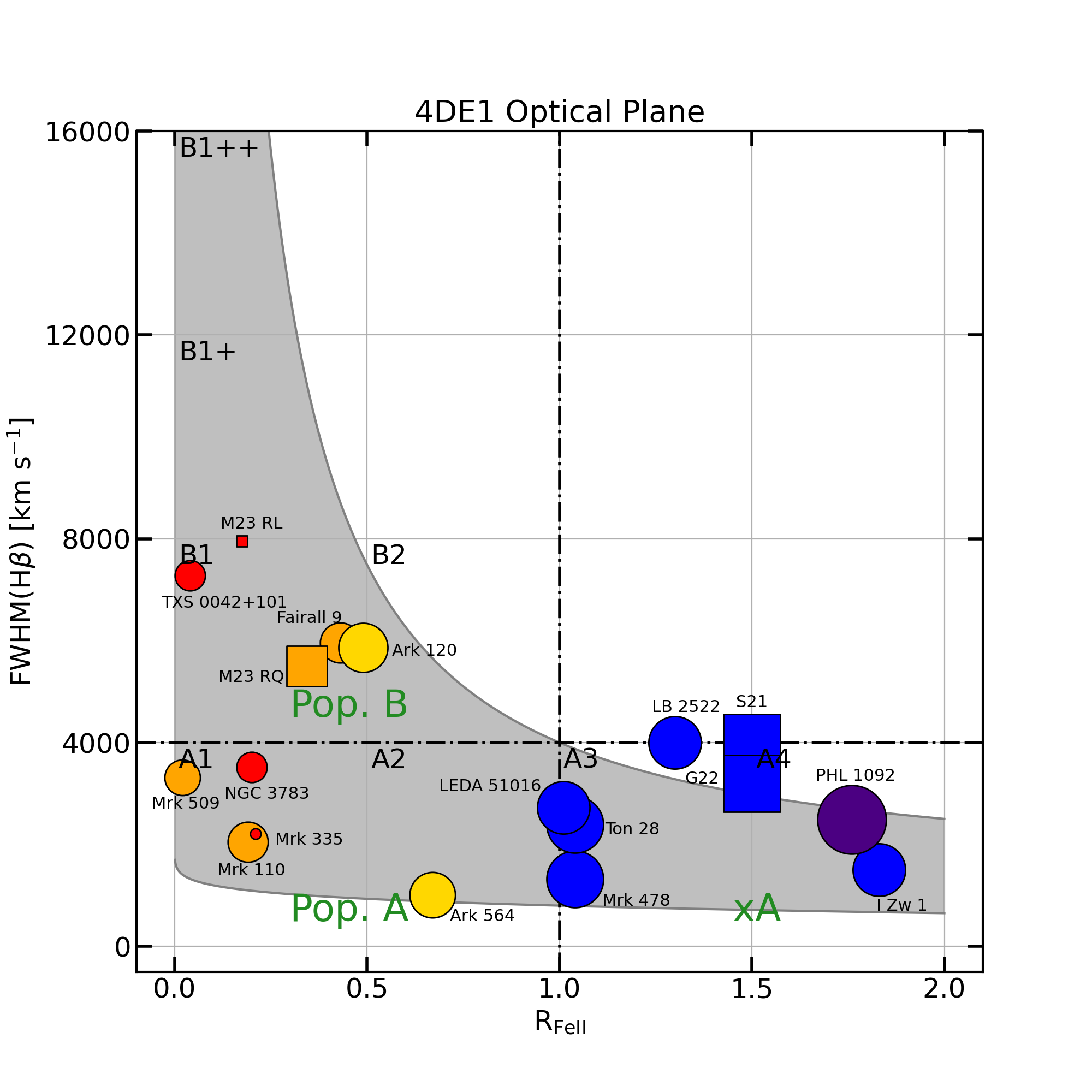}
    \caption{MS representation of our sample. Every object is placed according to its R$_{\rm FeII}$ and FWHM(H$\beta$). Greater metallicity $Z$ is expressed with increasing marker sizes and different colors. Objects with sub-solar metallicity are marked in red, objects with $1\ Z_\odot \leq Z \leq 2\ Z_\odot$ in orange, objects with $2\ Z_\odot < Z \leq 10\ Z_\odot$ are marked in gold, objects with $10\ Z_\odot < Z \leq 50\ Z_\odot$ are marked in blue, and objects with metallicity greater than 50 $Z_\odot$ are marked in indigo. Sources from this work are shown in circles. Squares are used as reference values for the objects in S21, G22, and M23. The latter with two squares, one for RL sources, and the other for RQ sources.}
    \label{fig:4DE1Sample}
\end{figure*}

Figure \ref{fig:4DE1Sample} depicts the MS representation of our sample, alongside representative values from \citetalias{Sniegowska2021}, \citetalias{Garnica2022}   and \citetalias{Marziani2023}.
The main result is that the highest-metallicity objects are clustered in the xA region of the MS. Conversely, all other objects with solar or subsolar metallicity are situated in the left portion of the MS, demonstrating a gradient of metallicity that peaks in the xA segment of the MS and gradually decreases towards RL Population B sources - an outstanding result confirming the theoretical predictions from \citet{Panda2019}. At one extreme, TXS 0042+101 comes close to the RL template of \citetalias{Marziani2023}, with subsolar $Z$, with $\log Z \sim -0.3 $ and $\sim -1.0$, respectively. For the sources of ST A1 and B1 with stronger \feii\ we encounter $Z$ around solar, which become supersolar for \rfe $\gtrsim$ 0.5. 
An important result is that these sources are at low-$z$, and reach $Z$ comparable to the samples of \citetalias{Sniegowska2021} and \citetalias{}{Garnica2022} that are high $z$, and associated with quasars of very high \mbh\ and extreme luminosity. Conversely, xA sources of the present sample are AGNs of moderate luminosity, Narrow Line Seyfert 1s (NLSy1s) with \mbh $\lesssim 10^8$ M$_\odot$. Since the high-$z$ objects are extreme radiators at high Eddington ratio \lledd $\sim 1 $, as are the xA of our sample, the enrichment phenomenon appears to be independent of cosmic epoch and from black hole mass, at least up to cosmic ages $\sim 3 $ Gyr after the Big Bang. 

The mechanisms that could explain the high $Z$ values and the presence of a gradient of metallicity will be briefly discussed in Section \ref{enrich}. 


\begin{table*}[h!]
\setlength{\tabcolsep}{3pt}
\caption{Metallicity and physical parameter estimates along the Quasar MS}
\label{tab:z}      
\centering                          
\begin{tabular}{l c c c c c c c c r}        
\hline\hline                 
Object & SED & Component & Minimum $\chi^2$ & $\log U$  &  $\Delta\log U$ & $\log Z$  &  $\Delta\log Z$ & $\log n_{\rm H}$  &  $\Delta\log n_{\rm H}$\\ 
& & & & & & [$Z_\odot$] & [$Z_\odot$] & [cm$^{-3}$] & [cm$^{-3}$]\\
(1) & (2) & (3) & (4) & (5) & (6) & (7) & (8) & (9) & (10)\\
\hline                       
Mrk 335 & A & BC & $12.68$ & $-1.75$ & $-1.75$ - $-1.50$ & $-1.00$ & $-1.00$ - $-1.00$ & $7.75$ & $7.00$ - $8.75$\\ 
Mrk 110 & A & BC & $0.67$ & $-0.50$ & $-0.75$ - $-0.50$ & $0.30$ & $0.30$ - $0.30$ & $8.50$ & $8.50$ - $8.75$\\ 
Mrk 509 & A & BC & $8.81$ & $-0.75$ & $-1.00$ - $-0.50$ & $0.00$ & $0.00$ - $0.00$ & $10.00$ & $9.75$ - $10.25$\\ 
Ark 564 & A & BC & $32.26$ & $-1.25$ & $-1.75$ - $-0.25$ & $0.70$ & $0.30$ - $1.00$ & $7.00^{\rm a}$ & $7.00$ - $10.00$\\
Mrk 478 & xA & BC & $84.01$ & $-0.75$ & $-2.75$ - $0.00$ & $1.70$ & $1.70$ - $2.30$ & $9.75$ & $7.00$ - $10.25$\\
Ton 28 & xA & BC & $1.40$ & $-0.50$ & $-1.00$ - $-0.50$ & $1.70$ & $1.30$ - $1.70$ & $11.25$ & $11.00$ - $13.25$\\
I Zw 1 & xA & BC & $36.73$ & $-0.75$ & $-1.75$ - $-0.50$ & $1.30$ & $1.00$ - $1.30$ & $10.75$ & $10.25$ - $11.75$\\
LB 2522 & xA & BC & $1.15$ & $-0.75$ & $-1.00$ - $0.00$ & $1.30$ & $1.00$ - $2.30$ & $11.75$ & $11.25$ - $14.00$\\
LEDA 51016 & xA & BC & $24.23$ & $-1.00$ & $-2.50$ - $-0.75$ & $1.30$ & $1.00$ - $1.70$ & $10.50$ & $10.00$ - $11.50$\\
PHL 1092 & xA & BC & $1.42$ & $-3.25$ & $-3.25$ - $-3.00$ & $3.00$ & $3.00$ - $3.00$ & $12.00$ & $11.75$ - $12.00$\\
& xA & BLUE & $2.20$ & $-0.25$ & $-0.75$ - $-0.25$ & $1.70$ & $1.30$ - $1.70$ & $11.50$ & $11.25$ - $12.50$\\
Fairall 9 & B & BC & $34.17$ & $-1.75$ & $-2.00$ - $-1.00$ & $0.30$ & $-1.00$ - $0.70$ & $10.50$ & $7.00$ - $11.00$\\ 
& B & VBC & $1.76$ & $-2.00$ & $-2.00$ - $-1.75$ & $0.00$ & $-0.30$ - $0.00$ & $10.50$ & $10.00$ - $10.50$\\
Ark 120 & B & BC & $32.26$ & $-1.25$ & $-1.75$ - $-1.00$ & $1.00$ & $-0.70$ - $1.30$ & $11.00$ & $7.00$ - $11.50$\\
 & B & VBC & $5.38$ & $-2.00$ & $-2.25$ - $-1.50$ & $-1.30$ & $-1.70$ - $-1.00$ & $9.00$ & $7.00$ - $9.75$\\
NGC 3783 & B & BC & $30.96$ & $-1.75$ & $-1.75$ - $-1.50$ & $-0.30$ & $-1.30$ - $0.00$ & $9.75$ & $7.00$ - $10.00$\\
& B & VBC & $9.67$ & $0.00$ & $-2.00$ - $0.25$ & $0.30$ & $-1.70$ - $0.30$ & $7.25$ & $7.00$ - $11.75$\\
TXS 0042+101 & B & BC & $5.12$ & $-2.00$ & $-2.00$ - $-1.00$ & $-0.30$ & $-1.30$ - $0.70$ & $9.75$ & $7.00$ - $10.00$\\
& B & VBC & $0.60$ & $-2.00$ & $-2.00$ - $-2.00$ & $-1.30$ & $-1.30$ - $-1.00$ & $9.00$ & $8.50$ - $9.50$\\
\hline
\end{tabular}
\tablefoot{(1) Object common name. (2) SED used for the computation, with A using the SED from \cite{Mathews1987}, xA using the xA SED computed from PHL 1092, and B using the Population B SED. (3) Component measured. (4) Minimum $\chi^2$ associated with the object and the component. (5) Decimal logarithm of the ionization parameter $U$ associated to the minimum $\chi^2$. (6) Logarithmic interval associated to the variation of the ionization parameter $U$. (7) Decimal logarithm of the metallicity $Z$ associated to the minimum $\chi^2$. (8) Logarithmic interval associated to the variation of the metallicity $Z$. (9) Decimal logarithm of the Hydrogen density $n_{\rm H}$ associated to the minimum $\chi^2$. (10) Logarithmic interval associated to the variation of the Hydrogen density $n_{\rm H}$. $^{\rm a}$: A significantly higher \nh\ ($\log n_\mathrm{H}\sim 11$ [cm$^{-3}$]) was obtained removing the diagnostic ratios connected to \ciii\ (see text on Ark 564 in Section \ref{ressources}).}
\end{table*}

\section{Validation}
\label{validat}

In the following sections, we will discuss the limits of our technique and the tests that we used to address these limits.

\subsection{Statistical errors}
\label{stat}

\begin{figure}[ht!]
    \centering
    \includegraphics[width=9.1cm]{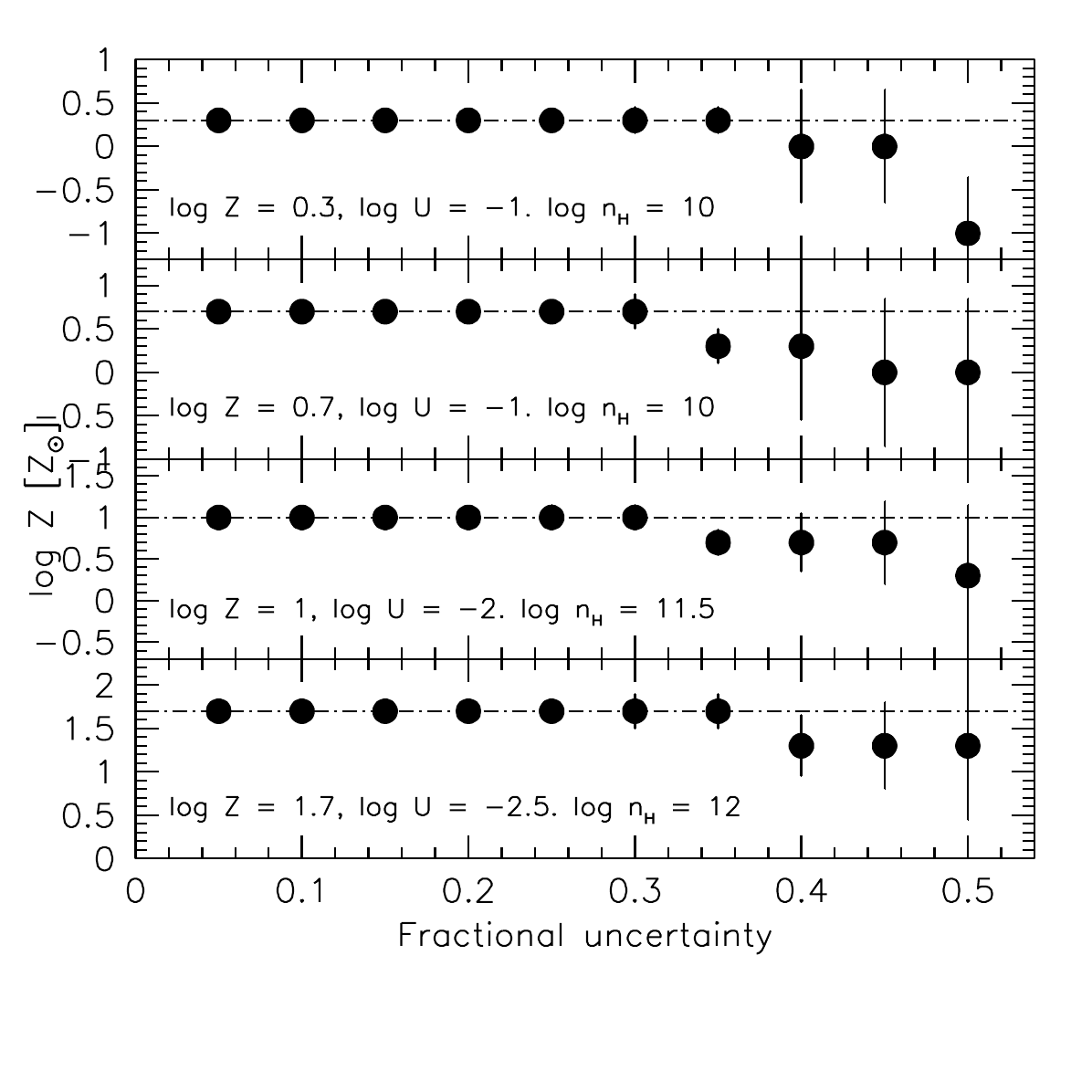}
    \vspace{-1cm}
    \caption{Median $Z$\ values for 200 replications of the intensity ratios deviating from the ones predicted by four models as a function of fractional uncertainties.}
    \label{fig:staterr}
\end{figure}

The minimum $\chi^2$ criterion and the $F$-test are not suited to estimate the influence of statistical errors. In this respect, bootstrap repetition of the $Z$, \nh\ and $U$\ estimate allows for deviation from a perfect solution (i.e., from the diagnostic ratio values predicted by a definite model $Z$, \nh\ and $U$) permit to assess how large the relative uncertainties in the intensity ratios can be until the correct solution in terms of metallicity is lost. Bootstrap simulations were carried out assuming random variations following a Gaussian distribution with $\sigma$\ corresponding to several fractional uncertainties. In practice, each replication assigned new values for the ratios, following Gaussian deviates with dispersion corresponding to the fractional uncertainty.  Fig. \ref{fig:staterr} shows that the $Z$ information is preserved for statistical errors up to $\approx 0.45$ for the somewhat idealized case of 10 diagnostic ratios having a constant fractional uncertainty, assuming diagnostic ratios from simulated cases with metallicity ranging from $\log Z = 0.3$ to 1.7, that could be fairly typical for Pop. A and xA. The statistical uncertainties do not affect the $Z$\ estimates up to 0.3 -- 0.35 fractional errors. Increasing the statistical uncertainties by 50\%\ for ratios  \civ/\heiiuv\ and \siiv/\heiiuv, because of the faintness of \heiiuv, lowers the highest uncertainty preserving the $Z$\ value by 0.05 (i.e., from 35\%\ to 30\%).      
The statistical errors tend to lower $Z$ (as well as $U$ and \nh, not shown). Therefore, statistical fluctuations are unlikely to induce systematically higher values of the metallicity to the ones derived from the model and to account for the very high $Z$\ deduced for the extreme Population  A sources. More relevant to the analysis of the result is the possibility of systematic discrepancies in emission line ratios to the expectation of a photoionization solution.

\subsection{Alternative $\chi^2$ evaluations}

Could the technique employed for the $\chi^2$ estimate be at the origin of the out-of-scale $Z$\ of PHL 1092, and of the very high values ($Z \lesssim 100$) encountered among xA sources? The $\chi^2$ expression of Eq. \ref{chi2} is weighting the contribution of  each measurement on the inverse of the uncertainty squared. This is expected to be effective in increasing the likelihood of the solution, but at the same time is giving minimum weight to ``bad" data that have large uncertainty. Several ratios  in Table \ref{tab:ratios} are between two detected lines, but have an uncertainty larger than the reported value. We therefore considered  two alternatives to the $\chi^2$ technique whose results are reported in Table \ref{tab:z}, and those are shown in Table \ref{tab:differr}. 

Instead of dividing each addendum by the error squared, the addenda could be divided by the value expected from the simulation for each ($U$, \nh, $Z$) grid element (different for each $U$, \nh, $Z$), or by the observed ratios that are constant. Especially this last option offers a valid alternative to the conventional $\chi^2$\ because it effectively quantifies the relative difference between any model and the observed set of ratios. Each ratio will have a weight independent from its associated uncertainty. If this exercise is carried out, the minimum $\chi^2$ \ solutions confirm the high $Z$ values for both LB 2522 and PHL 1092. The $Z$ derived for PHL 1092 BLUE is even increased to 20 -- 50 $Z_\odot$. Therefore, the metallicity estimates at high $Z$ do  not appear strongly sensitive to the different choices of $\chi^2$\ that we have considered.

\begin{table*}[h!]
\renewcommand{\arraystretch}{1.05}
\setlength{\tabcolsep}{3pt}
\caption{Different normalization for $\chi^2$ computation}
\label{tab:differr}      
\centering                          
\begin{tabular}{l c c c c c c c c r}        
\hline\hline                 
Object & SED & Component & Minimum $\chi^2$ & $\log U$  &  $\Delta\log U$ & $\log Z$  &  $\Delta\log Z$ & $\log n_{\rm H}$  &  $\Delta\log n_{\rm H}$\\ 
& & & & & & [$Z_\odot$] & [$Z_\odot$] & [cm$^{-3}$] & [cm$^{-3}$]\\
(1) & (2) & (3) & (4) & (5) & (6) & (7) & (8) & (9) & (10)\\
\hline
\multicolumn{10}{c}{Normalization over expected value} \\
\hline
LB 2522 & xA & BC & $19.62$ & $-3.50$ & $-3.50$ - $0.75$ & $3.00$ & $1.70$ - $3.00$ & $14.00$ & $10.75$ - $14.00$\\
PHL 1092 & xA & BC & $45.33$ & $-0.75$ & $-2.75$ - $-0.50$ & $3.00$ & $1.70$ - $3.00$ & $14.00$ & $12.50$ - $14.00$\\
& xA & BLUE & $1.93$ & $-0.25$ & $-1.00$ - $0.50$ & $1.70$ & $1.70$ - $2.70$ & $12.50$ & $11.25$ - $13.75$\\
\hline
\multicolumn{10}{c}{Normalization over observed value} \\
\hline
LB 2522 & xA & BC & $4.97$ & $0.00$ & $-2.50$ - $1.00$ & $1.70$ & $-2.00$ - $2.00$ & $7.00$ & $7.00$ - $11.25$\\
PHL 1092 & xA & BC & $2.78$ & $-3.25$ & $-3.25$ - $-3.00$ & $2.70$ & $1.70$ - $3.00$ & $11.50$ & $11.00$ - $12.00$\\
& xA & BLUE & $5.79$ & $-2.50$ & $-2.75$ - $1.00$ & $-2.00$ & $-1.00$ - $0.30$ & $10.25$ & $7.00$ - $14.00$\\

\hline                              
\end{tabular}
\tablefoot{(1) Object common name. (2) Adopted SED, with nomenclature used in Table \ref{tab:z}. (3) Component measured. (4) Minimum $\chi^2$ associated with the object and the component. (5) Decimal logarithm of the ionization parameter $U$ associated to the minimum $\chi^2$. (6) Logarithmic interval associated with the variation of the ionization parameter $U$. (7) Decimal logarithm of the metallicity $Z$ associated to the minimum $\chi^2$. (8) Logarithmic interval associated with the variation of the metallicity $Z$. (9) Decimal logarithm of the Hydrogen density $n_{\rm H}$ associated to the minimum $\chi^2$. (10) Logarithmic interval associated with the variation of the Hydrogen density $n_{\rm H}$.}
\end{table*}

\subsection{Influence of optical/UV mismatch}
\label{uvopt}

As discussed in Section \ref{ressources}, several sources are affected by an optical/UV flux mismatch, likely originated by observations conducted at different moments using both ground-based and space-based observatories. Optical observations are susceptible to uncontrollable light loss, significantly impacting the reliability of diagnostic ratios based on both optical and UV lines. However, diagnostic ratios dependent on lines found at similar wavelengths are less susceptible to this issue, as the intrinsic variability of the quasar should similarly affect them. The most likely candidate for this problem is the \civ/\hb\ ratio, as it is the sole diagnostic ratio involving a UV and an optical line. 

Indeed, when estimating the physical parameters associated with the BLR of our sample of quasars, the \civ/\hb\ ratio often disrupted the computation, returning combinations of $U$, $Z$ and $n_{\rm H}$ that were not compatible with our previous assumptions, or solutions strongly incompatible with any of the {\tt CLOUDY} simulations used in this work: very low \civ/\hb\ values are possible for low $U$, for most $Z$s (Fig. \ref{fig:c4hbzu}). The values of \civ/\hb\ along with the $\log U$ and $\log Z$ results obtained from the calculation are shown in Table \ref{table:civhbratio}. It is important to note that the \civ/\hb\ ratio was excluded in most calculations (Table \ref{tab:ratios}). 
The primary observation from Table \ref{table:civhbratio} is that the \civ/\hb\ value follows a clear pattern, steadily decreasing   towards the most extreme xA sources. This pattern has been observed in other works, and is a peculiar feature of xA quasars \citep{Sulenticetal17, Vietri2018}, exhibiting the bulk of their \civ\ emission in the BLUE. However, despite being a physical result, inappropriate estimation of this ratio due to the aforementioned problems strongly affects the measurements of $Z$ and $U$, rendering the estimation incorrect. The \civ/\hb\ ratio, being particularly sensitive to   $U$, loses its dependence on metallicity   (represented in Figure \ref{fig:c4hbzu}), returning a solution that is not physical, distorting the results associated with other diagnostic ratios.  

\begin{table*}[h!]
\caption{CIV/H$\beta$ results}
\label{table:civhbratio}      
\centering                          
\begin{tabular}{l c c c c r}        
\hline\hline                 
Object name & CIV/H$\beta$ & $\log U_{\rm CIV/H\beta}$ & $\log Z_{\rm CIV/H\beta}$ & $\log U_{\rm no\ CIV/H\beta}$ & $\log Z_{\rm no\ CIV/H\beta}$\\
 & & & [$Z_\odot$] & & [$Z_\odot$]\\
(1)  & (2) & (3) & (4) & (5) & (6)\\
\hline                     

Mrk 335 & $5.89\pm0.76$ & $-1.75$ & $-1.00$ & $-1.75$ & $-1.00$\\ 
Mrk 110 & $37.71\pm4.98$ & $-1.25$ & $0.00$ & $-0.50$ & $0.30$\\
Mrk 509 & $5.93\pm0.90$ & $-1.75$ & $-1.00$ & $-0.75$ & $0.00$\\
Ark 564 & $11.19\pm2.45$ & $-1.25$ & $0.70$ & $-1.25$ & $0.70$\\ 
Mrk 478 & $5.21\pm0.75$ & $-0.75$ & $1.70$ & $-0.75$ & $1.70$\\
Ton 28 & $1.78\pm0.75$ & $-3.00$ & $2.30$ & $-0.50$ & $1.70$\\ 
I Zw 1 & $2.28\pm0.44$ & $-2.75$ & $0.30$ & $-0.75$ & $1.30$\\ 
LB 2522 & $2.48\pm2.87$ & $-3.50$ & $3.00$ & $-0.75$ & $1.30$\\ 
LEDA 51016 & $7.60\pm1.56$ & $-0.75$ & $1.30$ & $-1.00$ & $1.30$\\ 
PHL 1092 & $0.41\pm0.24$ & $-3.50$ & $2.70$ & $-3.25$ & $3.00$\\ 
Fairall 9 & $12.60\pm1.71$ & $-1.75$ & $0.30$ & $-1.75$ & $0.30$\\ 
Ark 120 & $3.09\pm0.30$ & $-2.50$ & $-0.30$ & $-1.25$ & $1.00$\\ 
NGC 3783 & $7.64\pm0.68$ & $-1.75$ & $-0.30$ & $-1.50$ & $-0.70$\\ 
TXS 0042+101 & $6.01\pm2.24$ & $-2.00$ & $-0.30$ & $-1.00$ & $0.70$\\ 

\hline                                   
\end{tabular}
\tablefoot{(1) Object common name. (2) \civ/\hb\ ratio obtained from the multi-component fitting. (3) $\log U$ associated to the minimum $\chi^2$ between the observations and {\tt CLOUDY} simulations including the \civ/\hb\ ratio. (4) $\log Z$ associated to the minimum $\chi^2$ between the observations and {\tt CLOUDY} simulations including the \civ/\hb\ ratio. (5) $\log U$ associated to the minimum $\chi^2$ between the observations and {\tt CLOUDY} simulations excluding the \civ/\hb\ ratio. (6) $\log Z$ associated to the minimum $\chi^2$ between the observations and {\tt CLOUDY} simulations excluding the \civ/\hb\ ratio.}
\end{table*}




\subsection{Influence of SED}

\begin{figure*}[ht!]
    \centering
    \includegraphics[width=18cm]{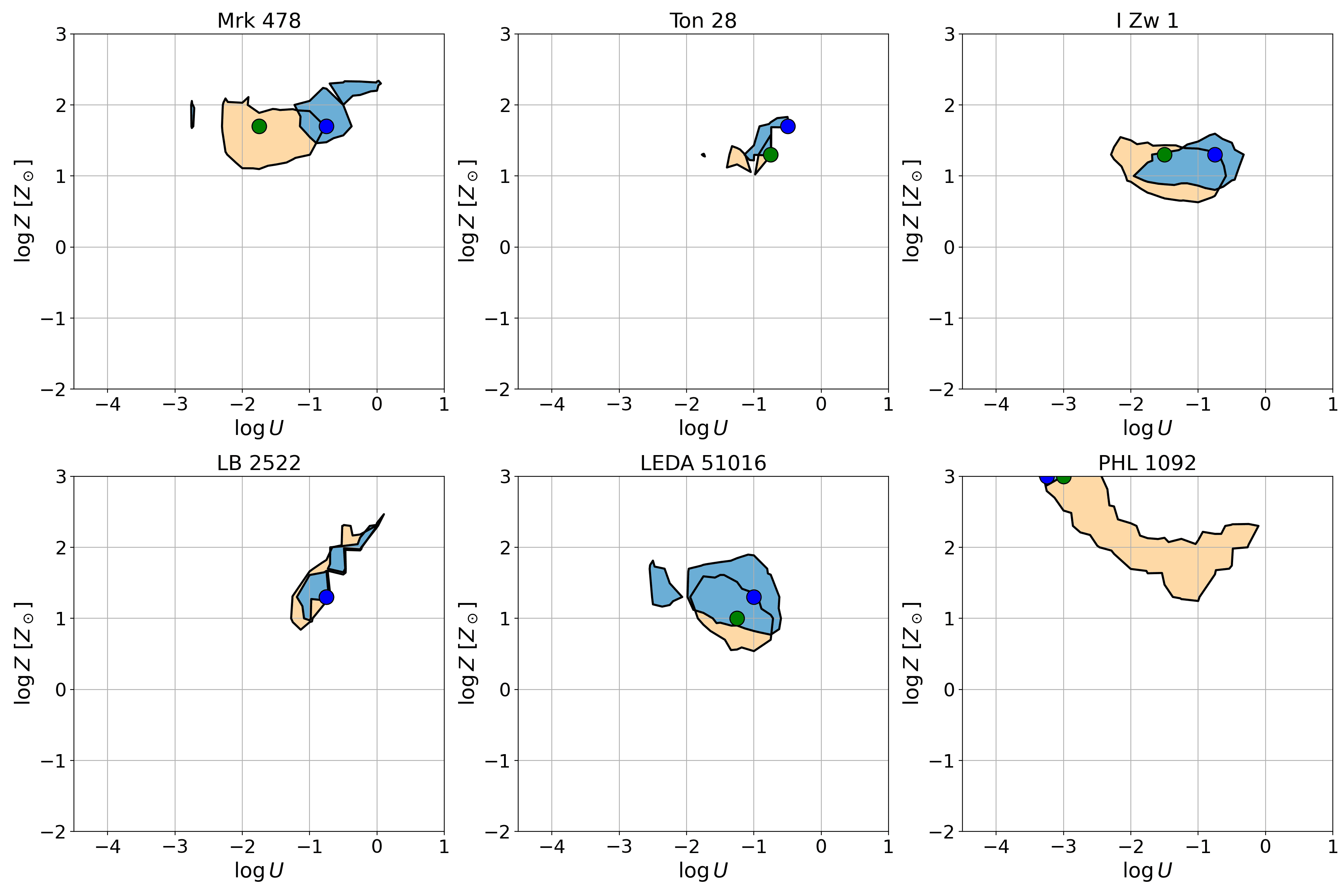}
    \caption{Two-dimensional parameter space of $\log Z$ - $\log U$ illustrating the physical conditions of the gas surrounding xA sources with different SEDs. In each plot the minimum $\chi^2$ computed between measured diagnostic ratios and {\tt CLOUDY} simulations is shown with dots: green for the \cite{Mathews1987} SED and blue for the xA SED. The orange region corresponds to the 1$\sigma$ accuracy range for the \cite{Mathews1987} SED, and the blue region for the xA SED.}
    \label{fig:doublesed}
\end{figure*}

The choice of the SED used for the photoionization simulations is expected to significantly influence the results obtained during the estimation of physical parameters in the BLR, as it dictates the luminosity of the central source.
Population A and B sources typically display distinct spectral features, leading to remarkably different SEDs \citep[see Figure 1 in][]{Panda2019}. However, the distinction between Population A and xA sources is less clear-cut, particularly between the A2 and A3 spectral bins,  as observed in some objects in our sample that exhibit borderline features. Furthermore, the Population A SED has been used in other works \citep{Marziani24}, since the xA SED employed in this work has often been considered too extreme for many Population A objects.
To address this issue, we recalculated the $\chi^2$ using the Population A SED for xA sources, keeping track of the systematic changes encountered (as summarized in Table \ref{tab:popAsed}). Additionally, in Figure \ref{fig:doublesed}, we illustrate the positions occupied by the results for both SEDs in the $\log U$ - $\log Z$ parameter space.

\begin{table*}[h!]
\setlength{\tabcolsep}{3pt}
\caption{Metallicity and physical parameter estimates using Pop.A SED}
\label{tab:popAsed}      
\centering                          
\begin{tabular}{l c c c c c c c c r}        
\hline\hline                 
Object & SED & Component & Minimum $\chi^2$ & $\log U$  &  $\Delta\log U$ & $\log Z$  &  $\Delta\log Z$ & $\log n_{\rm H}$  &  $\Delta\log n_{\rm H}$\\ 
& & & & & & [$Z_\odot$] & [$Z_\odot$] & [cm$^{-3}$] & [cm$^{-3}$]\\
(1) & (2) & (3) & (4) & (5) & (6) & (7) & (8) & (9) & (10)\\
\hline                       
Mrk 335 & A & BC & $12.68$ & $-1.75$ & $-1.75$ - $-1.50$ & $-1.00$ & $-1.00$ - $-1.00$ & $7.75$ & $7.00$ - $8.75$\\ 
Mrk 110 & A & BC & $0.67$ & $-0.50$ & $-0.75$ - $-0.50$ & $0.30$ & $0.30$ - $0.30$ & $8.50$ & $8.50$ - $8.75$\\ 
Mrk 509 & A & BC & $8.81$ & $-0.75$ & $-1.00$ - $-0.50$ & $0.00$ & $0.00$ - $0.00$ & $10.00$ & $9.75$ - $10.25$\\ 
Ark 564 & A & BC & $32.26$ & $-1.25$ & $-1.75$ - $-0.25$ & $0.70$ & $0.30$ - $1.00$ & $7.00$ & $7.00$ - $10.00$\\
Mrk 478 & A & BC & $68.22$ & $-1.75$ & $-2.25$ - $-1.00$ & $1.70$ & $1.30$ - $2.00$ & $10.25$ & $9.75$ - $10.75$\\
Ton 28 & A & BC & $1.05$ & $-0.75$ & $-1.75$ - $-0.75$ & $1.30$ & $1.30$ - $1.30$ & $11.25$ & $11.25$ - $13.50$\\
LB 2522 & A & BC & $1.11$ & $-0.75$ & $-1.25$ - $0.00$ & $1.30$ & $1.00$ - $2.30$ & $12.50$ & $11.25$ - $14.00$\\
LEDA 51016 & A & BC & $21.47$ & $-1.25$ & $-1.25$ - $-0.75$ & $1.00$ & $0.70$ - $1.30$ & $11.25$ & $10.25$ - $11.50$\\
I Zw 1 & A & BC & $30.84$ & $-1.50$ & $-2.25$ - $-0.75$ & $1.30$ & $0.70$ - $1.30$ & $11.50$ & $10.50$ - $12.00$\\
PHL 1092 & A & BC & $1.97$ & $-3.00$ & $-3.25$ - $0.00$ & $3.00$ & $1.30$ - $3.00$ & $11.75$ & $10.75$ - $12.50$\\
& A & BLUE & $2.28$ & $-0.75$ & $-1.25$ - $-0.50$ & $1.00$ & $1.00$ - $1.30$ & $11.75$ & $11.25$ - $12.25$\\

\hline                              
\end{tabular}
\tablefoot{(1) Object common name. (2) Adopted SED, with nomenclature used in Table \ref{tab:z}. (3) Component measured. (4) Minimum $\chi^2$ associated with the object and the component. (5) Decimal logarithm of the ionization parameter $U$ associated to the minimum $\chi^2$. (6) Logarithmic interval associated with the variation of the ionization parameter $U$. (7) Decimal logarithm of the metallicity $Z$ associated to the minimum $\chi^2$. (8) Logarithmic interval associated with the variation of the metallicity $Z$. (9) Decimal logarithm of the Hydrogen density $n_{\rm H}$ associated to the minimum $\chi^2$. (10) Logarithmic interval associated with the variation of the Hydrogen density $n_{\rm H}$.}
\end{table*}

The primary outcome of this test was a systematic decrease of the minimum $\chi^2$ obtained for all xA objects with the exception of PHL 1092 (which served as the basis for the xA SED). Furthermore, the solutions consistently indicate higher densities, typically moving upward by an order of magnitude or less. However, no significant changes to metallicity or $U$ have been observed, as depicted in Figure \ref{fig:doublesed}. Nonetheless, the parameter distributions obtained using the xA SED are generally more constrained.

\subsection{Micro-turbulence and highest $Z$ values}

\begin{figure}[ht!]
    \centering
    \includegraphics[width=4.25cm]{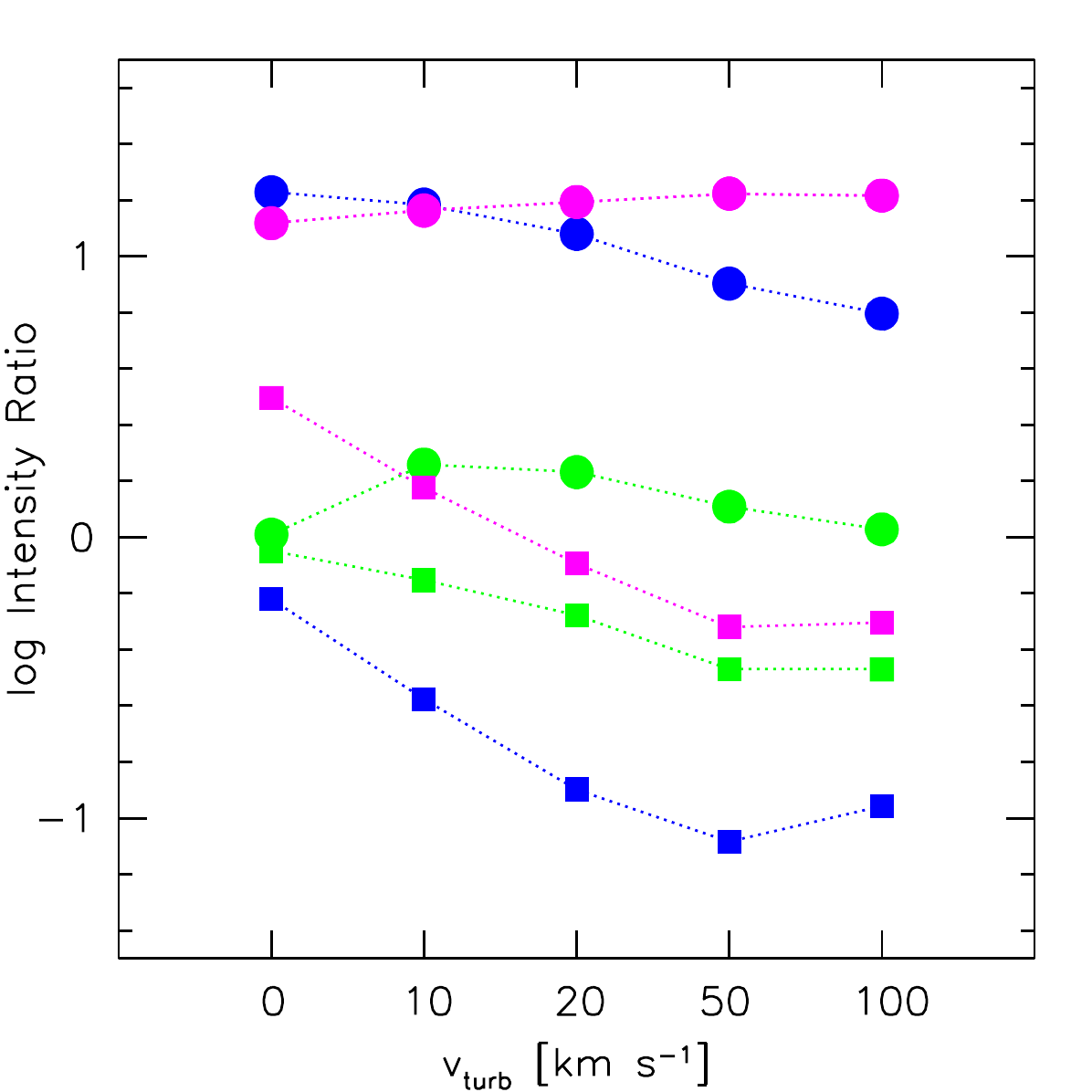}
    \includegraphics[width=4.25cm]{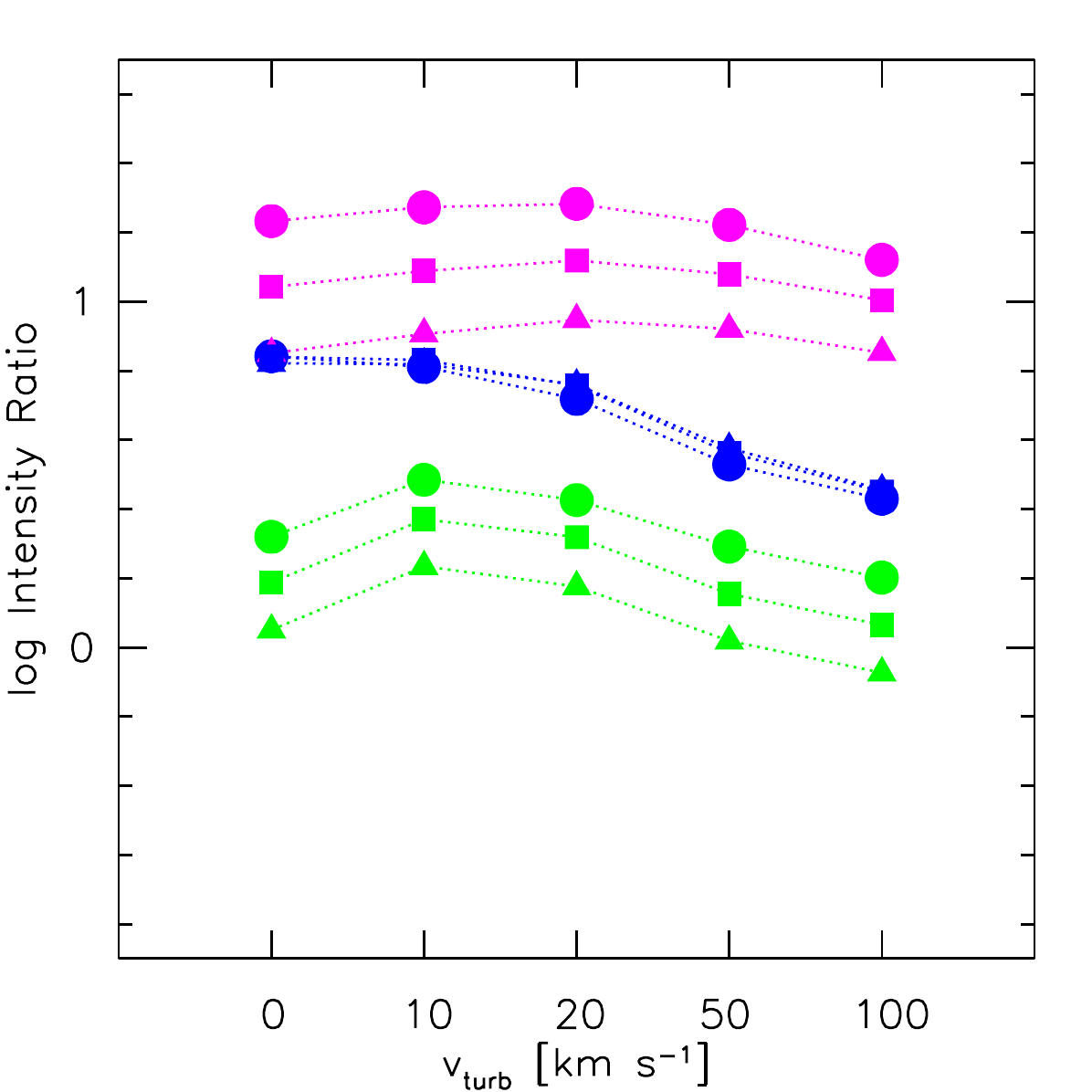}
    \caption{Behavior of the diagnostic ratio \civ/\heiiuv\ (magenta), \civ/\hb\ (blue), \rfe (green)  in logarithmic scale vs microturbulence velocity in \kms. Left: using the solution for I Zw 1 (squares) and PHL 1092 (circles); right: using a low-ionization, high-density solution appropriate for xA sources (see text for details), with three different $Z$\ values: 200 $Z_\odot$ (circles), 100 $Z_\odot$ (squares), 50 $Z_\odot$ (triangles).}
    \label{fig:turbo}
\end{figure}

Micro-turbulence is believed to increase the efficiency of metal line production. This is especially true for \feii\ blends \citep{verneretal99,verneretal04,bruhweilerverner08,Panda2018}, as micro-turbulence is expected to increase  \feii\ self-fluorescence and fluorescent excitation due to continuum and strong emission lines.  CLOUDY simulations were carried out starting from the ($U$, \nh, $Z$) values of I Zw 1 and PHL 1092. The left panel of Fig. \ref{fig:turbo}, where the  key intensity ratios \civ/\heiiuv, \civ/\hb, and \rfe\ are shown as a function of the microturbulent velocity. A microturbulent velocity $v_\mathrm{turn} = \sqrt{kT/m} \approx 10$ \kms\ corresponds to the thermal broadening of lines emitted by gas at the electron temperature of the BLR, $T \sim 10000$ K. The effect of microturbulence passing from 0 to 10 \kms\ is a net increase in \rfe\ for PHL 1092, while the effect is less clear for I Zw 1. The right panel of Fig. \ref{fig:turbo} shows the trends for the same ratios, starting from low ionization, and high-density solutions, for three very high $Z$ values, namely 50, 100, and 200 $Z_\odot$. The effect on \rfe\ is remarkable in all of the three cases. The increase in the emission efficiency of \feii\ features implies that less iron is needed to produce the same line luminosity. 

We recomputed the array of simulations for the \citet{Mathews1987} SED in the range $7\le \log$\nh\ $\le  14$, $-3.5 \le \log U \le 0$, and for 10 values of $Z$ between 1 and 1000 $Z_\odot$. The results for PHL 1092 indicate a greatly reduced value of the metallicity, $\log Z \approx 1.7^{+0.6}_{-0.4} $[Z$_\odot$]. The inclusion of turbulence aligns the estimate of the out-of-scale measurement to more likely values $Z \lesssim$ 100 $Z_\odot$. Most sources with $Z \ge 10 Z_\odot$ have their $Z$\ estimates unchanged or reduced by a factor $\approx 2$ as was first suggested in \citet{Panda2021}. It is interesting to note that this reduction seems to occur only for the highest values of $Z$, while for lower metallicity around solar, the effect of turbulence is not obvious.



\subsection{Stratification}

Evidence indicates trends for various parameters along the Quasar MS. Specifically, in the case of Population B, there is observable evidence of a radial stratification of properties within the BLR \citep{Baldwin1995, Korista1997}. As previously mentioned, this region is modeled by distinguishing between a BC and a VBC. However, it is not certain whether the assumption of constant physical parameters within the BLR is entirely accurate for Population A. On the one hand, the physical parameters appear to be associated with a much smaller uncertainty range in Pop. A than in Pop. B  (compare Figures \ref{fig:proj1}, \ref{fig:proj2}, \ref{fig:proj3} vs \ref{fig:proj4}, \ref{fig:proj5}). On the other hand, several Pop. A sources show prominent \ciii\ that may bias the photoionization solution toward lower \nh. To test whether the estimates of $Z$\ are stable to this potential bias, we repeated the $Z$\ estimates removing the diagnostic ratios involving \ciii\ for Mrk 335 and Ark 564, the objects which exhibit the lowest \nh. The computation yielded different results for the two objects: Mrk 335 retained a very low density ($\log$ \nh$\approx 7.5$ [cm$^{-3}$]), confirming the presence of low-density gas in its BLR, also supported by the fitting of a semi-broad component of \oiii, as described in Section \ref{ressources}. Ark 564, instead, returned a higher density solution, with $\log$\nh$\approx 11$ [cm$^{-3}$], suggesting that the absorption feature affecting \ciii\ led to an overestimation of its line intensity.

\begin{figure}[ht!]
    \centering
    \includegraphics[width=9.15cm]{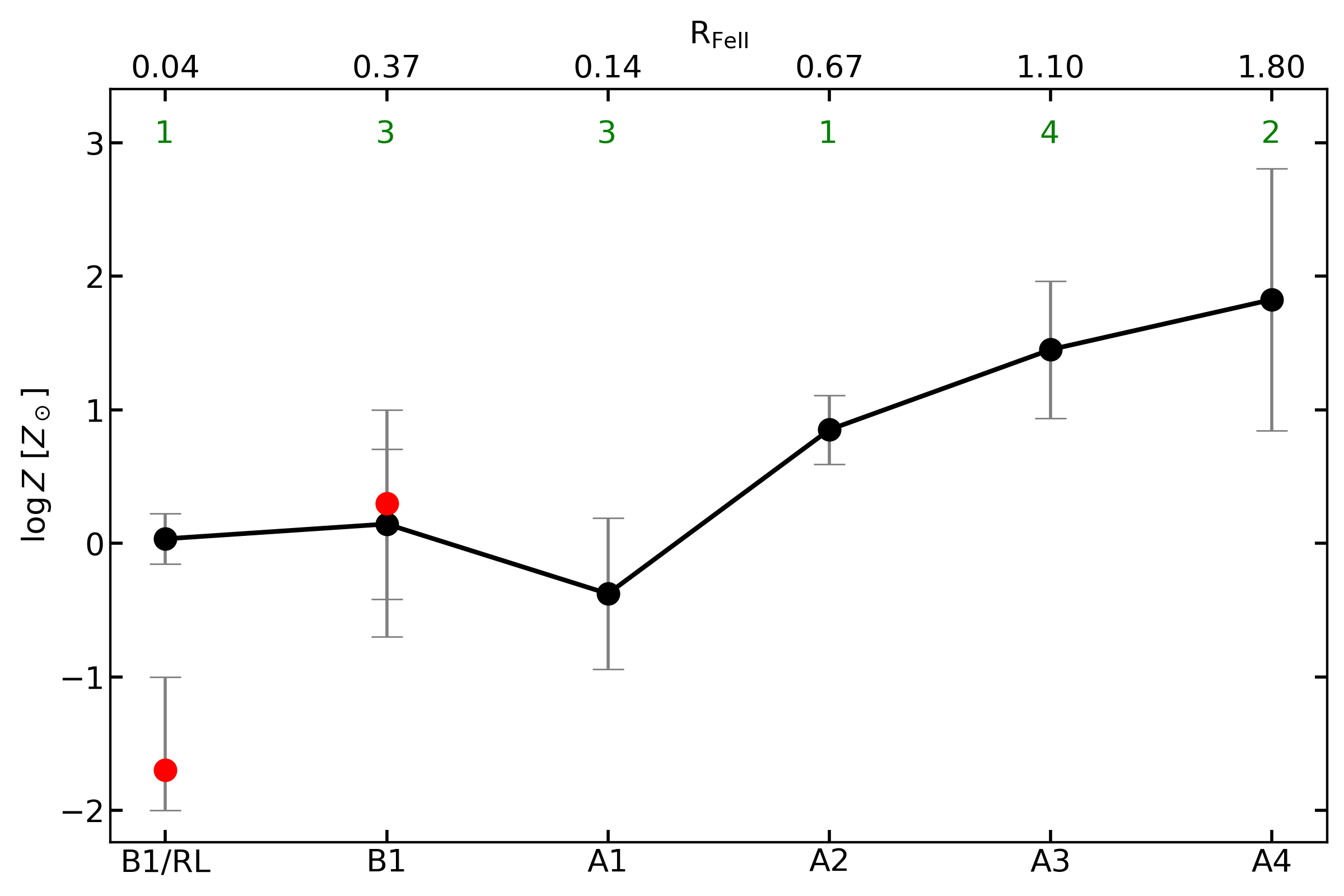}
    \caption{Diagram illustrating the trend of $Z$ across different spectral types, with the reference values for each spectral type obtained as the geometric mean of all the measurements assumed for all the objects belonging to a specific spectral type, including our method, the use of the \civ/\hb\ ratio, and the presence of microturbulence for all sources belonging to the A2, A3, and A4 spectral types. Confidence intervals are obtained as the standard deviation of our results. The number of objects contributing to each spectral type is shown in green, along with the average R$_{\rm FeII}$ of the sources considered. Two reference points (in red) are shown describing the RL and RQ Population B sources studied in M23.}
    \label{fig:zdisposition}
\end{figure}

Figure \ref{fig:zdisposition} displays the trend of $Z$ along the MS by analyzing the average value of $Z$ across specific spectral types, averaging all the values obtained with the methods that most affect the determination. The trend in metallicity is evident from this figure, with a dip in metallicity for the A1 spectral type being influenced by the selection of objects in our sample. In the A1 spectral type, the average R$_{\rm FeII}$ is lower than that measured for the adjacent bins: A1 sources are associated with lower R$_{\rm FeII}$ than typically measured in that spectral type, thus contributing to lower estimations of $Z$. Additionally, the absence of {\tt CLOUDY} simulations with $\log Z = -0.3, -0.7$ for Population A objects contributed to an underestimation of $Z$. Nevertheless, the trend in metallicity proceeds quite steadily from Population B RL quasars to xA quasars. 

\section{Discussion}
\label{disc}

\subsection{A recipe to compute BLR metal content}

Summarizing the procedure followed in this paper, we can define a recipe with several steps that could be easily applicable to larger samples of quasars:

\begin{enumerate}
    \item identification of the spectral type and radio loudness, or at least if an object belongs to Pop. B, Pop. A or extreme Pop. A. Note that RL xA AGN are extremely rare in Population A at low-$z$ \citep{gancietal19}; 
    \item  consideration of a maximum number of ratios for each component, avoiding ratios between two faint components;  
    \item     multi-component fitting, to separate  individual components associated with different associated physical conditions (Section \ref{empmodel});
    \item computation of the minimum $\chi^2$ between our measurements and the results obtained from {\tt CLOUDY} photoionization simulations to cover the parameter space in terms of SED, $Z$, \nh, $U$\ and microturbulence parameter. Ideally, SEDs should be different for at least Pop. B RQ and RL, Pop. A and xA. Microturbulence parameter could be 0 or 10 \kms; 
    \item definition of $1\sigma$\ confidence ranges via an F-test with the appropriate number of degrees of freedom. 
\end{enumerate}


In Section \ref{uvopt} we discussed that the presence of an optical-UV flux mismatch was a major obstacle encountered during this procedure. The effect of this mismatch, which plays a crucial role in nullifying the information of the \civ/\hb\ diagnostic ratio, mostly affects objects with non-contemporary observations. However, the \civ/\hb\ ratio is extremely effective for constraining $U$, so relinquishing this ratio has a negative impact on the accuracy of the measurement of the physical conditions in the BLR. 
We thus propose that the \civ/\hb\ ratio should be used with the highest confidence in the case of contemporary observations in the UV and optical range. In this case, the diagnostic ratio has a positive effect on the measurement and should help retrieve well-constrained results.

The \feii/H$\beta$\ ratio has often yielded discrepant information with respect to \civ/\hb. Indeed, this ratio is not typically predicted to be very high for the SEDs employed in this work. 
This raises the possibility that the optical ratios could be problematic for the estimation of the physical parameters, and should thus be excluded from the computation. 


\subsection{Estimation of accretion parameters}
\label{accrpar}

The bolometric luminosity ($L_{\rm bol}$) of the AGNs of the sample is estimated using the relation
\begin{equation}
\label{Lbol}
    L_{\rm bol} = 4\pi [\lambda \cdot f_\lambda](5100 \AA) d^2_\mathrm{P}  = L(5100 \AA) k_{\rm bol},
\end{equation}
where $d_\mathrm{P}$ is the luminosity distance divided by (1+$z$). In this paper, the values of the cosmological parameters are adopted as follows: $H_0$ =70 \kms\ Mpc$^{-1}$, $\Omega_\mathrm{M}$=0.3,  $\Omega_\Lambda$=0.7.  The bolometric correction factor $k_{\rm bol}$ is calculated using the formula from \cite{Netzer2019}:
\begin{equation}
\label{kbol}
    k_{\rm bol} = 40\times\left(\frac{L(5100 \AA)}{10^{42}\ {\rm erg}\ {\rm s}^{-1}}\right)^{-0.2},
\end{equation}
while $L(5100 \AA)$ is the luminosity of the quasar at 5100 \AA.
The mass of the black hole ($M_{\rm BH}$) is estimated using the relation:
\begin{equation}
\label{mbh}
\begin{split}
    \log(M_{\rm BH}/M_\odot) = 0.83\log({\rm FWHM}({\rm H}\beta)) + 0.45\log(l_{44}) + \\
    - 0.35\log({\rm R}_\mathrm{FeII}) + 5.22,
\end{split}
\end{equation}
with $l_{44} = L(5100 \AA)/10^{44} {\rm erg\ s}^{-1}$ obtained from \cite{duwang19}. The formula in Equation \ref{mbh} is the  \citet{Vestergaard2006} relation with two changes: (1) a correction on the FWHM following \citet{mejia-restrepoetal18a}, and   (2) the inclusion of the dependence of the BLR radius on \rfe\  \citep{duwang19}. In practice, Eq. \ref{mbh} rectifies the mass estimations for extreme accretors while keeping the results unchanged for other objects.

The Eddington ratio is estimated using the formula $L_{\rm bol}/L_{\rm Edd}$, with the Eddington luminosity $L_{\rm Edd}$ obtained as follows:
\begin{equation}
\label{Eddington}
    L_{\rm Edd} = 1.5\times10^{38} {\rm erg}\ {\rm s}^{-1} \frac{M_{\rm BH}}{M_\odot}.
\end{equation}
The derived properties of the sample are shown in Table \ref{table:objder}. 

\begin{table}[h!]
\caption{Derived properties of the sample}
\label{table:objder}      
\centering                          
\begin{tabular}{l c c r}
\hline\hline
Object Name  & $\log(L_\mathrm{bol})$ & $\log(M_\mathrm{BH})$ & $L/L_\mathrm{Edd}$ \\
 &   [erg s$^{-1}$] & [M$_\odot$] &    \\
(1) & (2) & (3) & (4)  \\
\hline
Mrk 335 		& $44.95\pm0.03$ & $7.78\pm0.05$ & $0.10\pm0.01$ \\
Mrk 110 		& $44.46\pm0.03$ & $7.48\pm0.05$ & $0.06\pm0.01$ \\
Mrk 509			& $45.34\pm0.03$ & $8.21\pm0.16$ & $0.09\pm0.03$ \\
Ark 564			& $44.25\pm0.03$  & $6.94\pm0.06$ & $0.14\pm0.02$\\
Mrk 478			& $45.32\pm0.03$ & $7.51\pm0.04$ & $0.43\pm0.05$ \\
Ton 28			& $46.26\pm0.03$ & $8.26\pm0.05$ & $0.67\pm0.10$ \\
I Zw 1 			& $45.62\pm0.03$ & $7.45\pm0.05$ & $0.98\pm0.12$ \\
LB 2522 		& $46.03\pm0.03$ & $8.22\pm0.06$ & $0.43\pm0.07$ \\
LEDA 51016		& $45.26\pm0.03$ & $7.75\pm0.05$ & $0.22\pm0.02$ \\
PHL 1092		& $45.72\pm0.03$ & $7.71\pm0.05$ & $0.67\pm0.12$ \\
Fairall 9 		& $45.08\pm0.03$ & $8.14\pm0.05$ & $0.06\pm0.01$ \\
Ark 120 		& $45.40\pm0.03$ & $8.29\pm0.04$ & $0.09\pm0.01$ \\
NGC 3783 		& $44.52\pm0.03$ & $7.71\pm0.05$ &  $0.04\pm0.01$\\
TXS 0042+101 	& $45.53\pm0.03$ & $8.60\pm0.23$ & $0.06\pm0.04$ \\
\hline
\end{tabular}
\tablefoot{(1) Object common name.  (2) Bolometric luminosity. (3) Mass of the SMBH at the center of the galaxy. (4) Eddington ratio associated with accretion onto the SMBH.}
\end{table}

\subsection{Correlation with accretion parameters}

We investigated the correlation of accretion parameters with metallicity and the R$_{\rm FeII}$ parameter that defines the MS. Utilizing the {\tt SLOPES} {\tt FORTRAN} code outlined in \cite{Feigelson1992}, we calculated the Pearson's correlation coefficient $r_{\rm P}$ between the quantities. Subsequently, we determined the probability $P$ of a spurious correlation, computed using the complementary error function erfc$(x) = \frac{2}{\sqrt{\pi}}\int_{x}^{\infty} e^{-t^2} \,dt$, where $x$ is defined as $x = \sqrt{\frac{N}{2}} |r_{\rm P}|$ with $N$ representing the number of objects \citep{Press1986}.

\begin{figure*}[ht!]
    \centering
    \includegraphics[width=18cm]{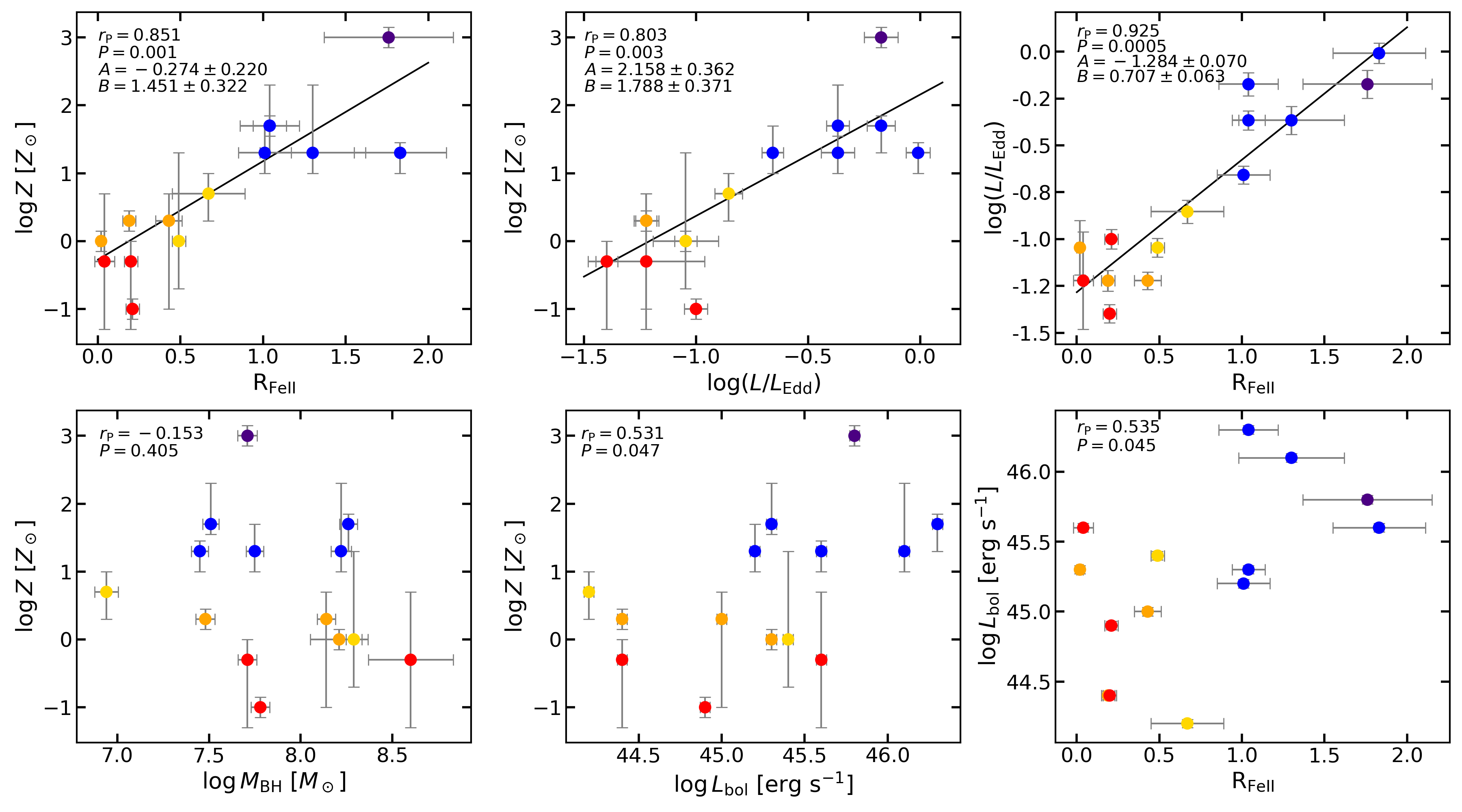}
    \caption{Figures showing correlation between various accretion parameters with $\log Z$ and R$_{\rm FeII}$. Points are displayed using the same color-coding as in Figure \ref{fig:4DE1Sample} with their errorbars. When the correlation is considered significant ($r_{\rm P} \gtrsim 0.7$, corresponding to a confidence level of $\approx 0.01$), the correlation line is shown, with its defining parameters intercept ($A$) and slope ($B$) shown with their respective errors. For each correlation are also shown the Pearson correlation coefficient $r_{\rm P}$\ from the {\tt SLOPES} program \citep{Feigelson1992} and the probability $P$\ of a spurious correlation.}
    \label{fig:corr}
\end{figure*}

The correlation analysis (Figure \ref{fig:corr})  yielded interesting results. The parameters correlating more strongly are R$_{\rm FeII}$ and $\log(L/L_{\rm Edd})$, with $r_{\rm P} = 0.925$. This result reinforces the hypothesis that the Eddington ratio is the physical parameter behind the R$_{\rm FeII}$, serving as the primary driver of the MS, as found by several works \cite[][see also the synopsis in \citealt{Donofrio2021}]{Marziani2001,sunshen15,duetal16a,Panda2019,Panda2023}. Additionally, very robust correlations were also observed in this sample between R$_{\rm FeII}$ and $\log Z$, and between $\log(L/L_{\rm Edd})$ and $\log Z$, confirming once again that extreme-accreting sources are characterized by high metallicities \citep{martinez-aldama18, Sniegowska2021, Garnica2022}. Moreover, the correlation in this work is extended to most spectral types along the MS, also including Population B sources, rather than being limited exclusively to xA sources as in \cite{Sniegowska2021} and \cite{Garnica2022}. The presence of such strong correlations is not trivial, as these quantities were derived independently and using different procedures.

Our results confirm the earlier finding concerning low-$z$ PG quasars that the BLR gas $Z$ correlates with Eddington ratio while $Z$ shows much weaker correlation
with \mbh\ \citep{Shin2013}. Typical values of the \siiv/\civ\ and \nv/\civ\ at low redshifts suggested values around solar or a few times solar. The range covered by the two diagnostic ratios were found consistent with those measured for a sample of intermediate redshift quasars of luminosity comparable to the one of luminous AGN at low-$z$\ \citep{sulenticetal14}. The trend with \lledd\ is due to the inclusion of Population B, and spectral type A1 that are radiating at lower Eddington ratio with respect to xA sources. The absence of a clear trend at intermediate-$z$ corresponding to the cosmic noon with a maximum  of nuclear activity and star formation \citep[e.g.,][]{madaudickinson14,florezetal20,forsterschreiberwuits20}, and  the large $Z$ estimated for  quasars at that cosmic epoch, might be just the result of a selection effect disfavoring the lowest \lledd\ for any given \mbh\ \citep{sulenticetal14}.

The three remaining correlations depicted in Figure \ref{fig:corr} were not found to be statistically significant. The absence of a significant correlation between $\log Z$ and $\log M_{\rm BH}$ was particularly puzzling, as these quantities have been found to correlate in other works \citep{Xu2018, Maiolino2019}, and larger SMBHs were found to be the ones with the highest metallicity, since their stronger gravitational potential is able to trap the more metallic matter in their virialized region.

\subsection{Comparison between the photoionization and reverberation solutions for $r_\mathrm{BLR}$}
\label{revphot}

\begin{figure}[ht!]
    \centering
    \includegraphics[width=9.15cm]{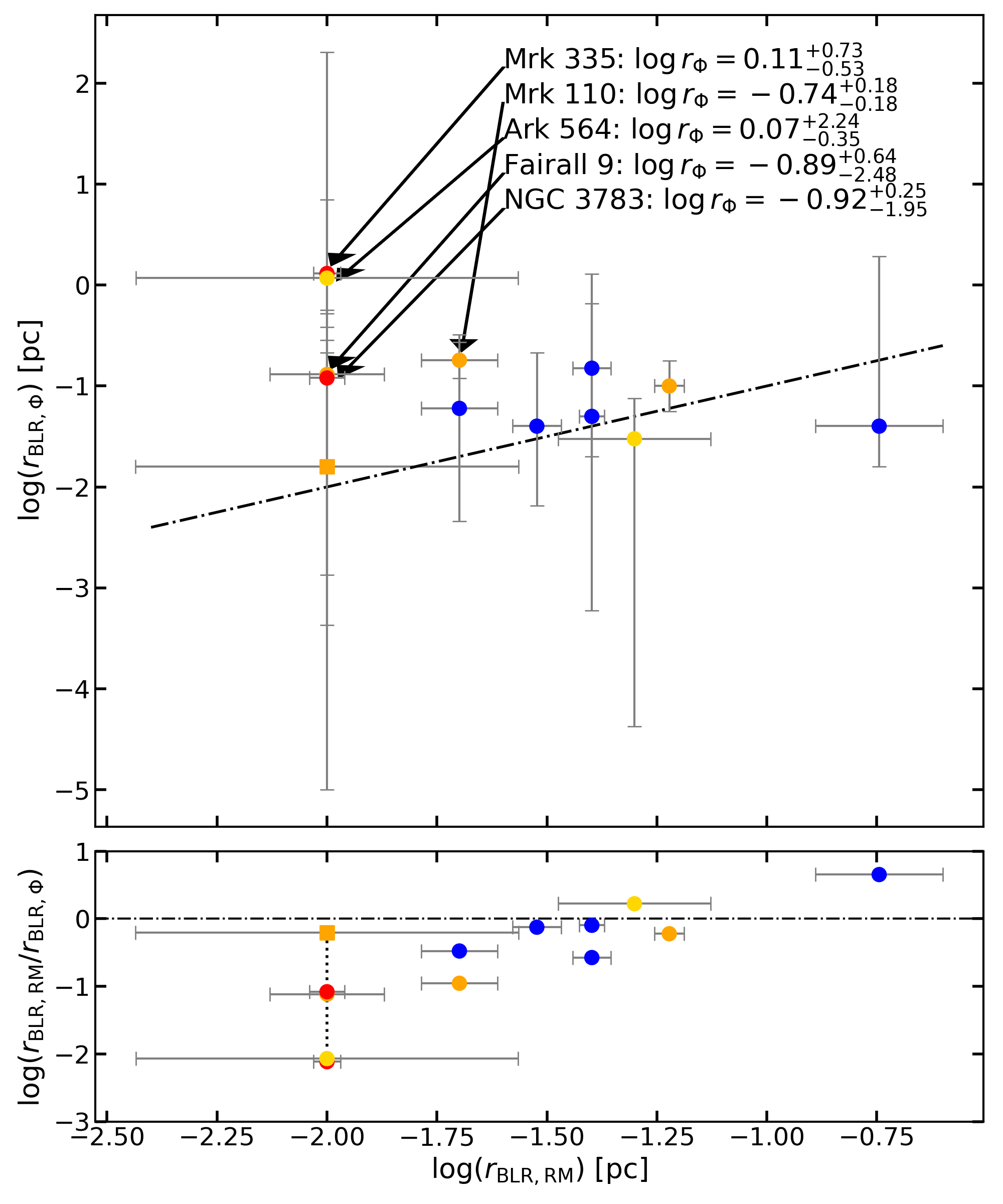}
    \caption{{\it Top}: Scatter plot showing the comparison of $r_\mathrm{BLR}$ estimated with reverberation mapping ($r_{\rm BLR,RM}$), and using {\tt CLOUDY} photoionization simulations ($r_{\rm BLR, \Phi}$), in the logarithmic plane. Points are displayed using the same color-coding as in Figure \ref{fig:4DE1Sample}, with their respective errorbars. The dotted-dashed line describes the position of objects for which the two independent determinations coincide. The values of $r_{\rm BLR, \Phi}$ for objects whose determinations do not coincide with $r_{\rm BLR,RM}$ by a margin are shown in the upper-right quadrant of the figure (expressed in pc), connected to their representative points through an arrow. An alternative result of the determination of physical parameters is shown for Ark 564 with a square, obtained removing the diagnostic ratios employing \ciii. {\it Bottom}: Plot showing the difference between the two estimations. The horizontal dotted-dashed line describes the position of objects for which the two independent determinations coincide. The dotted line connects the two different determinations for Ark 564 described above.} 
    \label{fig:rcomparison}
\end{figure}

The computation of physical parameters in the BLR, coupled with Equation \ref{ionization} for the ionization parameter, yields an independent measurement of $r_\mathrm{BLR}$, which could be compared with the results obtained from \hb\ reverberation mapping for our sources \citep[cf. ][]{Wandel2000,Negrete2013}. The reverberation $r_\mathrm{BLR}$ or time delays are reported whenever available in the presentation of the individual objects of the sample. The $r_\mathrm{BLR}$\ estimations are compared in Figure \ref{fig:rcomparison}. The two independent estimations yield similar results for most sources of our sample,  within a factor $\sim 3$ \citep{Vestergaard2006, Negrete2013,panda2022}. 

Three estimations of the BLR radius using the physical parameters obtained from our spectral measurements yield results incompatible with the reverberation mapping estimation: Mrk 335, Mrk 110 and Ark 564 (Fig. \ref{fig:rcomparison}).  This discrepancy arises from the  estimations returning a low-density solution, thus likely overestimating the extent of the H$\beta$-emitting region. The \ciii\ intensity of Ark 564 might be overestimated because of the correction needed to account for heavy absorption close to the line rest-frame (Fig. \ref{fig:fits1}). Removing the \ciii\ diagnostic ratios, moves the data point for this source in agreement with the \hb\ reverberation mapping estimate.  However, in the other two cases, the intensity of \ciii\ and the presence of other spectral features suggested the presence of low-density gas in the BLR, a stratification of properties that may be consistent with the presence of an extended BLR. This seems to be especially true for Mrk 110, where the redward asymmetry of a faint VBC might require an innermost radius of $\sim 100$\ gravitational radii. 

Nevertheless, the bulk of the measurements confirm the physical conditions obtained from the estimation associated with {\tt CLOUDY} photoionization simulations, providing an independent measurement of the $r_\mathrm{BLR}$.

\subsection{Enrichment mechanisms along the quasar MS}
\label{enrich}

The results obtained from the estimation of metallicity in the BLR of our sources suggest the presence of different enrichment mechanisms acting in the different objects of our sample, giving rise to the solar or subsolar metallicity that we observe in Population A and B sources, and the highest metallicity that we observe in extreme accretors. 
\cite{Matteucci93} discussed the possibility of metal enrichment in quasars, and predicted that the gas can reach solar abundances in less than $10^8$ years after star formation starts for a Salpeter-like Initial Mass Function (IMF). There are thus many possible mechanisms explaining the metal abundances observed in Population A and Population B quasars. 
We have, however, observed a remarkable number of objects displaying supersolar metallicity, with the totality of xA sources exhibiting $Z \gtrsim 10 Z_\odot$. 

One of the primary proposals to explain this phenomenon is the possibility of star formation in the vicinity of the active nucleus, starting from the models developed by \cite{Artymowicz93} and \cite{Collin1999}.
In this context, \cite{Wang2009} predicted the presence of highly enriched BLRs in quasars, with $Z$ up to 100 $Z_\odot$, suggesting the possibility of star formation in the vicinity of the accretion disk of the central nucleus as a means of chemical enrichment. The SMBH could then trap these metals in the region due to its gravitational potential. A similar mechanism involves Accretion Modified Stars (AMS): a stellar population characterized by stars accreting material from the disk and rapidly attaining high masses, subsequently exploding as supernovae and further enriching the surrounding material \citep{wangetal21,Cantiello2021,Wang2023}.
In our sample, I Zw 1 and PHL 1092 display blueshifts that dominate the emission of their HILs, particularly \civ. The presence of strong winds originating from the high-accretion phase experienced by these objects produce outflows capable of escaping the BLR and enriching the surrounding regions \citep[][\citetalias{Marziani2023}]{Vietri2016, Vietri2018}.

Furthermore,  there is a lack of redshift evolution of the metallicity of quasars \cite[][and references therein]{juarezetal09,Fan23}. Metallicity in the BLRs is higher than in the NLRs, suggesting that the chemical enrichment occurs in-situ, near the molecular torus \citep{Collin1999}. A top-heavy IMF can explain the observed metal abundances, that could produce a strong metal enrichment in a short timescale ($\tau < 10^9$ yr, even though accretion-modified stars could be longer living than isolated stars, see \citealt{chenlin24}).
An additional line of evidence supporting an in-situ enrichment is provided by the strong similarity observed between the objects in our sample, and those found in high-redshift samples \citep{Banados2016, dodorico23}, since at high redshift there is no time to produce the extreme amounts of metals observed \citep{Xu2018}.  

The similarity between low-$z$ quasars accreting at a high rate and most intermediate-to-high $z$ quasars \citep{sulenticetal00a} has been interpreted as a selection effect \citep{sulenticetal14}. In flux-limited surveys, for a given mass, the only observable objects are those of higher luminosity or, equivalently,  higher Eddington ratio. Therefore, the presence of a high percentage of high-accretors at high redshift should not be interpreted as them being the only quasars present at high redshift. Low-luminosity quasars observed at low redshifts, such as NGC 3783 in our sample, are yet unobservable at intermediate-to-high redshifts. If, as our results suggest, $Z$\ is correlated with \lledd, a population of quasars with metallicity around solar might be present up to $z \approx 4$, as indeed found from a small sample of deep observations of very faint quasars  \citep{sulenticetal14}.


The morphological types of the objects of the sample were retrieved from the NED extragalactic database and other works \citep{Mckernan10, Khorunzhev12, Kim2017, Zhao2021, Kim2021}. The sample is however too small to reach even a tentative conclusion. A reliable analysis of the non-ionizing part of the SED (especially radio, FIR, and MIR emission) that may yield information on the circumnuclear of host star-formation properties is deferred to an eventual work.

\section{Conclusions}
\label{conclusions}

We analyzed a sample of 14 Type-I AGNs distributed along the MS of quasars. We carried out a multi-component fitting, as described in Section \ref{methods}, to measure the intensity of major emission lines in the $\lambda 1300$ - $\lambda 2100$ and optical wavelength ranges, and confronted the results from approximately 10 diagnostic ratios with those obtained through {\tt CLOUDY} photoionization simulations, constraining the physical parameters that describe the BLRs of these sources. The outcomes of our analysis can be summarized as follows:
\begin{itemize}
    \item We identified a strong trend in metallicity along the MS, exhibiting an increasing gradient of metallicity from Population B quasars (with subsolar or solar metallicity) to xA sources (with supersolar metallicity $Z > 10 Z_\odot$). The hypothesis regarding $Z$ as one of the principal correlates of the MS \citep{Panda2019,Marziani24} is confirmed.

    \item We proposed a ``recipe” to estimate the metal content in the BLR, utilizing approximately 10 diagnostic ratios sensitive to $Z$, $U$, and $n_{\rm H}$. We recommend employing the \civ/\hb\ ratio only in the case of UV and optical spectra acquired simultaneously, as it was identified as a major source of error in this work. 

    \item We confirmed metallicity values of several tens of the solar metallicity, among sources classified as xA that are consistent on average with the values reported for the highly accreting quasars of \citet{Sniegowska2021,Garnica2022}.

    \item The highest $Z$ objects, might be significantly affected by the introduction of micro-turbulence, at the level expected for the line thermal broadening ($\sim$ 10 \kms).
\end{itemize}

The main caveats concerning the $Z$ estimates are related to the precision of the measurements (ideally, relative uncertainties should be less than $\sim 30$ \%), the lack of contemporaneity of optical and UV observations, and the simplification introduced by the single zone approximation in the photoionization modeling, even if the separation between BLR, VBLR and outflowing component takes into account major physical differences. In this respect, we plan to implement a locally optimized emitting cloud scheme \citep{Baldwin1995}. The application of the scheme to higher $z$\ would require photoionization computations employing SEDs appropriate for sources at high luminosity \citep{durasetal20,krawczyketal13}. However, estimations of metallicity at very high redshifts \citep[e.g.,][at $z\approx 7.24$]{rojas-ruizetal24} are now becoming possible from the combination of NIR and MIR observations and may offer an important constraint on the early evolution of the supermassive black holes and their host galaxies \citep[e.g.,][]{cattaneoetal09}. 

\begin{acknowledgements}
We thank Pu Du, Jian-Min Wang and Chen Hu for providing us with the tentative value of the time-delay of H$\beta$ for PG 1259+593. SP acknowledges the financial support of the \emph{Conselho Nacional de Desenvolvimento Científico e Tecnológico (CNPq)} Fellowships 300936/2023-0 and 301628/2024-6. The authors would like to thank Dr. Jaime Perea for his useful comments and for his help and generous allocation of computing time on {\tt hypercat} and {\tt idilico} at IAA (CSIC). A.d.O. acknowledges financial support from the Spanish MCIU through project PID2022-140871NB-C21 by “ERDF A way of making Europe” and the Severo Ochoa grant CEX2021- 515001131-S funded by MCIN/AEI/10.13039/501100011033.  
\end{acknowledgements}

%
%

\bibliographystyle{aa}
\bibliography{aanda.bib}

\begin{appendix} 
\onecolumn
\section{Fit spectra of the sample}
\label{app:fits}

\begin{figure*}[h!]
\includegraphics[width=4.53cm]{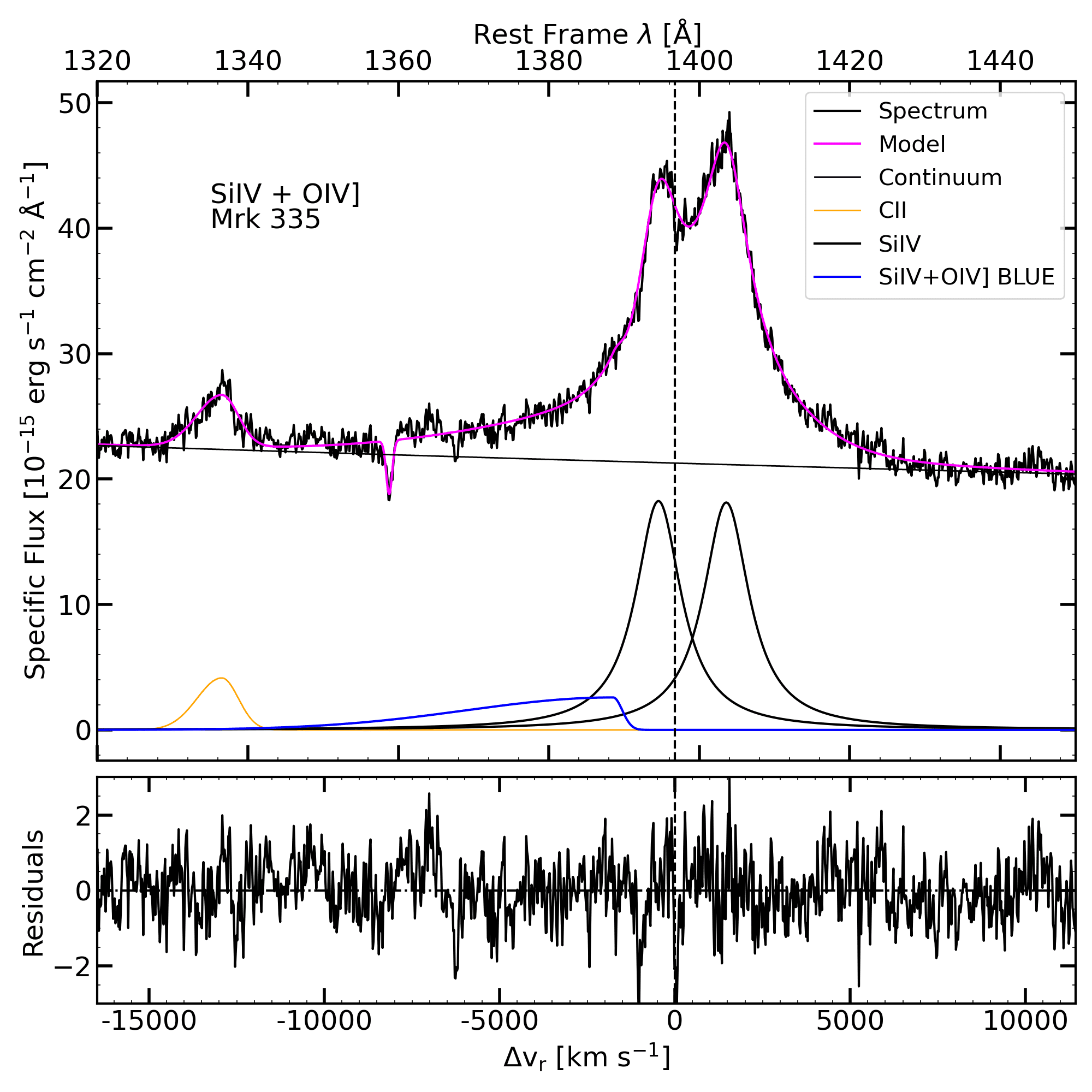}
\includegraphics[width=4.53cm]{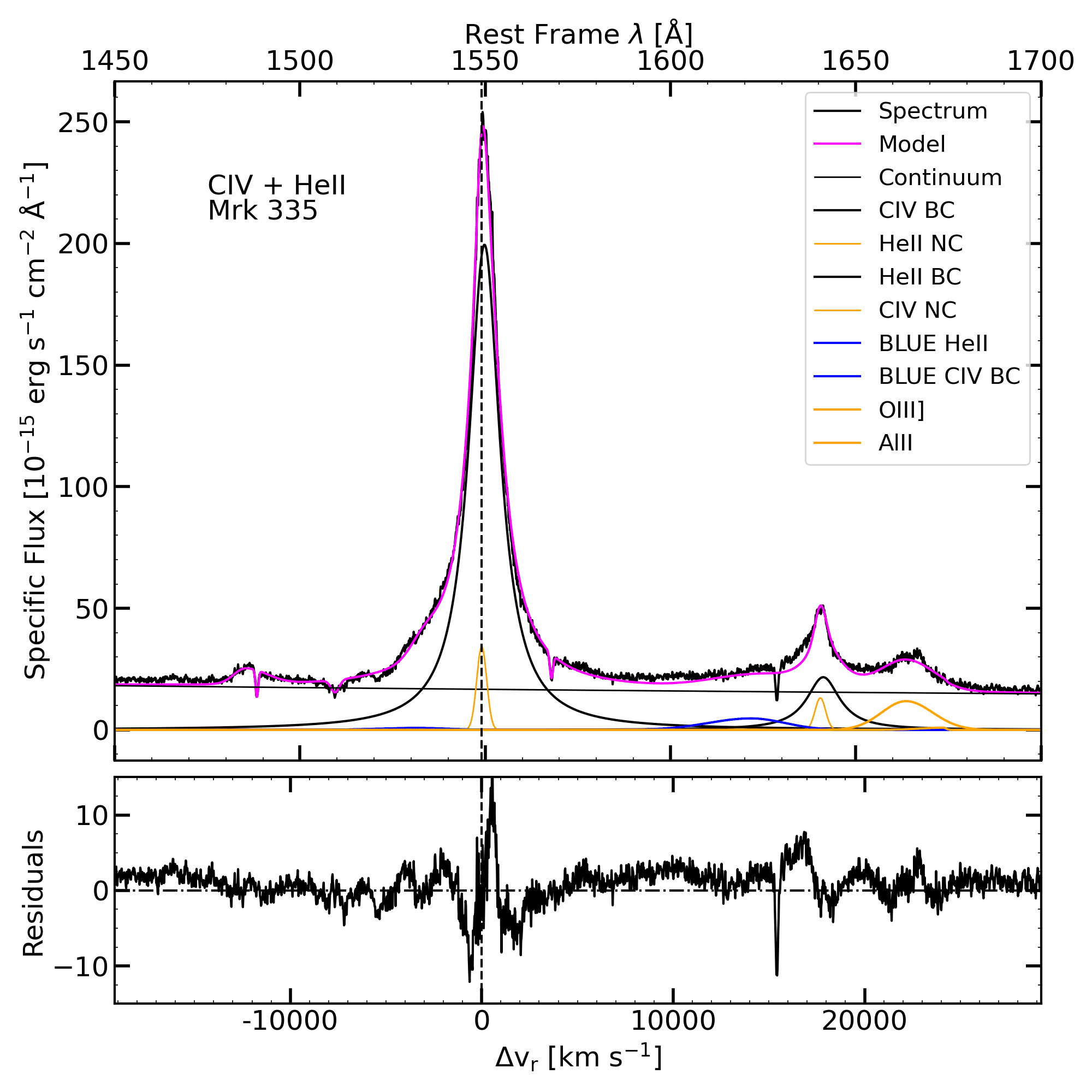}
\includegraphics[width=4.53cm]{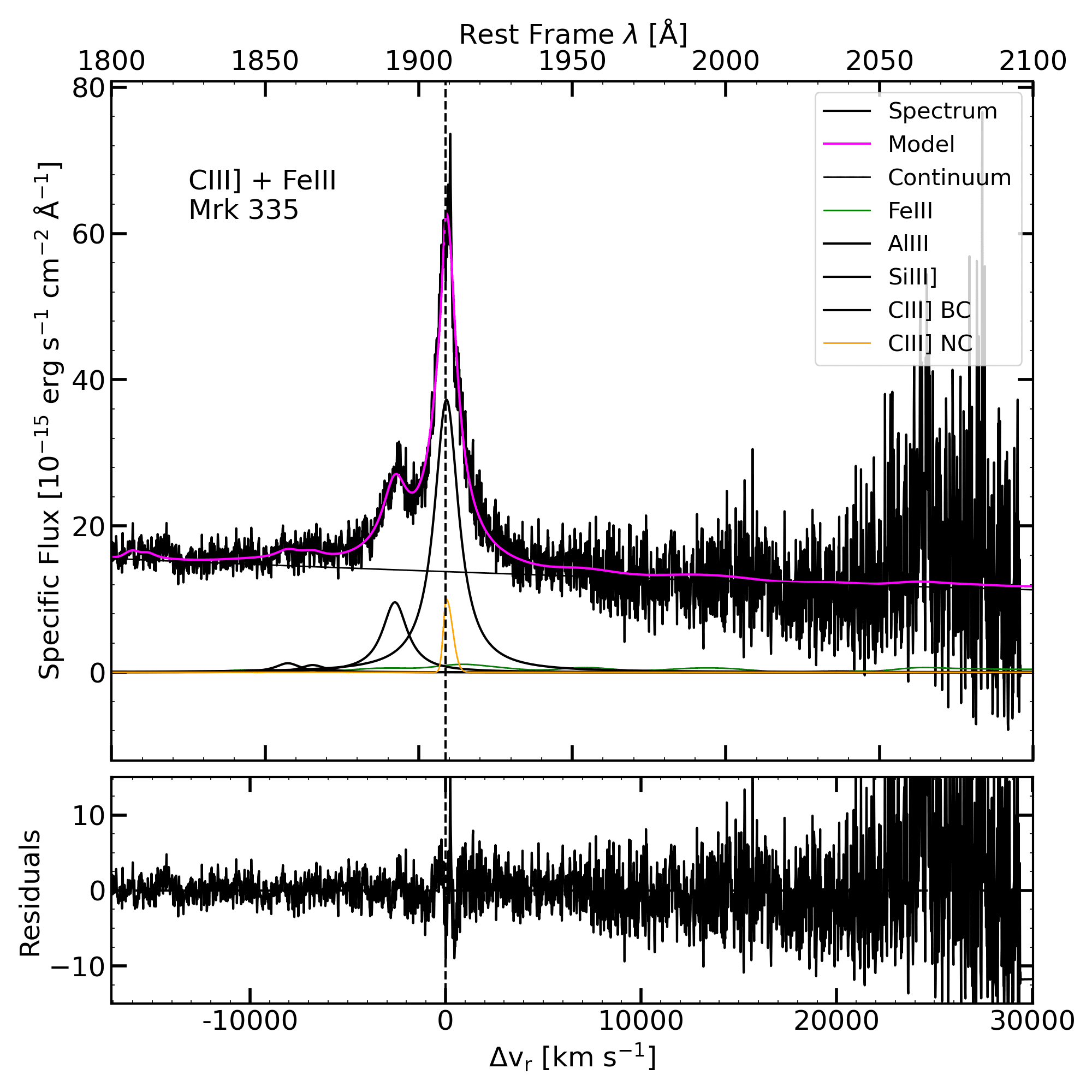}
\includegraphics[width=4.53cm]{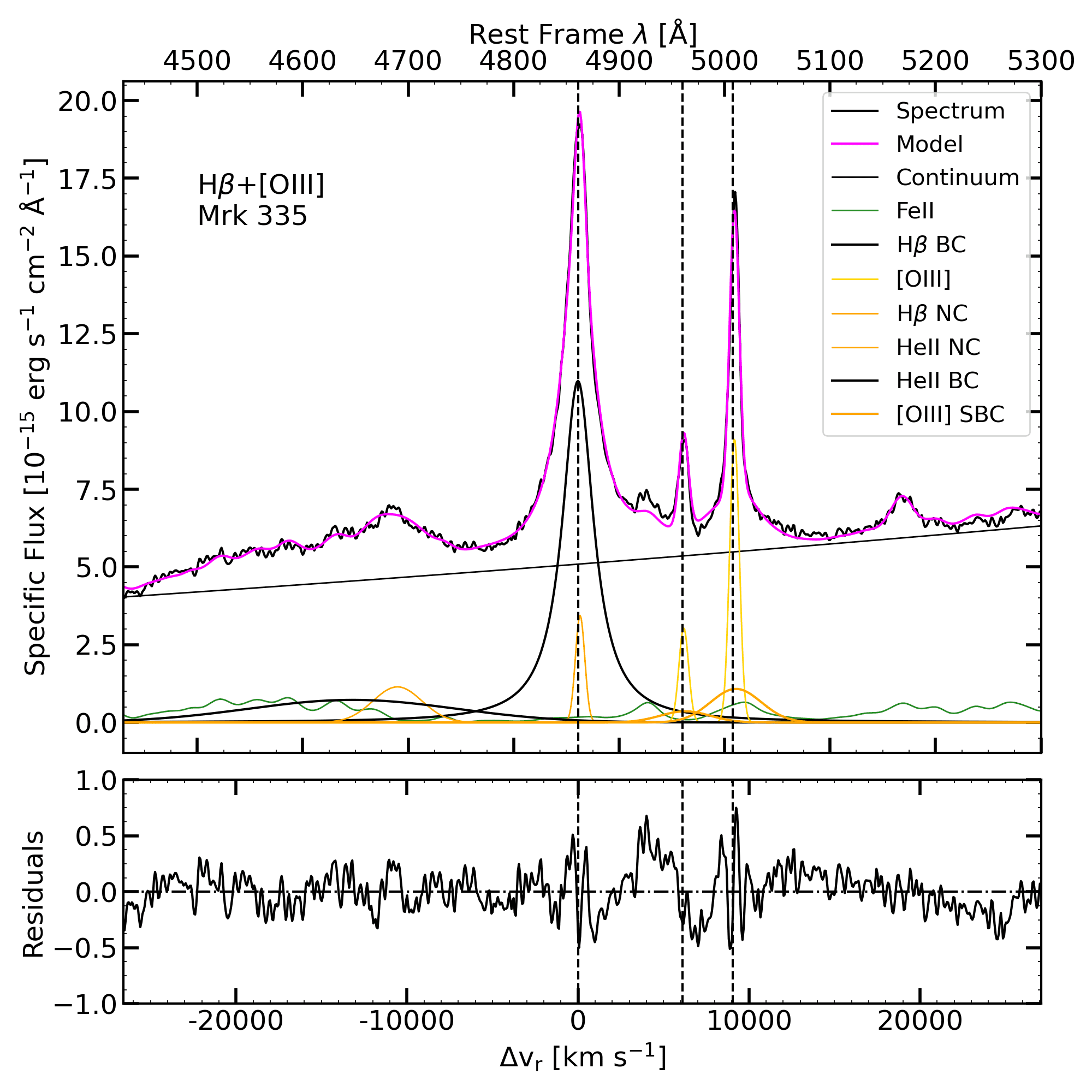}\\
\includegraphics[width=4.53cm]{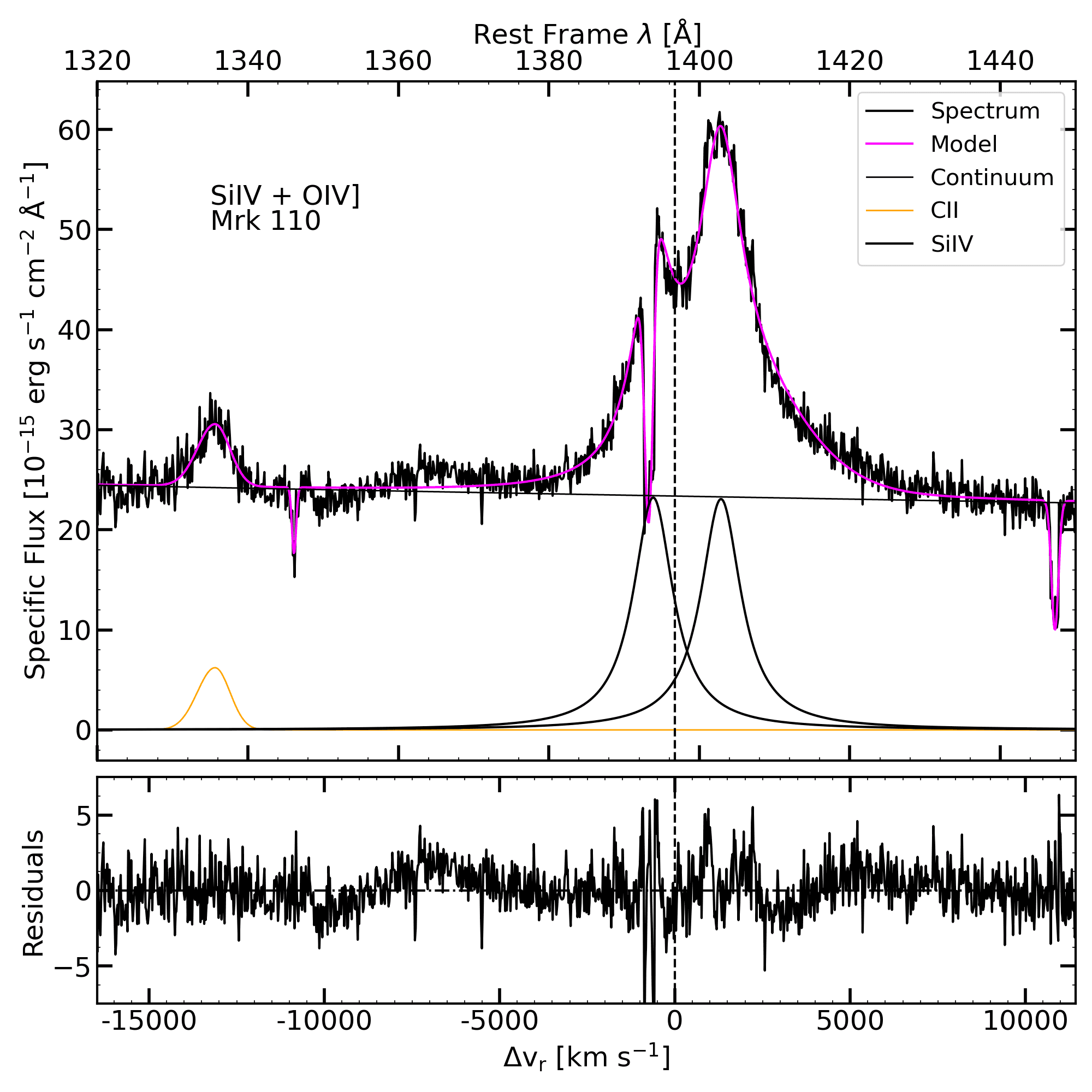}
\includegraphics[width=4.53cm]{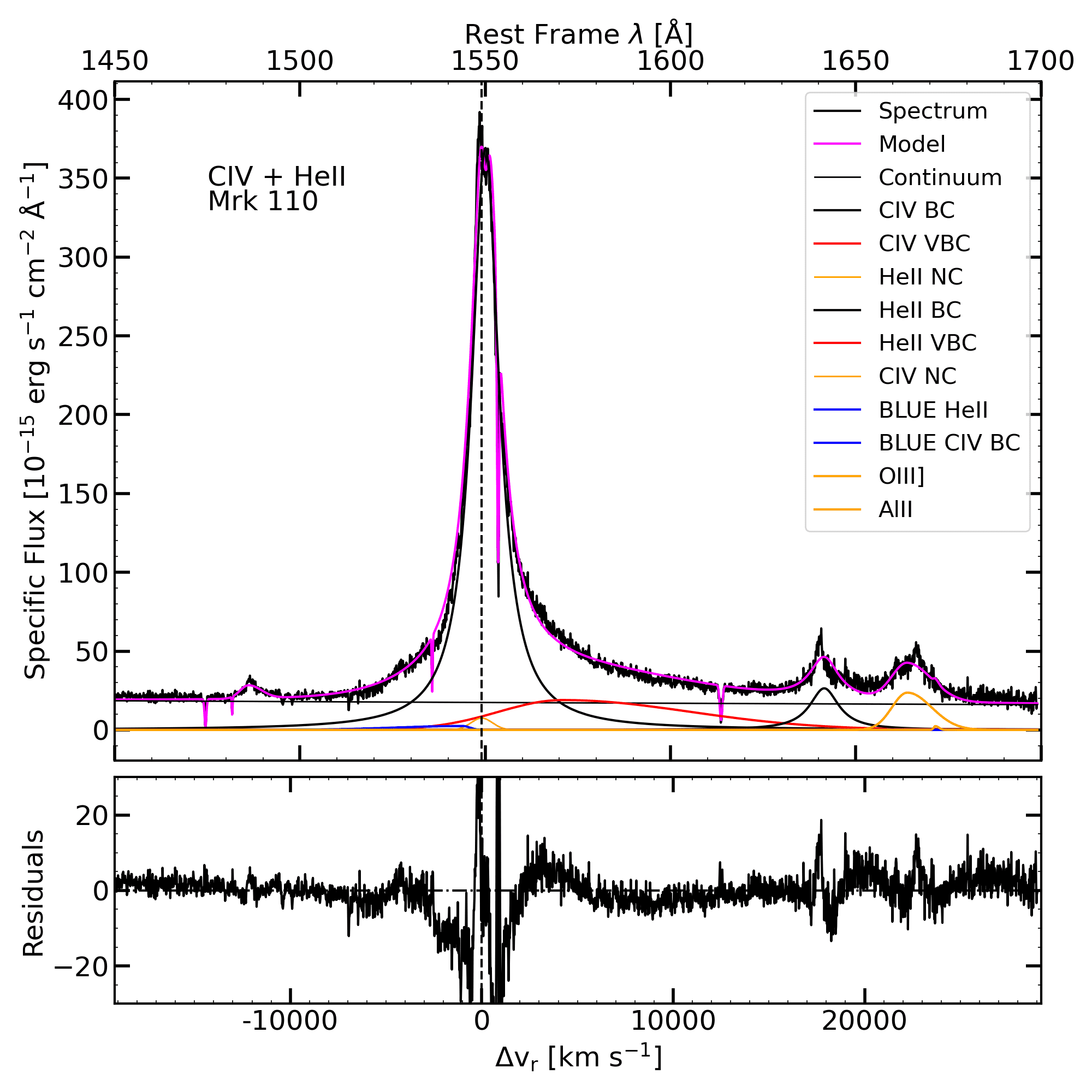}
\hspace{4.53cm}
\includegraphics[width=4.53cm]{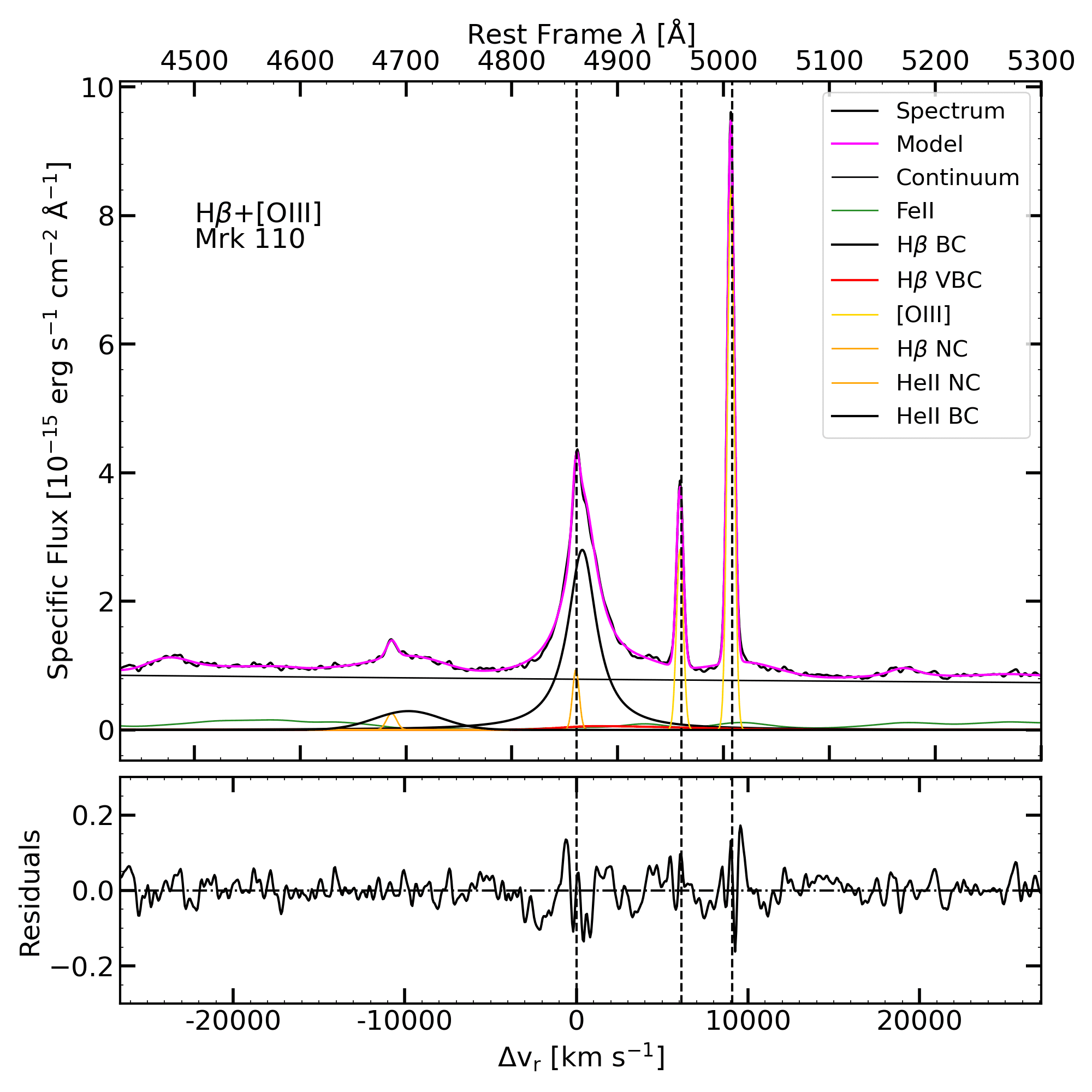}\\
\includegraphics[width=4.53cm]{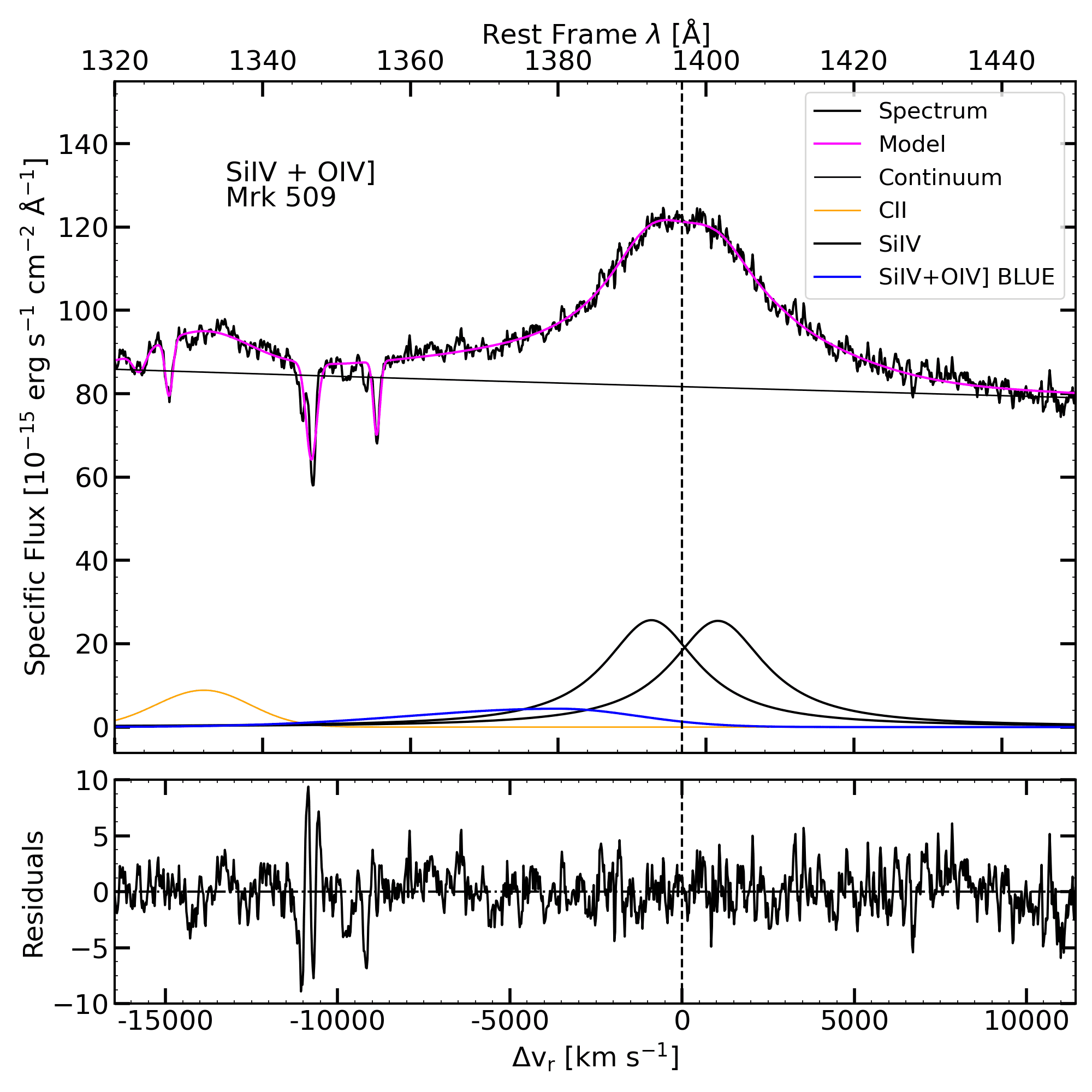}
\includegraphics[width=4.53cm]{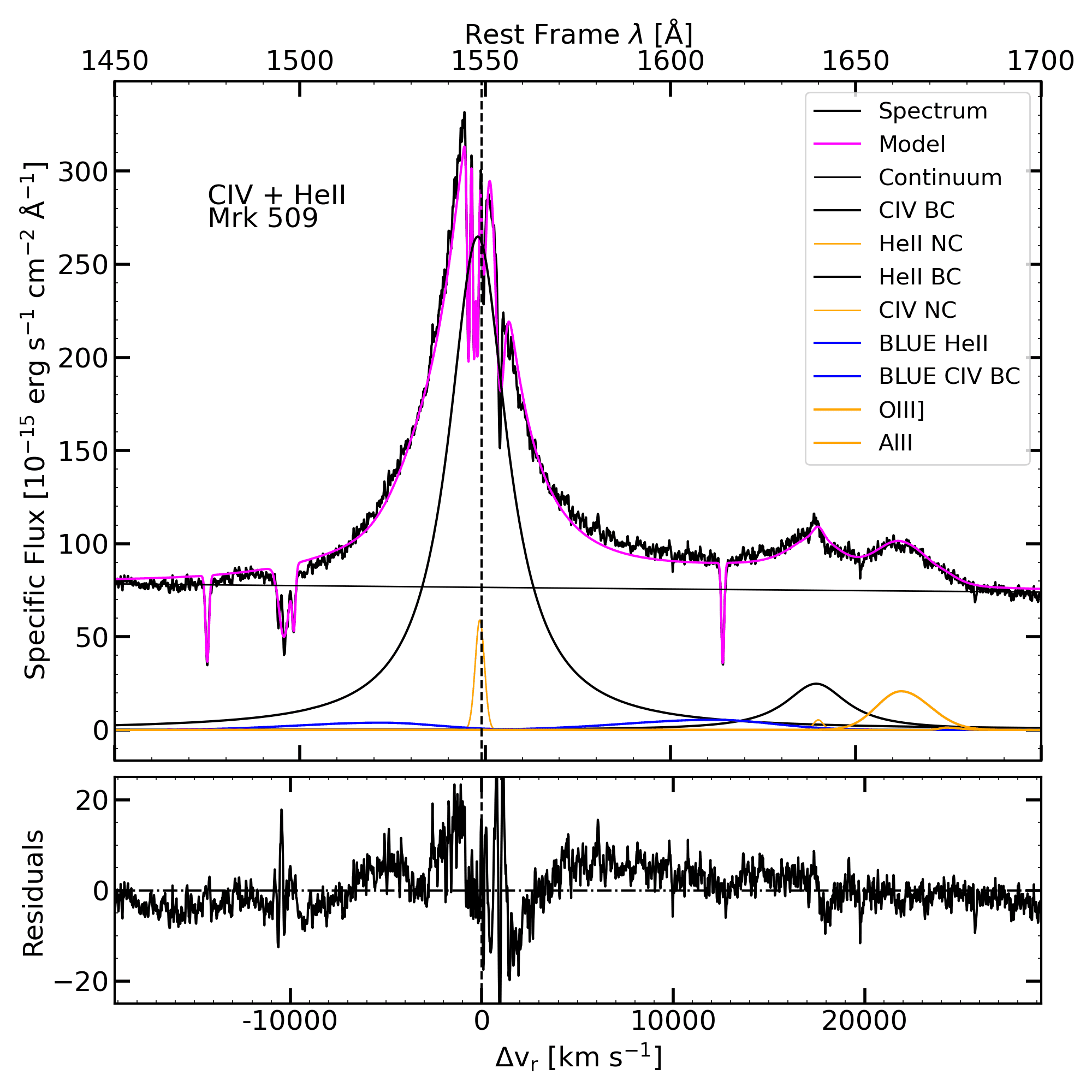}
\includegraphics[width=4.53cm]{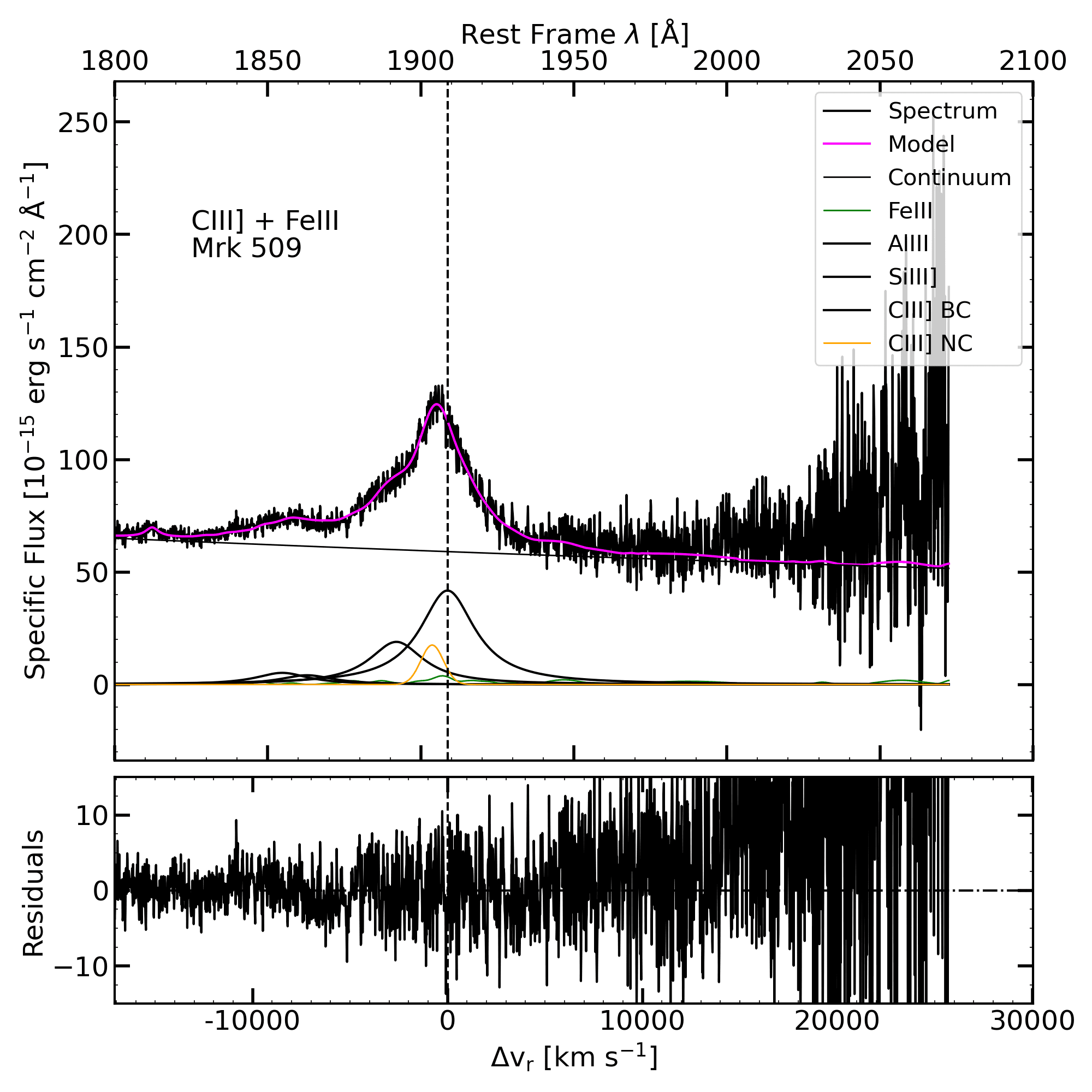}
\includegraphics[width=4.53cm]{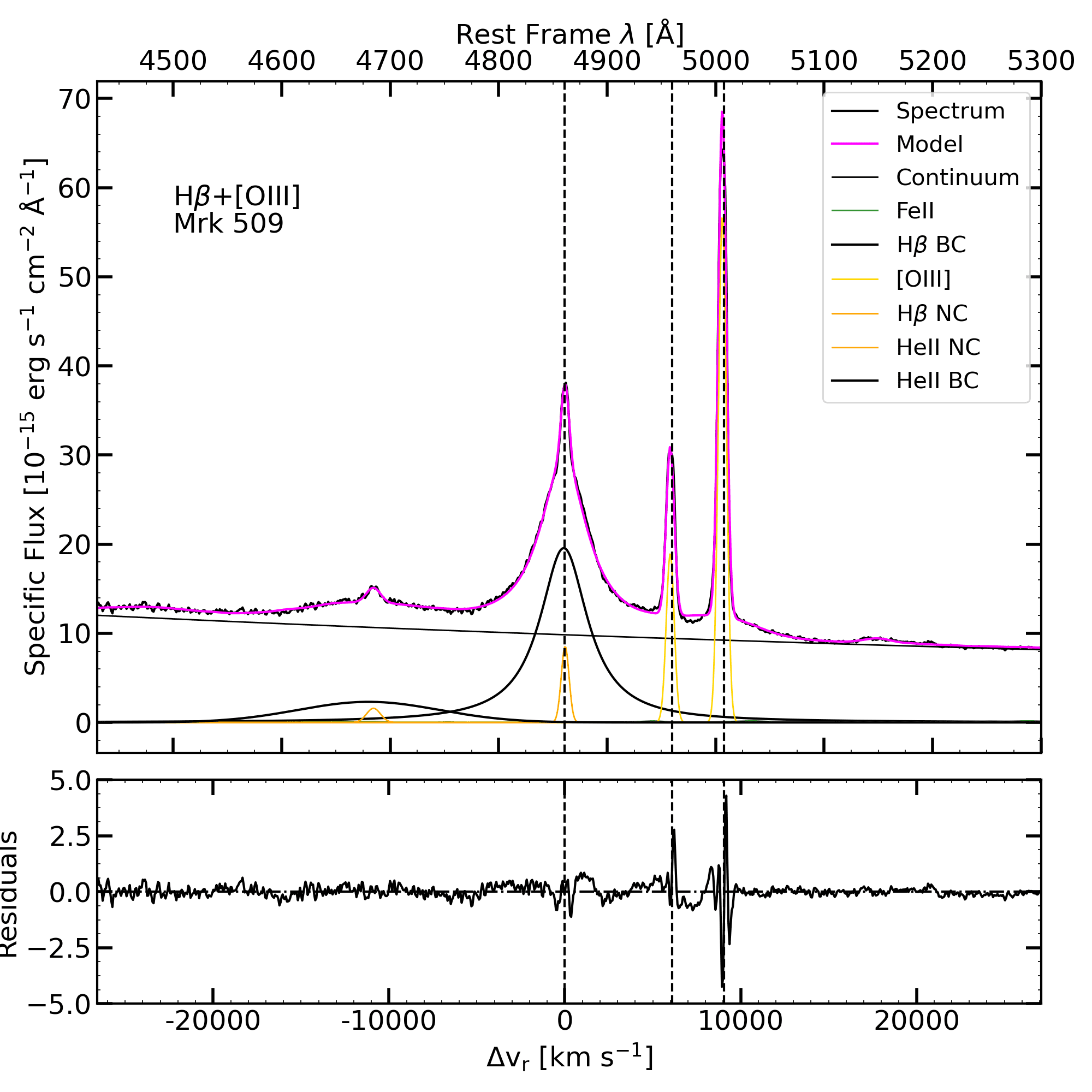}\\
\includegraphics[width=4.53cm]{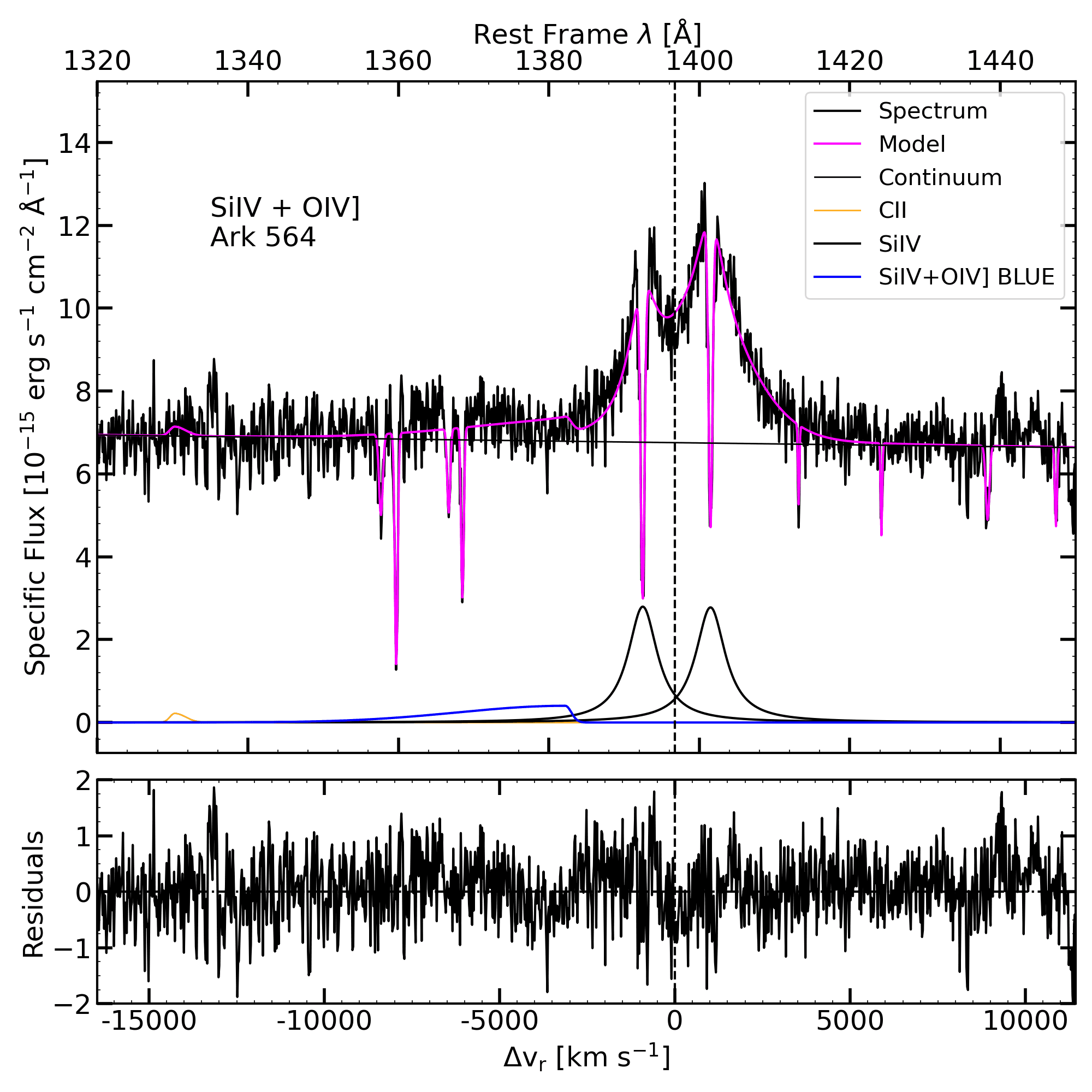}
\includegraphics[width=4.53cm]{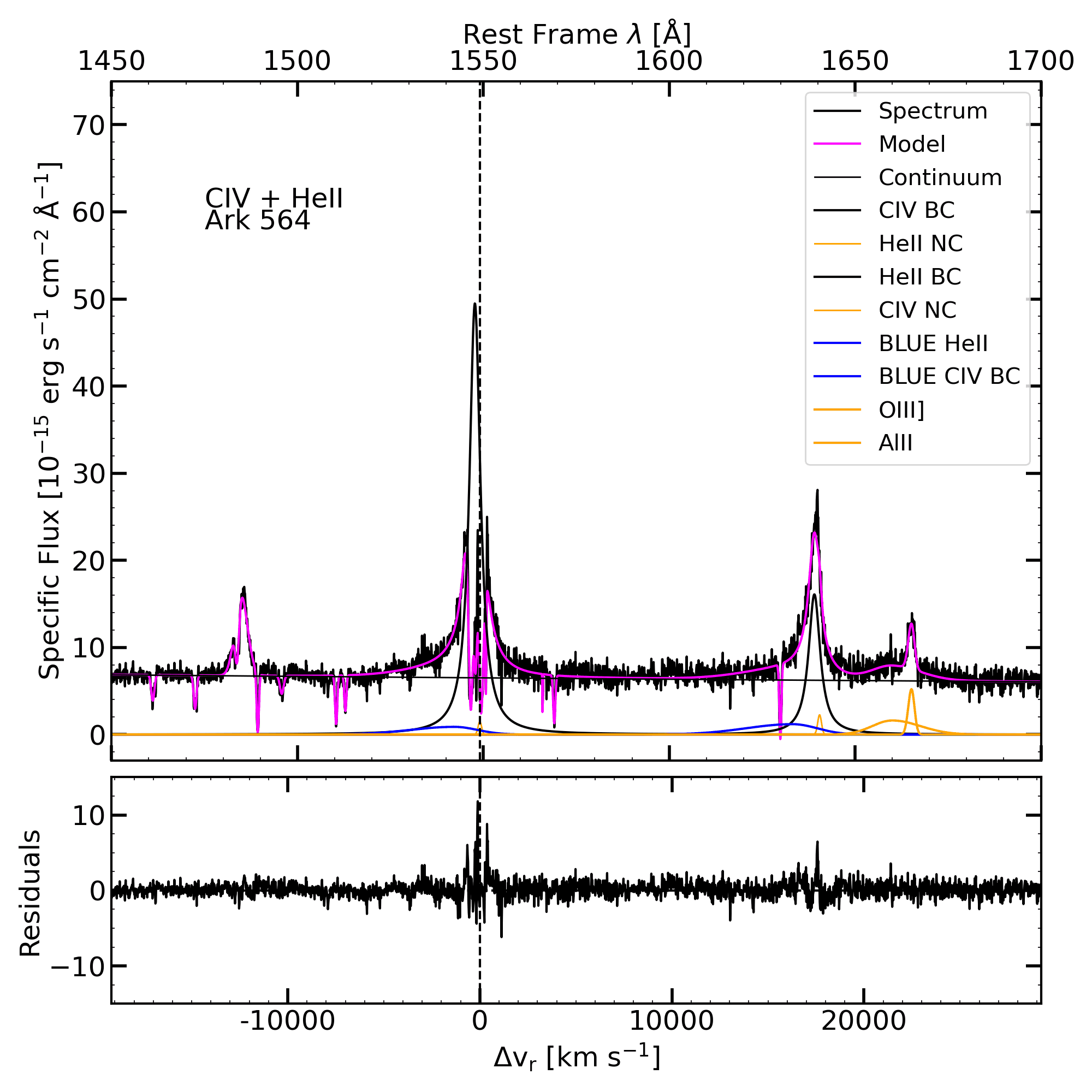}
\includegraphics[width=4.53cm]{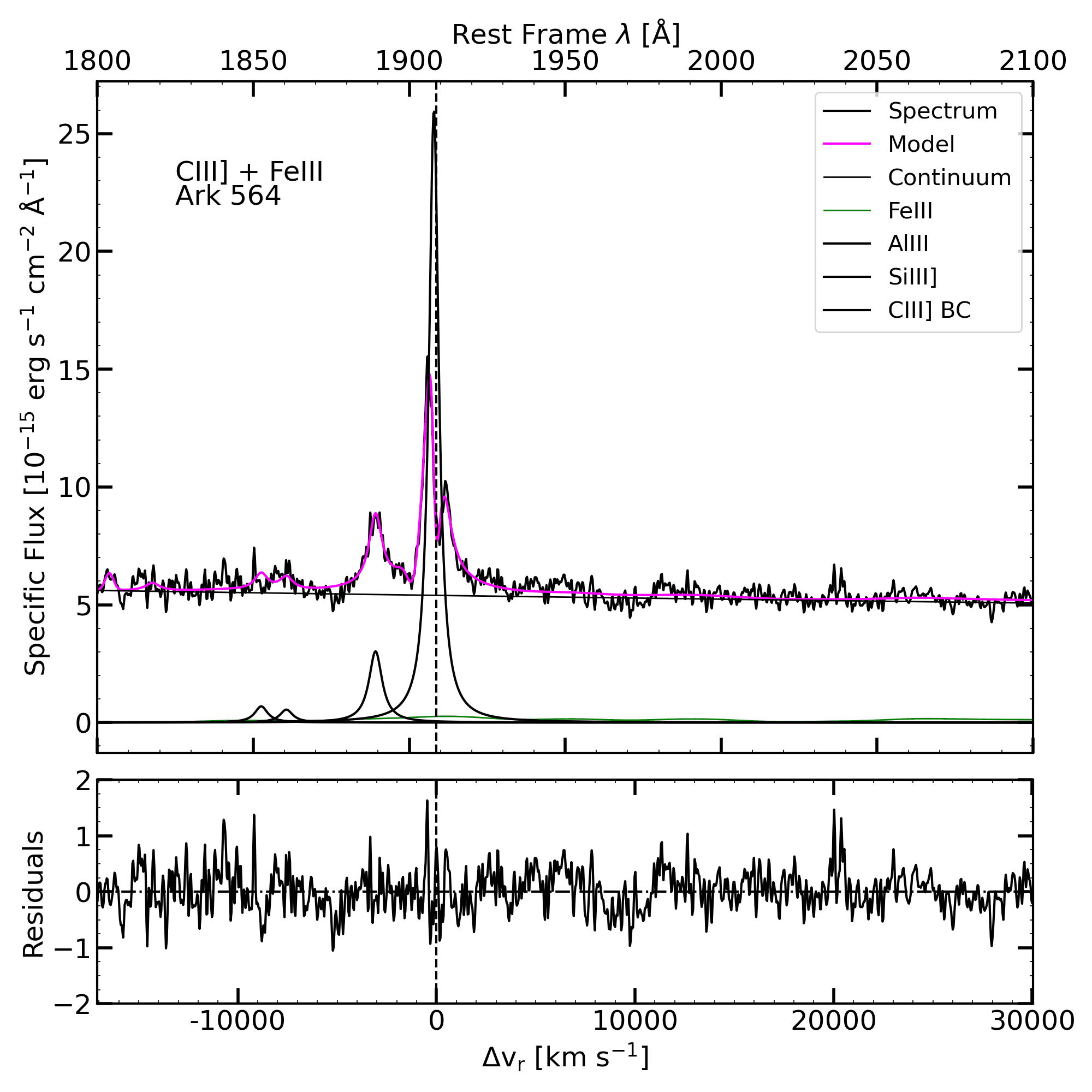}
\includegraphics[width=4.53cm]{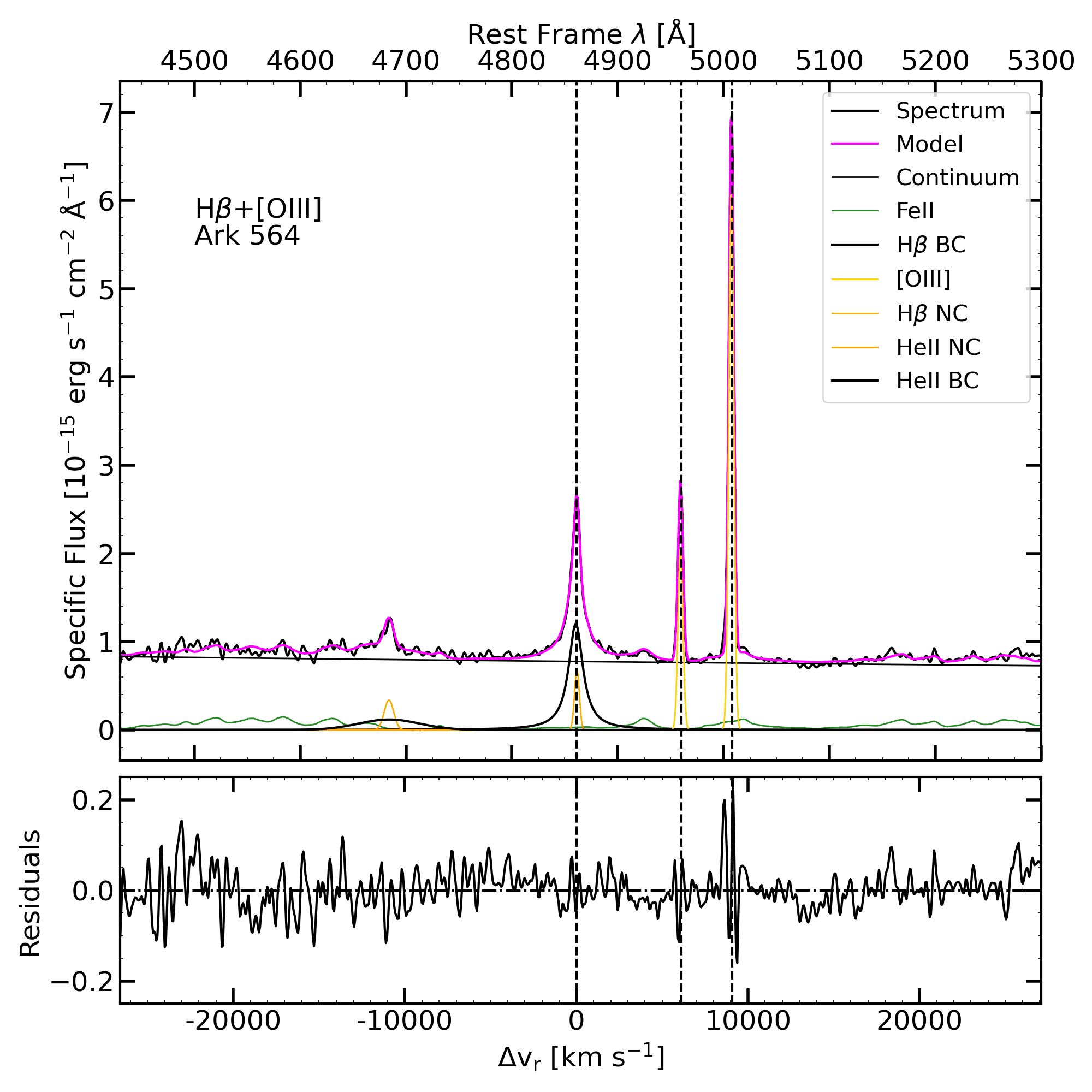}\\
\\
\caption{Spectral fits of (from top to bottom) Mrk 335, Mrk 110, Mrk 509 and Ark 564  in the wavelength ranges considered, with specific fluxes measured in units of $10^{-15}$ erg s$^{-1}$ cm$^{-2}$ ${\rm\AA}^{-1}$ and the wavelength measured in \AA. A lower box describing the residuals is associated with each spectral fit, with wavelength expressed in units of radial velocities (km s$^{-1}$). The original spectrum is plotted with a black line and the modeled spectrum is plotted in magenta. Each additional line component is plotted with a different colour, according to the legend. Dashed vertical lines represent the rest-frame wavelengths of the main lines of the selected spectral range. {\it Left}: $\lambda$1320-1450 \AA\ range. {\it  Middle-left}: $\lambda$1450-1700 \AA\ range. {\it Middle-right}: $\lambda$1800-2100 \AA\ range. {\it Right}: $\lambda$4430-5300 \AA\ range.}
\label{fig:fits1}
\end{figure*}

\begin{figure*}[ht!]
\centering
\includegraphics[width=4.53cm]{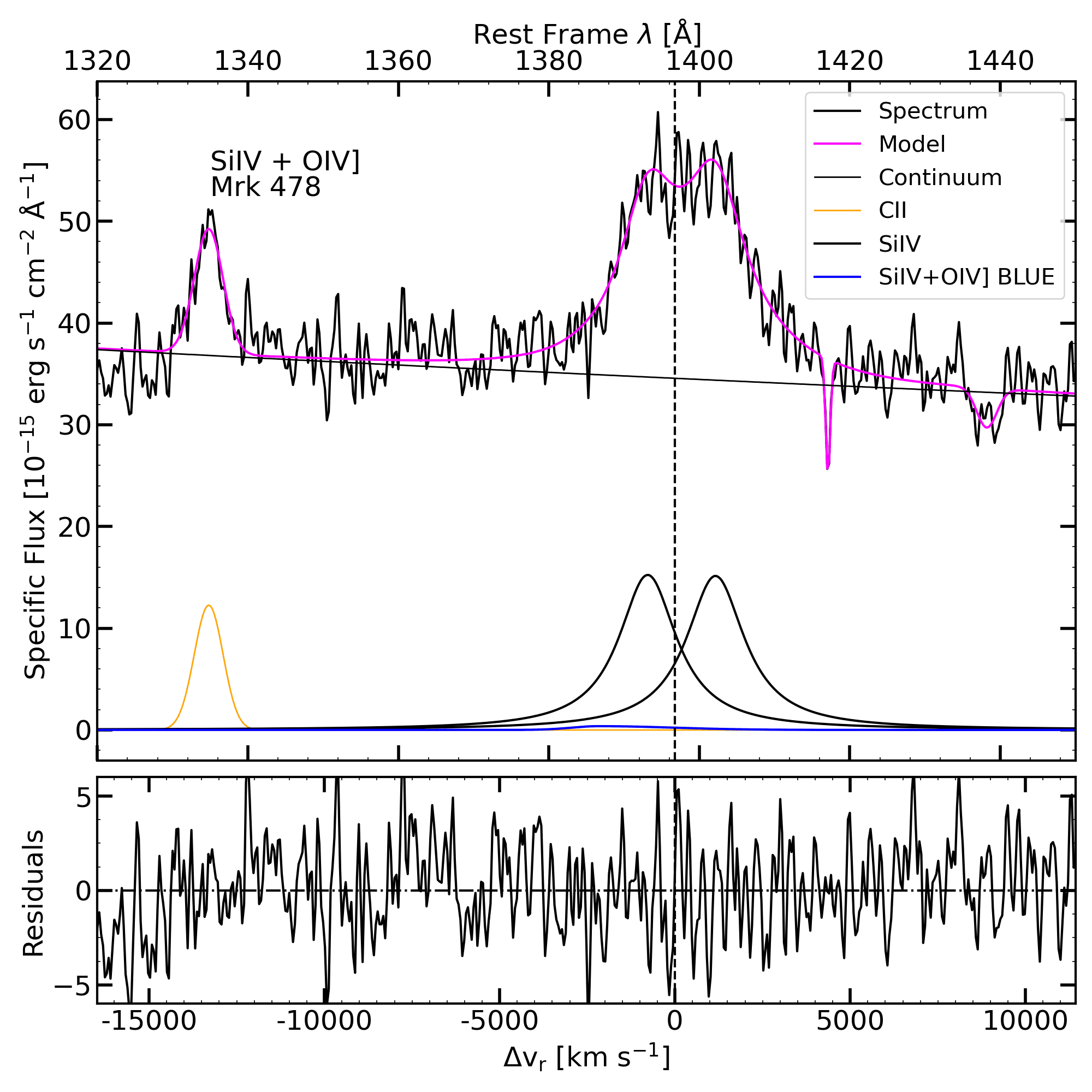}
\includegraphics[width=4.53cm]{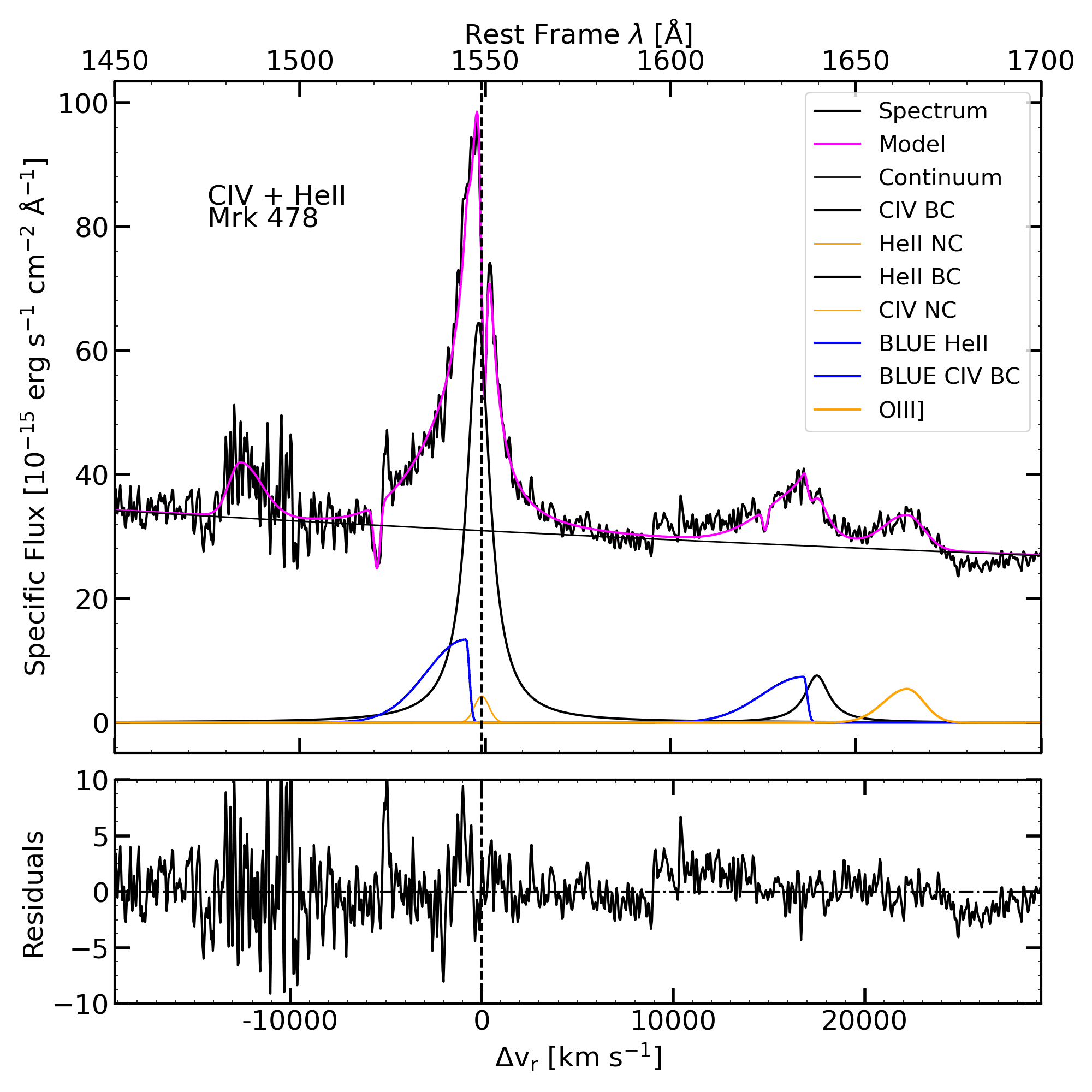}
\includegraphics[width=4.53cm]{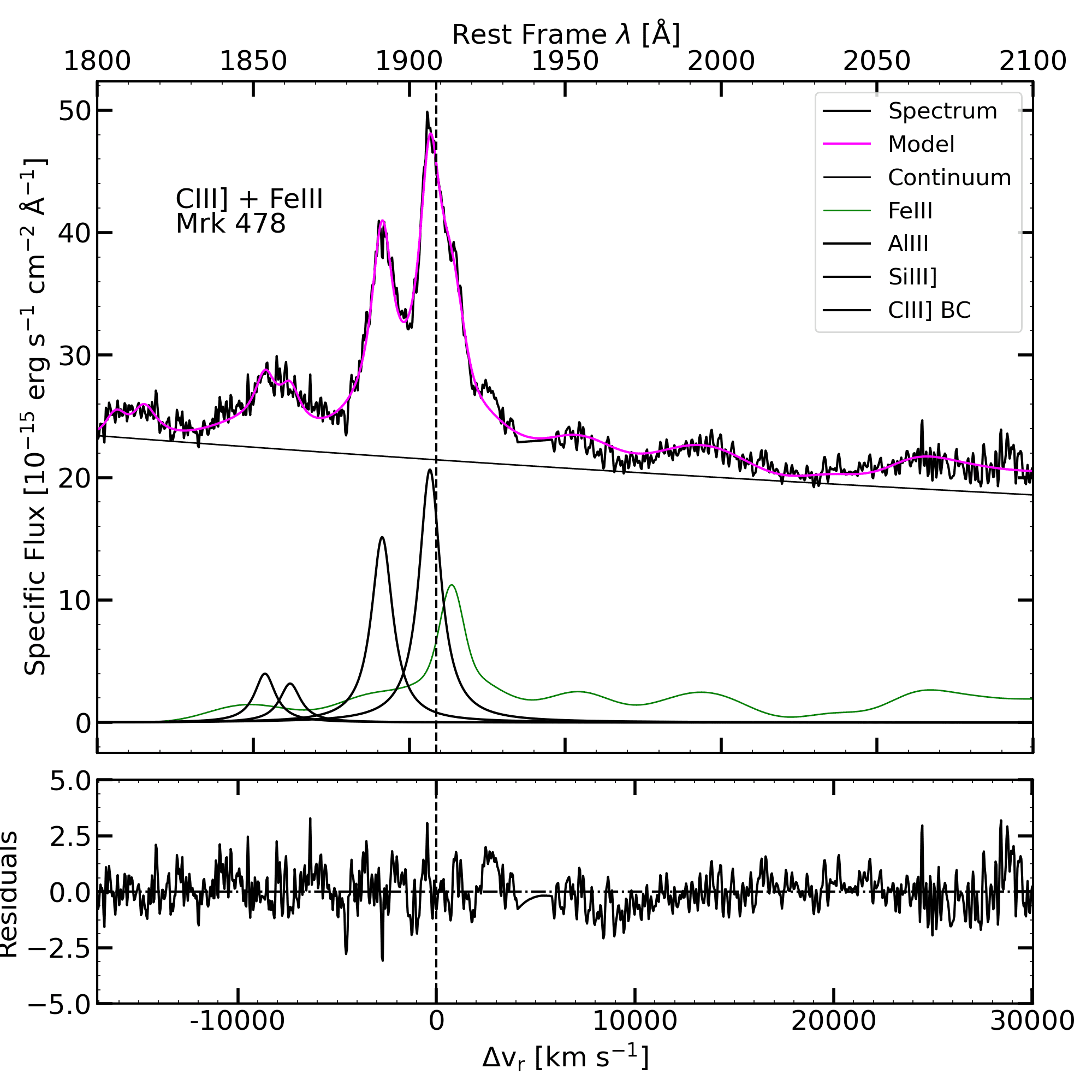}
\includegraphics[width=4.53cm]{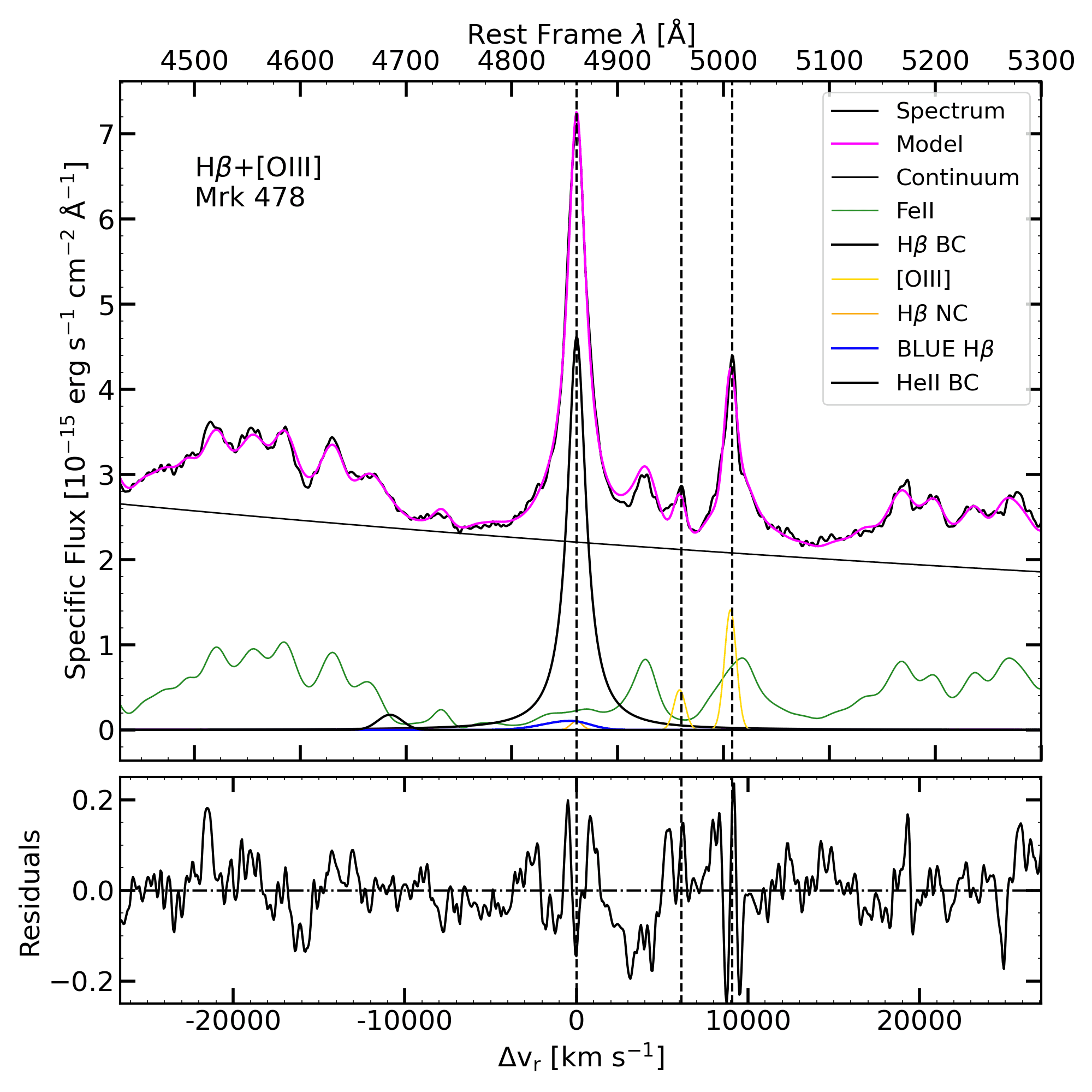}\\
\includegraphics[width=4.53cm]{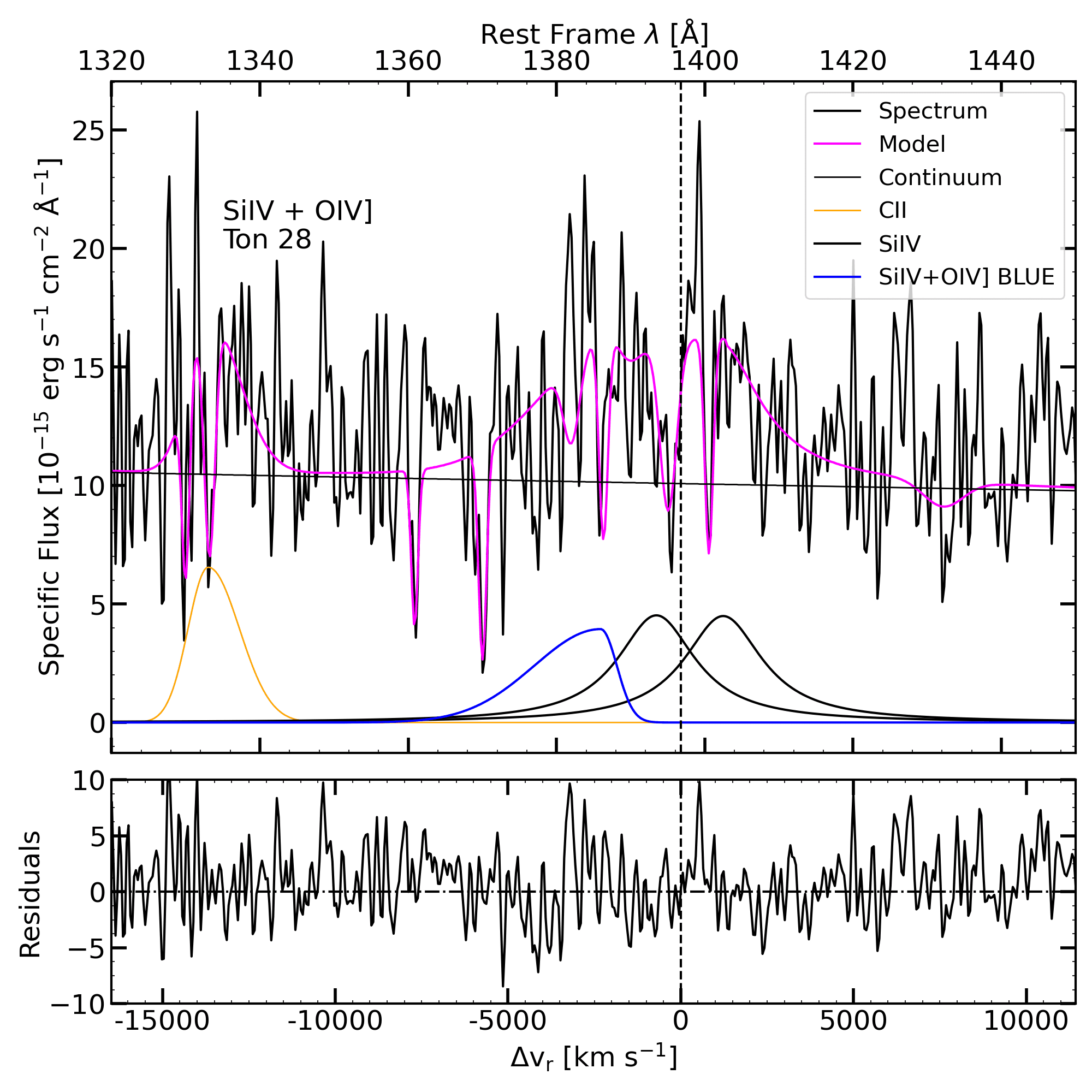}
\includegraphics[width=4.53cm]{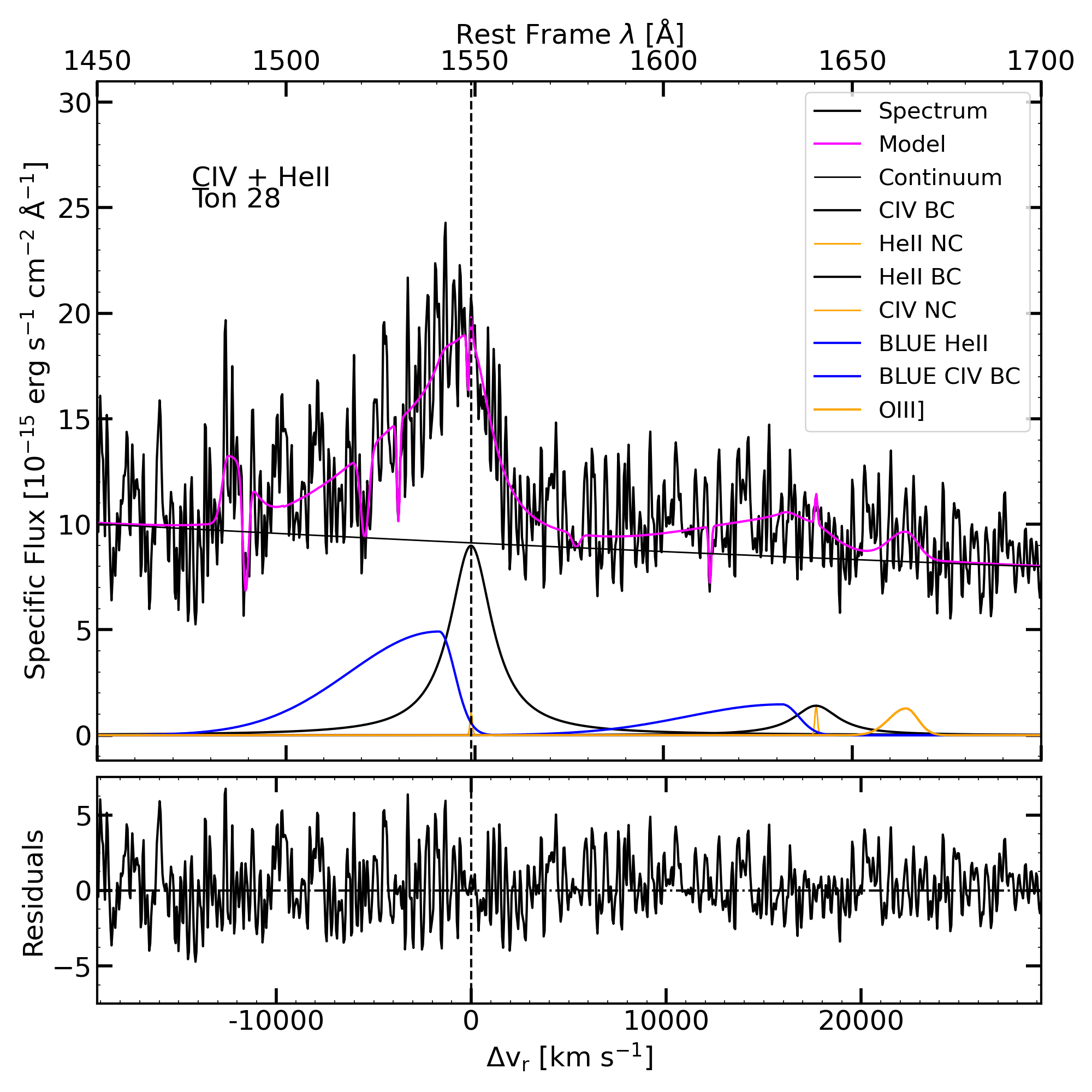}
\includegraphics[width=4.53cm]{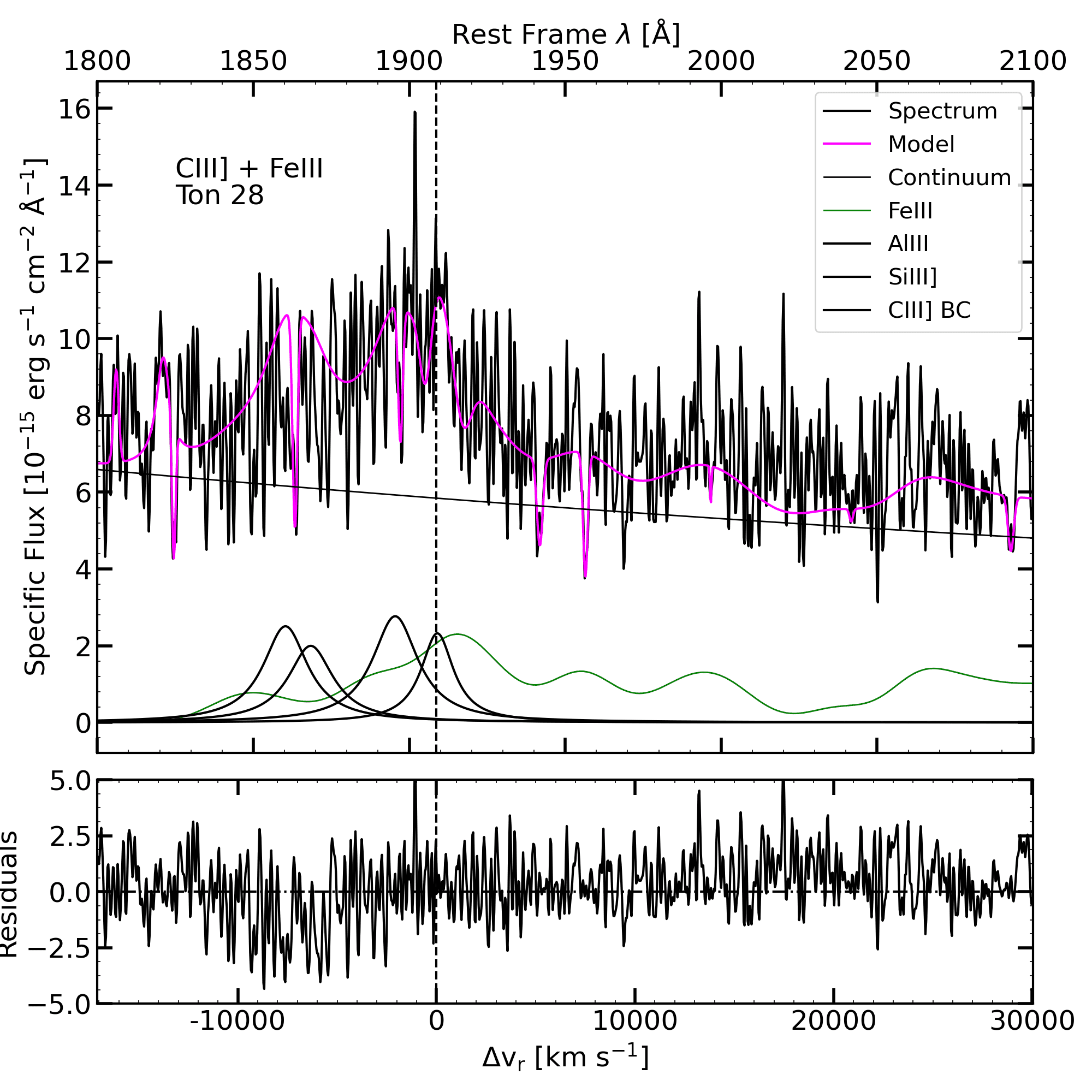}
\includegraphics[width=4.53cm]{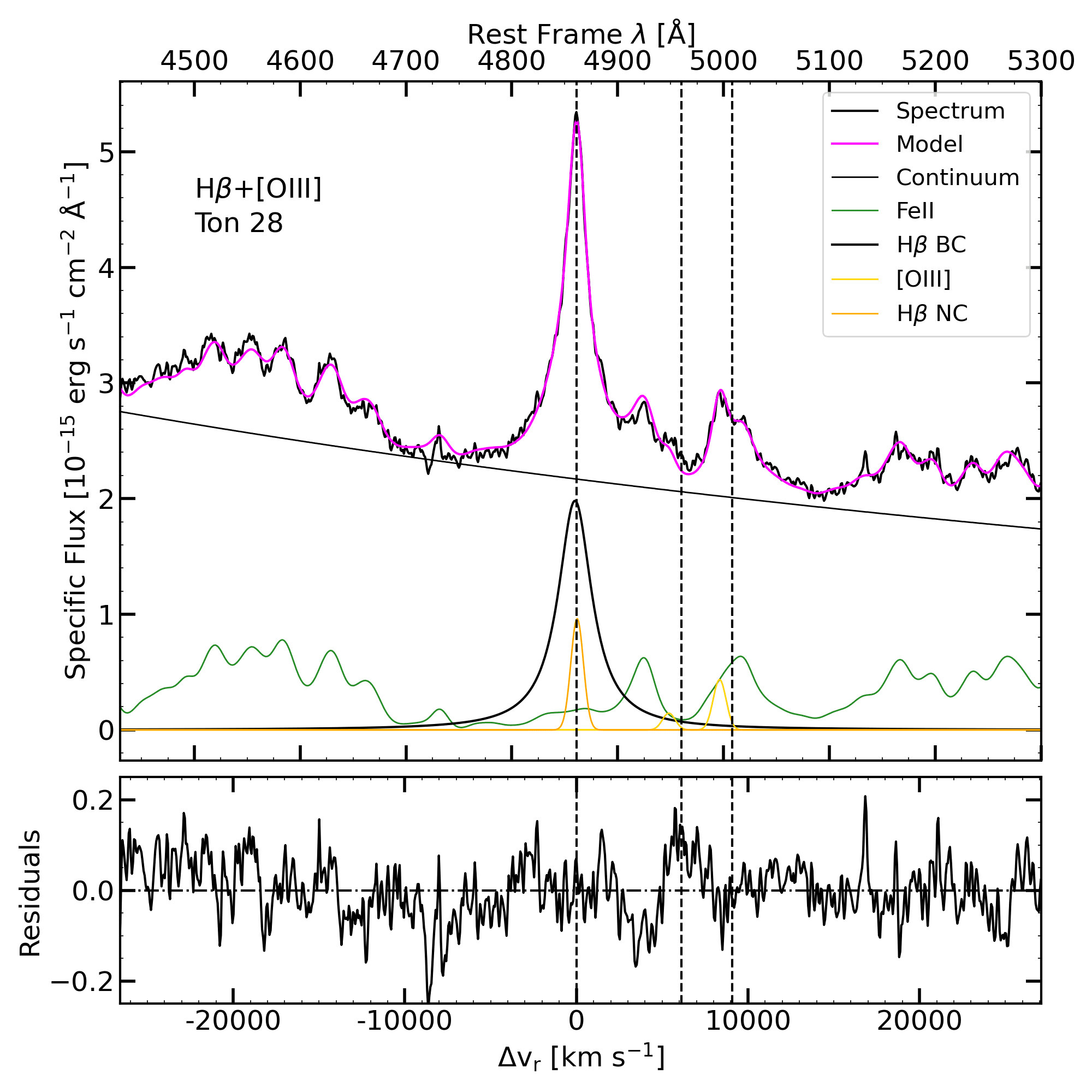}\\
\includegraphics[width=4.53cm]{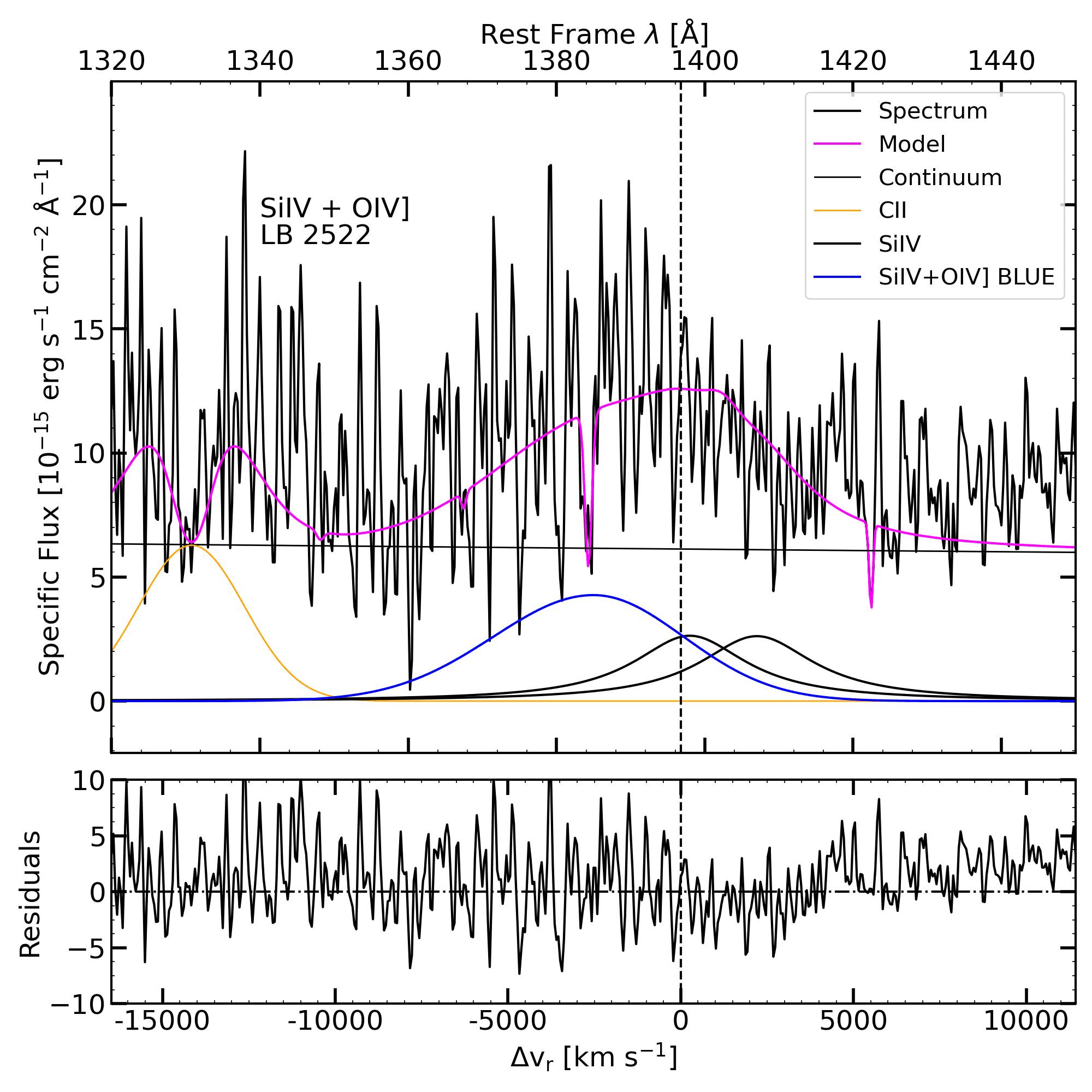}
\includegraphics[width=4.53cm]{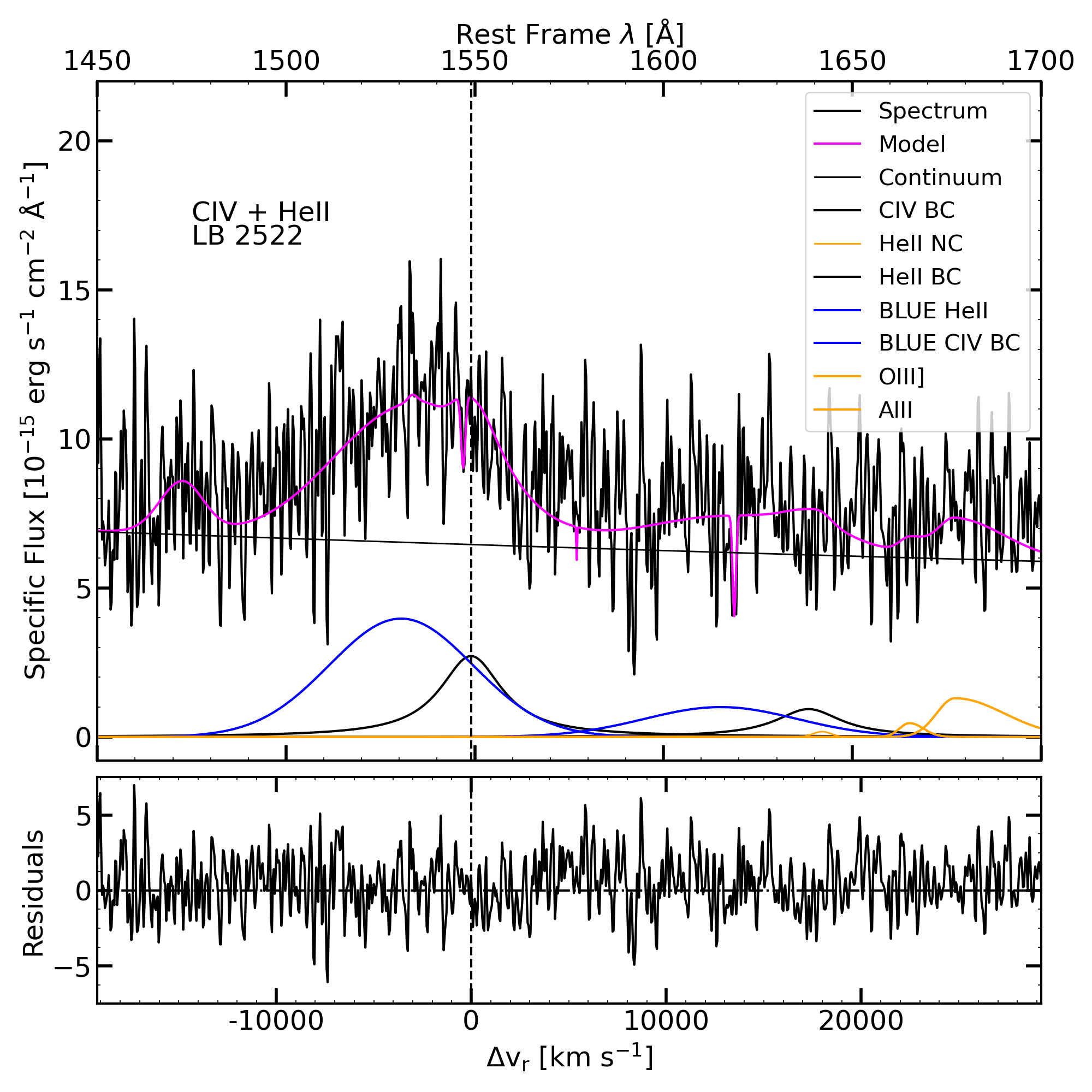}
\includegraphics[width=4.53cm]{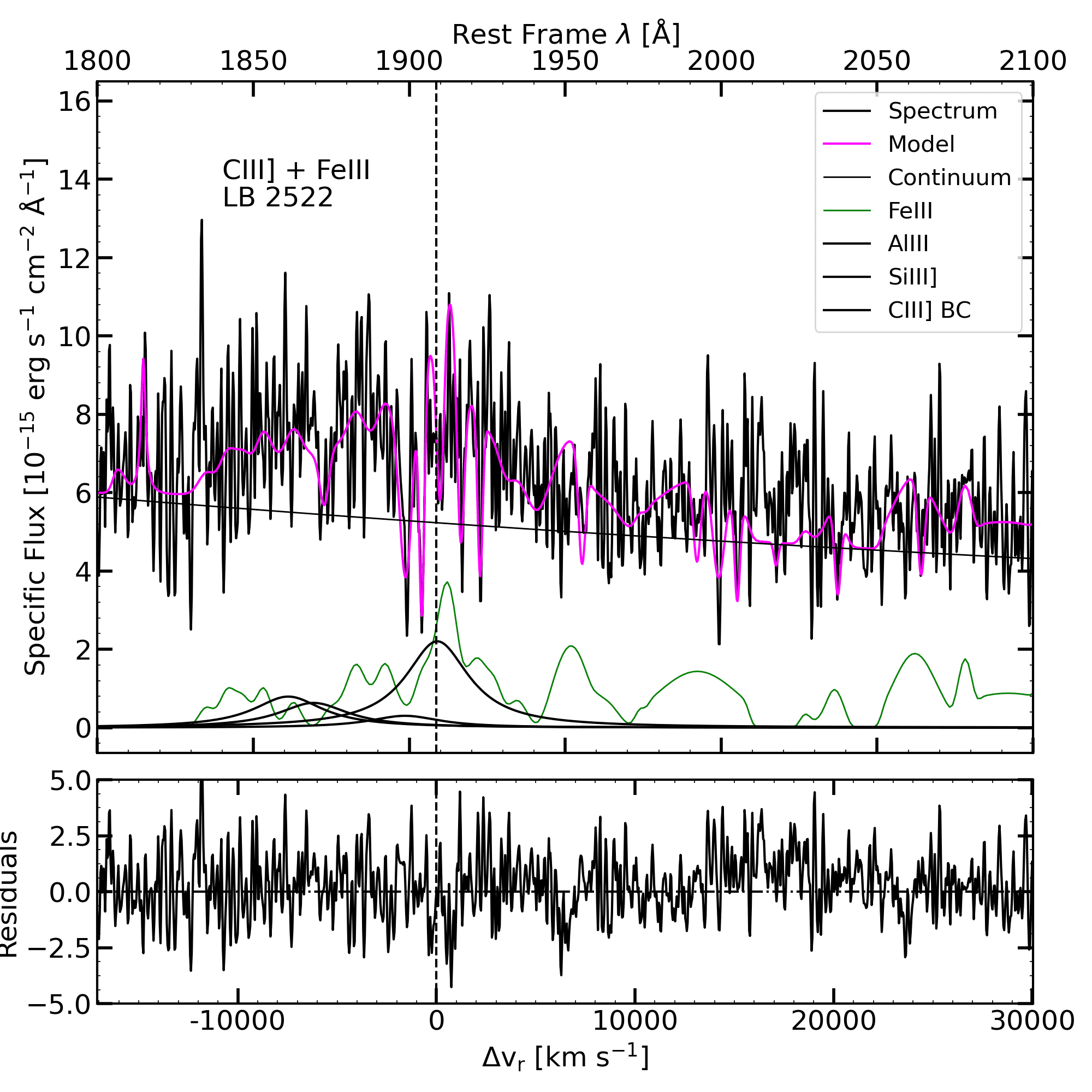}
\includegraphics[width=4.53cm]{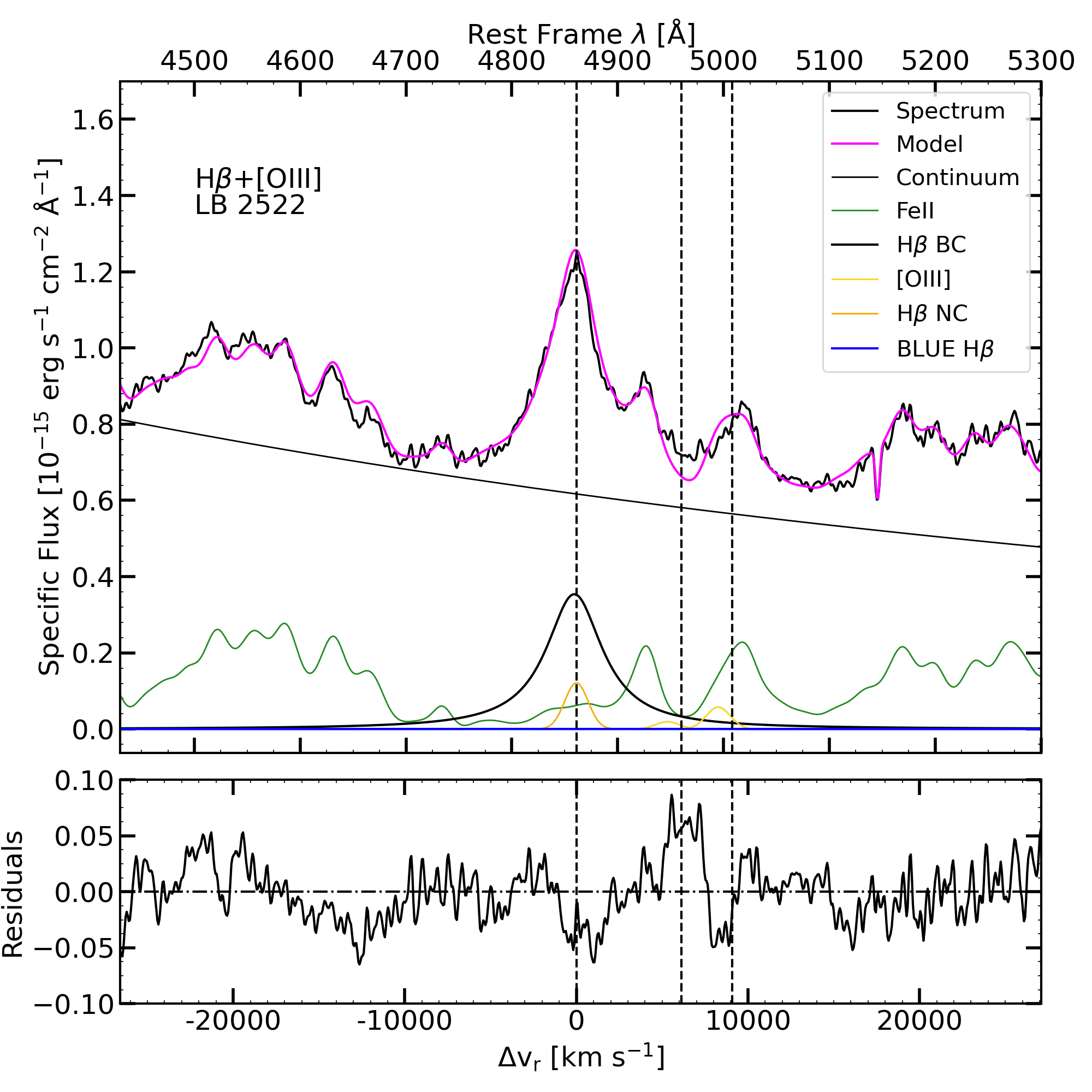}\\
\includegraphics[width=4.53cm]{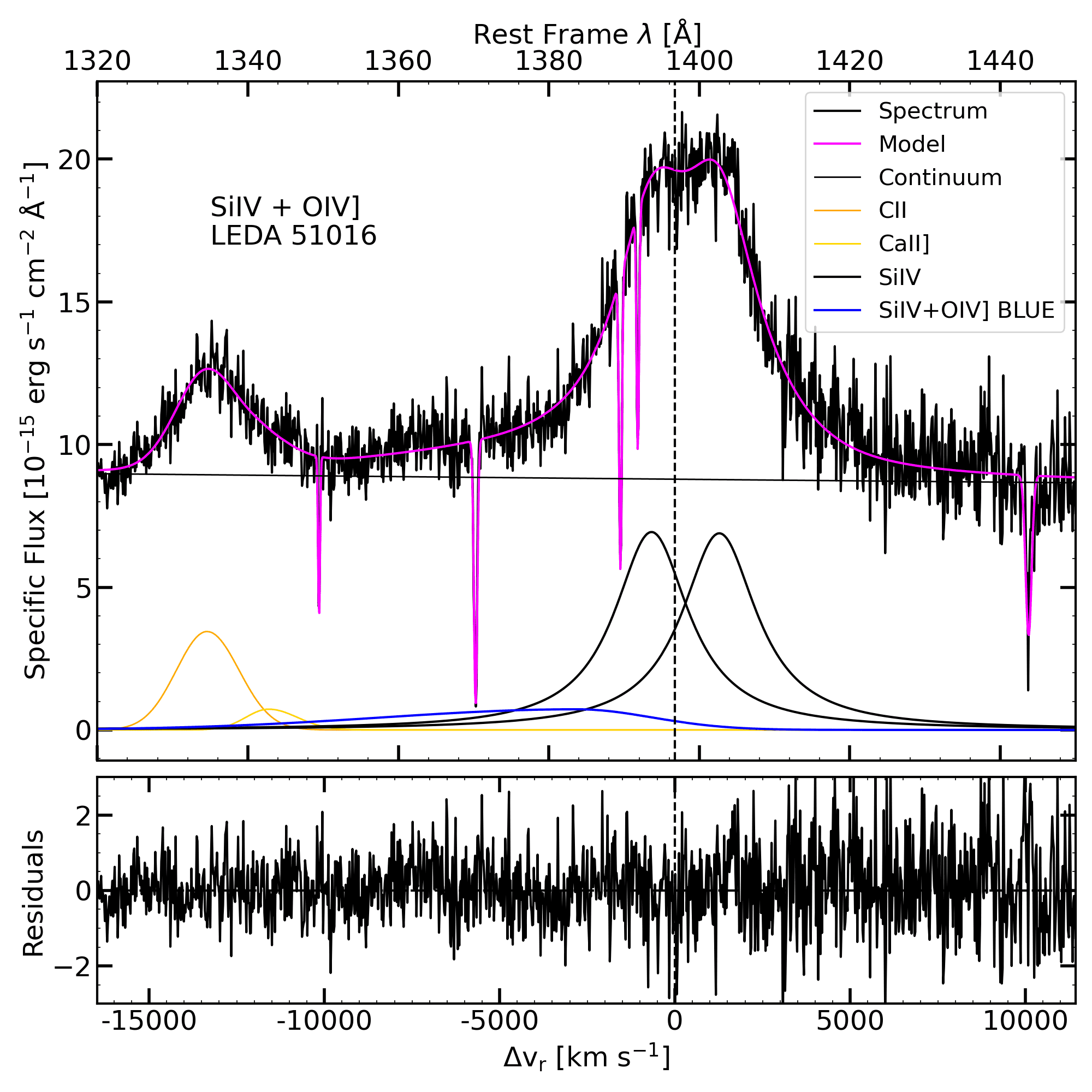}
\includegraphics[width=4.53cm]{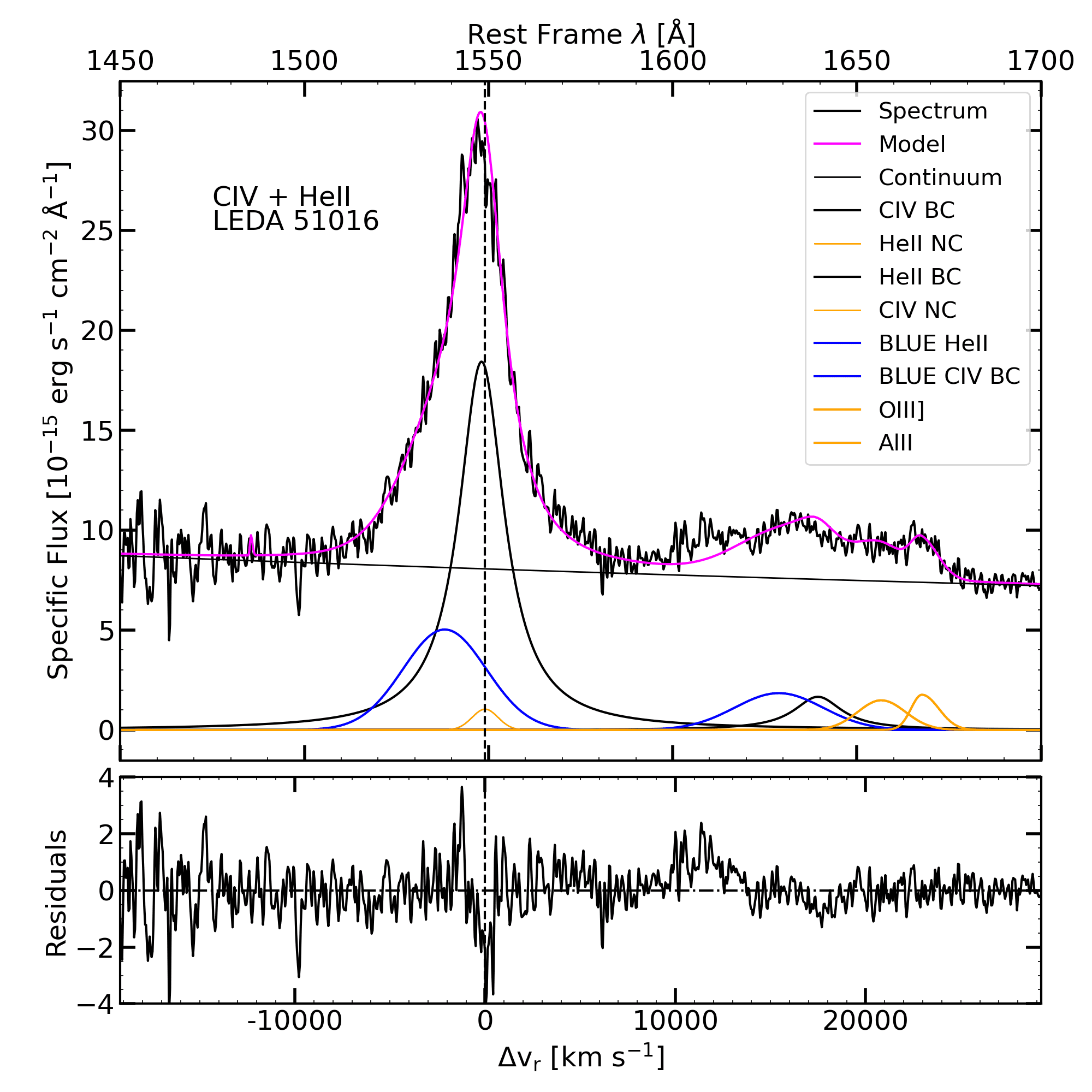}
\includegraphics[width=4.53cm]{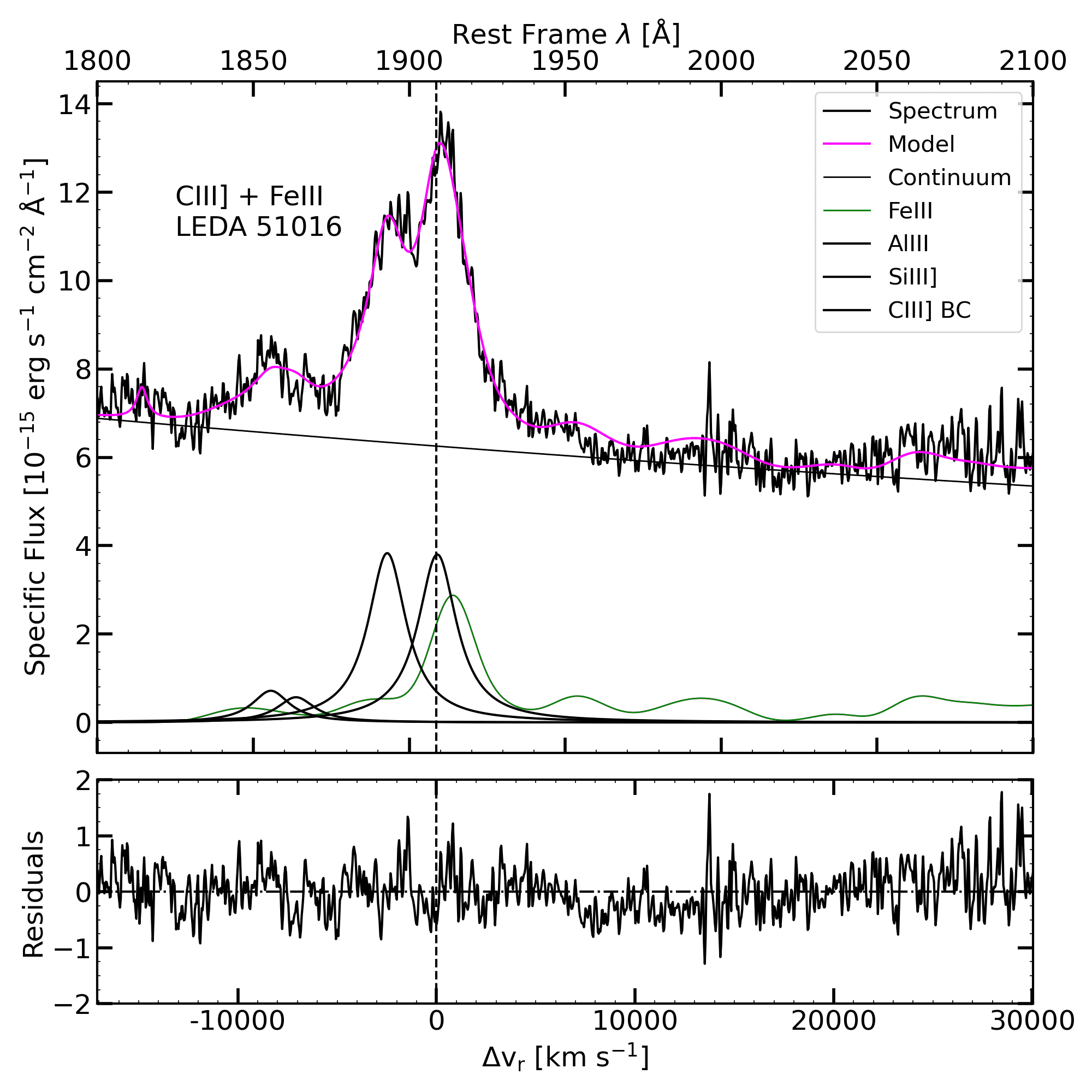}
\includegraphics[width=4.53cm]{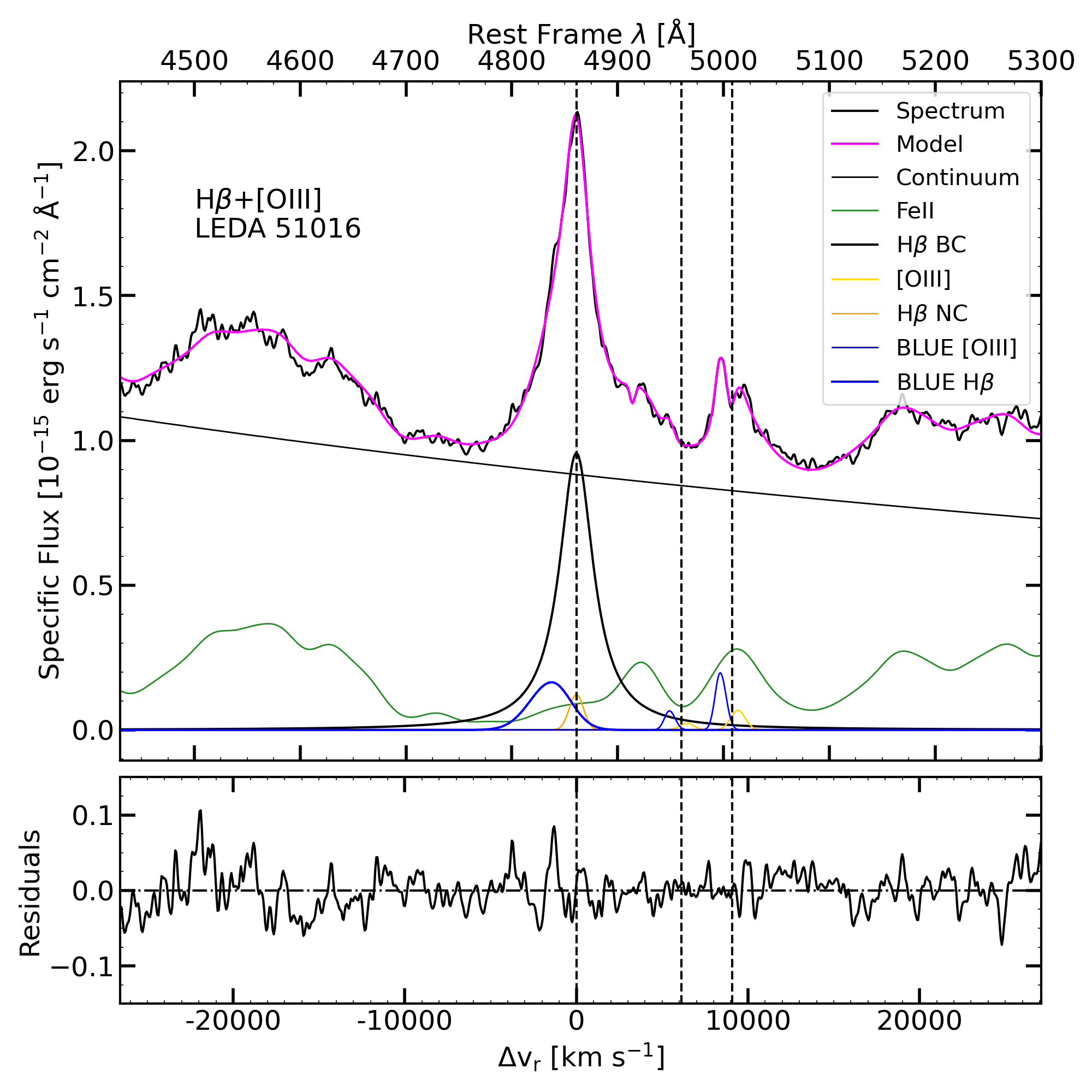}\\
\caption{Same as in Figure \ref{fig:fits1} but for Mrk 478, Ton 28, LB 2522 and LEDA 51016, respectively.}
\label{fig:fits2}
\end{figure*}

\begin{figure*}[ht!]
\centering
\includegraphics[width=4.53cm]{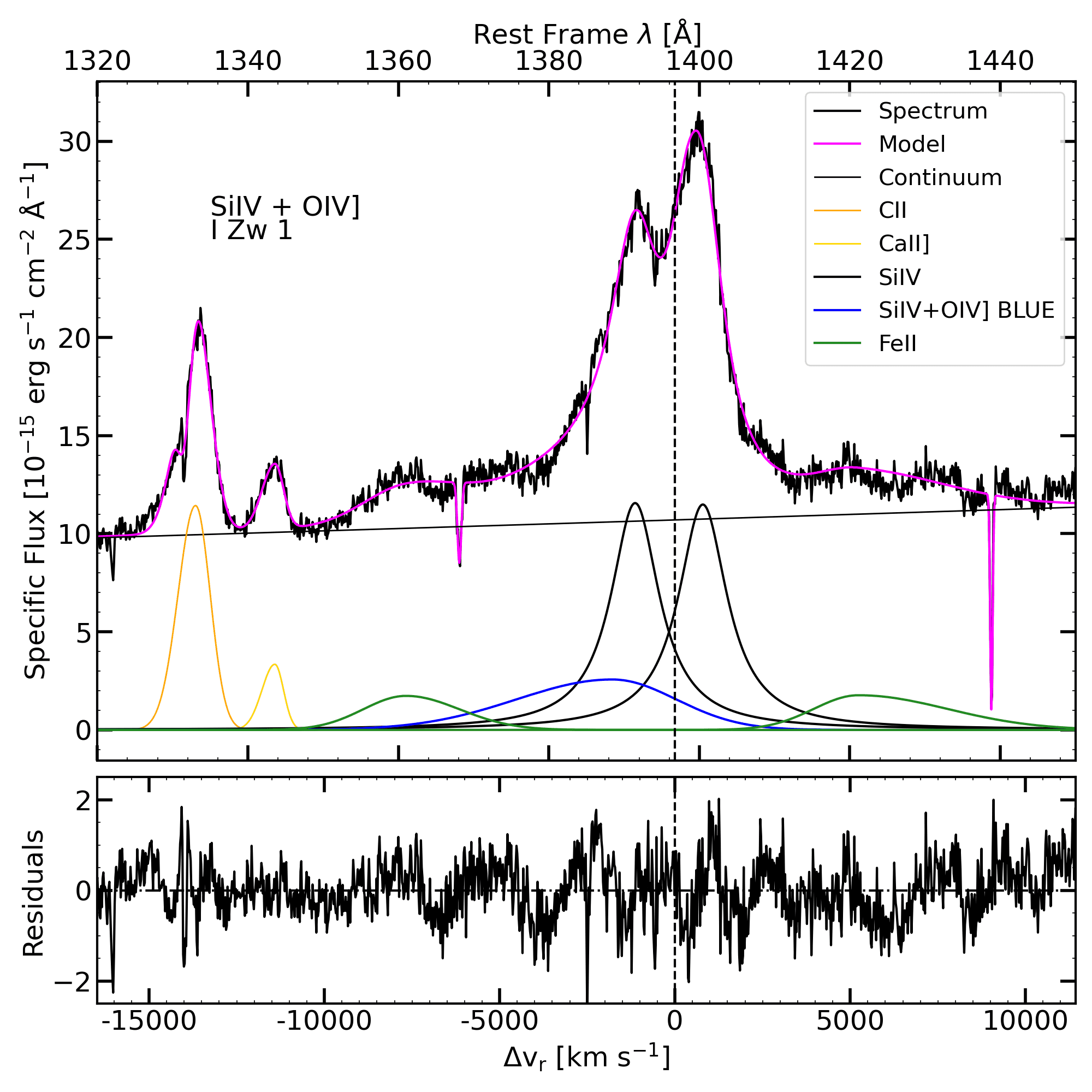}
\includegraphics[width=4.53cm]{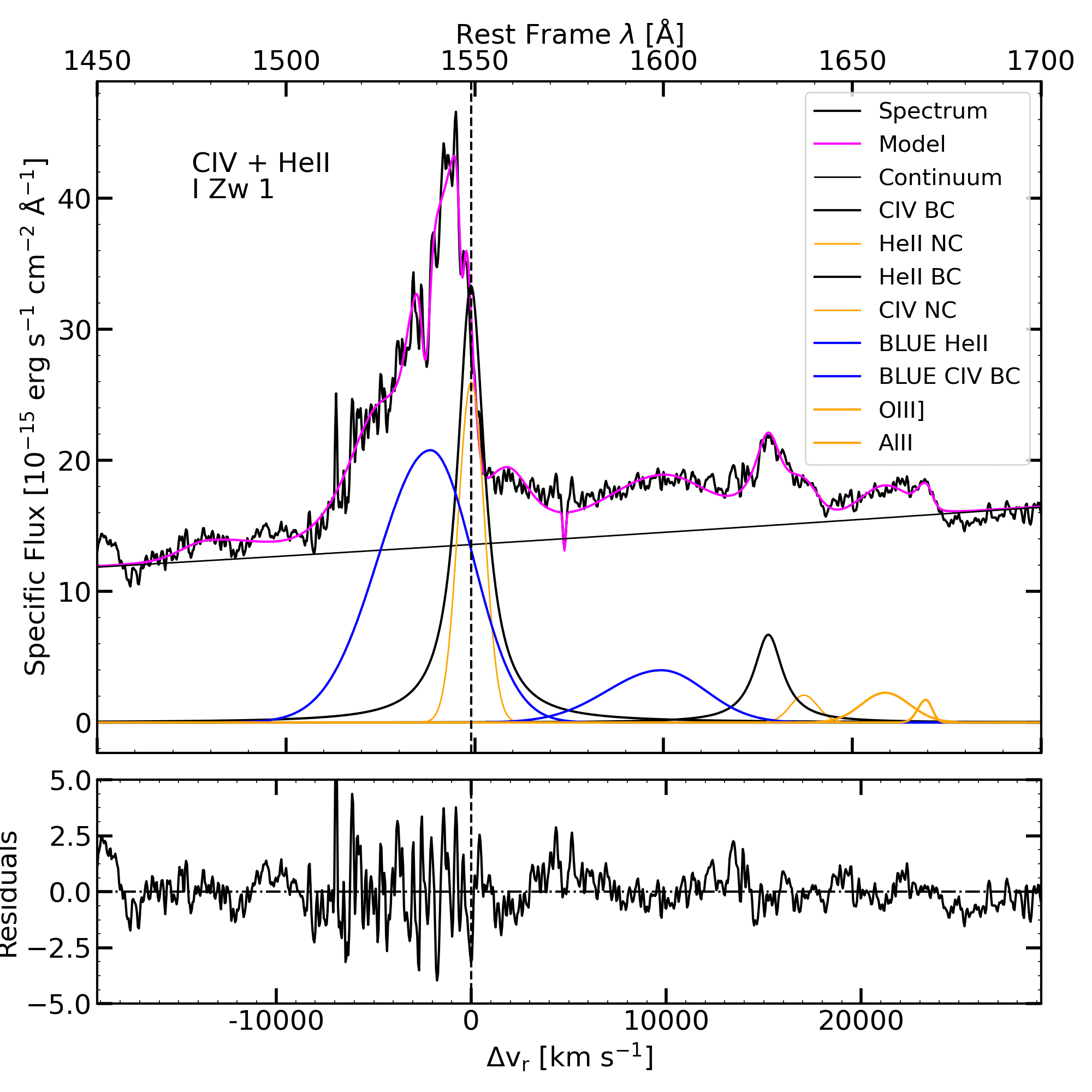}
\includegraphics[width=4.53cm]{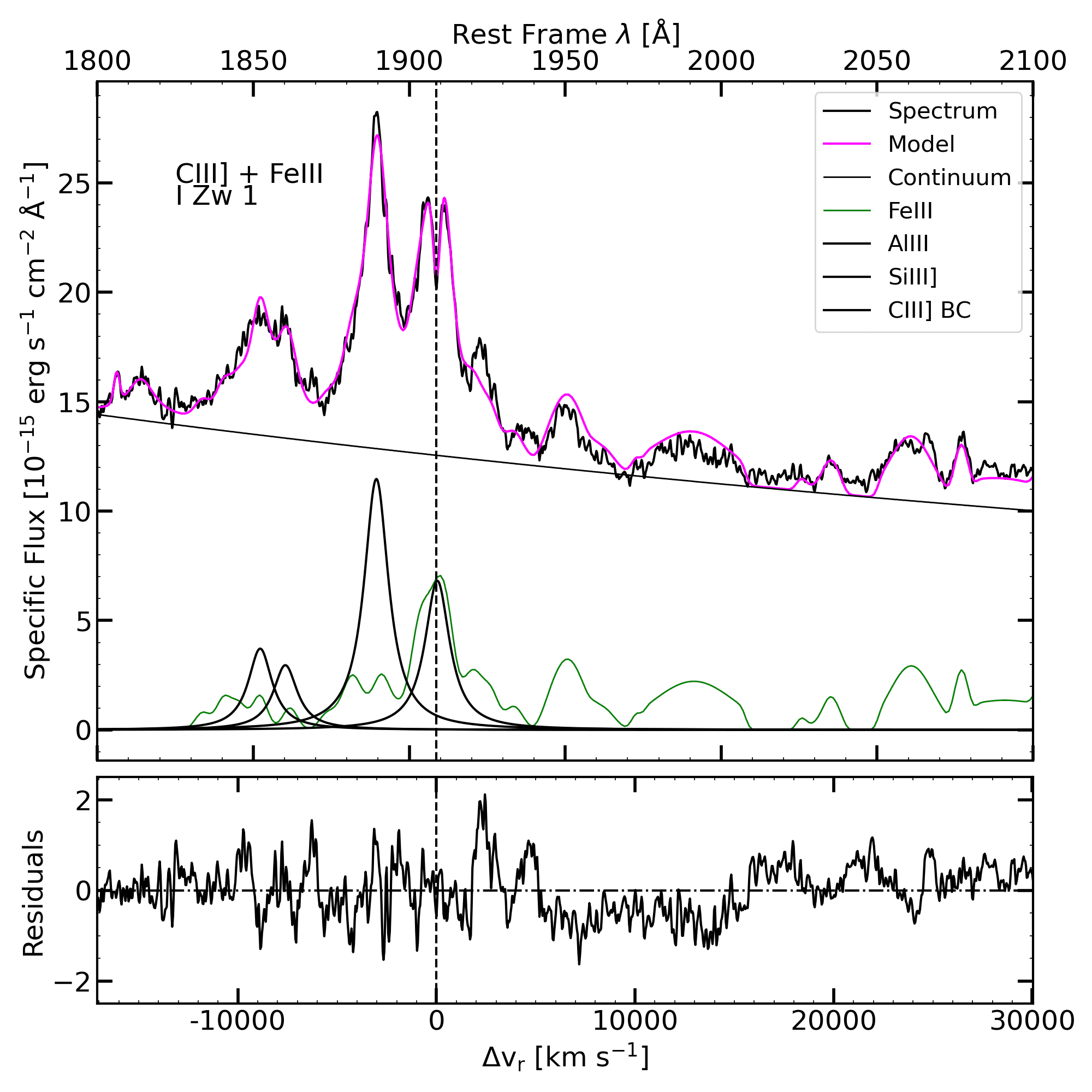}
\includegraphics[width=4.53cm]{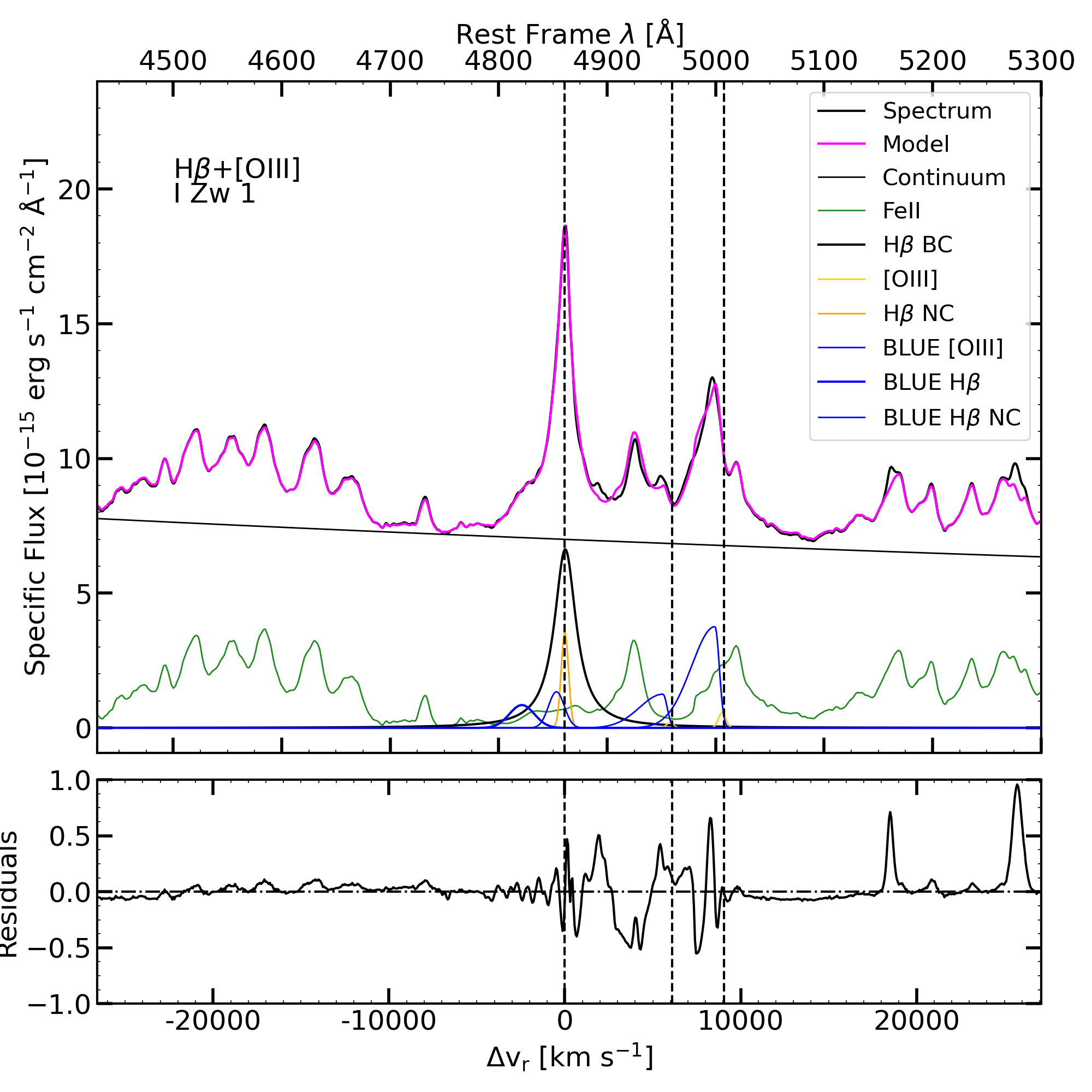}\\
\includegraphics[width=4.53cm]{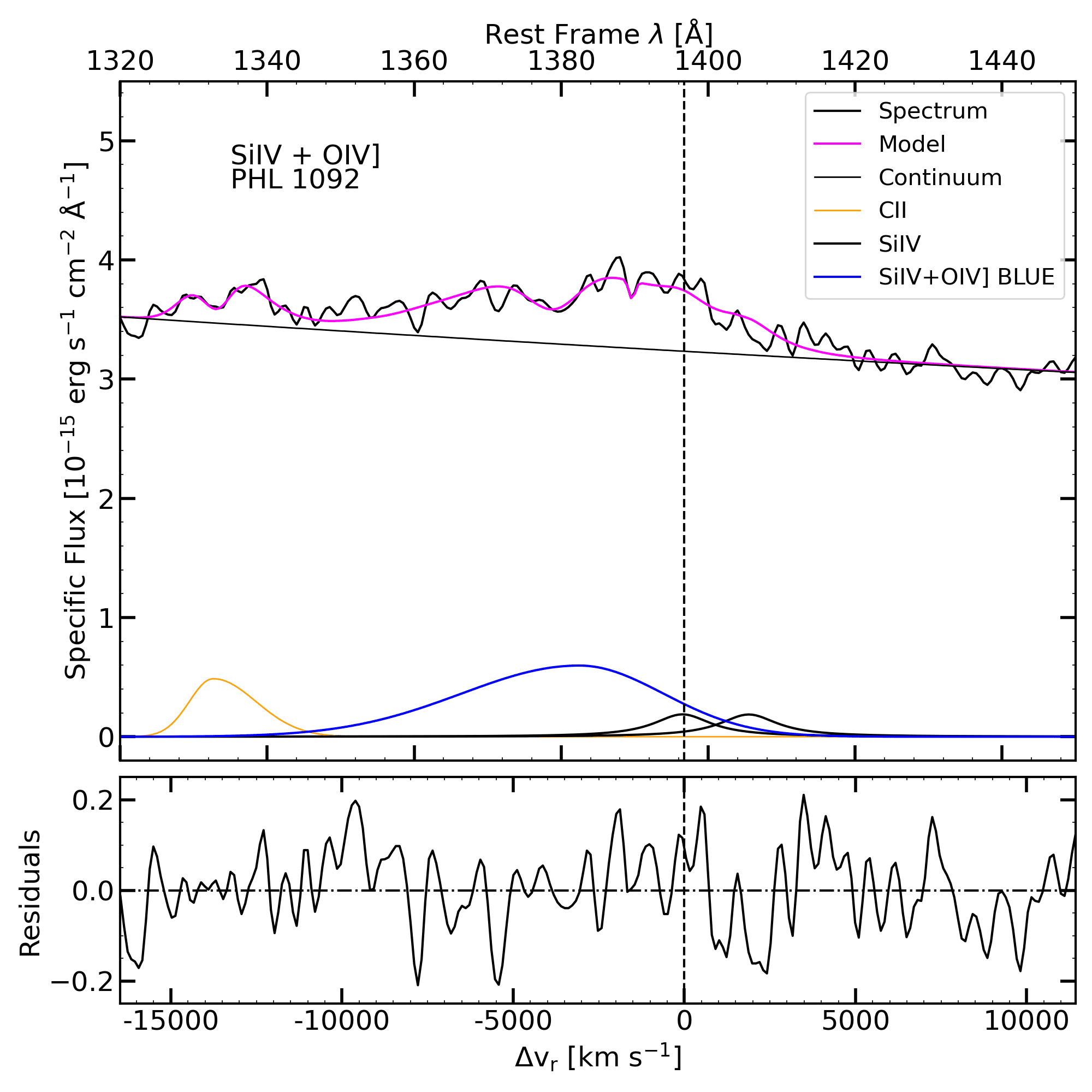}
\includegraphics[width=4.53cm]{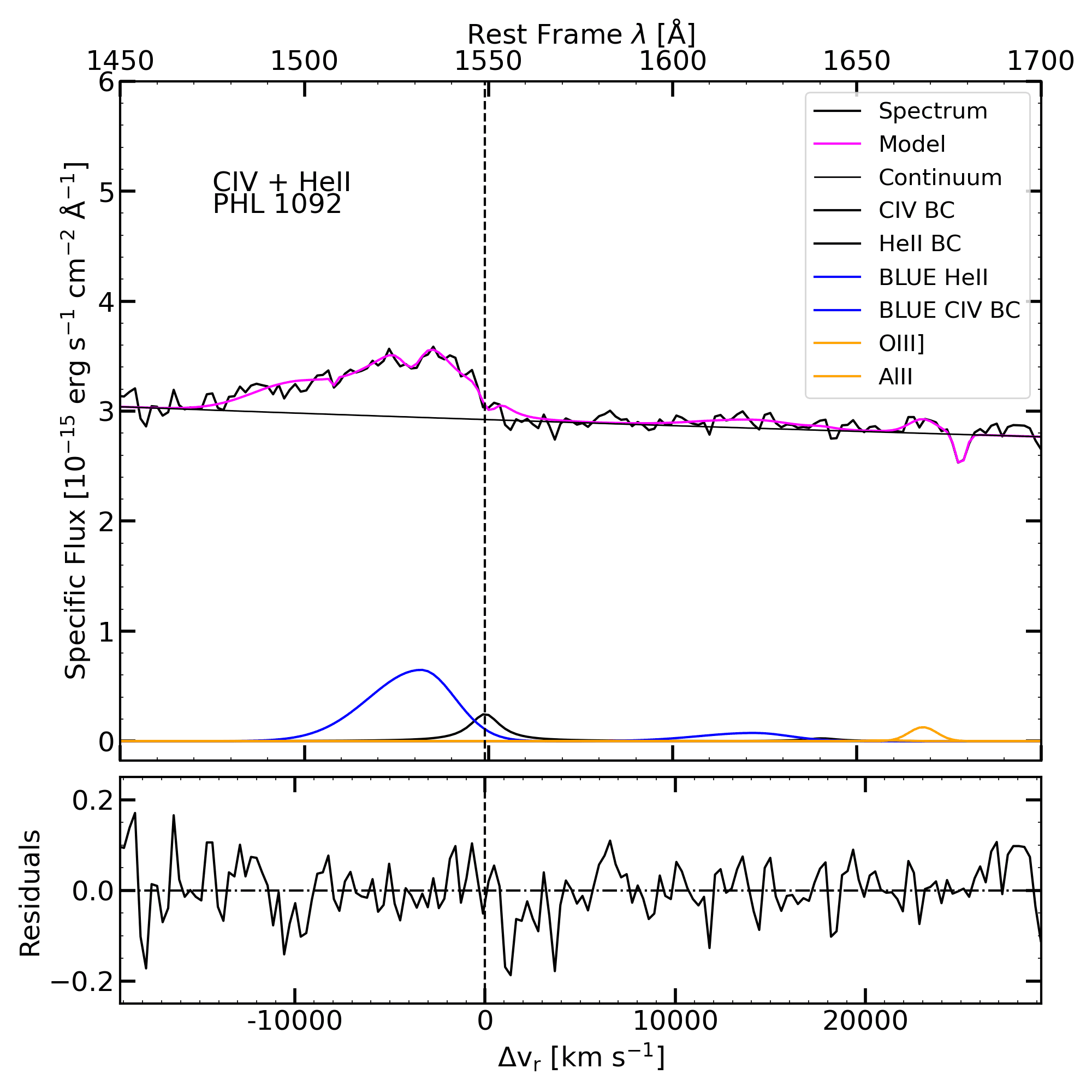}
\includegraphics[width=4.53cm]{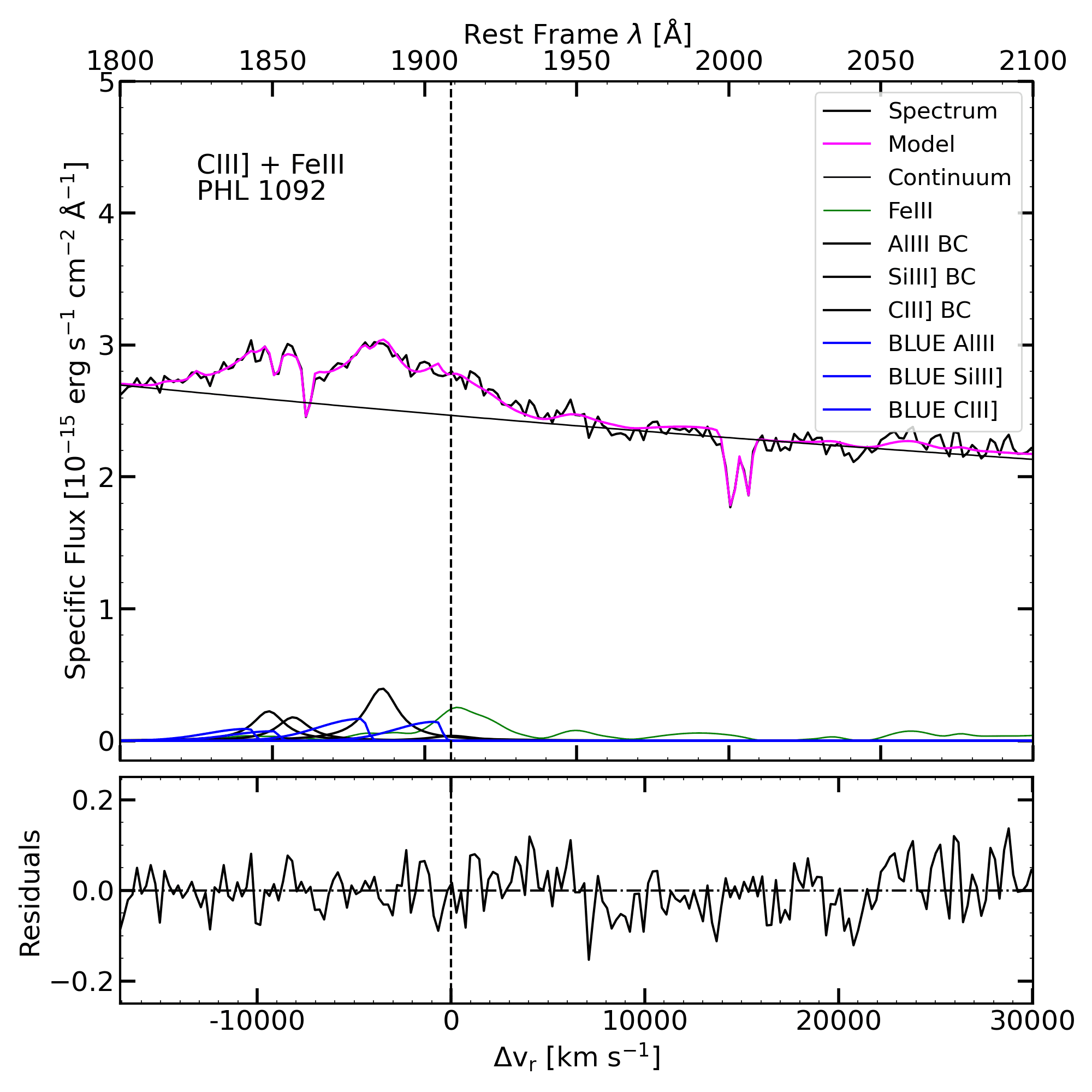}
\includegraphics[width=4.53cm]{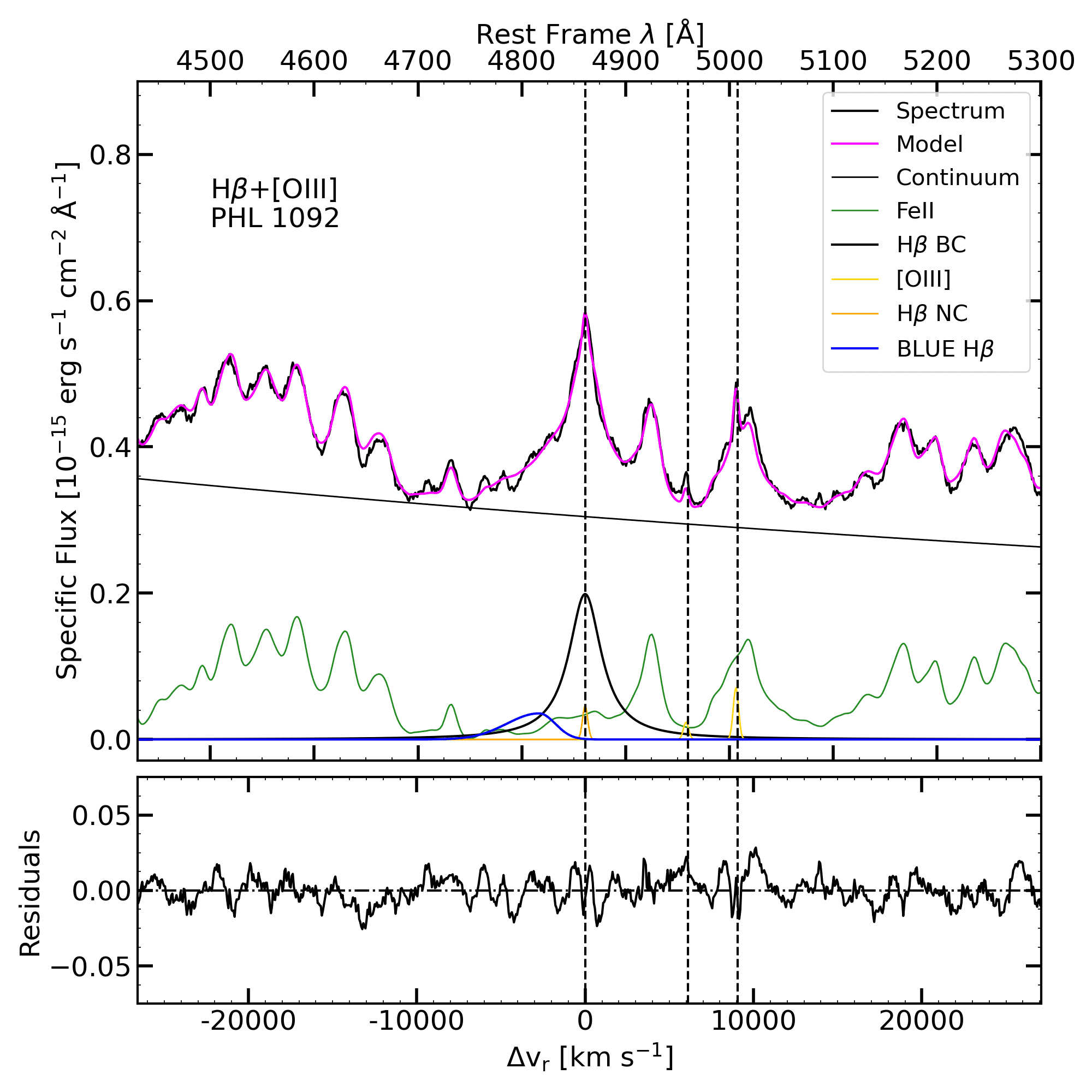}\\
\includegraphics[width=4.53cm]{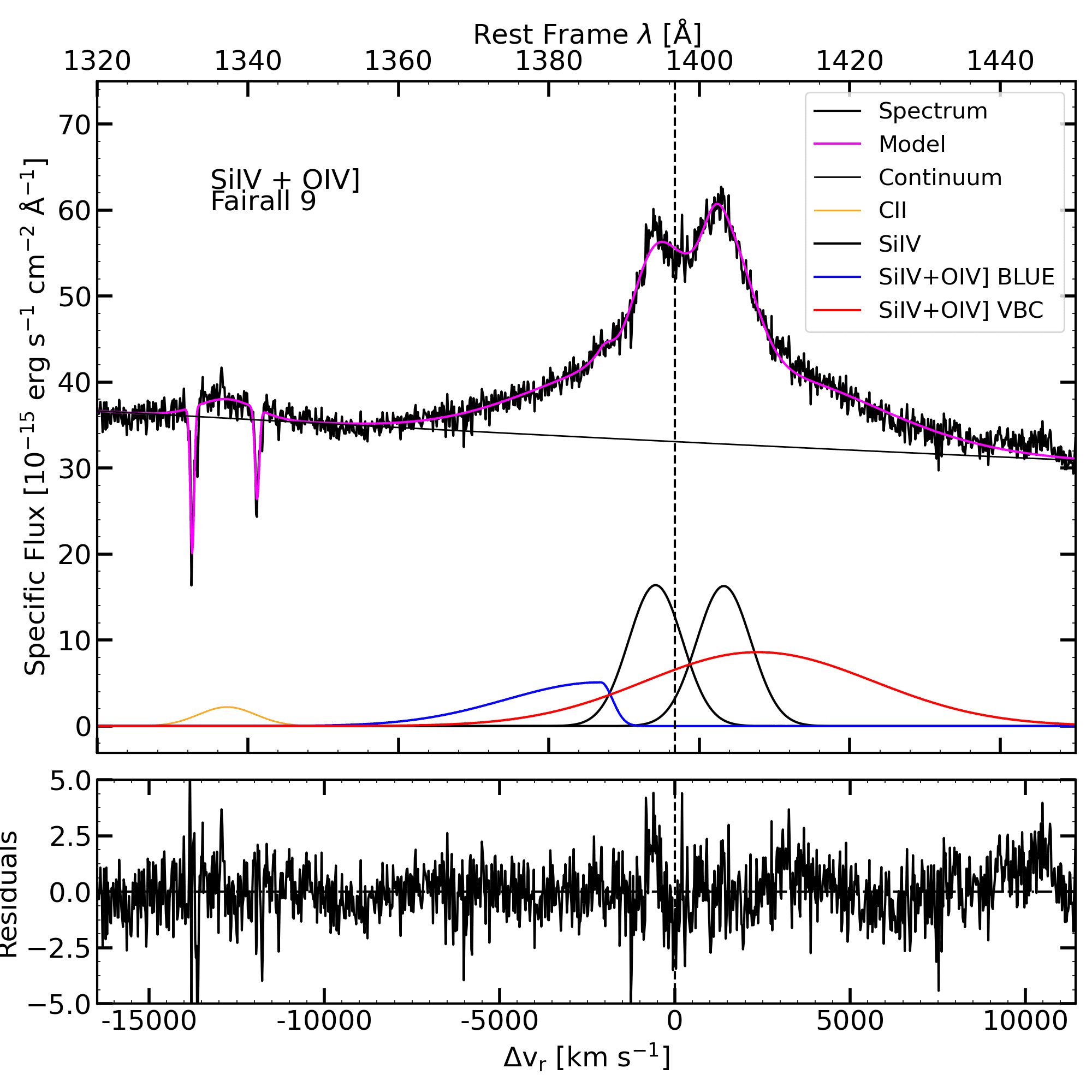}
\includegraphics[width=4.53cm]{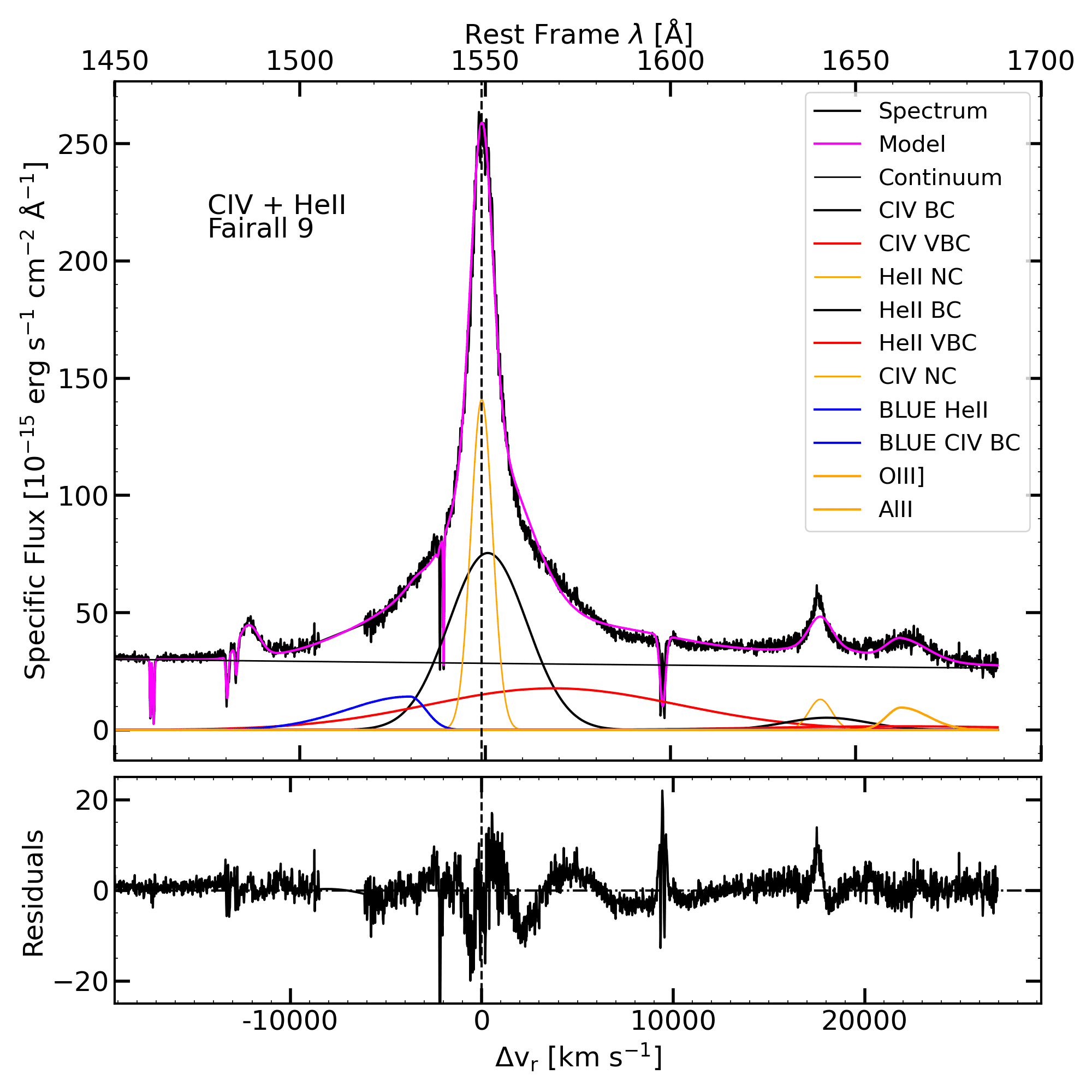}
\includegraphics[width=4.53cm]{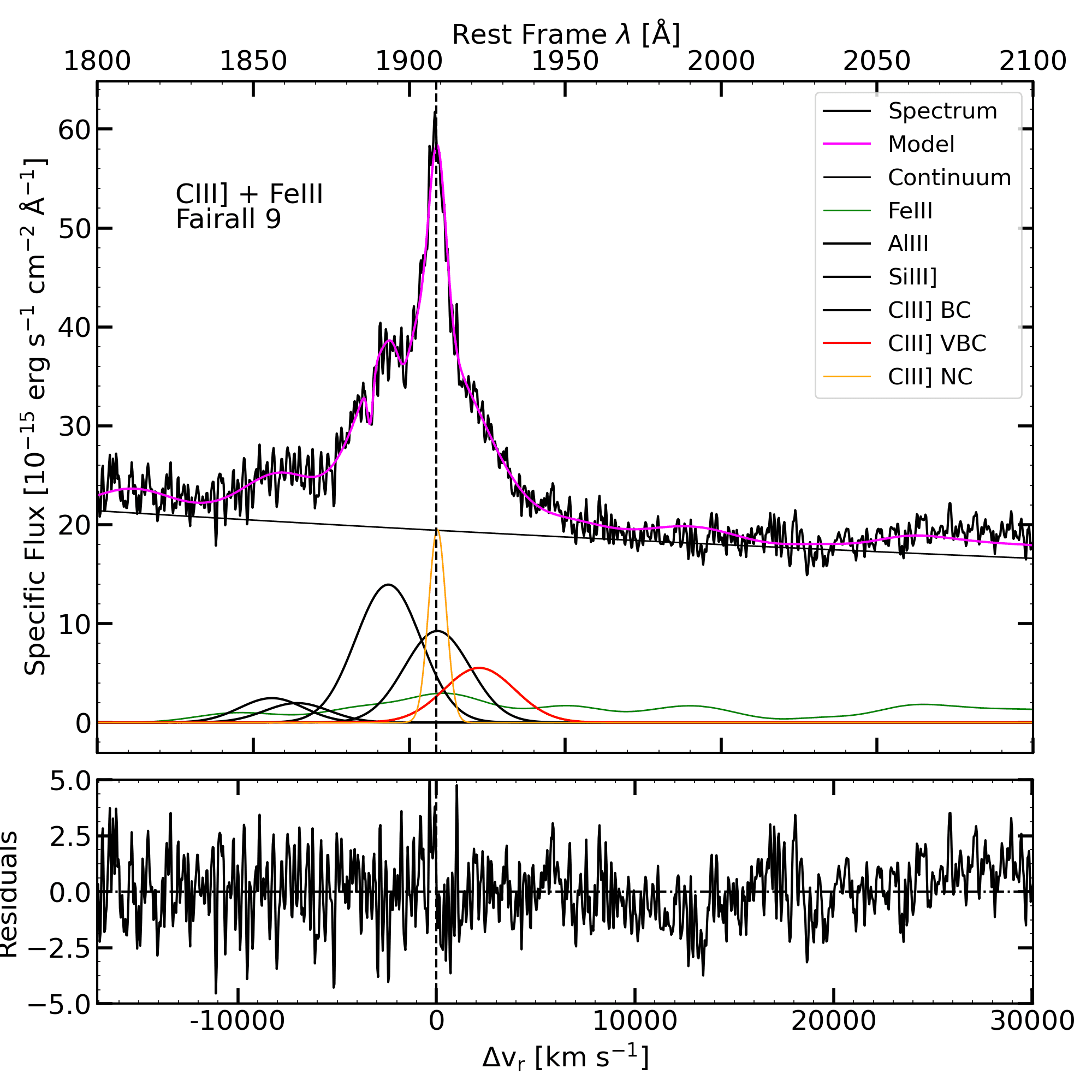}
\includegraphics[width=4.53cm]{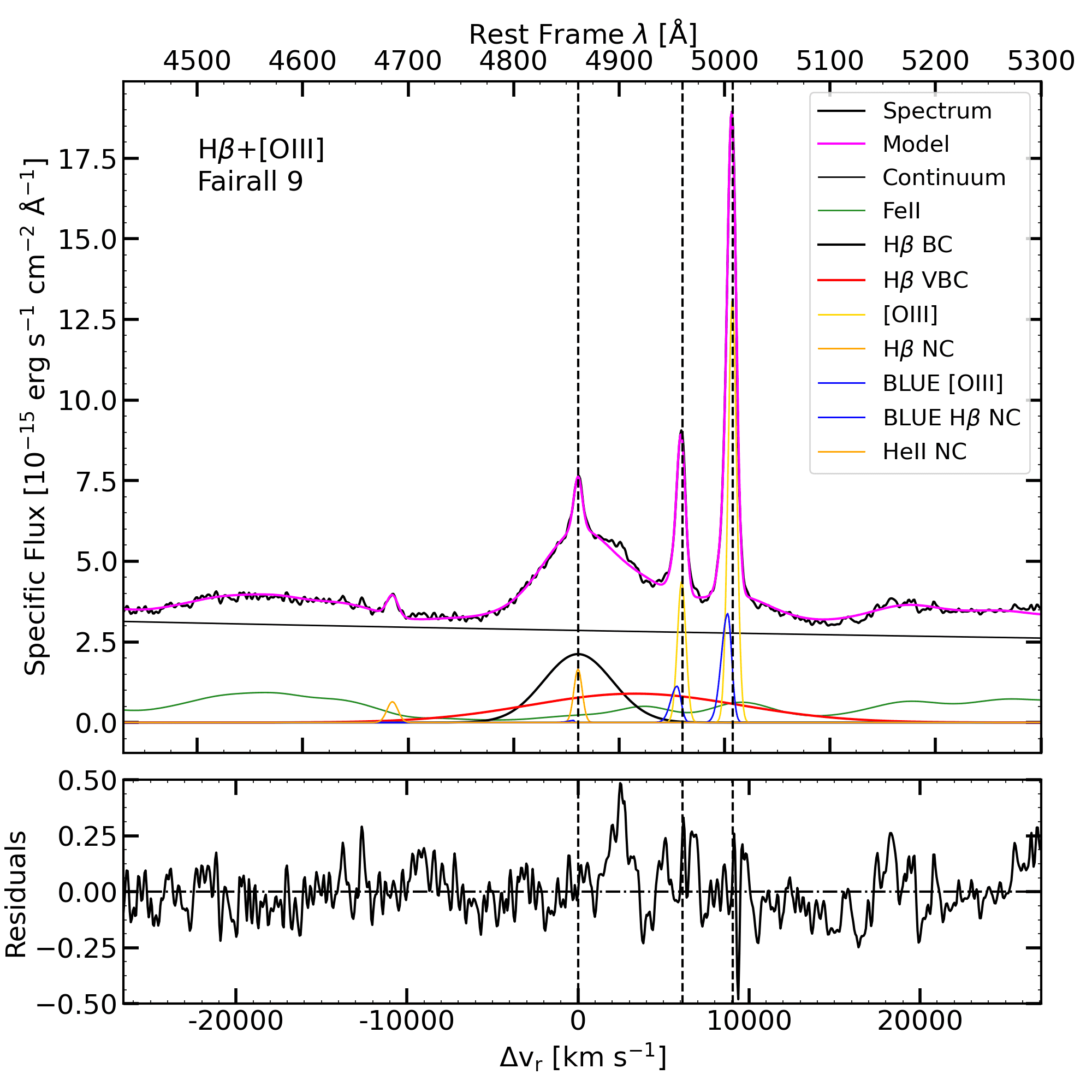}\\
\includegraphics[width=4.53cm]{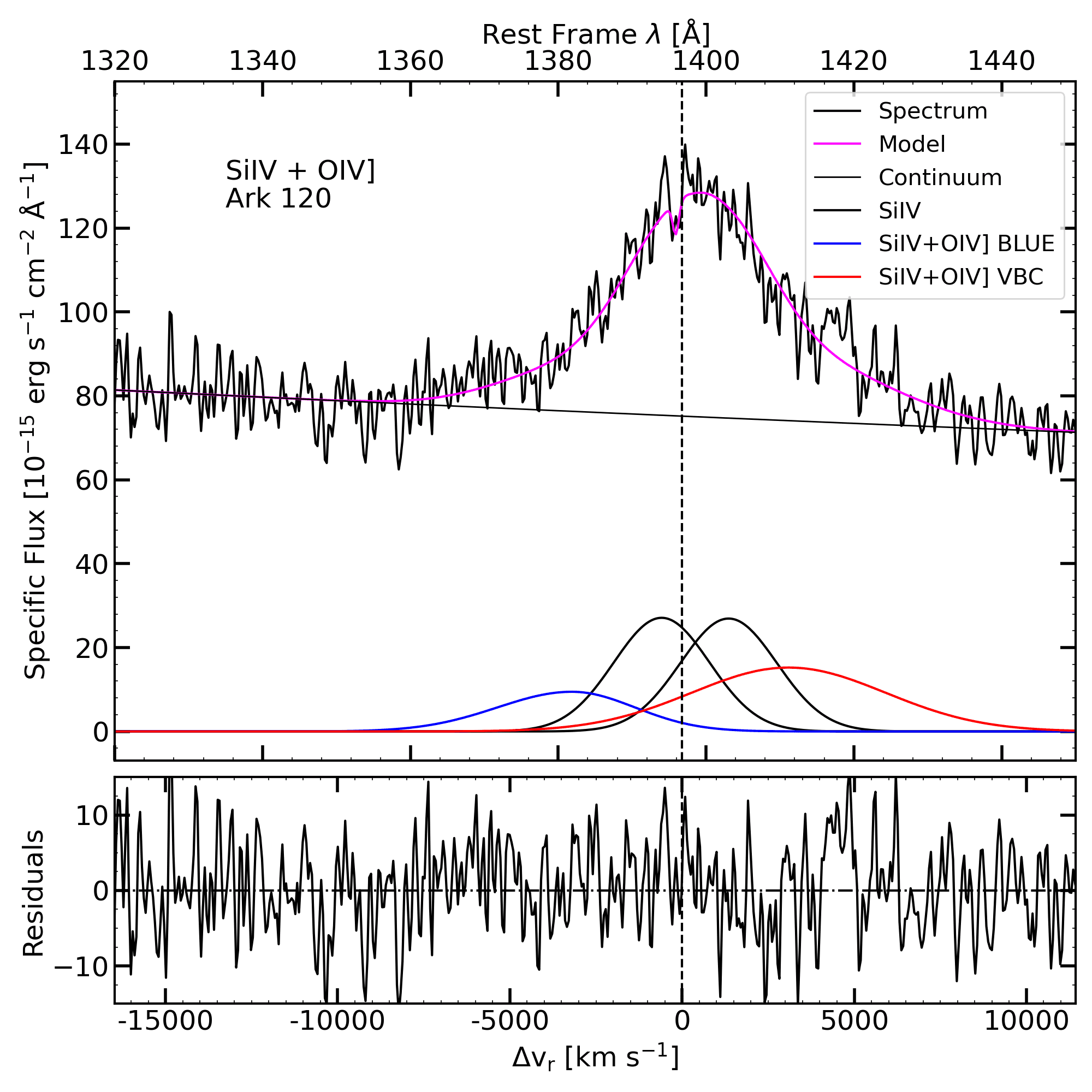}
\includegraphics[width=4.53cm]{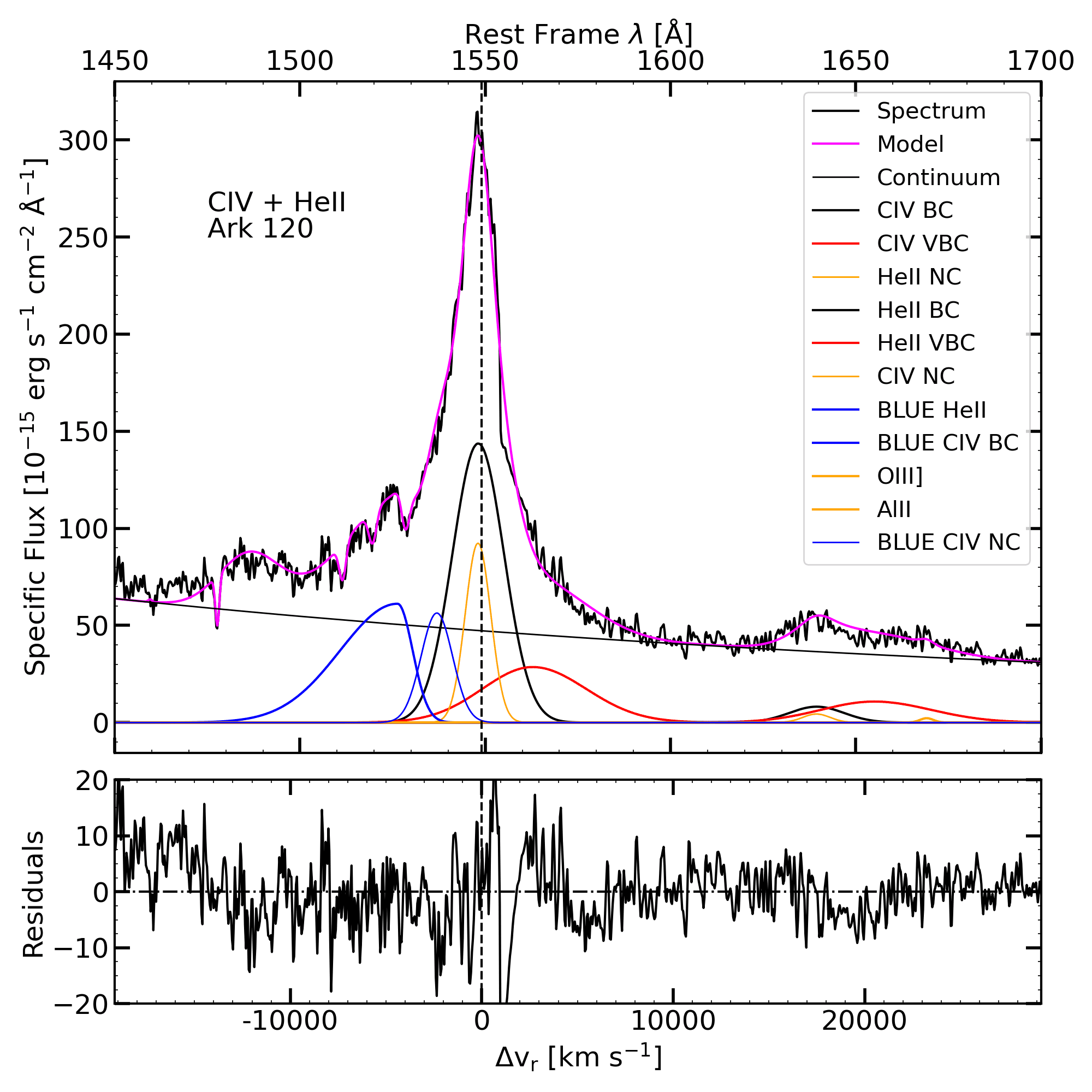}
\includegraphics[width=4.53cm]{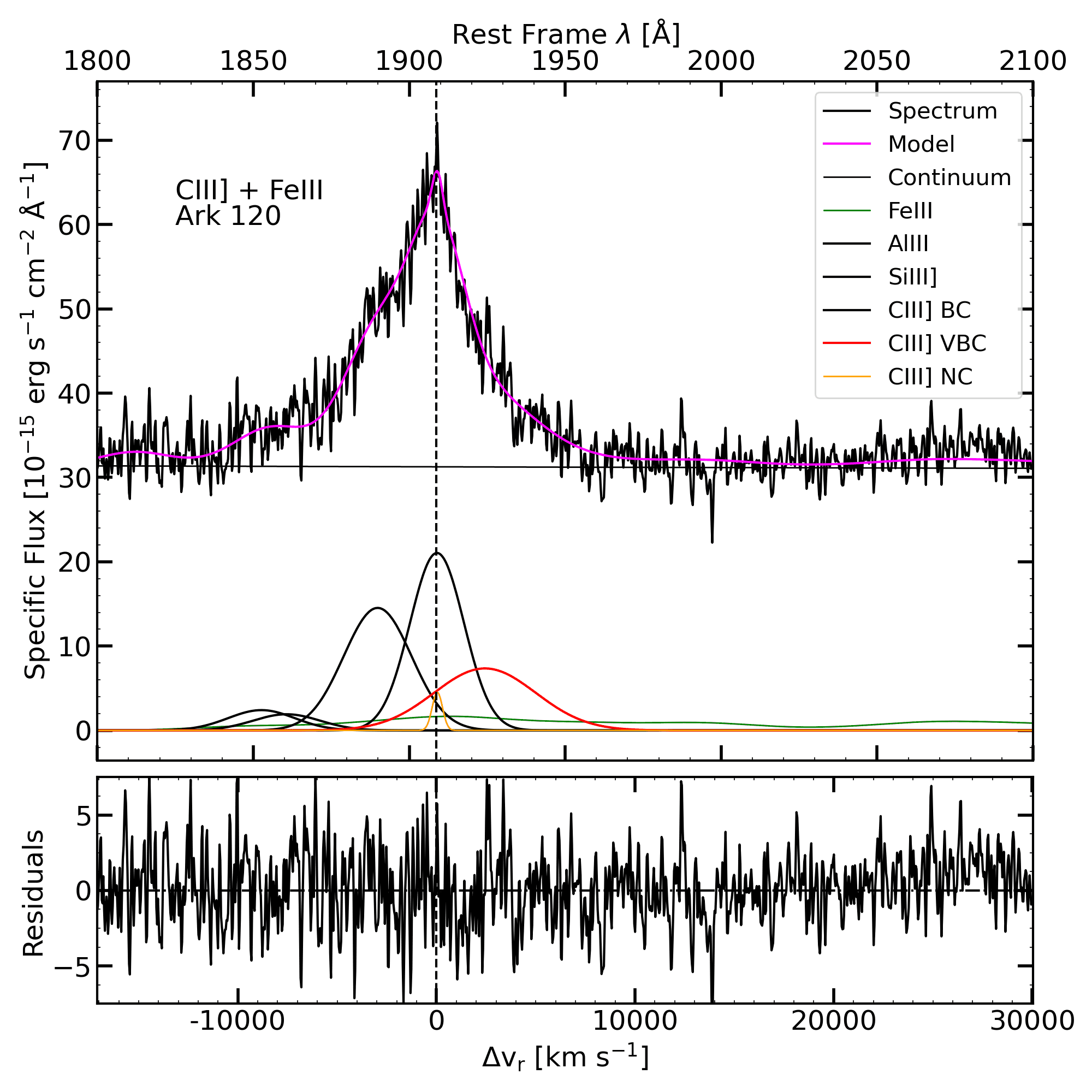}
\includegraphics[width=4.53cm]{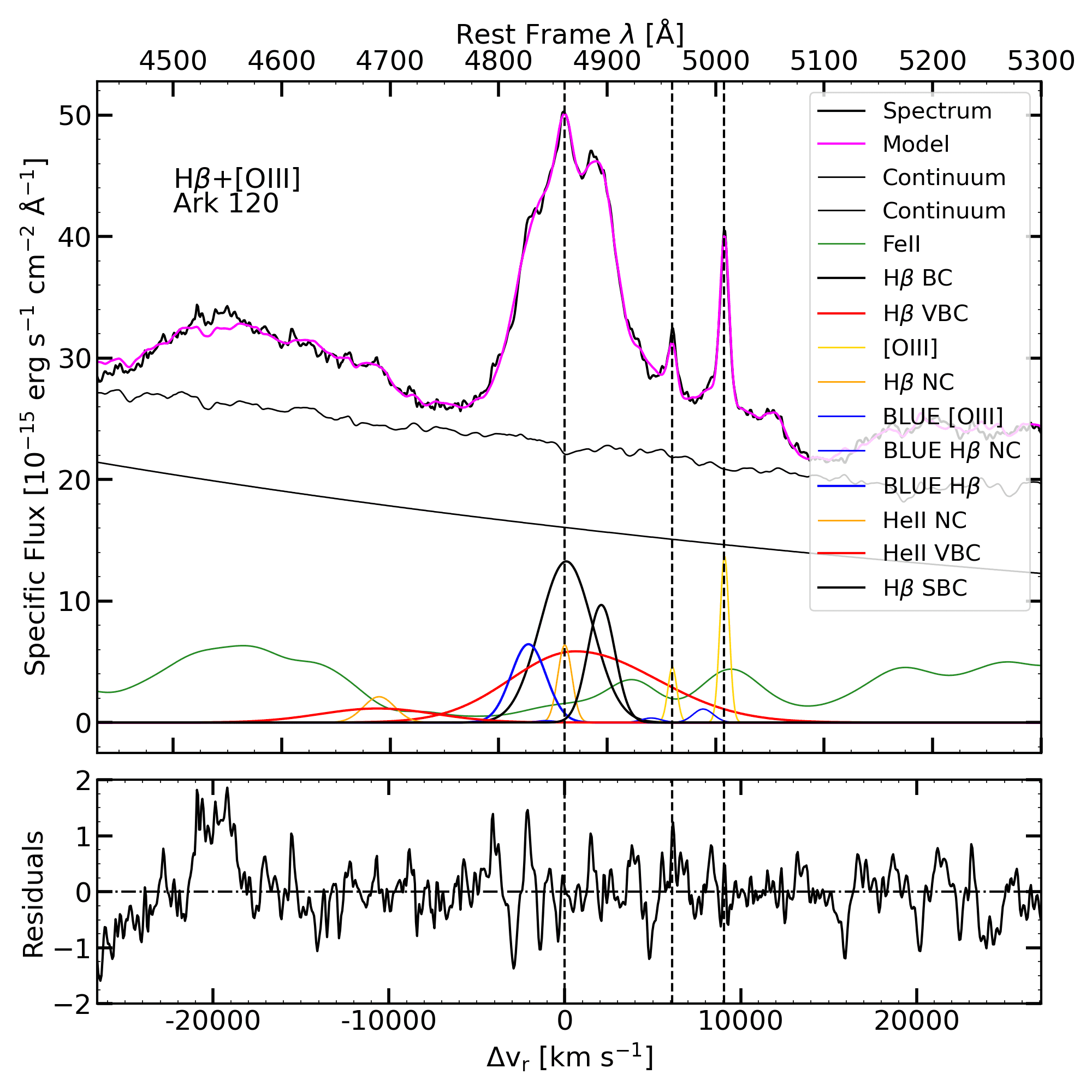}\\

\caption{Same as in Figure \ref{fig:fits1} but for I Zw 1, PHL 1092, Fairall 9 and Ark 120, respectively.}
\label{fig:fits3}
\end{figure*}

\begin{figure*}[ht!]
\centering
\includegraphics[width=4.53cm]{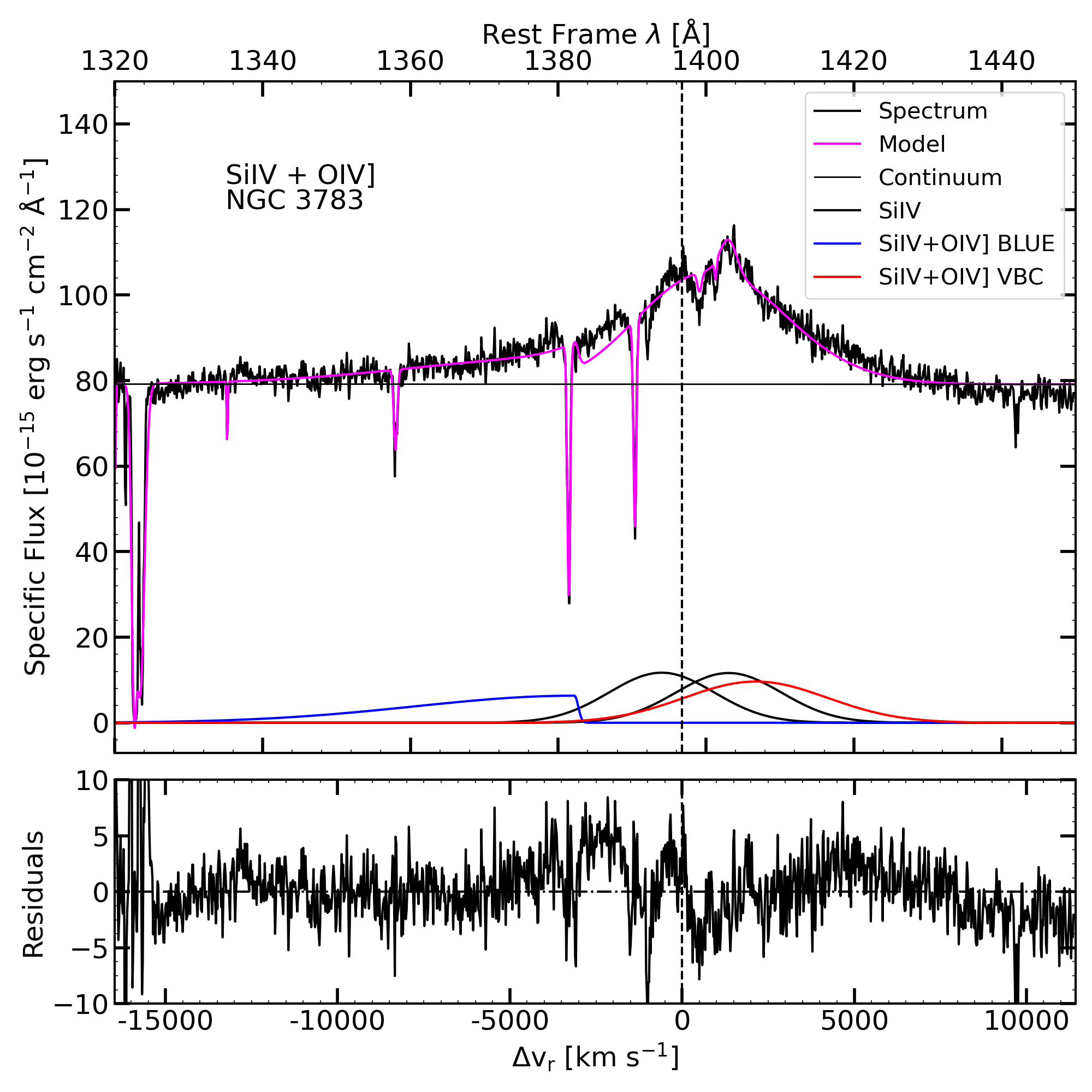}
\includegraphics[width=4.53cm]{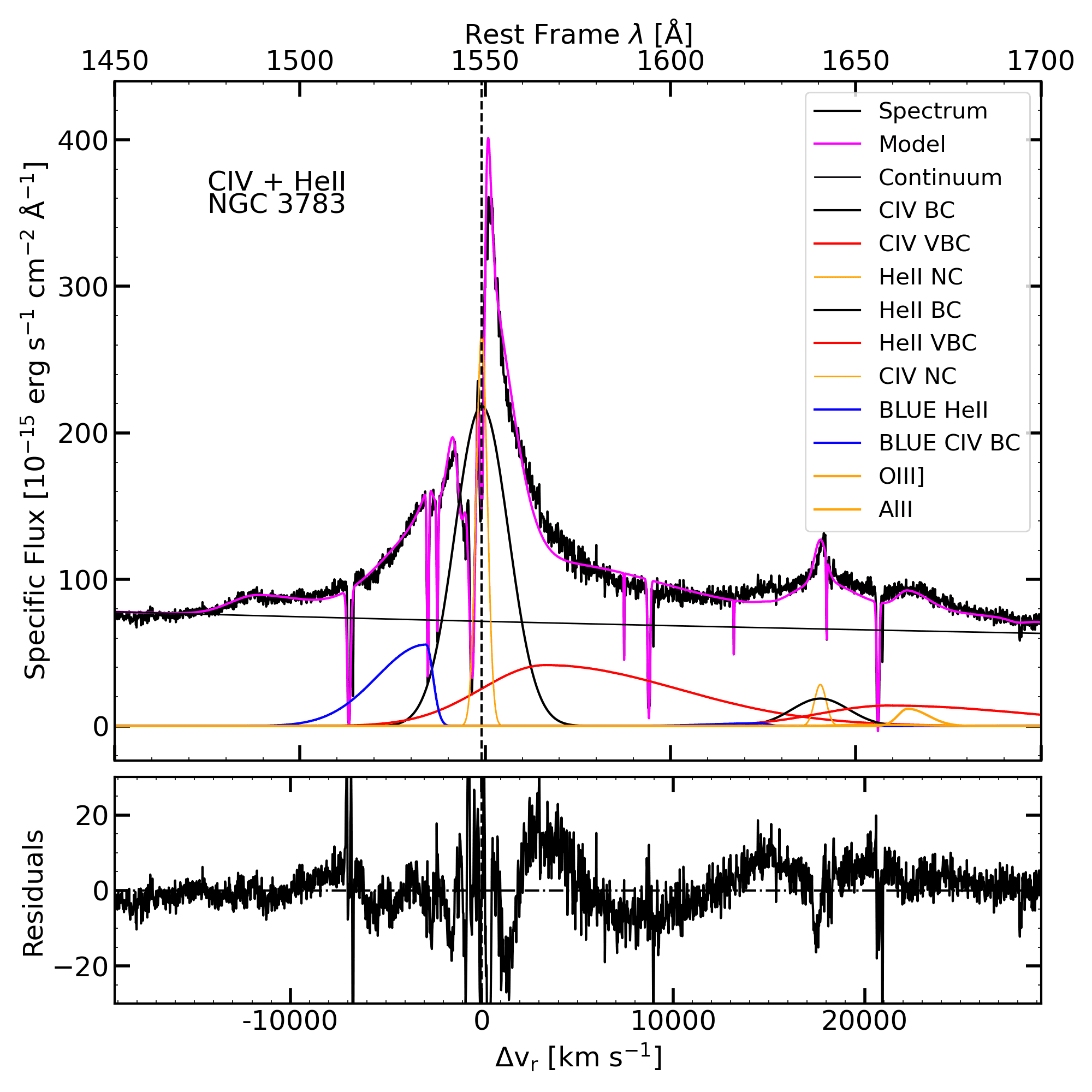}
\includegraphics[width=4.53cm]{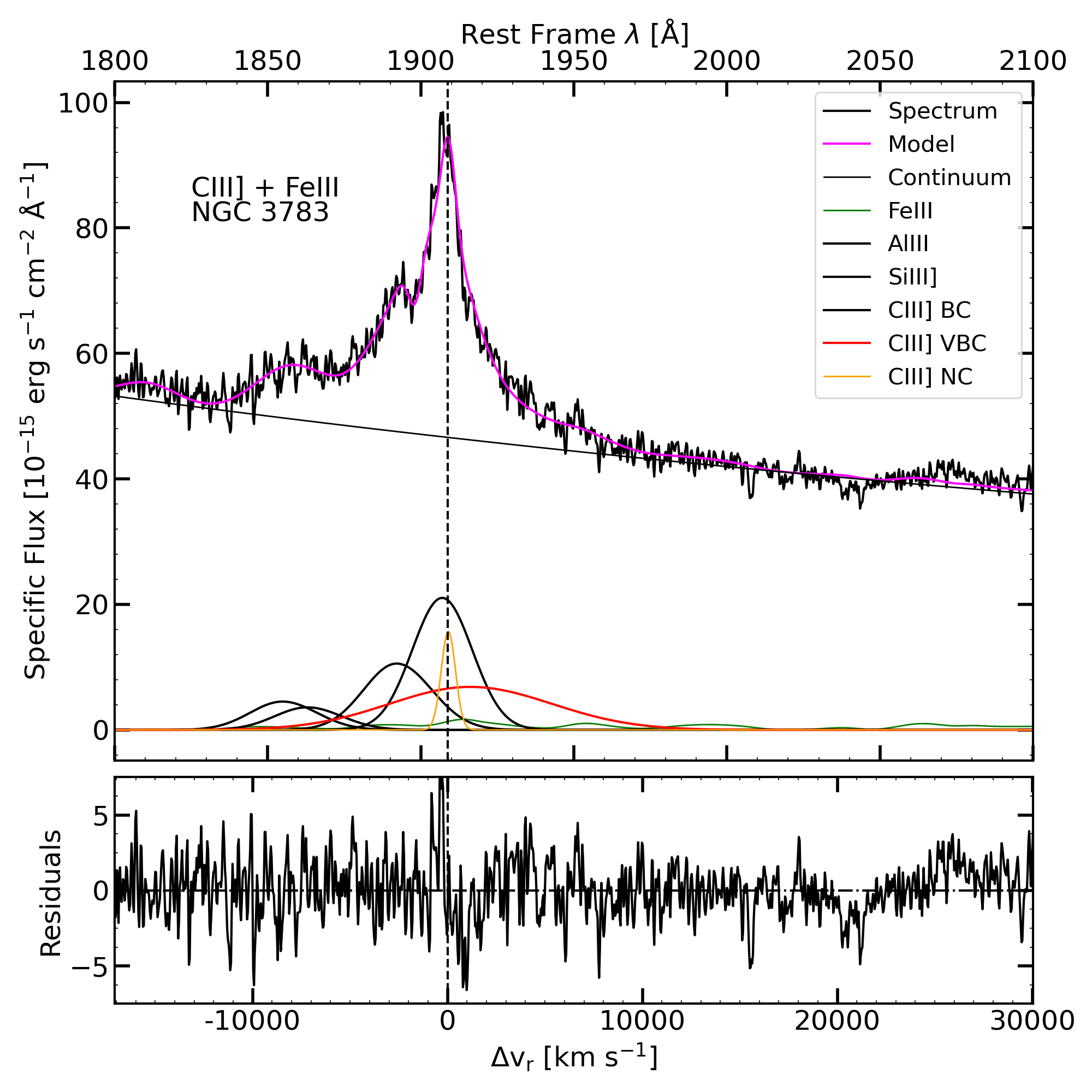}
\includegraphics[width=4.53cm]{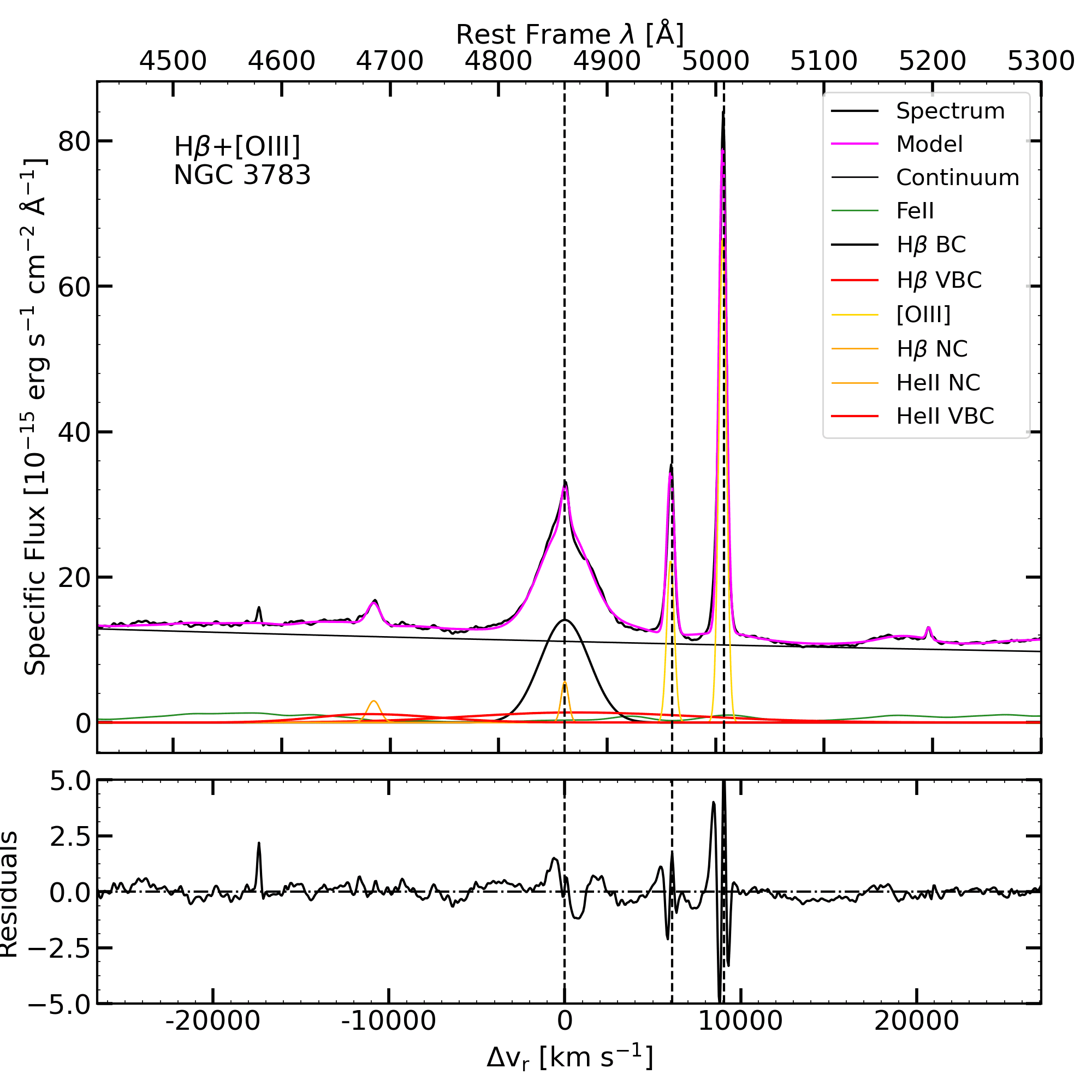}\\
\includegraphics[width=4.53cm]{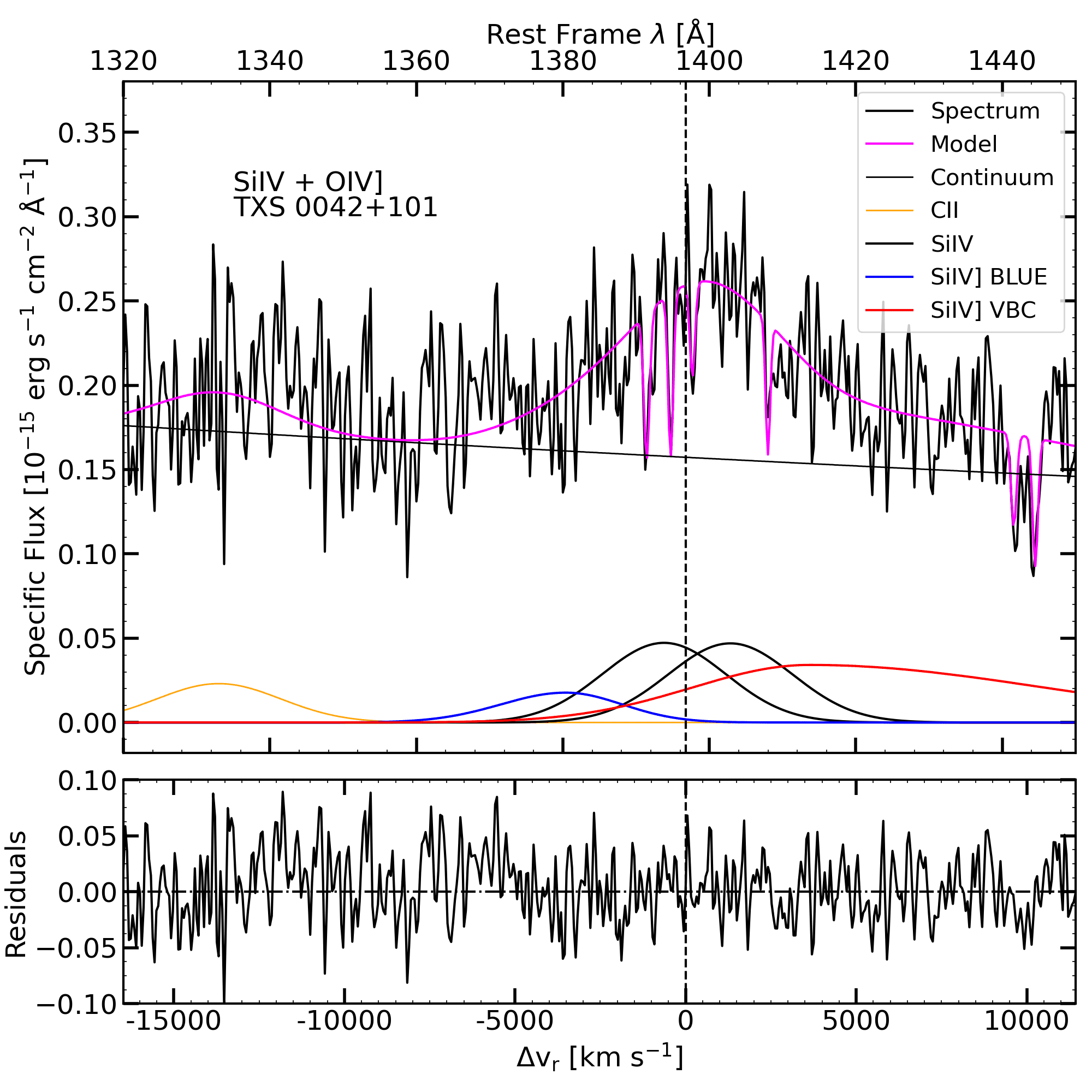}
\includegraphics[width=4.53cm]{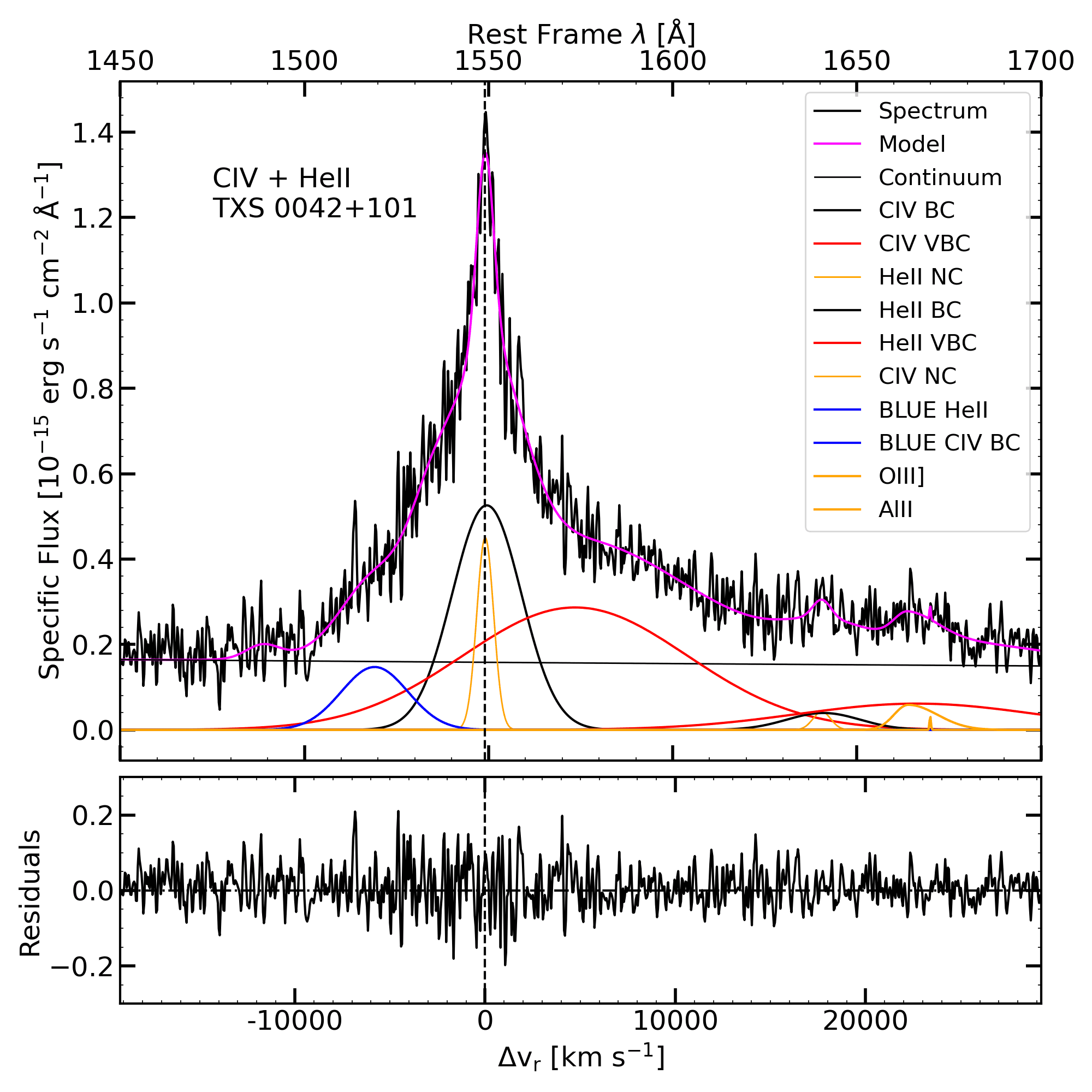}
\includegraphics[width=4.53cm]{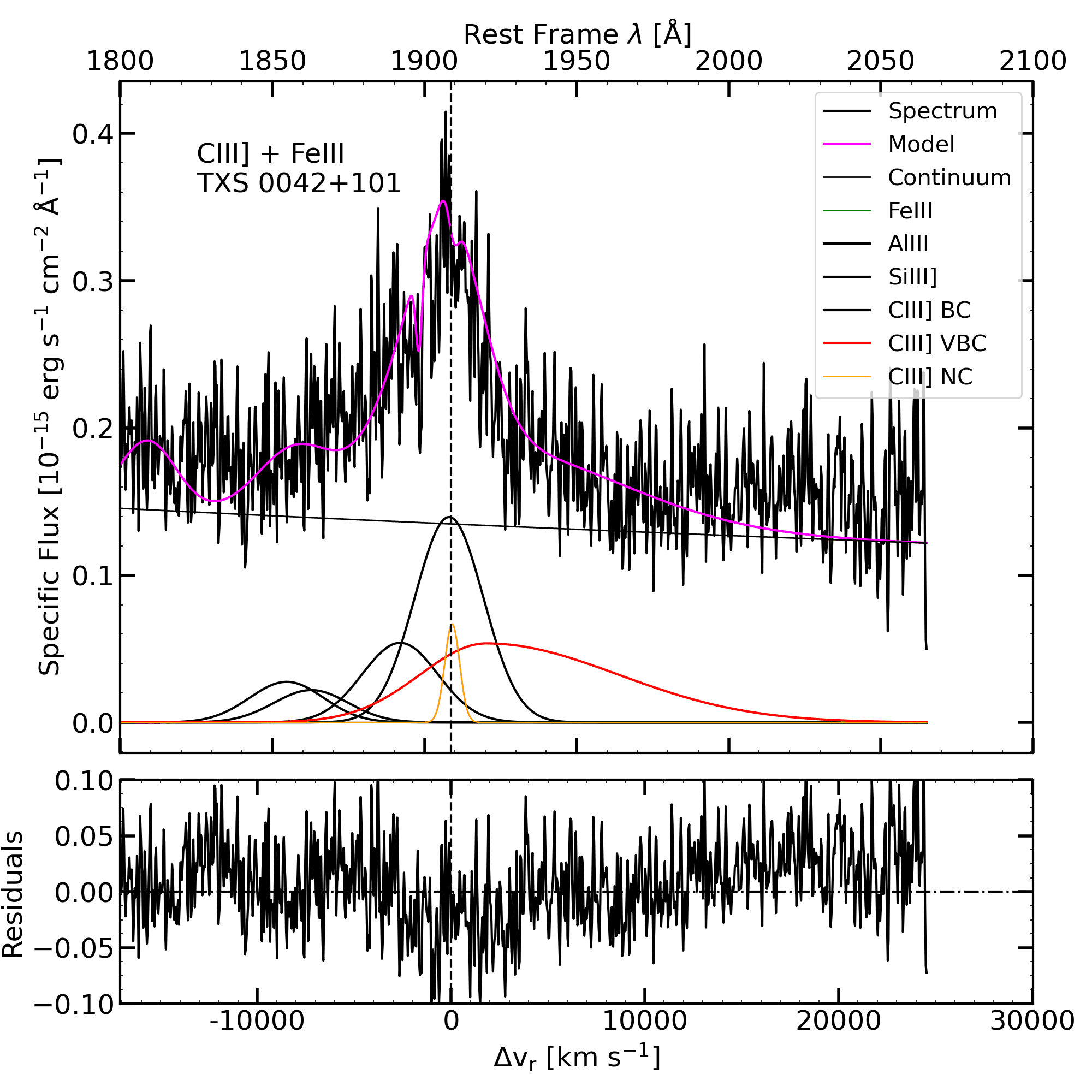}
\includegraphics[width=4.53cm]{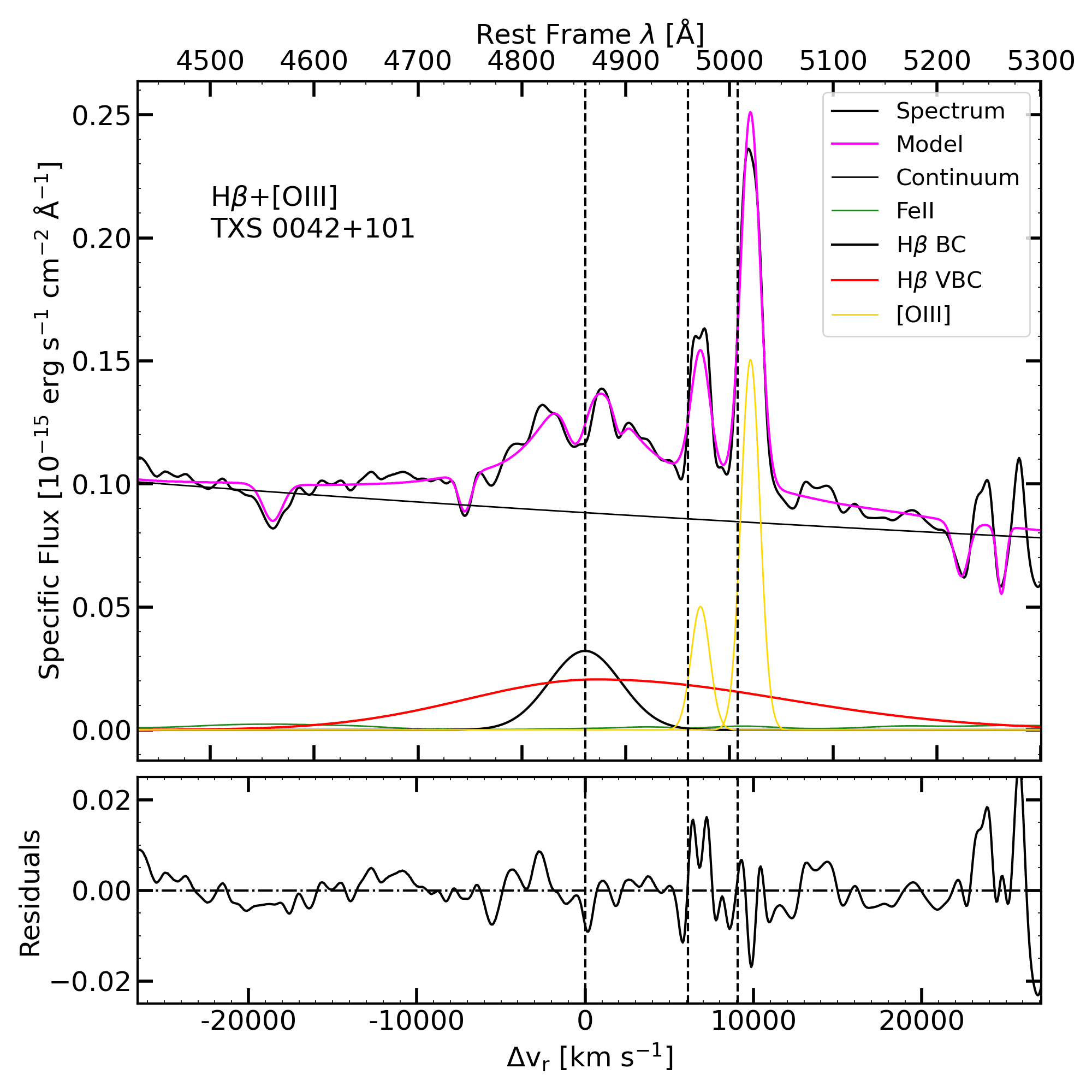}\\

\caption{Same as in Figure \ref{fig:fits1} but for NGC 3783 and TXS 0042+101, respectively.}
\label{fig:fits4}
\end{figure*}

The spectral analysis of the 14 objects of our sample is shown in Figures \ref{fig:fits1}-\ref{fig:fits4}. The spectral fit for the \ciii\ spectral range of Mrk 110 is absent because there are no \ciii\ data for this object.


\section{Distribution of $U$, $n_H$ and $Z$ for each object}
\label{app:isoph}
The distribution of the physical parameters of $U$, $n_H$ and $Z$ for the 14 objects of our sample is shown in Figures \ref{fig:proj1}-\ref{fig:proj5}.

\begin{figure*}[ht!]
\centering
\includegraphics[width=6.05cm]{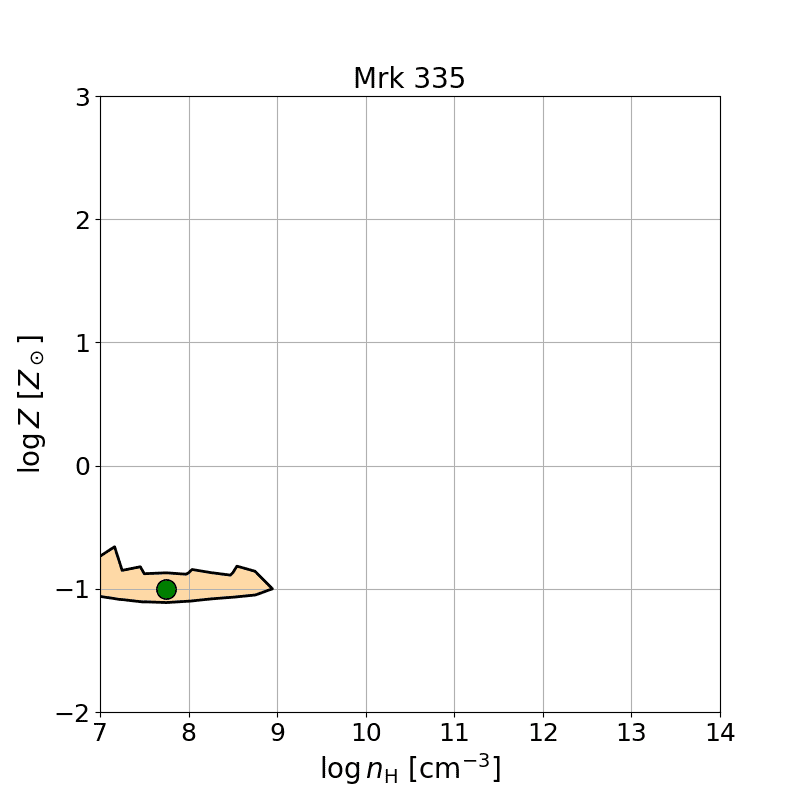}
\includegraphics[width=6.05cm]{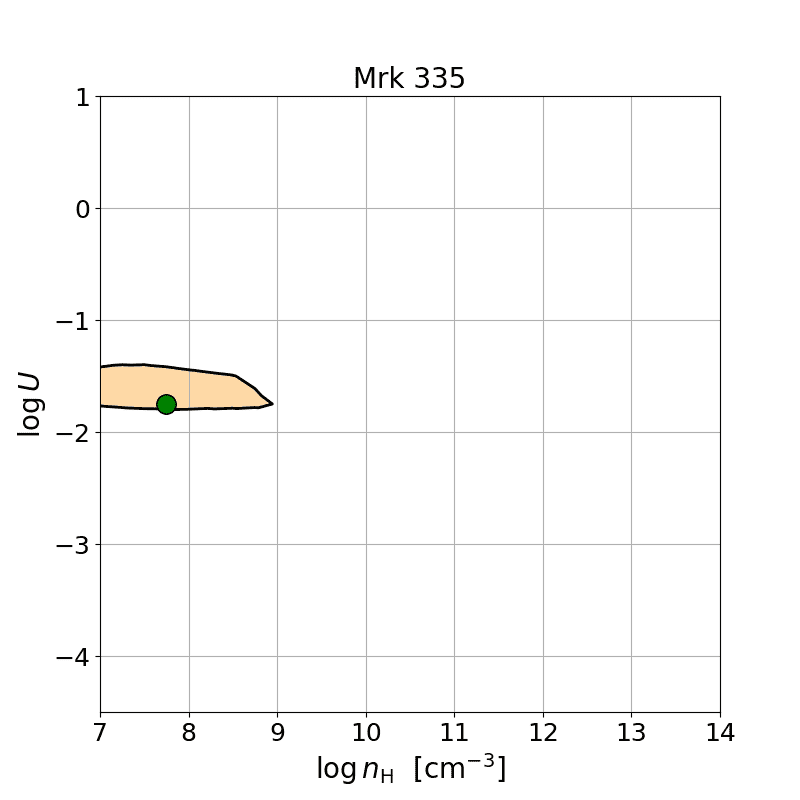}
\includegraphics[width=6.05cm]{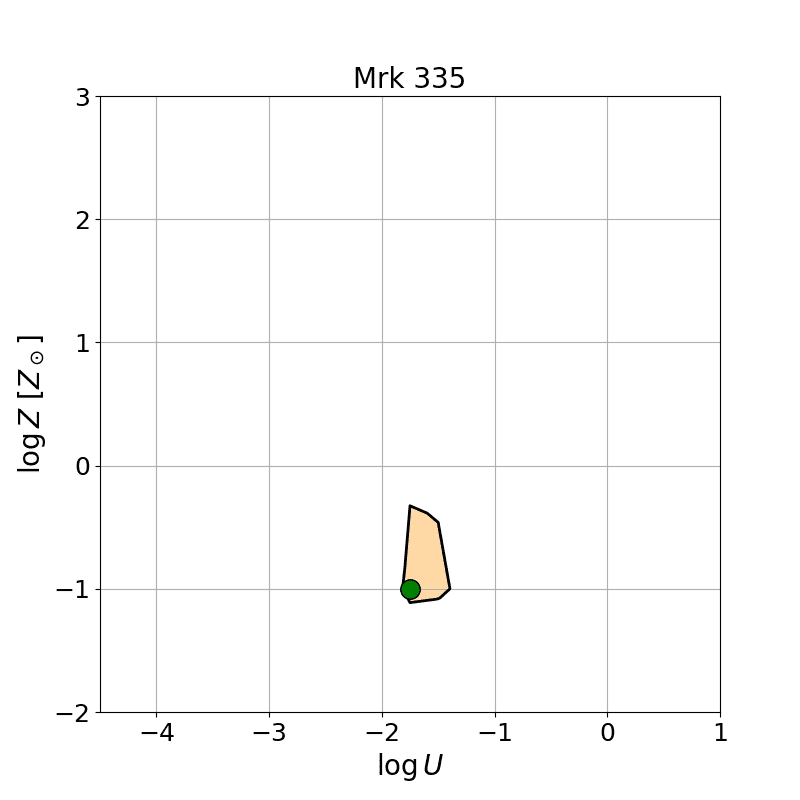}\\
\includegraphics[width=6.05cm]{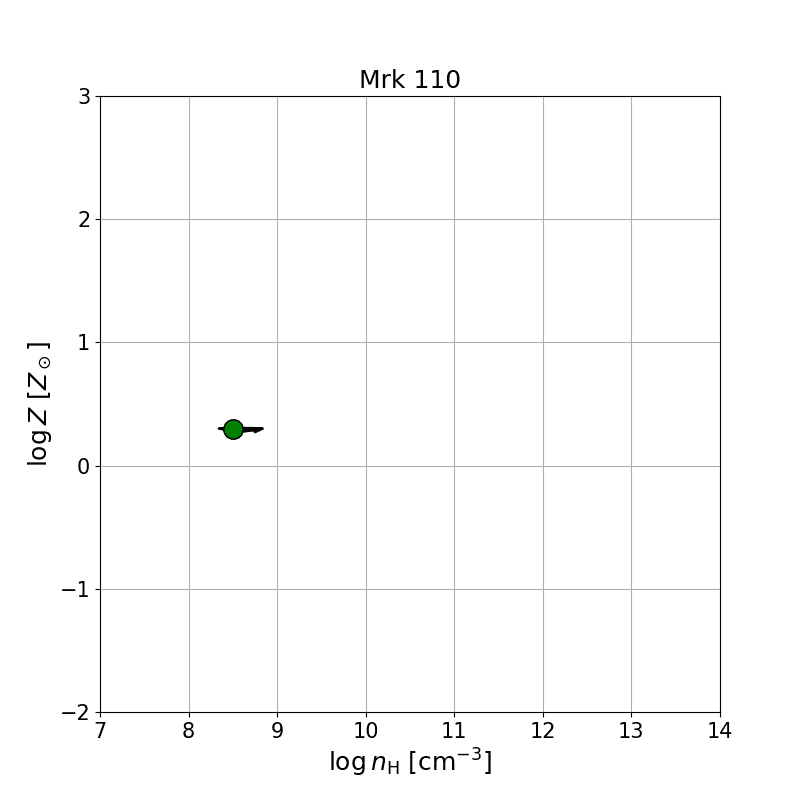}
\includegraphics[width=6.05cm]{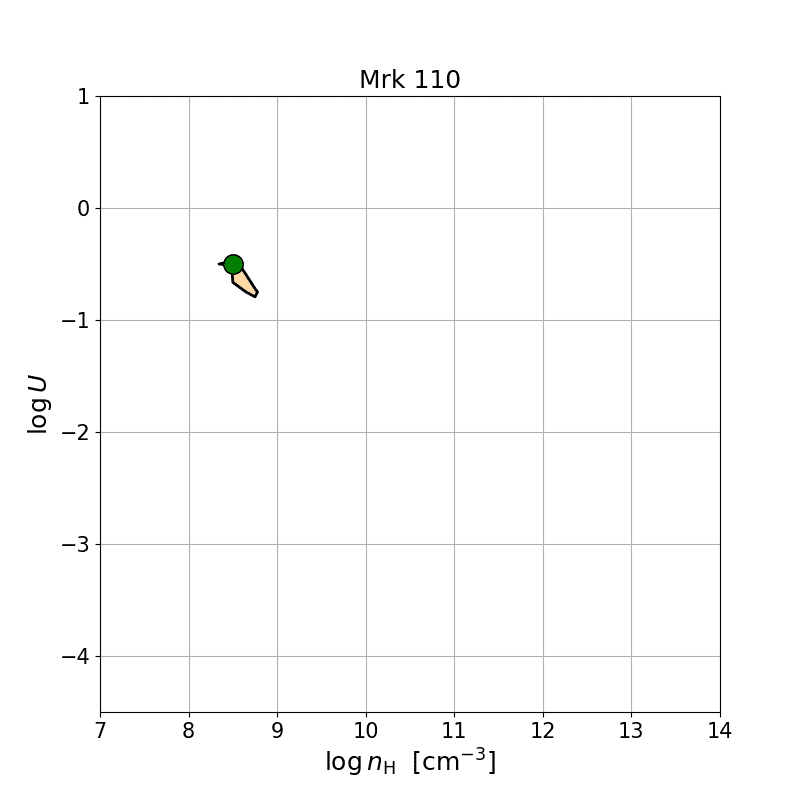}
\includegraphics[width=6.05cm]{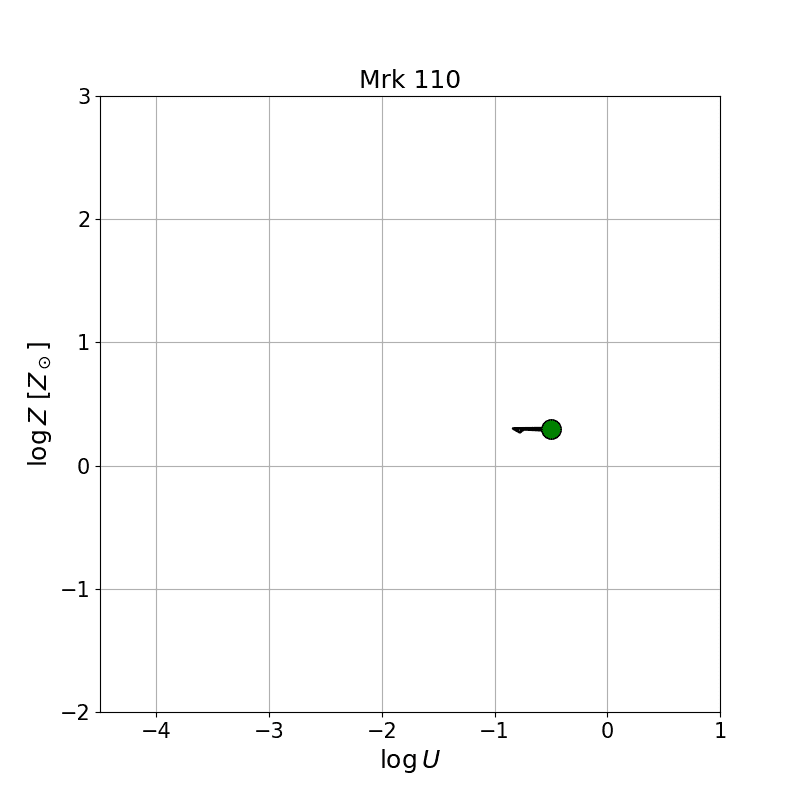}\\
\includegraphics[width=6.05cm]{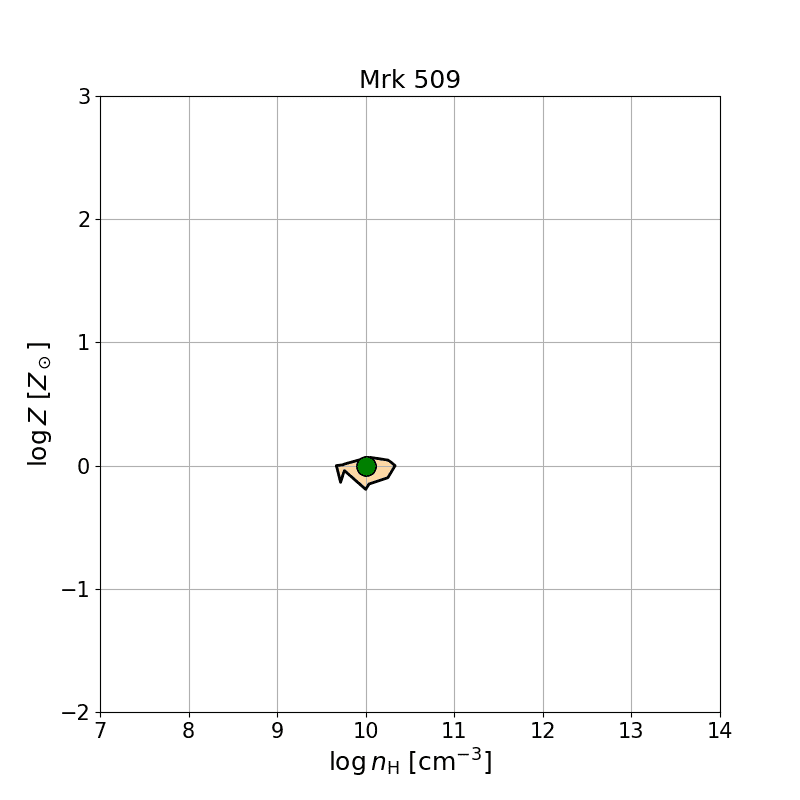}
\includegraphics[width=6.05cm]{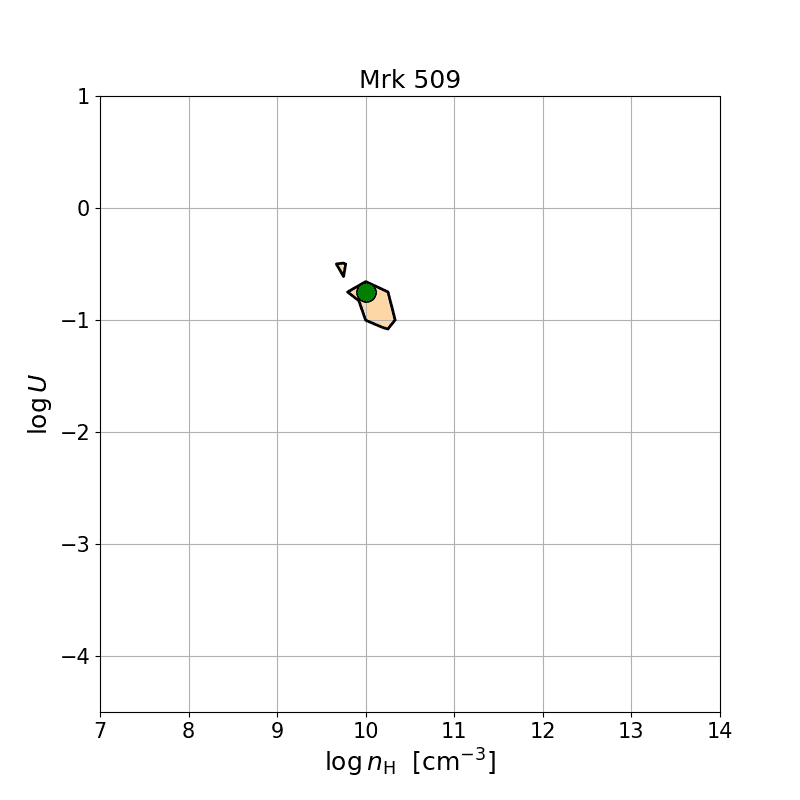}
\includegraphics[width=6.05cm]{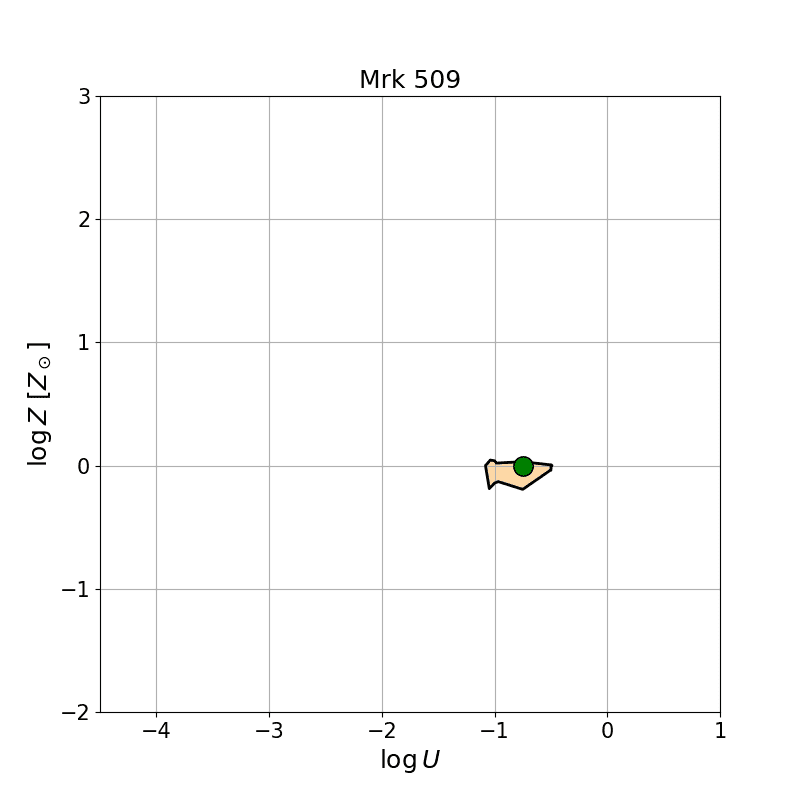}\\

\caption{Two-dimensional parameter spaces illustrating the physical conditions of the gas in the BLR of the objects. Each plot represents the BC with a green dot indicating the minimum $\chi^2$ computed between measured diagnostic ratios and {\tt CLOUDY} simulations. The orange region corresponds to 1$\sigma$ accuracy. Blue and red dots and shades represent the minimum $\chi^2$ and associated regions for the BLUE and VBC, respectively. {\it Left}: Logarithmic plane of $n_H$ and $Z$. {\it Middle}: Logarithmic plane of $n_H$ and $U$. {\it Right}: Logarithmic plane of $U$ and $Z$.}
\label{fig:proj1}
\end{figure*}

\begin{figure*}[ht!]
\centering
\includegraphics[width=6.05cm]{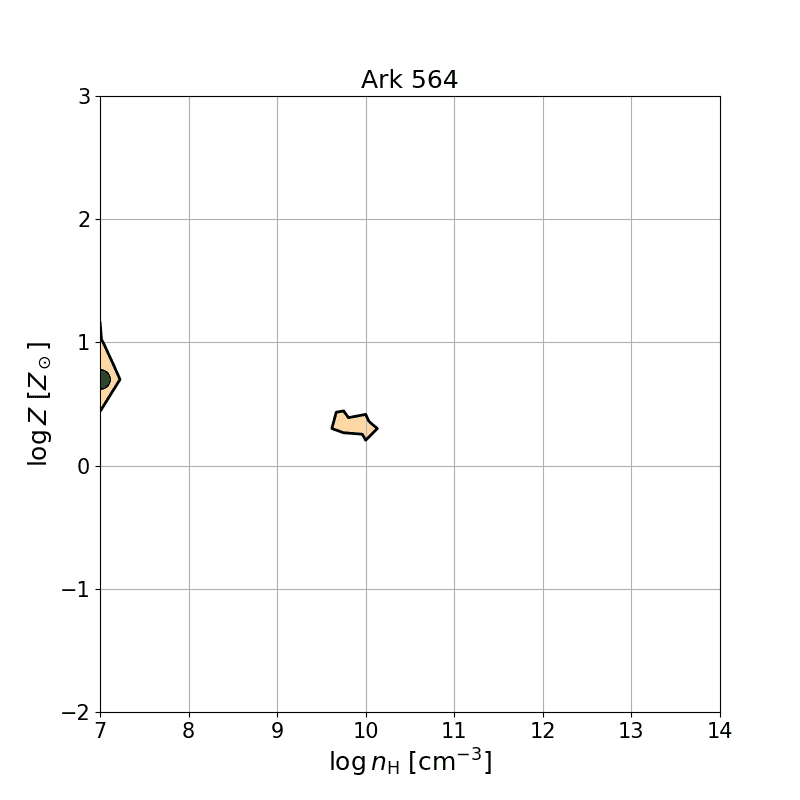}
\includegraphics[width=6.05cm]{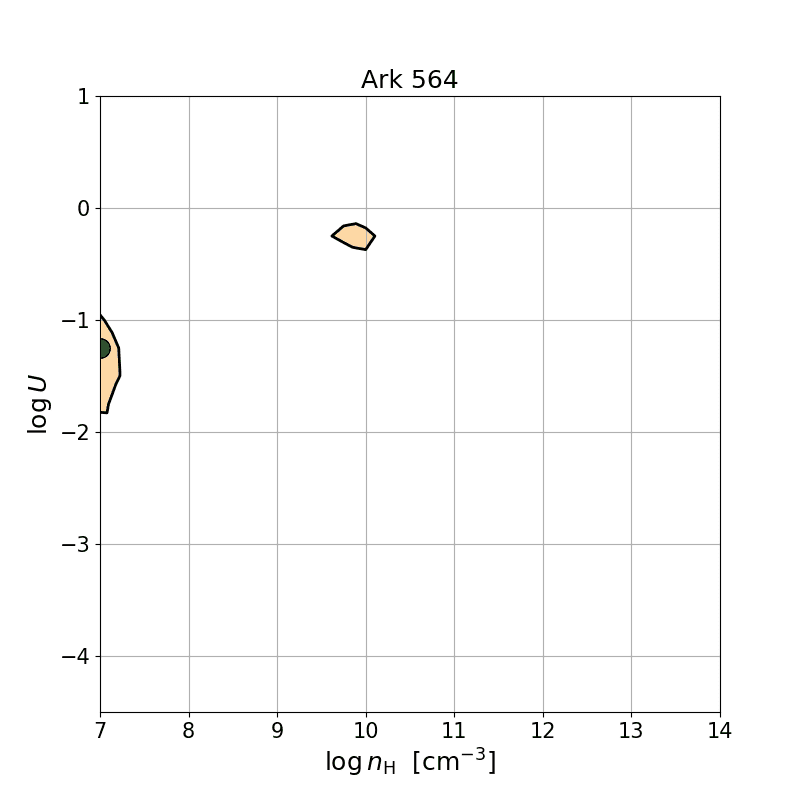}
\includegraphics[width=6.05cm]{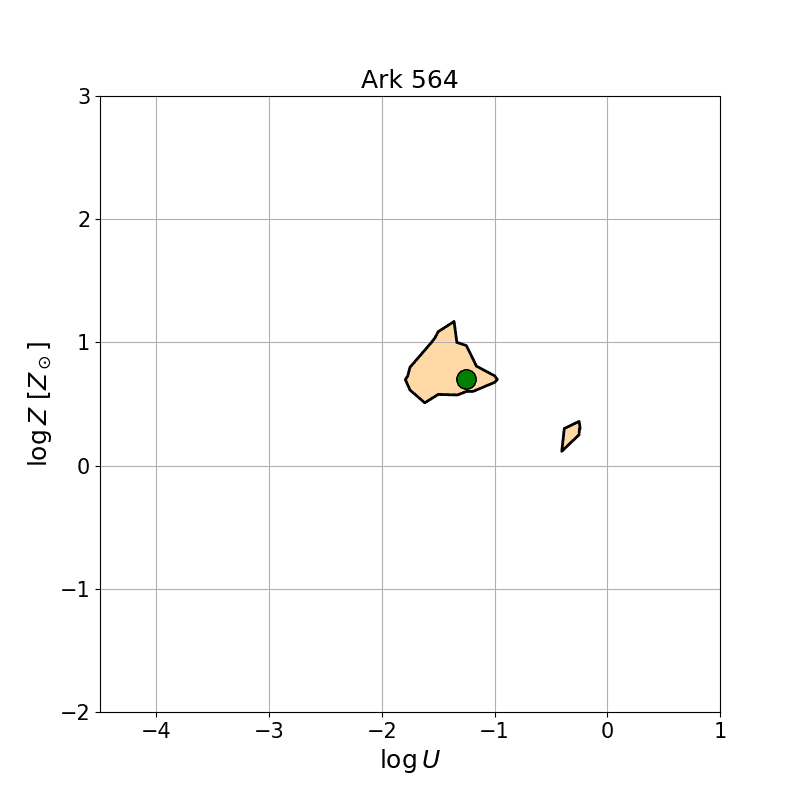}\\
\includegraphics[width=6.05cm]{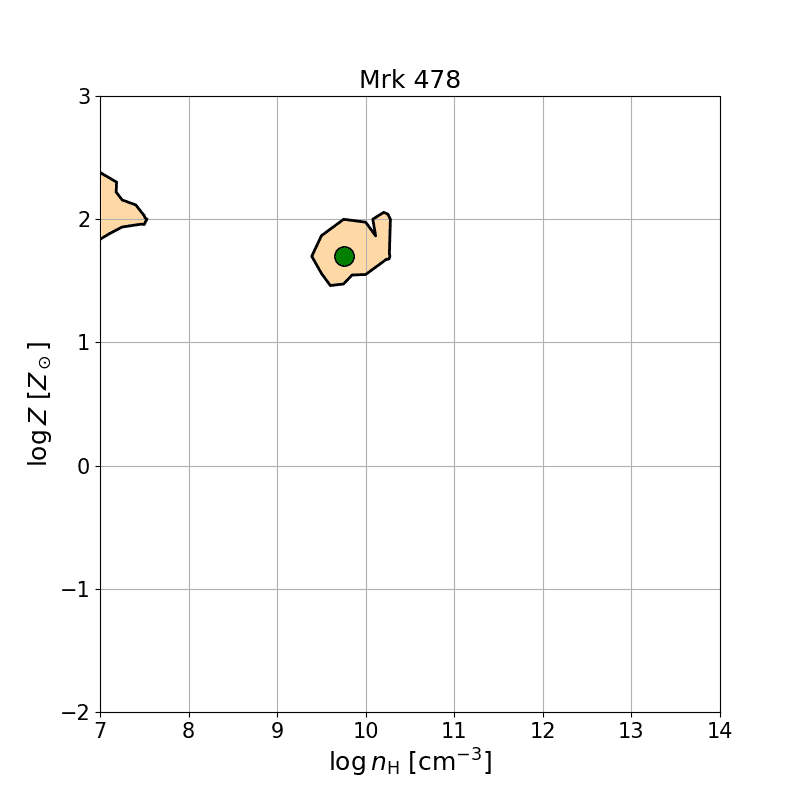}
\includegraphics[width=6.05cm]{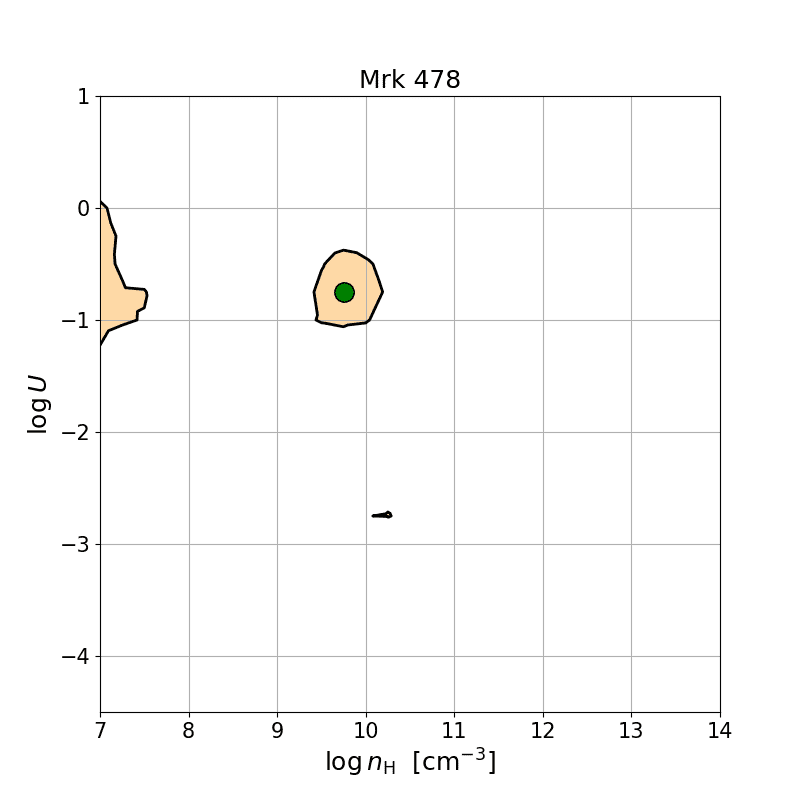}
\includegraphics[width=6.05cm]{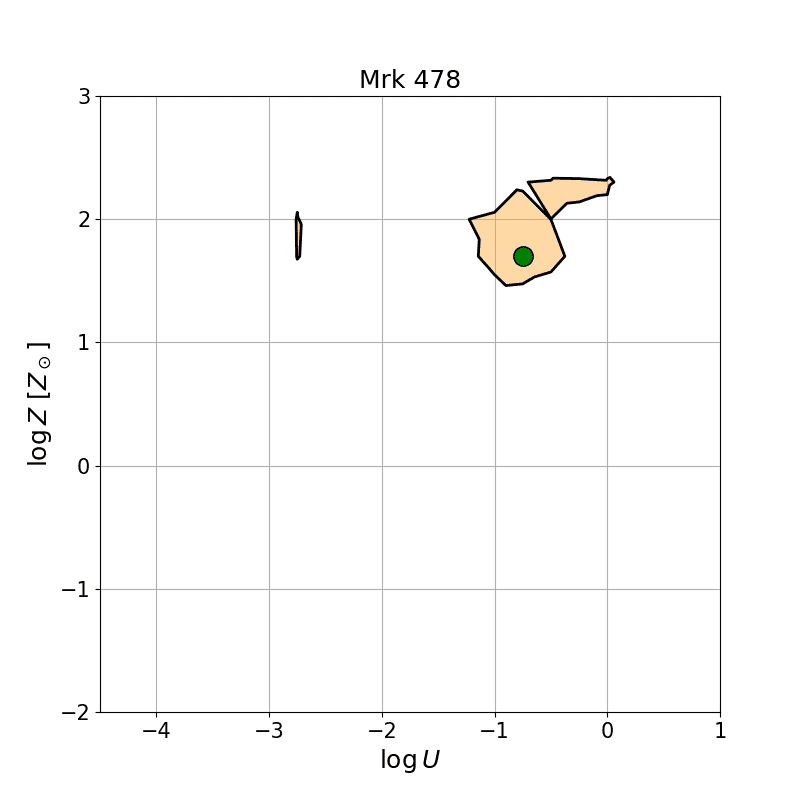}\\
\includegraphics[width=6.05cm]{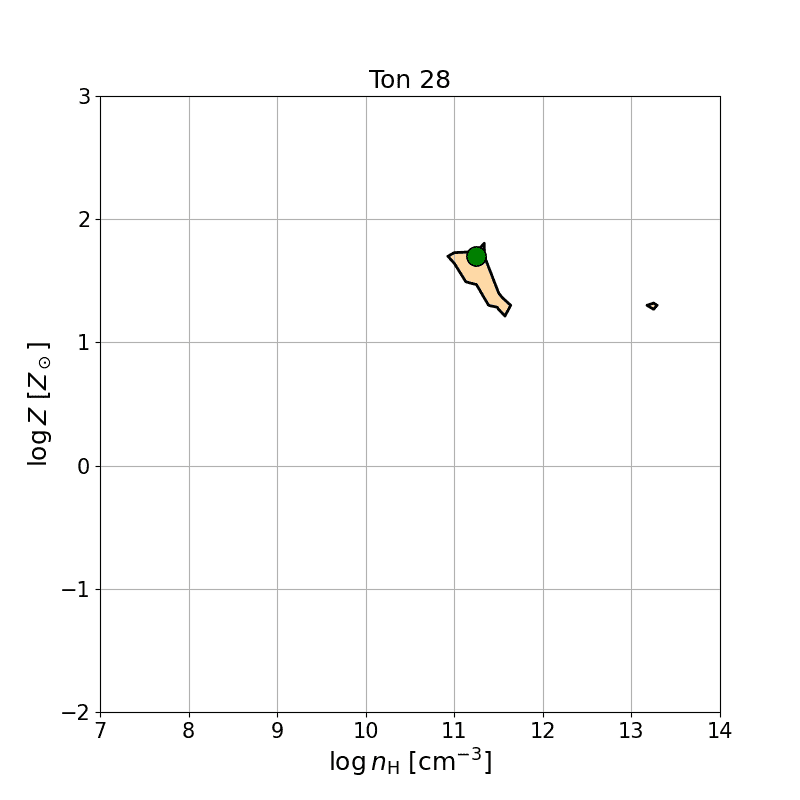}
\includegraphics[width=6.05cm]{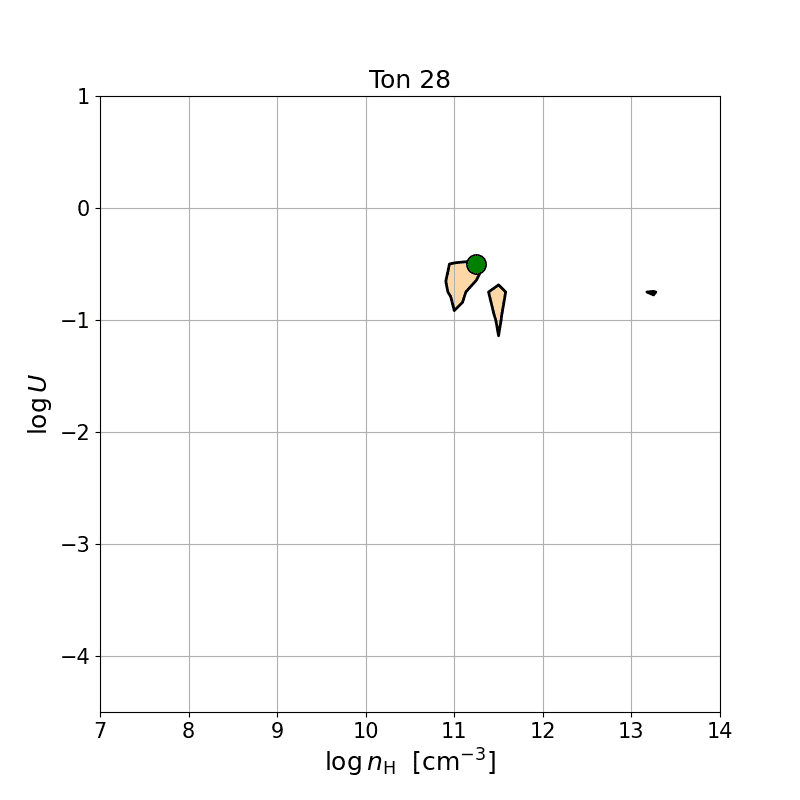}
\includegraphics[width=6.05cm]{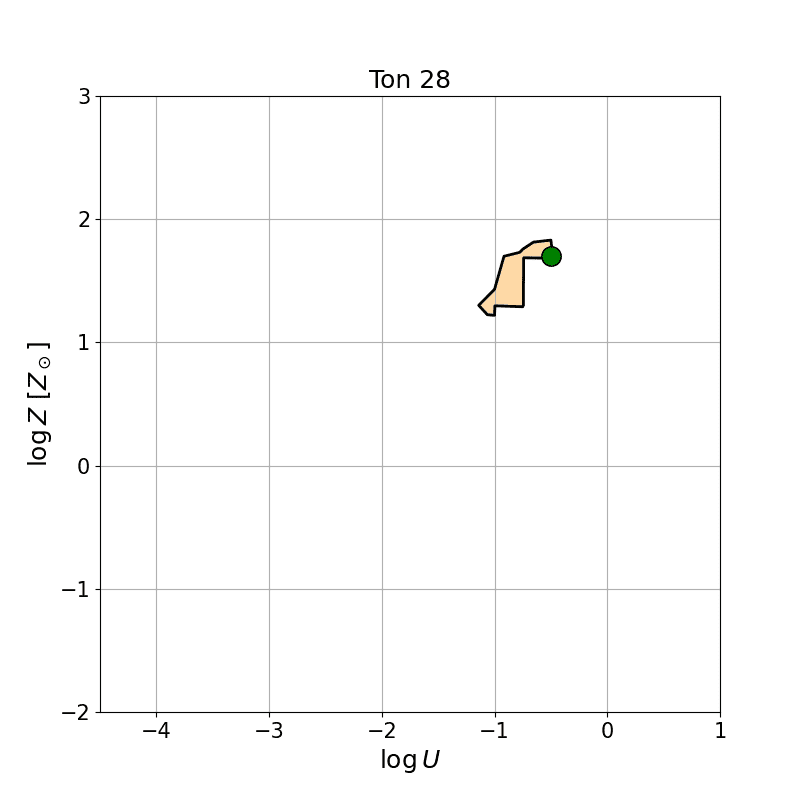}\\

\caption{Continued.}
\label{fig:proj2}
\end{figure*}

\begin{figure*}[h!]
\centering
\includegraphics[width=6.05cm]{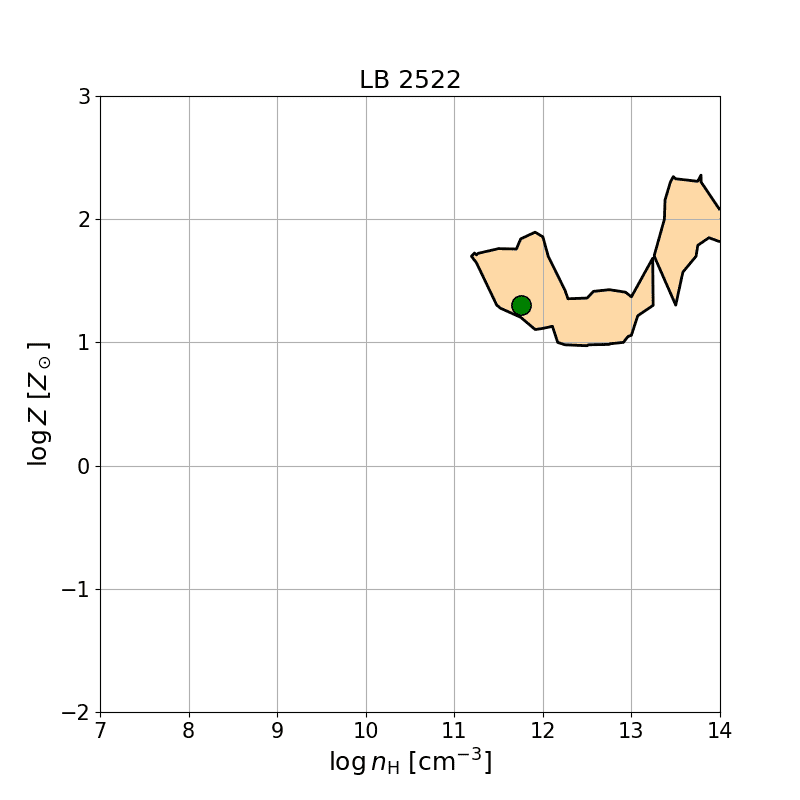}
\includegraphics[width=6.05cm]{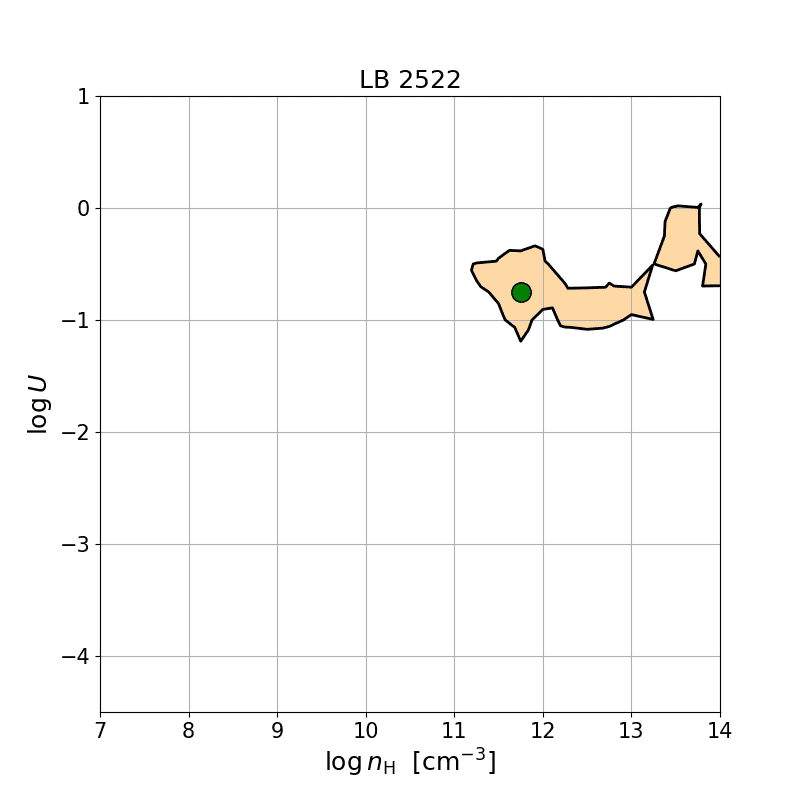}
\includegraphics[width=6.05cm]{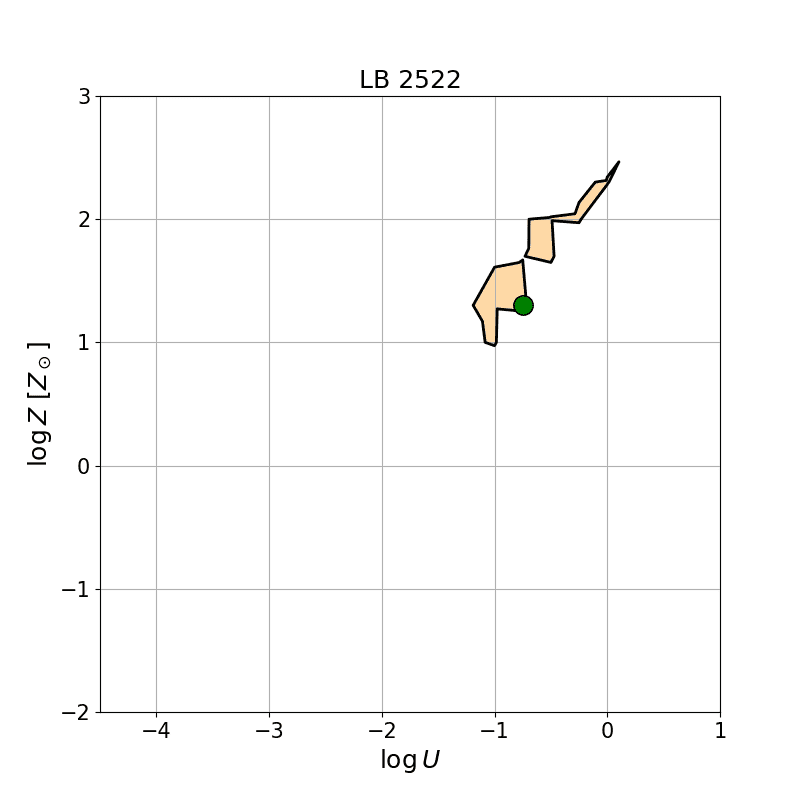}
\includegraphics[width=6.05cm]{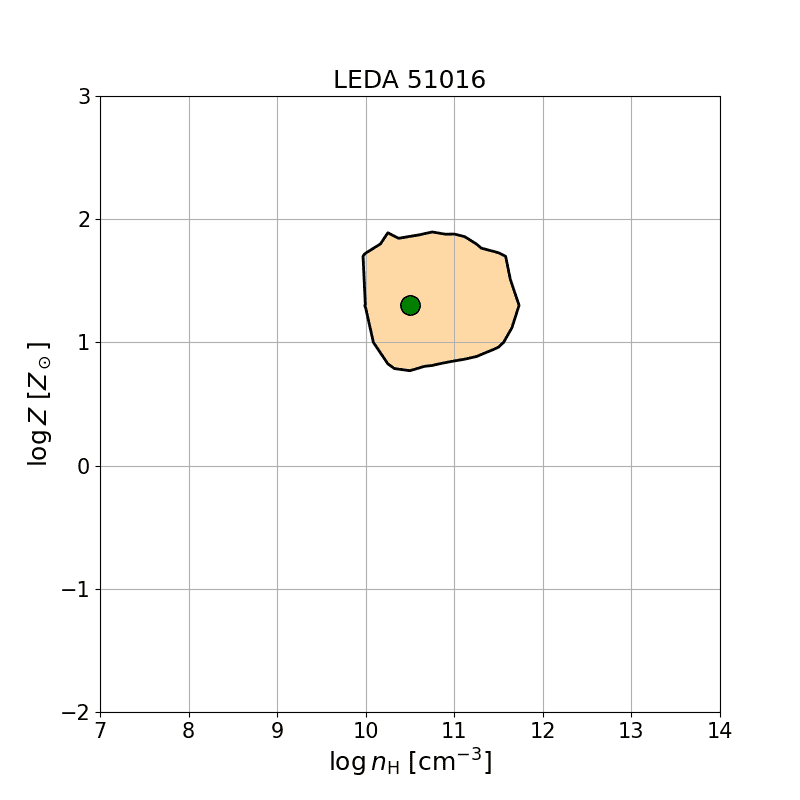}
\includegraphics[width=6.05cm]{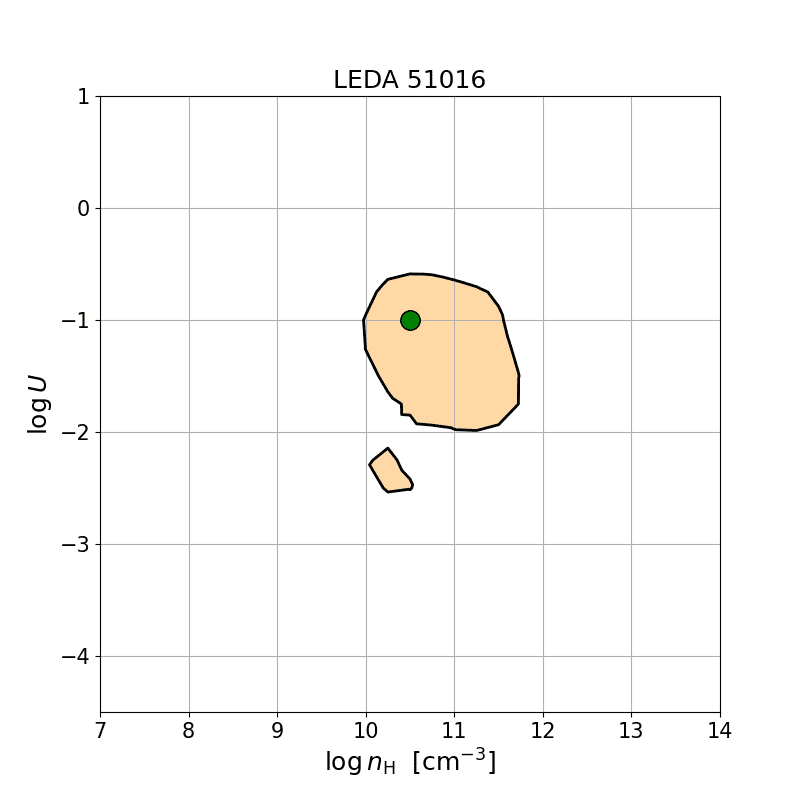}
\includegraphics[width=6.05cm]{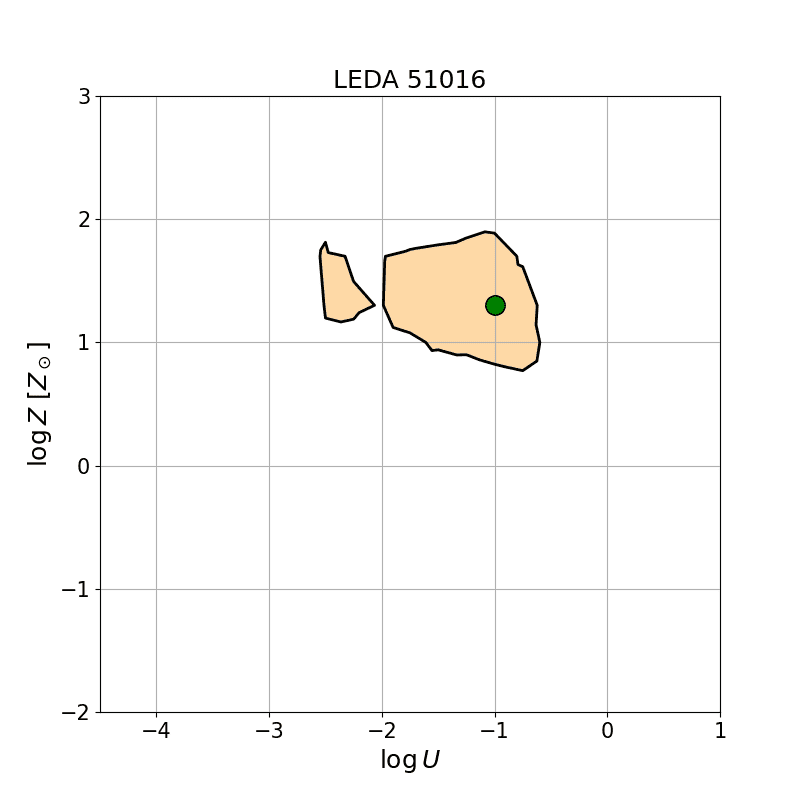}
\includegraphics[width=6.05cm]{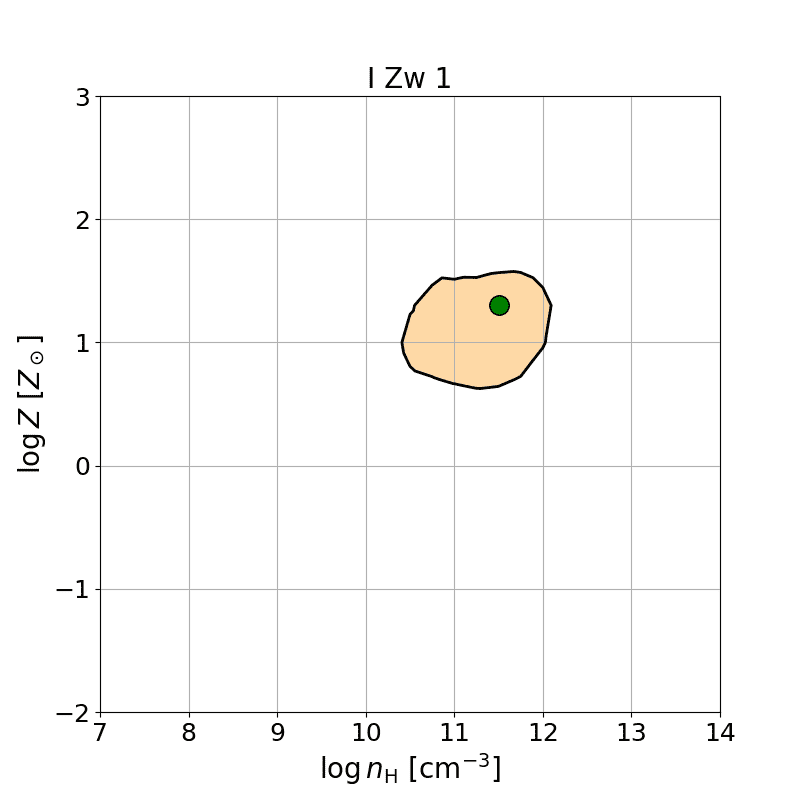}
\includegraphics[width=6.05cm]{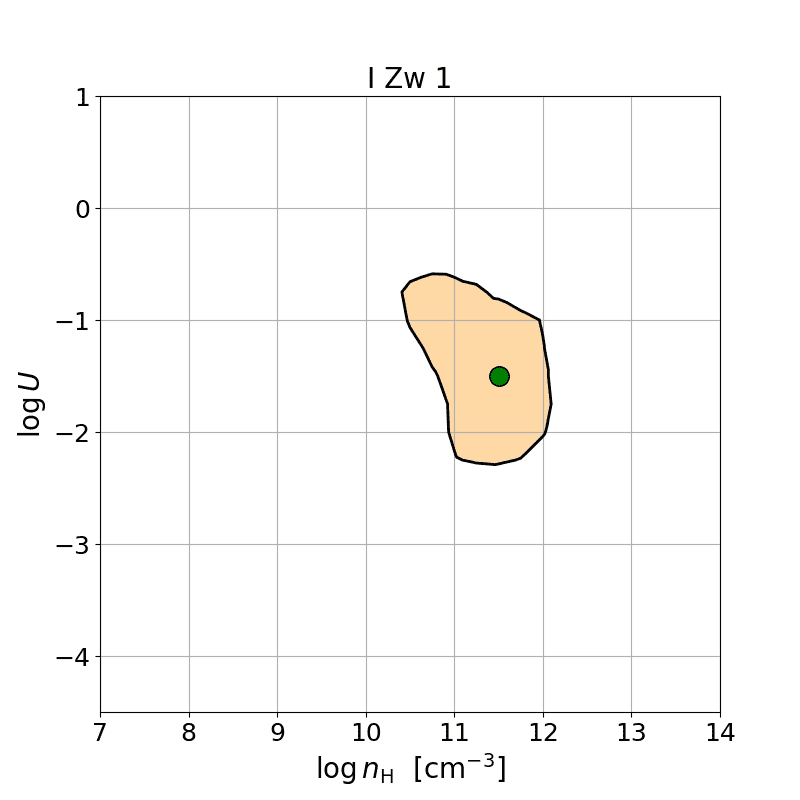}
\includegraphics[width=6.05cm]{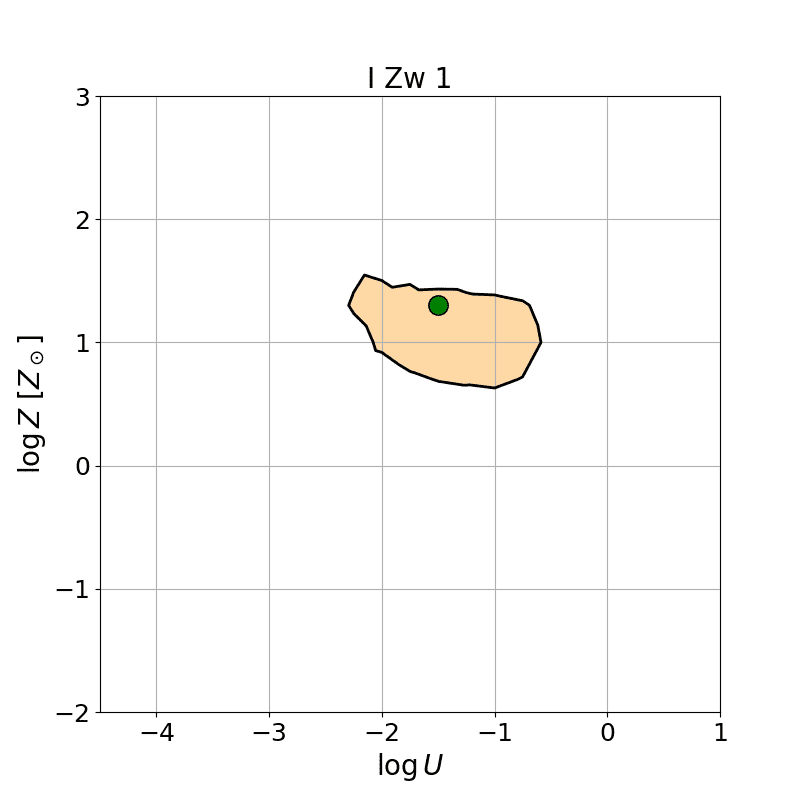}\\

\caption{Continued.}
\label{fig:proj3}
\end{figure*}

\begin{figure*}[ht!]
\centering
\includegraphics[width=6.05cm]{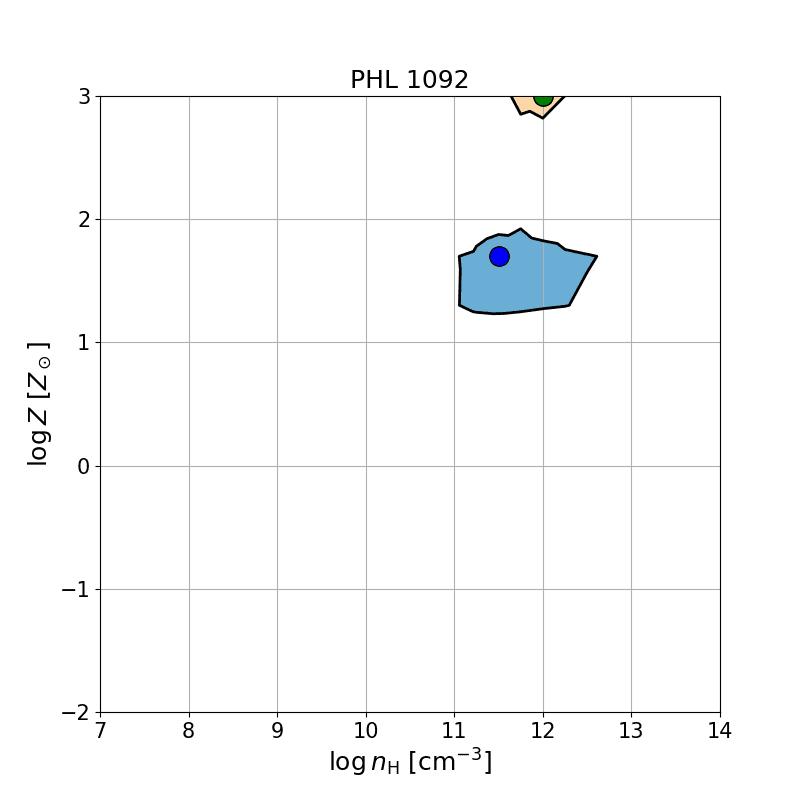}
\includegraphics[width=6.05cm]{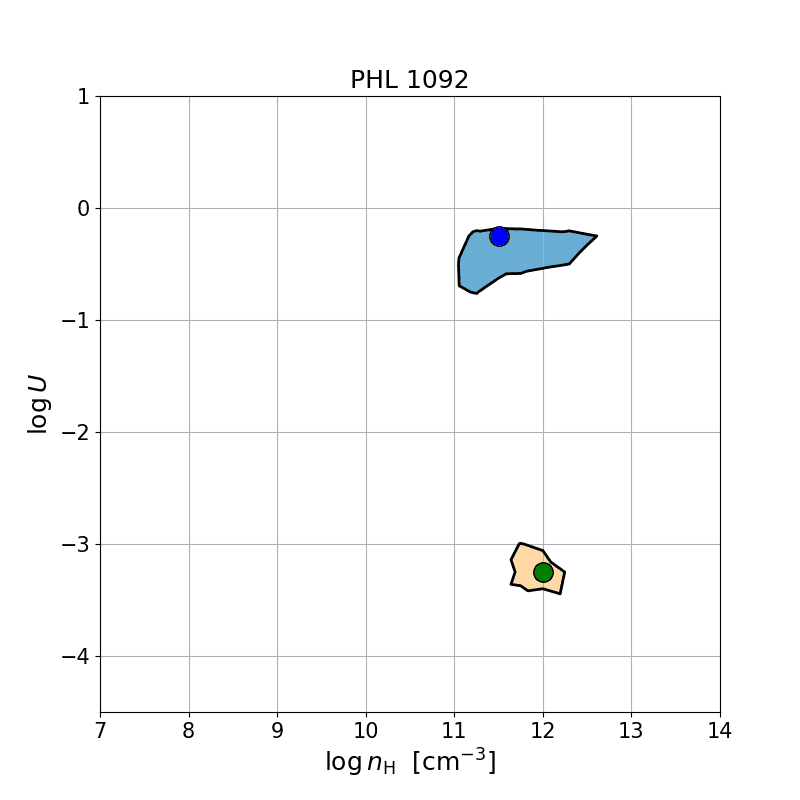}
\includegraphics[width=6.05cm]{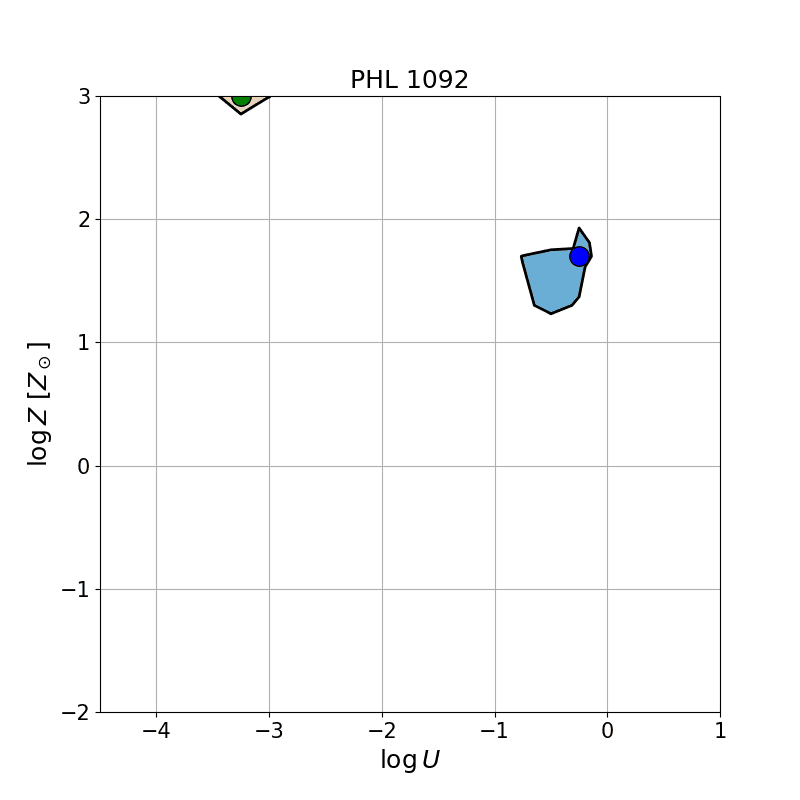}\\
\includegraphics[width=6.05cm]{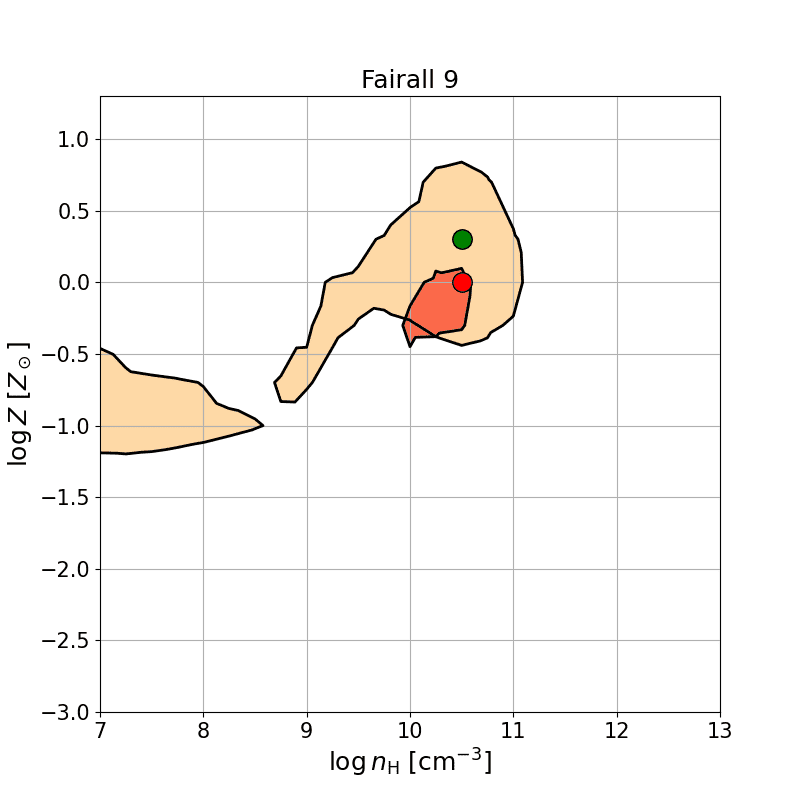}
\includegraphics[width=6.05cm]{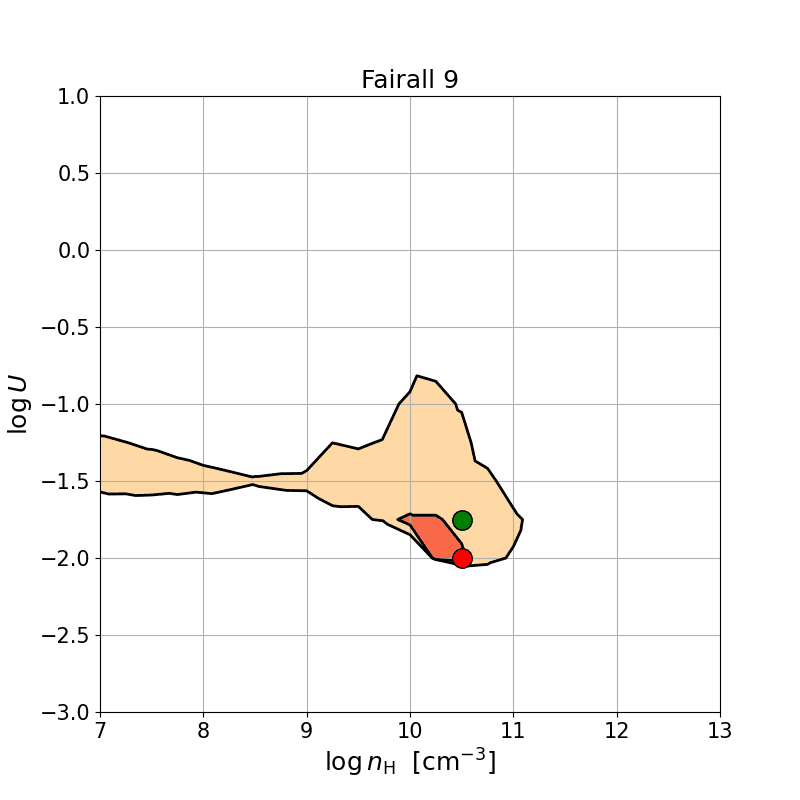}
\includegraphics[width=6.05cm]{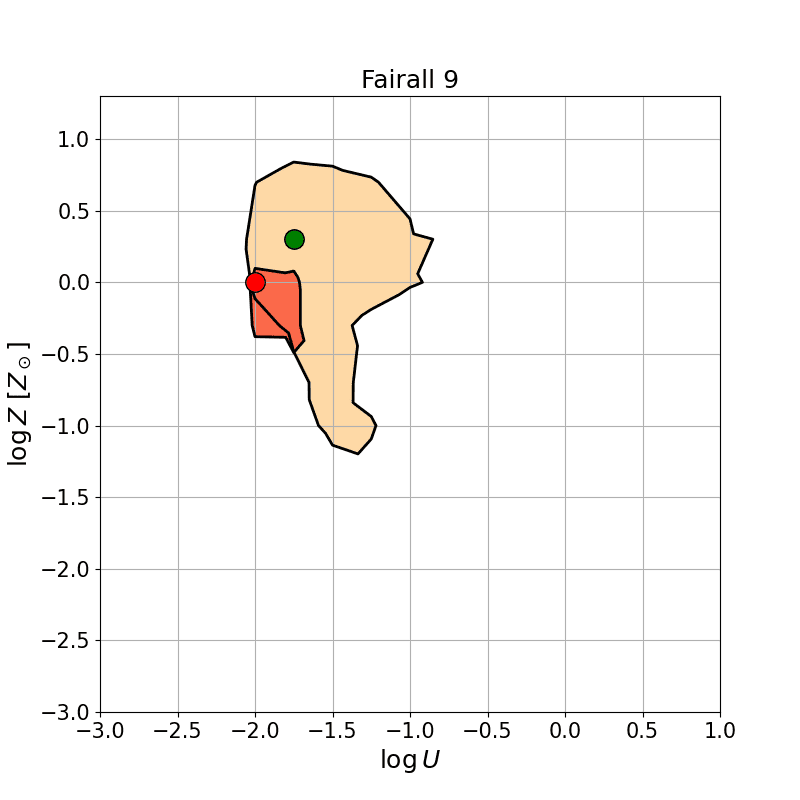}\\
\includegraphics[width=6.05cm]{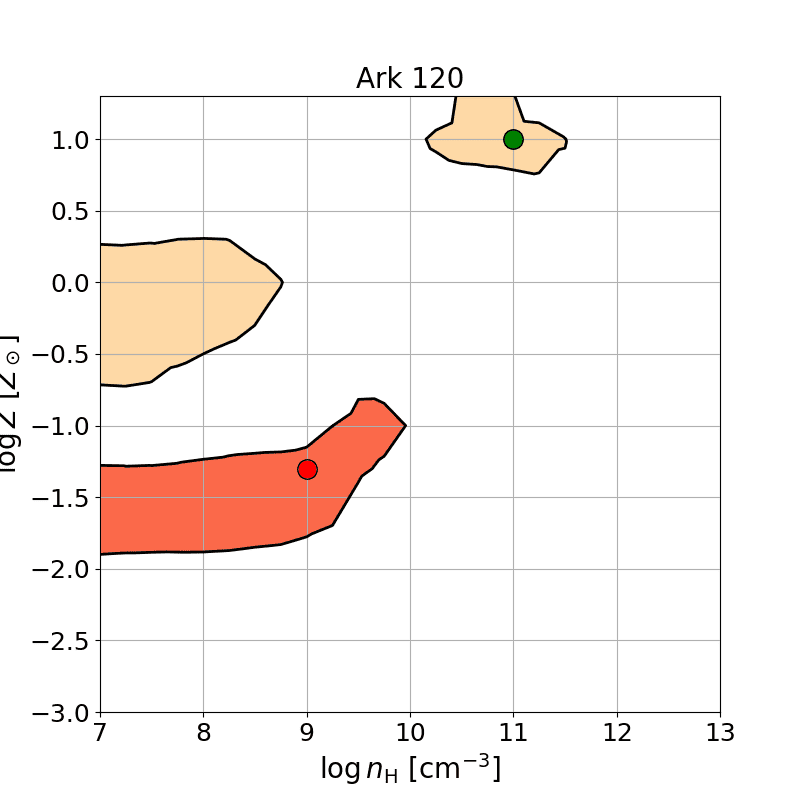}
\includegraphics[width=6.05cm]{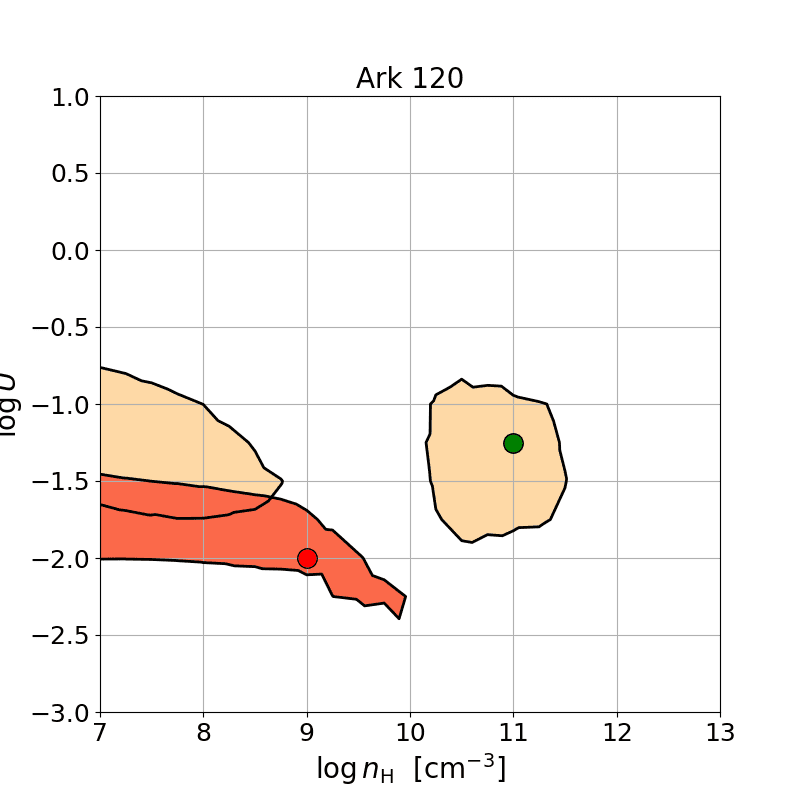}
\includegraphics[width=6.05cm]{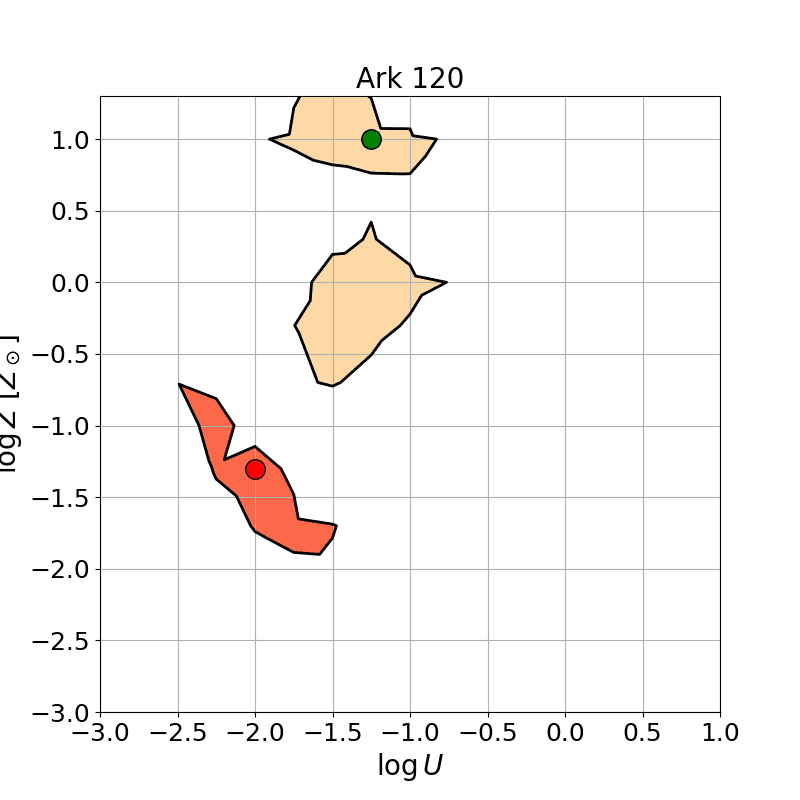}\\

\caption{Continued.}
\label{fig:proj4}
\end{figure*}

\begin{figure*}[ht!]
\centering
\includegraphics[width=6.05cm]{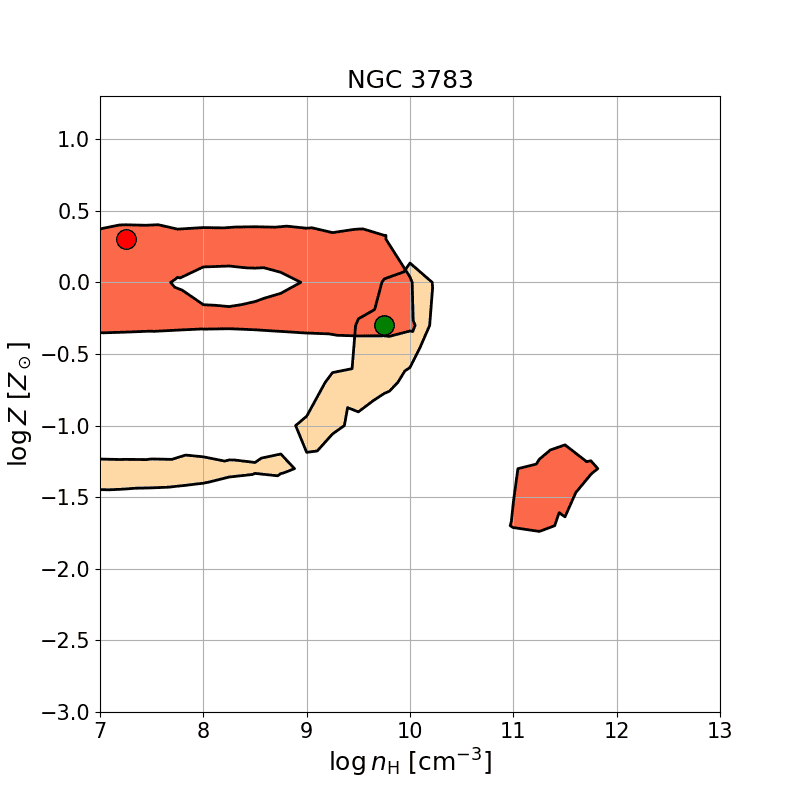}
\includegraphics[width=6.05cm]{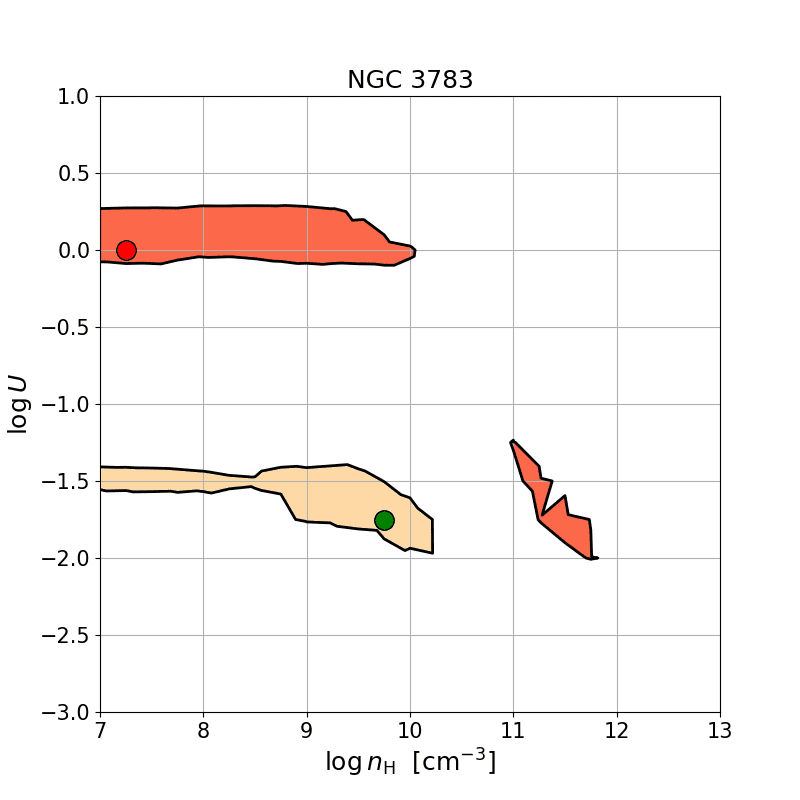}
\includegraphics[width=6.05cm]{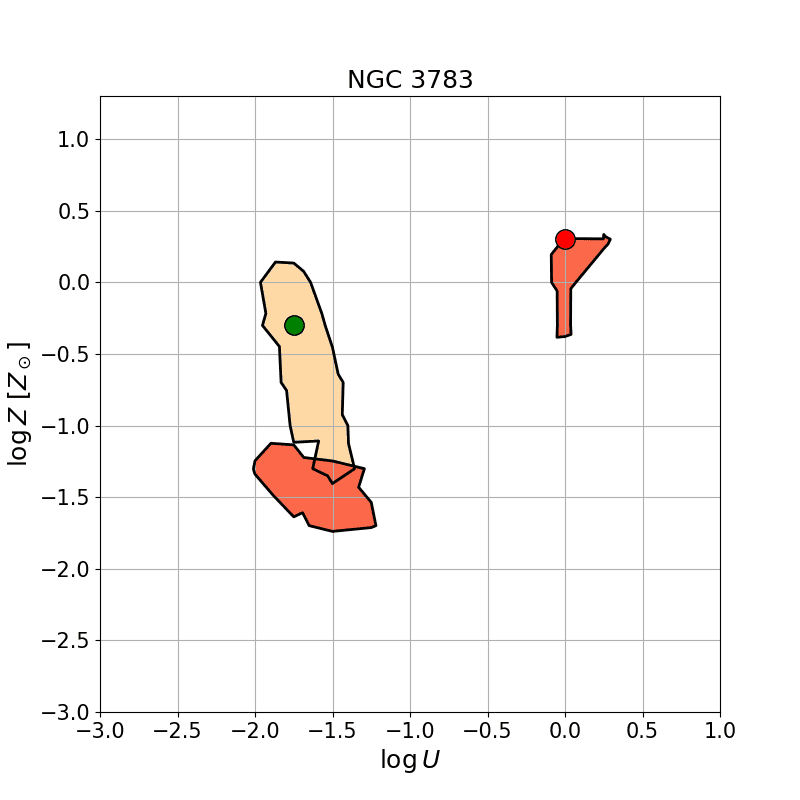}
\includegraphics[width=6.05cm]{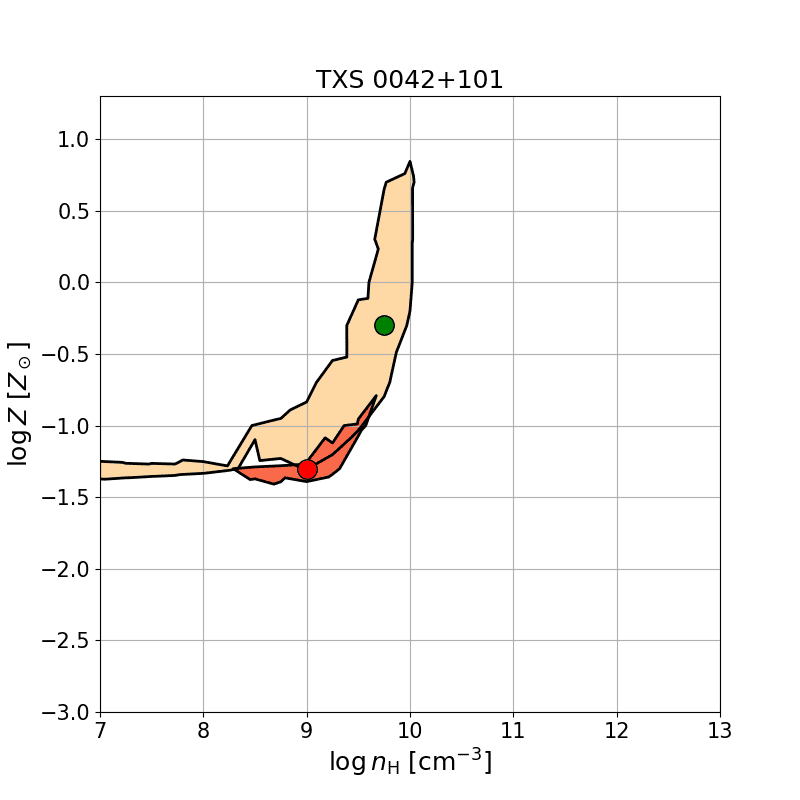}
\includegraphics[width=6.05cm]{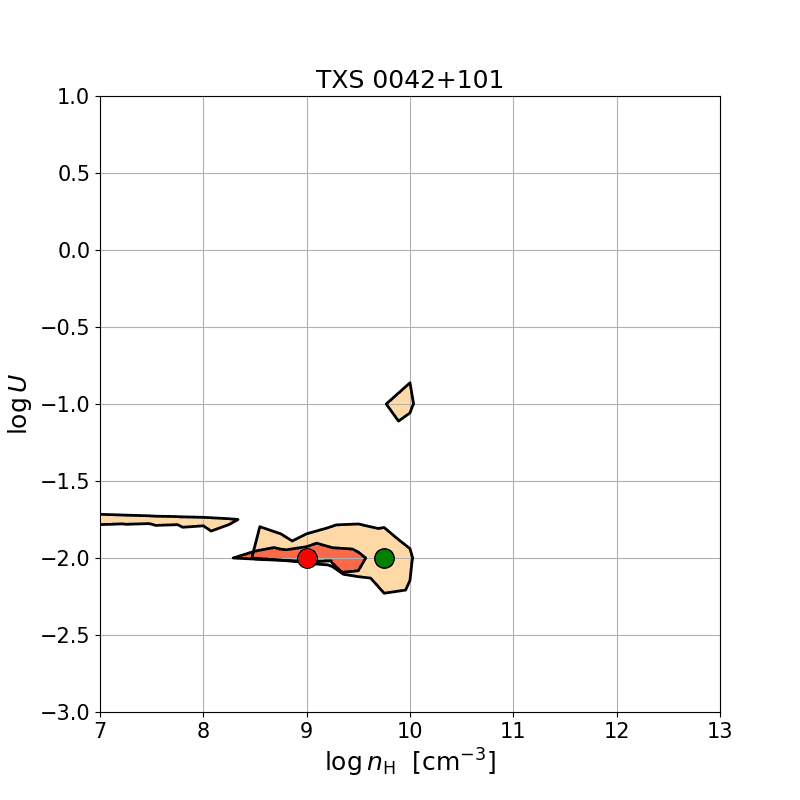}
\includegraphics[width=6.05cm]{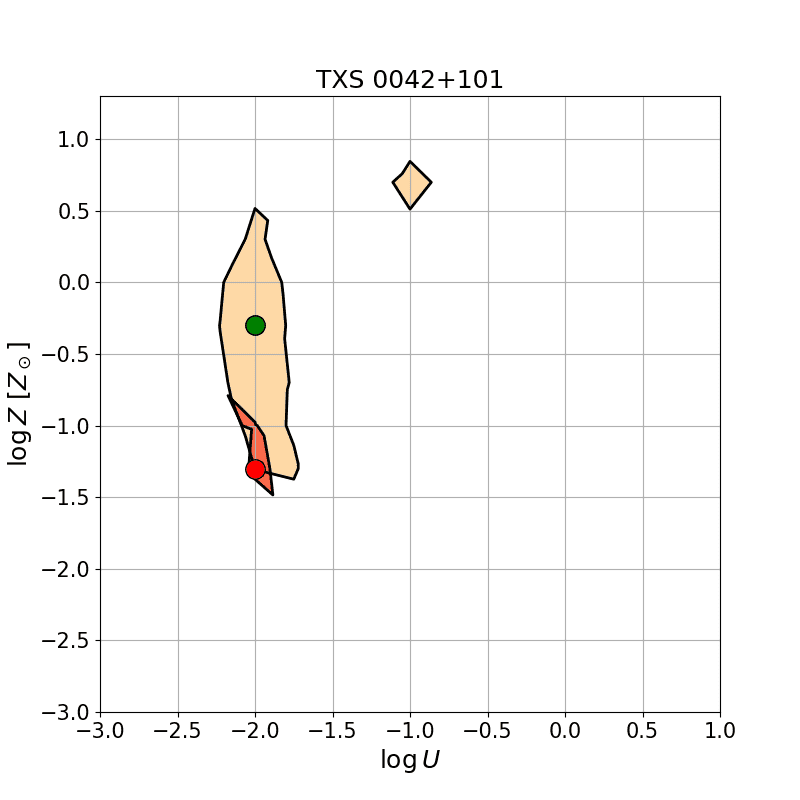}\\

\caption{Continued.}
\label{fig:proj5}
\end{figure*}

\end{appendix}

\end{document}